\documentclass[12pt]{JHEP3}
\input epsf.tex
\epsfclipon
\let\ifpdf\relax

\usepackage{epsfig}
\usepackage{epstopdf}
\usepackage{epsf}
\input{epsf.sty}
\usepackage{graphicx,amsmath,amssymb}
\usepackage{epsfig,multicol}
\allowdisplaybreaks

\usepackage{bbm,bm,amsmath,amssymb}

\def\bg{\begin{eqnarray}}
\def\nd{\end{eqnarray}}
\def\sin{{\rm sin}}
\def\cos{{\rm cos}}
\def\tan{{\rm tan}}

\title{Knot Invariants and M-Theory I: Hitchin Equations, Chern-Simons Actions, and the Surface Operators} 
\author{Keshav Dasgupta${}^{1}$, Ver\'{o}nica Errasti D\'{i}ez${}^{1}$, P. Ramadevi${}^{2}$, Radu Tatar${}^3$\\
\vskip.03in
${}^1$ Ernest Rutherford Physics Department, McGill University \\
~~3600 rue University, Montr\'{e}al, Qu\'{e}bec, Canada H3A 2T8\\
${}^2$ Department of Physics, Indian Institute of Technology Bombay\\
~~Mumbai-400076, India\\
${}^3$ Department of Mathematical Sciences, University of Liverpool\\
~~Liverpool, L69 3BX England, UK\\
{\tt dasgupta.keshav8@gmail.com, vediez@physics.mcgill.ca}
~~{\tt rama0072006@gmail.com, Radu.Tatar@liverpool.ac.uk}}

\date{\today}
\maketitle

\abstract{Recently Witten introduced a type IIB brane construction with certain boundary conditions to study knot invariants and Khovanov homology. The essential 
ingredients used in his work are the topologically twisted ${\cal N} = 4$ Yang-Mills theory, localization equations and surface operators.
In this paper we extend his construction in
two possible ways. On one hand we show that a slight modification of Witten's brane construction could lead, using certain well defined duality transformations, to the
model used by Ooguri-Vafa to study knot invariants using gravity duals. On the other hand, we argue that both these constructions, of Witten and of Ooguri-Vafa, lead to two different
seven-dimensional manifolds in M-theory from where the topological theories may appear from certain twisting of the G-flux action. The non-abelian nature of the 
topological action may also be studied if we take the wrapped M2-brane states in the theory.
We discuss explicit constructions of the seven-dimensional manifolds in M-theory, and show that both the localization equations and surface operators appear naturally from 
the Hamiltonian formalism of the theories. Knots and link invariants are then constructed using M2-brane states in both the models.} 

\begin{document}

\section{Introduction and summary}

Knot theory has attracted both mathematicians and physicists to tackle some of the challenging problems. There are various approaches of constructing invariants of knots and links. Mathematicians put forth skein/recursion relation \cite{skein} to evaluate the invariants. The skein method involves study of knots  projected onto two dimensions.  These invariants can also be obtained 
from braid group representations deduced from the two dimensional statistical mechanical models, rational conformal field theories and quantum groups. All these approaches show that the invariants are Laurent polynomials in variable $q$  with integer coefficients. That is, for any knot ${\bf K}$:  
\bg\label{laurie}
J({\bf K}, q) = \sum_n a_n q^n,
\nd
where $a_n$ are integers. 

On the other hand, Chern-Simons gauge theory based on any compact group $G$ provides a natural framework to study knots  and their invariants
\cite{witten89}. In particular, this approach gives a three-dimensional definition for knots and links.
For any knot ${\bf K}$ carrying representation $R$ of gauge group $G$, the expectation value of Wilson loop operator $W({\bf K}, R) = {\rm Tr}_R P~ \exp\left(\oint_{\bf K} A\right)$ 
gives the knot invariants:
\bg\label{ramCS}
 J({\bf K}, R, q) &= & \langle W({\bf K}, R)\rangle \\
 &= &  \int {\cal D} A ~{\rm exp}\left[ik  \int_{{\bf R}^3} {\rm Tr}\left(A \wedge dA + 
{2 \over 3} A\wedge A\wedge A\right)\right] 
{\rm Tr}_R P~{\rm exp}\left(\oint_{\bf K}A\right), \nonumber 
\nd
with the first trace being in the adjoint representation, and the second trace ${\rm Tr}_R$ being in the representation $R$ of $G$;  
and $k$, an integer giving the coupling constant that we can use to write $q$ in the following way:
\bg\label{qexpr}
q = {\rm exp}\left({2\pi i \over k+ h}\right), \nd
where $h$ is the 
 dual coxeter number for group $G$. The Jones and HOMFLY-PT polynomials  correspond to placing defining representations of $SU(2)$ and $SU(N)$ respectively. 
Additionally, 
the skein relation obtained from $SU(N)$ Chern-Simons theory resembles skein relation of Alexander polynomial when $N = 0$. Similarly
for the defining representation of $SO(N)$, we get Kauffman polynomials. Besides the well known polynomials, we can obtain many new generalised knot invariants \cite{kaul}. 
Within this theory  having manifest  three-dimensional symmetry, it is not obvious as to why these knot invariants have to be Laurent polynomials with {\it integer} 
coefficients. 
Giving a topological interpretation to these integer coefficients is one of the challenging problem which
has been addressed by both mathematicians and physicists during the last 17 years.

An understanding of this issue came from the 
works on homological invariants initiated by Khovanov \cite{kovan0}. In this interesting work, Khovanov argued 
that the integer coefficients can be accounted as dimensions of vector spaces. 
This imples, for any knot ${\bf K}$, Khovanov polynomial  will be:
\begin{equation}
{\rm Kh} ({\bf K}, q, t) = \sum_{i,j} t^i q^j {\rm dim}~ H_{i, j},
\end{equation}
where ${\rm dim} ~H_{i,j}$ is the dimension of the bigraded homological chain complex. Taking $t=-1$, the above invariant is the $q$-graded Euler characteristic of the homology which gives
 Jones polynomial (for $G = SU(2)$), namely:
\begin{equation}
 J({\bf K}, \square, q) = \sum_{i,j} (-1)^i q^j {\rm dim} ~H_{i,j}.
\end{equation}
Generalisations of the bigraded homological theory for $sl_3$ \cite{khovan1}, $sl_N$ \cite{kovrosan} and arbitrary colors which are referred to as 
{\it categorifications} of knot polynomials leading to vector spaces have been extensively studied. 

Parallel development from topological string duality conjecture proposed by Gopakumar-Vafa \cite{GoV} followed by Ooguri-Vafa \cite{OV} conjecture for knots have shown 
that these invariants and their reformulations can be interpreted as counting of BPS states in string theory. 
Interestingly, this approach led to various checks of integrality properties of generalised knot invariants \cite{integ}. Further the works on categorifications  
motivated the study of triply graded polynomials discussed in \cite{sgv} succinctly within the string theory context. 

More recently, 
with the aim of interpreting Khovanov homology within intersecting brane model,
Witten considered the NS5-D3 brane system to study four dimensional
gauge theory on  ${\bf W} \times {\bf R}^+$ with knots ${\bf K}$ stuck on the three 
dimensional boundary ${\bf W}$ \cite{wittenknots1}. 
Interestingly, the number of solutions $a_n$ to the Hitchin equation in the four-dimensional gauge theory, for a given instanton number 
$n$, now give topological meaning to the integer coefficients in the Laurent polynomials \eqref{laurie}. 
The homological invariants involve one more variable $t$ besides the already existing variable $q$,  
and require study of the  surface operators in a {\it five} dimensional theory.

A relation between Witten's brane setup \cite{wittenknots1} and the Ooguri-Vafa \cite{OV} approach with intersecting D4-branes has been studied in section 5 of \cite{wittenknots1}. However a more
generic construction that relates the four-dimensional ${\cal N} = 4$ model of Witten to the ${\cal N} = 1$ set-up of Ooguri-Vafa has not been spelled out in full 
generalities\footnote{The actual comparison will be between two ${\cal N} = 1$ models as we discuss in section \ref{coknots}.}.  
In this detailed paper, we will study a unified setting in low energy supergravity description of M-theory where we relate the brane setup of Witten with the Ooguri-Vafa string theory background. 
Specifically we focus on reproducing all the results of Witten in the supergravity picture. Further, we also detail the construction of oper 
equation useful for the study of knots stuck at the three-dimensional boundary.

\subsection{Organization and summary of the paper}

This paper is organized in two broad topics. On one hand, we analyze in details the model studied by Witten in \cite{wittenknots1}. On the other hand we discuss, albeit briefly, the 
model studied by Ooguri-Vafa \cite{OV}, pointing out some of the key ingredients that may link various aspects of the two models \cite{wittenknots1} and \cite{OV}. 

We start 
section \ref{gwov} by introducing the two models in question. In section \ref{witbra} we discuss the brane constructions associated with the two models, and argue how
they can stem from similar brane configurations. This is of course a first hint to show that the two pictures in \cite{wittenknots1} and \cite{OV} may not be so different as they appear on
first sight. However subtlety lies in the construction of the Ooguri-Vafa \cite{OV} model because there are at least {\it two} possible realizations of the model $-$ one in type IIB and the 
other in type IIA. Additionally, because of the large $N$ nature of \cite{OV}, there are also {\it gravity duals} in each pictures that may be used to study the model. This is illustrated 
in section \ref{TRC}, where certain issues related to knot configurations are being pointed out. 

Section \ref{nonkah2} is dedicated completely to analyzing the physics of Witten's model \cite{wittenknots1} using a {\it dual} configuration in M-theory that has only geometry and fluxes and no other 
branes except the M2-branes. The technique considered in our work is very different from what is utilized 
in \cite{wittenknots1}. Witten uses mostly brane configurations and tactics of four-dimensional 
${\cal N} = 4$ gauge theory, along with its topological twist,  
to discuss the physics of knots in the three-dimensional boundary ${\bf W}$. In fact in the notation of \cite{wittenknots1}, the four-dimensional space will be 
denoted by ${\bf V}$ such that ${\bf V} = {\bf W} \times {\bf R}^+$, where ${\bf R}^+$ is a half-line. Our approach  
will be to use eleven-dimensional M-theory to study similar physics on the boundary ${\bf W}$. Question naturally arises as to how could two wildly different methods lead to the same physics on 
${\bf V}$ as well on the boundary ${\bf W}$. Elaborating this is of course 
one of the purposes of section \ref{nonkah2}, but before we summarize the story, let us discuss Witten's model in some details below.     

The work of Witten \cite{wittenknots1} utilizes certain crucial ingredients useful in studying knots on the boundary ${\bf W}$. The first is the topological theory on ${\bf W}$. In \cite{wittenknots1}
this is achieved in two steps using an intersecting NS5-D3 brane configuration shown in {\bf Table \ref{wittenbranes}}. The details are discussed in section \ref{flook1}.   

The second is the localization equations that are not only responsible in simplifying the path integral formalism of the theory, but also helpful in fixing the boundary terms discussed above. 
We will call these localization equations as BHN equations, the acronym being related to Bogomolnyi, Hitchin and Nahm. A derivation of the BHN equations, using techniques different from 
what is being used in \cite{wittenknots1}, is presented in section \ref{bhuc}. It turns out, and as explained in \cite{wittenknots1} and \cite{wittenknots2}, the {\it number} of solutions of the 
BHN equations, for a given instanton number, determines the coefficient of the knot polynomial. In other words, if we express the Jones' polynomial as \eqref{laurie}, then $a_n$ is the number of 
solutions to the BHN equation with instanton number $n$. This accounts for the integer coefficients in the knot polynomials.

The knots appear as Wilson loops in the boundary theory. In the S-dual picture the knots are given by 'tHooft loops. There are some advantages in discussing the S-dual story, particularly in 
connection with solving the BHN equation, and 
this forms the third crucial feature of Witten's work \cite{wittenknots1}. In section \ref{sdualt} we use our technique to analyze the S-dual picture, putting special emphasis on the {\it form} of
the BHN equations. 

There is yet another way to study the knots in the theory involving co-dimension two operators, both in the boundary ${\bf W}$ as well as in the bulk ${\bf V}$. These are called the surface 
operators, and is the fourth crucial ingredient in Witten's work \cite{wittenknots1}. We discuss the surface operators in section \ref{bathle}, and as before show that most of the results studied 
in \cite{wittenknots1} do also appear from our analysis. 

Finally, Witten discusses a possible realization of the Ooguri-Vafa model \cite{OV} given in terms of intersecting D4-branes. Similar analysis is also studied by Walcher \cite{walcher}. Our 
study in section \ref{nonkah} differs from both Witten and Walcher analysis as we discuss the D6-branes' realization of the Ooguri-Vafa model using the brane set-up in {\bf Table \ref{modbranes}}.
Although this is intimately connected to the minimally supersymmetric four-dimensional gauge theory, the specific realization of knots in this picture is more subtle. This is 
elaborated in sections \ref{toptwist} and \ref{coknots}.  

{}From the above discussions we see that the general picture developed by Witten and Ooguri-Vafa in \cite{wittenknots1} and \cite{OV} respectively, may be addressed in a 
different, albeit unified, way by dualizing the
brane configurations of {\bf Table \ref{wittenbranes}} and {\bf Table \ref{modbranes}} to M-theory. 
The duality proceeds via an intemediate configuration in type IIB involving wrapped five-branes on two-cycles of certain
non-K\"ahler manifolds. The choice of the non-K\"ahler manifolds remain specific to the model that we want to analyze. For example, Witten's model dualizes to a configuration of 
D5-${\overline{\rm D5}}$ branes wrapped on a warped Taub-NUT space as shown in section \ref{32}. This Taub-NUT space, or more appropriately a warped ALE space, is very different from the 
ALE space that may appear from T-dualizing the NS5-brane in {\bf Table \ref{wittenbranes}}. The latter creates problem in path integral representation because of the lack of a {\it global} 
one-cycle rendering it useless to study Khovanov homology. The Taub-NUT that we study here is different as discussed in section \ref{32} and we do not use it to study Khovanov homology. Instead 
our configuration is {\it only} used to study knots in the three-dimensional boundary ${\bf W}$. 

However, restricting the knots to the three-dimensional boundary is non-trivial. In Witten \cite{wittenknots1} this is achieved by switching on the gauge theory $\theta$ angle. In our 
supergravity approach in type IIB, as we show in sections \ref{aluraj} and \ref{mothdaug}, this may be achieved by switching on a non-commutative or a RR deformation on the wrapped five-branes. 
Interestingly, as we argue in section \ref{mothdaug}, these two deformations have similar four-dimensional physics when it comes to restricting the knots to the boundary ${\bf W}$.  

The M-theory uplift of the type IIB configuration is then elaborated in section \ref{jahangir}. This is the dual description of Witten's model in the absence of the knots (knots will be inserted later),
and consists of only geometry and fluxes with no branes other than the M2-branes. In this section we argue how the precise geometric information is essential to derive the harmonic two-form 
which is normalizable and unique. This two-form is essential to derive the $U(1)$ gauge theory on ${\bf V}$. This is elaborated in section \ref{fstep}, first by ignoring certain backreactions, and
then in section \ref{inclu}, by including all possible backreactions. 

The $U(1)$ theory is of course only a toy model, and what we need is the full non-abelian theory in four-dimensional space ${\bf V}$. This is achieved in section \ref{NAE}, where the first 
appearance of the M2-branes wrapped on the two-cycles of certain warped multi Taub-NUT space occurs. All these lead to the non-abelian theory on ${\bf V}$, whose details are analyzed in the 
subsequent sections. In section \ref{dynamo} we introduce the boundary dynamics.

In sections \ref{actionf} and \ref{actionR} we present our first set of major computations, related to the four-dimensional scalar fields. The complete interacting lagrangian is derived from M-theory
dimensionally reduced over a seven-dimensional manifold of the form \eqref{pwhite}. It turns out that the dynamics of three scalar fields that are dimensional reduction of the seven-dimensional gauge 
fields are somewhat easier to derive than the other three scalar fields that are fluctuations of the multi Taub-NUT space. The two sections \ref{actionf} and \ref{actionR} are elaborations on this. 

We then combine everything and write the complete four-dimensional action as \eqref{stotal}. The action contains two pieces: a topological piece and a non-topological piece. 
This is the start of section \ref{bhuc}, being one of the important section of the paper. The 
action computed in \eqref{stotal} now leads succinctly to the total Hamiltonian \eqref{hamilbeta}. This is the central result of the paper, from where all other results are derived by minimization 
and other techniques.
For example the BPS equations from the Hamiltonian \eqref{hamilbeta} may be studied by minimizing. 
The first set of BPS equations appear in \eqref{sbtaaja} for the gauge choice \eqref{gaugec}. As we showed in details, for 
example in \eqref{pacchor}, the coefficients computed in sections \ref{inclu}, \ref{actionf}, and \ref{actionR} solve {\it all} the BPS equations \eqref{sbtaaja} precisely! 

The second set of BPS equations also follow easily from the Hamiltonian \eqref{hamilbeta}. Our analysis proceeds by first ignoring the topological piece of the action \eqref{stotal}. The BPS equations 
turn out to be the BHN equations studied in \cite{wittenknots1}. The BHN equations are given by \eqref{f12bhn} and \eqref{meyepita}, with \eqref{meyepita} being further expressed in terms of 
component equations as \eqref{bhishon}. Incidentally, if we change our gauge choice from \eqref{gaugec} to \eqref{gaugec2}, the first and the second set of BPS equations change to 
\eqref{sbtaaja2} and \eqref{pharmacy} respectively, perfectly consistent to what one would expect from \cite{wittenknots1}. 

Among all the crucial ingredients of Witten's model \cite{wittenknots1}, one that we did not emphasize earlier is the appearance of the parameter $t$. This parameter has appeared before 
in describing the geometric Langland programme using supersymmetric gauge theories in \cite{langland}. In the work of \cite{wittenknots1}, $t$ appears once we try to express the BHN equations in terms of 
topologically twisted variables. In section \ref{flook} we show how $t$ appears naturally in our set-up too, {\it although} all informations that may be extracted from \cite{wittenknots1} using $t$
may appear from our supergravity analysis without involving $t$. This is to be expected as supergravity data contains all information and there is no need to add new parameters. Nevertheless, 
as we elaborate in section \ref{flook}, one may use supergravity to define $t$ and then use this to extract informations similar to \cite{wittenknots1}. One immediate advantage of this 
procedure is for finding the BHN equations once the topological piece in the action \eqref{stotal} is switched on. For example the BHN equation \eqref{bhorthi} appears easily now, and the 
full background equations, including the constraint equations plus the BHN equations, can be presented succinctly as \eqref{prolpart}. As mentioned above, all these could be done directly using 
supergravity without involving $t$, but the use of $t$ avoids certain technical challenges. 

We have now assimilated all the ingredients, namely the constraint equations and the BHN equations, to construct the theory on the boundary ${\bf W}$.
The crucial ingredients are the electric and the magnetic charges ${\bf Q_E}$ and ${\bf Q_M}$ respectively that appear in the Hamiltonian \eqref{horsho} which is the modified version
of the Hamiltonian \eqref{hamilbeta} once the topological term in the action \eqref{stotal} is switched on. In section \ref{boundth}  
we compute the two charges and show that the electric charge vanish due to our gauge choice \eqref{gaugec}, and the magnetic charge is given by \eqref{QoM}. After {\it twisting}, the magnetic charge
combines with the topological piece, now reduced to the boundary ${\bf W}$, to give us the boundary theory. This is easier said than done, because a naive computation yields an 
{\it incorrect} boundary action of the form \eqref{bandu}. There are numerous subtleties that one needs to take care of before we get the correct boundary action. These are all explained 
carefully in section \ref{boundth}, and the final topological action on ${\bf W}$ is given by \eqref{plaza}. This is a Chern-Simons action but defined with a modified one-form field ${\cal A}_d$, 
given by \eqref{newgf}, and {\it not} with the original gauge field ${\cal A}$. This is one of our main results, and matches well with the one derived in \cite{wittenknots1} using a different technique. The 
story can be similarly reproduced in the S-dual picture, and we elaborate this in section \ref{sdualt}. Various subtleties in the S-dual description discussed in \cite{wittenknots1} also show 
up in our description.  

So far we have managed to reproduce the complete boundary topological theory on ${\bf W}$. Question is, where are the knots in this picture? Section \ref{bathle} is dedicated to answering this question. 
It turns out, 
one of the key player is the {\it surface operator} that will be used to explore the knots and knot invariants in the boundary theory. In this section we 
start by discussing how the surface operators modify the 
BHN equations that we studied in section \ref{bhuc}. The surface operators are M2-branes in the theory, but their orientations are different from the M2-branes used earlier in 
section \ref{NAE} to enhance the gauge symmetry from abelian to non-abelian. In fact the M2-brane surface operators are co-dimension two singularities both in the bulk ${\bf V}$ and 
in the boundary ${\bf W}$, and their configurations are presented in {\bf Table \ref{surfaceoper}} and in {\bf Table \ref{m2m2}} for type IIA and M-theory respectively.  
   
In the language of {\bf Table \ref{surfaceoper}}, the supersymmetry preserved by the surface operator is ($4, 4$). The (4, 4) supersymmetric representation contains a vector multiplet, containing 
vectors and four scalars all in the adjoint representations of the gauge group, and a hypermultiplet, containing four scalars. If we concentrate only on the hypermultiplet sector then, in the 
absence of the surface operator, the BHN equations satisfy \eqref{may6baz} which are exactly the Hitchin's equations that one would expect from  
\cite{gukovwitten2}, \cite{gukovwitten}, \cite{gukovsurf}. In the presence of the surface operators \eqref{may6baz} changes to \eqref{niceg}, again consistent with 
\cite{gukovwitten2}, \cite{gukovwitten}, \cite{gukovsurf}. Interestingly, comparing \eqref{niceg} with \eqref{may6baz} we see that the RHS of the three equations in \eqref{may6baz} are now no longer
zeroes but proportional to certain source terms parametrized by the triplets ($\alpha, \beta, \gamma$). These triplets can be expressed in terms of supergravity parameters as given in 
\eqref{f685}, which in our opinion is a new result. 

One might also ask how the full BHN and the constraint
equations appear in the presence of the surface operators when we consider both the vector and the hypermultiplet of (4, 4) supersymmetry. The results are presented 
in \eqref{karfish}, and \eqref{mktan94} for the BHN equations and \eqref{rodmoore} for the constraint equations. 

Having got all the background equations and constraints, our next question is the {\it form} of the boundary theory. We follow similar steps as before, and express the Hamiltonian, in the presence of 
the surface operators, as \eqref{fracture}. The Hamiltonian again can be expressed as sum of squares plus the magnetic charge ${\bf Q_M}$. However now it turns out, and as explained in 
section \ref{bathle}, that the non-abelian case is in reality  much harder to study in the presence of the surface operators. To simplify, we then go to the abelian case and express the BHN and the 
constraints equations as \eqref{kings}. The magnetic charge is not too hard to find now $-$ it is presented in \eqref{abandon}; and from here the boundary theory on ${\bf W}$ is given by 
\eqref{rambof} by taking care of similar subtleties as encountered in section \ref{boundth}. 

Construction of knots on the boundary ${\bf W}$ using surface operators now easily follow using the configuration depicted in {\bf fig \ref{fig2}} and as given at the start of section \ref{knotty}. 
More precisely, the Wilson loop structure that we will 
consider is as given in \eqref{krisisjun25}. i.e using gauge fields {\it parallel} to the $x_1$ axis. This way we are able to trace all the computations with the same rigor as of the earlier sections. 

The next set of computations rely on three crucial steps for the Wilson line configurations. First is the Heegaard splitting \eqref{heegaard}
as shown in {\bf fig \ref{fig4}}. Second is the monodromy 
identifications \eqref{mormath}, as shown in {\bf fig \ref{fig6}}; and third is the braid group action, as shown in {\bf fig \ref{fig5}}. These three steps form the building blocks 
for all the knot configurations 
that we study here. We represent them as operators ${\bf A}_k$, ${\bf B}_k$ and ${\bf C}_{(2, \sigma_j)}$ respectively acting on the Wilson line state $\vert n_k \rangle$, 
where the subscript $k$ denotes the number of Wilson lines; and $\sigma_j$ is the braid group action on the $j$-th set of two consecutive Wilson lines.   
Using the three operators, for example the {\it unknot} may be represented 
as {\bf fig \ref{fig8}} and we can use them to compute the knot invariant for this case. However the steps leading to 
the actual computation of the invariant are riddled with numerous subtleties $-$ dealing with monodromies and framing anomalies to name a few $-$ 
that we discuss in details in section \ref{knotty}. The 
final knot invariant, or more appropriately the {\it linking number} for the unknot is given by \eqref{unknot}. Similar analysis is presented for the trefoil knot, torus ($2, n$) knots, 
figure 8 knot and 
$5_2$ knot in \eqref{trefoil}, \eqref{torusknot}, \eqref{figure8} and \eqref{52} respectively. These knot configurations easily follow the three-steps building blocks mentioned above, as shown in 
{\bf fig \ref{fig9}}, {\bf fig \ref{fig10}}, {\bf fig \ref{fig11}} and {\bf fig \ref{fig12}} respectively, and we discuss how this generalizes to {\it all} knot configurations that may be built 
in our model. 

In fact other invariants, beyond the linking numbers, may also be studied for the knot configurations that we discuss here. These invariants have been addressed in \cite{surfknot} and may be 
constructed using the monodromies ${\cal M}_k$ in \eqref{iamlegend}, implying that our analysis is generic enough not only to include all the constructions of \cite{surfknot} but also give them 
appropriate supergravity interpretations. Despite the success, a non-abelian extension of this picture is harder, and we do not attempt it here leaving a more detailed elaboration for the sequel. 
Instead however
we dedicate the last section, i.e section \ref{opar}, albeit briefly on {\it opers} that may generalize more easily to the non-abelian case.  

Section \ref{nonkah} is dedicated completely in exploring the physics of the Ooguri-Vafa \cite{OV} model. From start, there are many points of comparison with section \ref{nonkah2} dealing with 
the physics of Witten's model \cite{wittenknots1}. For example, 
the absence of a Coulomb branch, the location of the knots on the {\it internal} ${\bf S}^3$ and the existence of a gravity dual might suggest that the 
Ooguri-Vafa \cite{OV} model is very different from Witten's model \cite{wittenknots1}. In section \ref{nonkah} we argue that this is {\it not} the case. In spirit, these two models are far closer
in many respects than one would expect from naive comparison.  

The first hint already appears from the discussion in section 5 of \cite{wittenknots1} and in \cite{walcher}, 
where the intersecting D4-branes' construction of the Ooguri-Vafa model is discussed from the brane set-up of 
{\bf Table \ref{wittenbranes}}. However we want to emphasize the connection using the brane set-up of {\bf Table \ref{modbranes}} that directly relates the four-dimensional ${\cal N} = 4$ model of 
Witten to the ${\cal N} = 1$ set-up of Ooguri-Vafa.  

Our starting point is then multiple D5-branes wrapped on a two-cycle of a non-K\"ahler resolved conifold. We take $N$ five-branes so that IR gauge group for the minimally supersymmetric four-dimensional 
gauge theory becomes $SU(N)$. The geometry can be worked out precisely as we show in section \ref{second}, which in turn is based on the recent work \cite{DEM}. However existence of a similar picture 
as in section \ref{aluraj} without dipole deformation, doesn't mean that the physics remains similar now. The absence of the Coulomb branch changes the story a bit, and this is discussed in details 
in section \ref{toptwist}. However the two models, despite the small difference, are identical in some respect regarding the four-dimensional picture, even when we go to the mirror type IIA side. The 
Ooguri-Vafa model is then realized from the mirror picture by first Euclideanizing the geometry, so that the four-dimensional physics is defined on ${\bf S}^3_{(1)} \times {\bf R}^+$, 
and then performing a flop \eqref{sebeast} that exchanges the ${\bf S}^3_{(1)}$ with ${\bf S}^3_{(2)}$, the three-cycle of the mirror deformed conifold. The flop transfers the physics to the three-cycle
of the deformed conifold, and this way we can get \cite{OV} from \cite{wittenknots1}. 

The discussion in section \ref{toptwist} leaves a few questions unanswered. The first is related to the physics on ${\bf S}^3_{(1)}$, namely, what is the precise topological theory 
on ${\bf S}^3_{(1)}$ that we eventually transfer to ${\bf S}^3_{(2)}$? The second is related to the knots, namely, what about the knot configurations and the knot invariants? 
In the remaining part of the paper we answer these two questions.  

To answer the first question we will require the precise supergravity background in type IIB, before mirror transformation. This is studied in section \ref{5brane}, where the fluxes are worked out in 
section \ref{5brane1} and the warp-factors, in the type IIB metric, are worked out in section \ref{5brane2}. The M-theory lift of this configuration is studied in section \ref{g21}, where we show 
that the seven-dimensional manifold is again a warped Taub-NUT fibered over a three-dimensonal base. This time however the warping of the base and fibre in the seven-dimensional manifold \eqref{g2} is 
different from what we had in section \ref{jahangir} such that the four-dimensional supersymmetry can be minimal. Of course the right comparison with section \ref{jahangir} can only be done {\it after} 
we make a dipole deformation to the type IIB background. It turns out, and as expected, dipole deformation doesn't break any supersymmetry, but does break the four-dimensional Lorentz symmetry  
to three-dimensional Lorentz symmetry. This is good because we can localize the knots in the three-dimensional space where there is a manifest Lorentz invariance. Details on this are presented 
in section \ref{g22}. 

Once we have the full geometry and fluxes in M-theory, with dipole deformation, it is easy to follow similar procedure as in sections \ref{jahangir}, \ref{fstep}, \ref{inclu} and \ref{NAE} to work 
out the normalizable harmonic forms, and non-abelian enhancement to study the gauge theory in four-dimensional space. This is the content of section \ref{g23}, where we discuss the 
vector multiplet structure, leaving the study of chiral multiplets for the sequel. The vector multiplet structure leads to a non-abelian gauge theory in four-dimensions whose coupling constant, much 
like \eqref{c11theta} before, may to traced to the underlying supergravity variables in M-theory. 

The above discussions then brings us to the second question related to the knot configurations and knot invariants. In fact the story is already summarized in section \ref{toptwist}, and in 
section \ref{coknots} we elaborate on individual steps. The first step is related to the topologically twisted theory on the three-dimensional boundary ${\bf W}$. This time, because of the absence 
of the Coulomb branch, the boundary theory is simpler than the one in Witten's model, namely \eqref{plaza}. It is now given by \eqref{ajapple}, which is again a Chern-Simons theory but the 
coupling constant is not the one that we naively get from the topological piece \eqref{vietduck} in M-theory,  rather it is a combination that appears from both the G-flux kinetic and the topological 
pieces in M-theory. This is identical to what we had in section \ref{boundth} related to Witten's model. We now see that a similar structure, yet a bit simpler from \cite{wittenknots1}, 
is played out for the Ooguri-Vafa model \cite{OV} too. 

All these are defined on ${\bf S}^3_{(1)}$, and once we take the mirror, the theory on ${\bf S}^3_{(1)}$ remains identical. The second step is to perform a flop operation \eqref{sebeast}, so that 
we can  transfer the physics to the three-cycle ${\bf S}^3_{(2)}$ of the non-K\"ahler deformed conifold, giving us \eqref{pasoli}. 
For this case, the knots may now be introduced by inserting co-dimension two singularities   
as depicted in {\bf fig \ref{fig14}}. Again, the picture may look similar to what we discussed in sections \ref{bathle} and \ref{knotty}, but there are a few key differences. One, we cannot study 
the abelian version now as the model is only defined for large $N$. This means all the analysis of the knots using operators ${\bf A}_k$, ${\bf B}_k$  and ${\bf C}_{(2, \sigma_j)}$ may not be 
possible now. Two, similar manipulations to the BHN equations that we did in section \ref{bathle} now cannot be performed. 

What can be defined here? In the remaining part of section \ref{coknots} we give a brief discussion of how to study knots in the Ooguri-Vafa model, leaving a more detailed exposition for the sequel. 
We summarize our findings and discuss future directions in section \ref{disco}. In a companion paper \cite{veronica}, and for the aid of the readers, we provide {\it detailed} proofs and derivations 
of all the results here including, at times, alternative derivations of some of the results.  

\subsection{What are the new results in this paper?}

In this paper we construct two different configurations in M-theory, which consist of only geometry, fluxes and M2-branes
(the latter provide a non-abelian enhancement of the underlying gauge group).
We refer to them as {\bf Model A} and {\bf Model B}.
These are dual to the models in \cite{wittenknots1} and \cite{OV}, respectively.
An important new result is that we show the exact duality transformations that relate {\bf Models A} and {\bf B}.
Consequently, we make explicit the direct connection between the seemingly very distinct models in \cite{wittenknots1} and \cite{OV}.
In other words, we provide a unifying picture of the two existing physics approaches to compute knot invariants from the counting of solutions to BPS equations.

The present work focuses on the study of {\bf Model A}, that dual to the model in \cite{wittenknots1}.
We first obtain the complete four-dimensional gauge theory Lagrangian \eqref{stotal}, appropriatedly compactifying {\bf Model A}.
Then, we derive its associated Hamiltonian \eqref{hamilbeta}.
Clearly, the coefficients appearing in \eqref{hamilbeta} are expressed in terms of supergravity parameters, by construction.
All our results for {\bf Model A} stem from this Hamiltonian and are mapped in exquisite detail to the results in \cite{wittenknots1}.
We thus conclude that another major outcome of our analysis is that it allows for a precise physical interpretation of \cite{wittenknots1}
in the conceptually simple and long-known classical Hamiltonian formalism.
For example, the BPS equations follow from the minimization of \eqref{hamilbeta} for static configurations of the gauge theory fields.
We refer to them as BHN equations, the acronym standing for Bogomolnyi-Hitchin-Nahm equations.

The four-dimensional space $\mathbf{V}$ where our Hamiltonian is defined naturally decomposes as $\mathbf{V}=\mathbf{W}\times{\bf R}^+$.
After minimization of the energy and topologically twisting our theory, we show that the action on the three-dimensional boundary $\mathbf{W}$ of
$\mathbf{V}$ is topological.
This is a Chern-Simons action for a modified gauge field,
which is a certain linear combination of the original gauge fields and some of the scalar fields in our theory.
Then, the inclusion of surface operators in this set up provides an inherent framework for realizing knot invariants,
as argued in \cite{surfknot}. A key result in our work is
the  realization of surface operators as M2-branes, different from the ones used for the non-abelian enhancement of {\bf Model A}.
Upon restricting to the abelian case for simplicity, this allows us to work out the
linking numbers for the most well-known knots:
unknot, trefoil, torus $(2,n)$, figure 8 and $5_2$, given by \eqref{unknot}-\eqref{52}.

Finally, it is interesting to note that we have not yet exploited most of the immense potential of the constructed {\bf Models A} and {\bf B}.
To mention a few possibilities, we hope to learn about the Jones, Alexander and HOMFLY polynomials and Khovanov homology in the sequel. 


\section{Brane constructions and Knots  \label{gwov}}

In this section we will study the knots first from a brane construction proposed by Witten \cite{wittenknots1, wittenknots2} 
and argue how this could be mapped to the geometric transition
picture of Ooguri-Vafa \cite{OV, AV}. We will argue that certain fourfolds along with specific configurations of surface operators 
are useful in making the connections between the two scenarios. 

\subsection{Brane constructions for Knots \label{witbra}}

In the original Witten's construction \cite{wittenknots1} of knot theory in type IIB theory, we will call this\footnote{Not to be confused with A-model and B-model that appear in the topologically 
twisted version of our constuction.} 
{\bf Model A}, the branes were arranged as in {\bf Table \ref{wittenbranes}},  
\begin{table}[h!]
 \begin{center} 
\begin{tabular}{|c||c|c|c|c|c|c|c|c|c|c|}\hline Directions & 0 & 1 & 2
& 3 & 4 & 5 & 6 & 7 & 8 & 9 \\ \hline
NS5 & $\surd$  & $\surd$   & $\surd$   & $\surd$ & $\ast$  & $\ast$ & $\ast$ & $\ast$ & $\surd$
& $\surd$ \\  \hline
D3 & $\surd$  & $\surd$   & $\surd$   & $\ast$ & $\ast$  & $\ast$ & $\surd$ & $\ast$ & $\ast$  & $\ast$
\\  \hline
  \end{tabular}
\end{center}
  \caption{The orientations of various branes in the intersecting branes set-up. The notation $\surd$ is the
direction along which the branes are oriented.}
  \label{wittenbranes}
\end{table}
 \noindent with an additional source for IIB axion, $C_0$, switched on such that the knots are localised along the $2+1$ dimensional intersection parametrised by 
$x_{0, 1, 2}$. 

Let us now modify the original set-up of Witten by converting the direction $x_6$ along which the D3-brane is stretched into a finite interval. This is achieved by 
introducing another NS5-brane oriented along $x_{0, 1, 2, 3, 4, 5}$. This crucial step will be useful for us to relate the configuration of Witten to the 
configuration of Ooguri-Vafa \cite{OV}, as we will soon see. For later convenience we will call this, and the subsequent modification of this, as {\bf Model B}.

The type IIB configuration can be modified further by T-dualizing along $x_3$ direction. This T-duality leads us to the well-known configuration in type 
IIA theory \cite{DOT1, DOT2} as depicted in {\bf Table \ref{modbranes}}.
\begin{table}[h!]
 \begin{center}
\begin{tabular}{|c||c|c|c|c|c|c|c|c|c|c|}\hline Directions & 0 & 1 & 2
& 3 & 4 & 5 & 6 & 7 & 8 & 9 \\ \hline
NS5 & $\surd$  & $\surd$   & $\surd$   & $\surd$ & $\ast$  & $\ast$ & $\ast$ & $\ast$ & $\surd$
& $\surd$ \\  \hline
NS5 & $\surd$  & $\surd$   & $\surd$   & $\surd$ & $\surd$ & $\surd$ & $\ast$ & $\ast$ & $\ast$
& $\ast$ \\  \hline
D4 & $\surd$  & $\surd$   & $\surd$   & $\surd$ & $\ast$  & $\ast$ & $\surd$ & $\ast$ & $\ast$  & $\ast$
\\  \hline
  \end{tabular}
\end{center}
  \caption{The orientations of various branes in the T-dual of the modified Witten set-up.}
  \label{modbranes}
\end{table}
\noindent In addition to the required branes we will have a background type IIA gauge field $A_3$, that will have a pull-back on the D4-brane and furthermore 
introduce a non-trivial complex structure on the ($x_3, x_6$) torus. The latter operation will help distinguish the non-compact world-volume directions $x_{0, 1, 2}$ with the 
compact toroidal directions even in the limit of large size of the torus. However although supersymmetry of the background still remains valid,  
the localization of the knots in the $x_{0, 1, 2}$ directions is not: we have lost the Coulomb branch, so the discussion of 
knots should be taken with care here. We will study this soon. 

Finally let us make yet another modification to the set-up studied above: introduce large $N$ number of D4-branes. Such a modification will help us to study the 
{\it gravity} dual of this set-up, in other words will connect us directly to the model studied by Ooguri-Vafa \cite{OV}
or more recently to Aganagic-Vafa \cite{AV}! This is because an 
appropriate T-duality to the above brane configuration will convert the two NS5-branes to a singular conifold
 and the $N$ D4-branes to $N$ wrapped D5-branes on the 
vanishing two-cycle of the conifold. We can then blow-up the two cycle to convert the singular conifold to a resolved 
conifold\footnote{We will see that the metric on this will be a non-K\"ahler one.}. 
The D5-branes will then wrap the resolution two-cycle.
To see how this works, let us discuss this in some details. 

\subsection{T-duality, resolved cone and a geometric transition \label{TRC}}

We begin by  introducing  a circle action on the conifold and extend it to the 
resolved conifold in a compatible manner.  Consider an action  $S_c$ on the conifold 
$xy-uv=0$, where ($x, y, u, v$) are complex coordinates, in the following way: 
\bg\label{sc}
S_c:~~(e^{i\theta}, x) \to x ,~~ (e^{i\theta}, y) \to y, ~~(e^{i\theta}, u) \to 
e^{i\theta} u ,~~ (e^{i\theta}, v) \to
e^{-i\theta} v. \nd 
The orbits of the action $S_c$ degenerates along the 
union of two intersecting complex
lines $y=u=v=0$ and $x=u=v=0$ on the conifold. Now, if we take a T-dual along
the direction of the orbits of the action, there will be NS branes along these
degeneracy loci as argued in \cite{bvs}. So we have two NS branes which are
spaced along $x$ (i.e. $y=u=v=0$) and $y$  directions (i.e. $x=u=v=0$)
together with non-compact direction along the Minkowski space
 which will be denoted by $NS_x$ and $NS_y$. 

One may lift this action so as to define a resolved cone.
To do that, let us start with two ${\bf C}^3$ with coordinates ($Z, X, Y$) and ($Z', X', Y'$) respectively, where ($Z, Z'$) are the coordinates of ${\bf P}^1$ in the two ${\bf C}^3$'s respectively, and 
the rest form the coordinates of the fiber.
Then the manifold ${\cal O}(-1) \oplus {\cal O} (-1)$ over ${\bf P}^1$ can be obtained by gluing the two copies of ${\bf C}^3$, parametrized above,  
by the following identification: \bg
\label{-1-1} ZZ' = 1 ~, ~\quad X^{-1} X' = Z ~, ~\quad Y^{-1} Y' = Z. \nd 
The blown-down map from the resolved conifold to the 
conifold ${\cal C}$ is given by eq (7) of \cite{DOT1}, from where one may infer the 
action $S_r$ on the resolved conifold to be an extension of the 
action $S_c$ (\ref{sc}) given by by eq (8) of \cite{DOT1}.
The rest of the discussion after eq (8) of \cite{DOT1}, till the end of section 2 in \cite{DOT1}, details how 
the T-dual picture becomes the following brane configuration: a D4 brane along the
interval with two NS branes in the {\it orthogonal} direction at the ends of
the interval exactly as illustrated in {\bf Table \ref{modbranes}}.
Here the length of the interval is the same as the
size of the two-cycle of the resolved conifold.  As the two-cycle shrinks to zero, the brane construction of a resolved conifold approaches the brane construction of a 
conifold\footnote{In the first version of the paper some of the details presented here overlapped with \cite{DOT1}. Here we remove all the overlap and the readers are instead referred to 
section 2 of \cite{DOT1}.}. 

In the language of branes, the two NS5 branes are along directions $x_{4, 5}$ and $x_{8, 9}$ and fill simultaneously the spacetime directions $x_{0, 1, 2, 3}$. 
This means the T-duality was done along direction $x_6$, or in the language of a conifold, along $\psi$. The conifold geometry is parametrized by ($\theta_i, \phi_i$) 
with $i = 1, 2$ with the $U(1)$ direction
$\psi$ and the non compact radial direction $r$.   
In the following let us clarify some subtleties related to the T-duality. 
First let us consider the wrapped D5-brane on a conifold geometry. 
A standard T-duality along an orthogonal direction should convert this to a wrapped D6-brane. The $C_7$ source charge of the D6-brane decomposes
in the following way:
\bg\label{c7}
C_7(\overrightarrow{\bf x}, \psi, \theta_1, \phi_1) ~ = ~ 
C_5(\overrightarrow{\bf x}, \psi) \wedge \left({e_{\theta_1} \wedge e_{\phi_1} \over \sqrt{V_2}}\right)\nd
where $V_2$ is the volume of the two-sphere that is being wrapped by the D6-brane and whose cohomology is represented by the term in the 
bracket\footnote{The representative of second cohomology for a two-cycle of a conifold is $e_{\theta_1} \wedge e_{\phi_1} -
e_{\theta_2} \wedge e_{\phi_2}$ as both ${\bf P}^1$ vanish at the origin \cite{DM2}. 
For resolved conifold we will take \eqref{c7}, as
geometrically the D5-brane wraps a two-sphere parametrized by ($\theta_1, \phi_1$). This makes sense as one of the sphere
will be of vanishing size at $r = 0$.}. 
In the limit where 
the size of the two-sphere is vanishing (i.e for the T-dual conifold), the term in the bracket in \eqref{c7} will behave as a delta-function, and
consequently $C_7$ will decompose as $C_5$ i.e as a D4-brane. It will take infinite energy to excite any mode along the 
directions of the vanishing two-sphere, and 
therefore for all practical purpose a T-dual of the wrapped D5-brane on a conifold will be a D4-brane stretched along $\psi$ direction. This is 
of course the main content of \cite{uranga, DM1, DM2}. Similarly if the wrapped two-sphere is of {\it finite} size, i.e the D5-brane
wraps the two-cycle of a resolved conifold, then at energy lower than the inverse size of the two-sphere the T-dual will effectively behave again as a 
D4-brane \cite{DOT1, DOT2}. Once the energy is bigger than this bound $-$
the size of the two-cycle is much bigger than the string scale $-$ then
the intermediate energy physics will probe the full D6-brane. Our analysis in this paper will be related to this case only, i.e we will explore
the classical dynamics of a wrapped D6-brane on a four-cycle parametrized by ($\theta_1, \phi_1, \psi$) and $x_3$. 

The above discussion tells us that, under appropriate T-duality, we should get the IR picture of the geometric transition model studied by Ooguri-Vafa \cite{OV}. There are 
of course few {\it differences} that we need to consider before making the equivalences. The first is the existence of a $B_{\rm NS}$ field with one of its components 
along the D5-branes and another orthogonal to it\footnote{In general we expect both $B_{\rm NS}$ and $B_{\rm RR}$ to appear here. The latter however is more non-trivial to deal with, so
we will relegate the discussion for later.}.
This $B_{\rm NS}$ field should give rise to the {\it dipole} deformations of the D5-branes' gauge theory \cite{GB, bergman1, bergman2}.
This deformation should also be responsible for preserving supersymmetry in the model. It is however {\it not} clear that the knots in this model should again be 
restricted to $x_{0, 1, 2}$ directions, although naively one could argue that  
the two directions of the D5-branes are wrapped on the ${\bf P}^1$ of the resolved conifold, and the dipole deformation with a 
$B_{\rm NS}$ field $B_{3\psi}$ should restrict the knots further to the $x_{0, 1, 2}$ directions. The reason is of course the absence of the Coulomb branch which is a crucial
ingredient in \cite{wittenknots1, wittenknots2}. 

There is another reason why this should not be the case. We can ask the following question:
what will happen if we make a geometric transition to two-cycle on which we have wrapped D5-branes? From standard argument we 
know that the D5-branes will disappear and will be replaced by fluxes. In this flux picture, or more appropriately the gravity dual, it will be highly non-trivial to 
get the information about the knots from the fluxes on a deformed conifold background (as there are no branes on the dual side). One might think that a T-dual 
of this gravity dual could bring us back to branes in type IIA, but this doesn't help as the original D4-branes on which we had the knot configurations {\it do not} 
appear even on the brane side. To see this, consider the
following circle action $S_d$: 
\bg \label{sd} 
S_d:~(e^{i\theta}, x) \to x ,~~ 
(e^{i\theta}, y) \to y,~~ (e^{i\theta},
u) \to e^{i\theta} u ,~~ (e^{i\theta}, v) \to e^{-i\theta} v, 
\nd  on the 
deformed conifold $xy -uv =\mu$, where $\mu$ is the deformation parameter.
Then $S_d$ is clearly the extension of $S_r$ discussed in eq (8) of \cite{DOT1}, and the orbits of 
the action degenerate along a
complex curve  $u = v = 0$ on the deformed conifold. If we take a T-dual of the 
deformed conifold along the orbits of
$S_d$, we obtain a NS brane along the curve $u=v=0$ with non-compact direction 
in the Minkowski space which is
given by $xy = \mu$ in the x-y plane.
Topologically, the above curve is ${\bf R}^1 \times {\bf S}^1$.
Thus in the T-dual picture, the large $N$ duality implies a transition from 
the brane
configuration of $N$ coincident D4-branes between two orthogonal NS5-branes to
the brane configuration of a {\it single} NS5-brane wrapped on ${\bf R}^1 \times {\bf S}^1$
with appropriate background fluxes.  The D4-branes have disappeared in the dual brane configuration too, apparently along with our knot configuration!

The solution to the above conundrum is non-trivial and we will discuss this soon. But first let us discuss how to study {\bf Model A} using the approach of wrapped branes on
certain non-K\"ahler manifolds. This will lead us to a more unified approach to discuss both the models.

\section{Model A: The type IIB dual description and warped Taub-NUT \label{nonkah2}}

The situation for {\bf Model A} is slightly different as it is directly related to \cite{wittenknots1} and therefore to the Chern-Simons theory along $x_{0, 1, 2}$ directions for
the brane configurations given in {\bf Table \ref{wittenbranes}}. The claim is that the knot polynomial $J(q; {\bf K}_i, R_i)$ for any knot ${\bf K}_i$ 
is given in the Chern-Simons theory via the
following path integral: 
\bg\label{pathoi}
J(q; {\bf K}_i, R_i) &=& \langle {W}({\bf K}_i, R_i)\rangle = \langle {\rm Tr}_{R_i} P~{\rm exp}\oint_{{\bf K}_i} A\rangle \nonumber\\
 & = & {\int {\cal D}A ~{\rm exp}\left(iS_{cs}\right) \prod_i {\cal W}(K_i, R_i)\over \int {\cal D}A ~{\rm exp}\left(iS_{cs}\right)}, \nd
that is a generalization of \eqref{ramCS}, and 
where $q$ is the variable which is used to express the knot polynomial as a Laurent series, $R_i$ is the compact representation of the gauge group $G$ appearing in the Chern-Simons action
$S_{cs}$:
\bg\label{csa}
S_{cs} = {k\over 4\pi}\int_{\bf W} {\rm Tr} \left(A \wedge dA + {2\over 3} A \wedge A \wedge A\right). \nd
As discussed in the introduction, $k$ is an integer used to express $q$ as in \eqref{qexpr}. The denominator appearing in \eqref{pathoi} is in general non-trivial 
function of $k$. For example for $SU(2)$ group with ${\bf W} = {\bf S}^3$, as shown in \cite{witten89} and \cite{wittenknots1}, the denominator becomes:
\bg\label{atwork}
\int {\cal D} A ~{\rm exp}\left(iS_{cs}\right) = \sqrt{2\over k + 2} ~\sin\left({\pi\over k +2 }\right), \nd
but if we take ${\bf W} = {\bf R}^3$, this can be normalized to 1 and so \eqref{ramCS} and \eqref{pathoi} become identical. 
This is the case we will study in this section.   
The above two expressions \eqref{pathoi} and \eqref{qexpr} serve as dictonary that maps the knot polynomial $J$ and the knot parameter 
$q$ in terms of the variables of Chern-Simons theory.

\subsection{First look at the gravity and the topological gauge theory \label{flook1}}

We will discuss the knots appearing from this construction soon, but first let us modify {\bf Table \ref{wittenbranes}} slightly by first restricting the direction $x_6$ to an interval,
and secondly, T-dualizing along $x_3$ direction to convert the configuration to D4-branes between two parallel NS5-branes. 
T-dualizing
further along $x_6 \equiv \psi$ direction will convert the D4-branes to fractional D3-branes at a point on a {\it warped} Taub-NUT space. In particular, we will have a geometry like:
\bg\label{sugra1}
&&ds^2 = e^{-\phi} ds^2_{0123} + e^\phi ds_6^2\nonumber\\
&& {\cal F}_3 = e^{2\phi} \ast_6 d\left(e^{-2\phi} J\right), \nd
where $\phi$ is the dilaton and the Hodge star and the fundamental form $J$ are wrt to the dilaton deformed metric $e^{2\phi} ds_6^2$. The metric $ds_6$ will be given by:
\bg\label{chote}
ds^2_6 = F_1 dr^2 + F_2 (d\psi + {\rm cos}~\theta_1 d\phi_1)^2 + F_3 (d\theta_1^2 + \sin^2\theta_1 d\phi_1^2) + F_4 (dx_8^2 + dx_9^2), \nd
with $F_1 = F_1(r), F_2 = F_2(r)$ and $F_3 = F_3(r)$ as functions of $r$ only and $F_4 = F_4(r, x_8, x_9)$, as the simplest extension of the case with only radially dependent warp factors. 
Note also that the fractional D3-branes cannot be interpreted as wrapped D5 - ${\overline {\rm D5}}$
branes along ($\theta_1, \phi_1$) directions. Instead the fractional D3-branes will be
interpreted here as D5 - ${\overline{\rm D5}}$ pair wrapping direction $\psi$ and stretched along the radial $r$ direction.

We can also change the topology along the $x_{8, 9}$ directions from ${\bf R}^2$ to ${\bf T}^2$ or ${\bf P}^1$ without violating Gauss' law. Before elaborating on this story, let us
clarify few issues that may have appeared due to our duality transformation. First, one would have to revist the supersymmetry of the model, which seems to have changed from 
${\cal N} = 4$ to ${\cal N} = 2$. This still allows a Coulomb branch, but we need more scalars to complete the story. One way to regain the lost supersymmetry is to assume that the 
$x_6$ circle is large, so that essentially, for the half space $x_6 > 0$, we have the same physics explored in \cite{langland, wittenknots1}. 

Secondly, Witten discusses the possibility of T-duality along orthogonal $S^1$ for the D3-NS5 system and argues that, because of the absence of a topological one-cycle in the 
T-dual configuration, the path integral in this framework cannot be taken as a trace. Our configuration differs from this conclusion in the following way. The T-dual will lead us to 
a non-K\"ahler metric on the Taub-NUT space (we call this as a deformed Taub-NUT) and although the Taub-NUT circle will shrink to zero size, 
we will not be using the Taub-NUT configuration to compute the path integral. Rather a different Taub-NUT will feature later in our study of 
the gauge theory on the wrapped D5-branes. 

Thirdly, converting the D3-branes to D4-branes wrapped along direction $x_3$ would seem to give us only two scalars ($x_8, x_9$). But this is not quite the case as the fluctuation of
the gauge field along the $x_3$ direction will appear as an extra scalar field when we look at the three dimensional gauge theory along directions ($x_0, x_1, x_2$). These are therefore
exactly the scalar $\overrightarrow{X}$ in \cite{wittenknots1}. The other three scalar fields, namely ($x_4, x_5, x_7$), are related to $\overrightarrow{Y}$ in \cite{wittenknots1}. 

Below a certain energy scale, related to inverse radius of the $x_3$ circle, the theory on the D4-branes can be studied at the intersection space of NS5-D4 system. The boundary 
action is then given, for the Euclidean three dimensional space, by \cite{wittenknots1}:
\bg\label{eqn1}
S^{(1)}_b = {1\over g^2_{YM}} \int_{x_6 = 0} d^3x\left[l_1 \epsilon^{abc}{\rm Tr}~X_a[X_b, X_c] + l_2 \epsilon^{\mu\nu\rho}{\rm Tr}\left(A_\mu \partial_\nu A_\rho + {2\over 3} A_\mu A_\nu 
A_\rho\right)\right], \nonumber\\ \nd
where ($l_1, l_2$) are constants related to the background gauge field $\langle A_3\rangle$ (see also \cite{wittenknots1}) and the superscript is for later convenience. 

\subsubsection{On the topologically twisted theory \label{311}}

Constructing a topological field theory using R-symmetry twist to ${\cal N} = 4$ theory is well known, and could be easily applied to our configuration. The wrapped D4-branes on 
$x_3$ has a $SO(5)$ symmetry broken to $SO(4) \times U(1)$. The one-form associated with the $U(1)$ symmetry can be combined with the {\it twisted} scalar fields, i.e scalar fields
associated with ($x_8, x_9$) converted to one-forms $\phi_\mu dx^\mu$. The fluctuation of the gauge field along $x_3$ direction\footnote{Not to be confused with the type IIA $U(1)$ 
gauge field with expectation value $\langle A_3 \rangle$.} contributes another one-form. Finally the fourth one-form may appear from one component of the fluctuations of the D4-branes 
along orthogonal direction. Together these one-forms could be expressed (in Euclidean space) as:
\bg\label{flacu}
\phi \equiv \sum_{\mu = 0}^3 \phi_\mu dx^\mu, \nd
which captures the concept of R-symmetry twist (see \cite{langland, wittenknots1} for more details). 
Using these we can rewrite \eqref{eqn1} as the following topological theory \cite{wittenknots1}:
\bg\label{toptheory}
S^{(1)}_b = {1\over g^2_{YM}} \int_{x_6 = 0} d^3x \epsilon^{\mu\nu\rho} {\rm Tr}\left[2l_1 \phi_\mu \phi_\nu\phi_\rho  + l_2 \left(A_\mu \partial_\nu A_\rho 
+ {2\over 3} A_\mu A_\nu A_\rho\right)\right], \nd
where the coefficients $l_1$ and $l_2$ are defined\footnote{We thank Ori Ganor for explaining the coefficient $l_1$ of the cubic term in \eqref{toptheory} using bound state 
wavefunction of a F1-string with a NS5-brane \cite{gan0}.} 
in terms of $t \equiv \pm {\vert \tau \vert \over \tau}$, where $\tau$ is the standard definition for four-dimensional gauge theory,
namely $\tau = {\theta \over 2\pi} + {4\pi i\over g^2_{YM}}$, as:
\bg\label{c1c2}
l_1 \equiv  - {t + t^{-1}\over 6}, ~~~~~~~ l_2 \equiv {t + t^{-1} \over t - t^{-1}}. \nd
The derivation of the above relations are given in \cite{wittenknots1}, assuming the $\theta$ angle in the definition of $\tau$ to be related to the YM coupling $g^2_{YM}$. 

The topological theory that we got above in \eqref{toptheory} is however {\it not} complete. There are other terms that require a more detailed study to derive. The derivation has been
beautifully presented in \cite{wittenknots1}, so we will just quote the results. The idea is to take the five-dimensional action on the D4-branes:
\bg\label{5daction}
S_{\rm D4} = {1\over g^2_5}\int d^5 x \sqrt{g^{(5)}}{\cal L}^{(5)}_{kin} + T_4\int \epsilon^{\mu\nu\rho\sigma\alpha} {\cal A}_\mu {\rm Tr}~F_{\nu\rho} F_{\sigma\alpha}, \nd
where $T_4$ is the tension of the D4-brane, and reduce over the compact direction $x_3$. The expectation value of ${\cal A}_\mu$, alongwith $T_4$, 
will give rise to the $\theta$ angle in the dimensionally reduced four-dimensional ${\cal N} = 4$ SYM theory with the YM coupling determined by the length $R_3$ of the compact $x_3$ 
direction (assuming flat $g^{(5)}$):
\bg\label{ymcoupling}
{1\over g^2_{YM}} = {R_3 \over g^2_5}. \nd
The kinetic piece of the five-dimensional action of the D4-branes can now be represented as:
\bg\label{kinpiece}
&&{1\over g^2_5}\int d^5x \sqrt{g} {\cal L}^{(5)}_{kin} ~ \to ~ {1\over g^2_{YM}} \int d^4x \sqrt{g} {\cal L}_{kin} = \{Q, ....\} \\
&&~~~~~ +{1\over g^2_{YM}}\int d^3x {1\over \sqrt{1+ w^2}}\left[-w \Omega(A) + \epsilon^{\mu\nu\sigma} {\rm Tr}\left(\phi_\mu F_{\nu\sigma} + w \phi_\mu D_\mu \phi_\sigma - 
{2\over 3} \phi_\mu \phi_\nu\phi_\sigma\right)\right]\nonumber \nd 
where the bounday integral has to be defined at both ends of $x_6$, namely $x_6 = 0$ and $x_6 \to \infty$, or to the point along $x_6$ where we have moved the other NS5-brane. Of course,
as mentioned earlier, to preserve maximal supersymmetry, the other NS5-brane has to be kept far away so that near $x_6= 0$ we restore ${\cal N} = 4$ supersymmetry. We have also related
$t$, appearing in \eqref{c1c2}, and $w$ as:
\bg\label{tnw} t ~ = ~ w - \sqrt{1+w^2}. \nd 
The other parameters appearing in \eqref{kinpiece} are defined in the following way: $Q$ is the standard supersymmetric operator such that in the absence of any boundary, the kinetic piece
would only be given by the first line of \eqref{kinpiece} i.e as an anti-commutator with $Q$. The other parameter $\Omega(A)$ is the standard Chern-Simons term in three-dimension, such 
that:
\bg\label{csa}
\int d^3 x~ \Omega(A) = \int {\rm Tr}\left(A \wedge dA + {2\over 3} A \wedge A \wedge A\right). \nd
It is now easy to see that once we combine the boundary term of \eqref{kinpiece} with the bounday action \eqref{toptheory}, the final action takes the following simple form:
\bg\label{actions}
S = -{4\pi\over g^2_{YM}} \cdot {1\over w\sqrt{1+w^2}} \int_{x_6=0} d^3x \int {\rm Tr}\left(A_w \wedge dA_w + {2\over 3} A_w \wedge A_w \wedge A_w\right), \nd
as one may verify from \cite{wittenknots1} too. The above action is 
essentially $\Omega(A_w)$, with $A_w \equiv A + w\phi$. This tells us that we could insert a generalized one-form, given by $A_w$, into the Chern-Simons action and get the 
corresponding topological field theory! This generalized one-form, as we will argue soon, should appear from our M-theory analysis. Note also that the path integral description
should remain similar to \eqref{pathoi} as:
\bg\label{fath}
\int {\cal D}A_w {\rm exp}\left[{4\pi\over g^2_{YM}} \cdot {1\over w\sqrt{1+w^2}} \int d^3x ~\Omega(A_w)\right]  ~ =~ 
\int {\cal D}A ~{\rm exp}\left[{4\pi\over {g}^2} \int d^3x ~\Omega(A)\right] \nonumber\\ \nd
where we assume that the path integral is evaluated at the usual boundary $x_6 = 0$. Thus the $S_{cs}$ appearing in \eqref{pathoi} should then be identified with \eqref{csa} 
except with a scaled coupling ${g}^2$ defined as:
\bg\label{scacop}
{g} = g_{\rm YM} \sqrt{w\sqrt{1+w^2}}. \nd 
It is important to recall that, for our case, only 
the low energy dynamics is given by the Chern-Simons theory at the boundary. By low energy we mean the energy scale smaller than the inverse radius of
the $x_3$ direction. Using the language of \cite{wittenknots1} our five-dimensional Euclidean space is given by ${\bf V} \equiv {\bf W} \times {\bf S}^1 \times {\bf R}_+$, 
where ${\bf S}^1$ is parametrized by 
$x_3$ and ${\bf R}_+$ is parametrized by $x_6$. This ${\bf S}^1$ should not be confused with the ${\bf S}^1$ of \cite{wittenknots1} used in studying Khovanov Homology.   

There is one subtlety that we always kept under the rug: the physics at the other boundary associated with the existence of the second parallel NS5-brane. We had assumed that the 
second NS5-brane can be moved far away so that near $x_6 = 0$ we have the full ${\cal N} = 4$ supersymmetry. Although this description is roughly correct, this is not the full picture as
this $x_6$ circle will become the Taub-NUT circle in the dual type IIB framework. Therefore it is then necessary to determine the behavior of the following Chern-Simons form:
\bg\label{csform}
\Omega(A^{(1)}_w) - \Omega(A_w^{(2)}), \nd
where $A^{(1)}_w = A_w$ is gauge field we studied earlier for the boundary $x_6= 0$. As discussed by Witten in \cite{wittenknots1}, if we view $A_w^{(2)}$ to be trivial, then
the path integral can be represented as \eqref{fath}. We will elaborate on this later.

\subsection{Non-abelian extension and Chern-Simons theory \label{32}}

Having developed the basic strategy to study Chern-Simons theory from our brane construction, let us now analyze the geometry \eqref{sugra1}. The $x_6$ circle on the brane side will appear 
as a $S^1$, parametrized by $\psi$, 
fibered over the radial direction. The topology of this space is a $P^1$ and it will be assumed that 
the D5-branes wrap this two-cycle. The $\overline{\rm D5}$-branes are moved away along the Coulomb-branch. 

The fundamental form ${\cal J}$ can be computed from \eqref{chote} using 
standard procedure, and is given by:
\bg\label{funfor}
{\cal J} = \sqrt{F_1 F_2}~(d\psi + \cos~\theta_1 d\phi_1)\wedge dr + F_3 ~\sin~\theta_1 d\theta_1 \wedge d\phi_1 + F_4 ~dx_8 \wedge dx_9. \nd
One can plug $J \equiv e^{2\phi} {\cal J}$
in \eqref{sugra1} to determine the RR three-form flux using Hodge duality. Assuming non-zero background dilaton, this is given by the following expression:
\bg\label{jotat}
{\cal F}_3 = e_\psi \wedge \left(k_1~ e_{\theta_1} \wedge e_{\phi_1} + k_2 ~dx_8 \wedge dx_9\right), \nd
where due to the wedge structure there would be no $F_{48}= {\partial F_4 \over \partial x_8}$ or $F_{49}$ factors. This is reflected in the  
coefficients ($k_1, k_2$) which are given in terms of the warp factors of \eqref{sugra1} as:
\bg\label{k1k2d}
k_1 = {e^{2\phi} F_{4r} F_3 \sqrt{F_2}\over F_4 \sqrt{F_1}}, ~~~~~~~~~ k_2 = {e^{2\phi} \left(\sqrt{F_1 F_2} - F_{3r}\right) F_4 \sqrt{F_2}\over F_3 \sqrt{F_1}}, \nd
even if we keep $\phi$ as an arbitrary, but well defined, function of the internal coordinates.
Note that if the metric on the space \eqref{sugra1} is K\"ahler, then our formula would have yielded vanishing RR three-form flux. Thus when the D5-branes 
wrap the two-cycle of a 
blown-up Taub-NUT space, the metric has to be non-K\"ahler to preserve supersymmetry. 

\subsubsection{Generalized deformation and type IIB background \label{aluraj}}

It is now time to see what effect would the introduction of type IIA complex structure on the ($x_3, \psi$) torus have on our type IIB background.  This will {\it not} be a dipole 
deformation, rather it will be a non-commutative (NC) deformation of the wrapped five-brane theory, the non-commutativity only being along the ($x_3, \psi$) directions. 
Essentially the simplest non-commutative deformation amounts to switching on a NS B-field with both components along the brane.
The B-field for our case will have component $B_{3\psi}$ as we mentioned before, which of course
has the required property in the presence of a D5-brane along ($x_{0, 1, 2, 3}, r, \psi$). Since the warp factors are $r$ dependent, this B-field component will be a 
constant along the ($x_3, \psi$) directions but will depend on the radial coordinate $r$. 
This case is unlike anything that has been studied so far, although from an effective three-brane point of view this will be a 
dipole deformation. Thus this is not the standard NC deformation but we will continue calling it one. 

We now expect a field strength of the form $dB$. This field strength will then back-react on our original type IIB background \eqref{sugra1} and change the metric to the following:
\bg\label{metudip}
ds^2 & = & e^{-\phi}\left[-dt^2 + dx_1^2 + dx_2^2 + {dx_3^2\over \cos^2\theta + F_2 ~\sin^2\theta}\right] \\
& + & e^\phi\left[F_1 dr^2 + {F_2\left({d\psi\over \cos~\theta} + \cos~\theta_1 d\phi_1\right)^2\over 1 + F_2 ~\tan^2\theta} + F_3\left(d\theta_1^2 + \sin^2\theta_1 d\phi_1^2\right)
+ F_4(dx_8^2 + dx_9^2)\right],
\nonumber\nd
where $\theta$ is related to the NC deformation. It is easy to see how the Lorentz invariance along the compact $x_3$ direction is broken by the NC deformation. This is one reason (albeit not the most important one)
for the knots to be restricted along the Euclidean three directions. 

Coming now to the fluxes, it is interesting to note that the RR three-form flux remain mostly unchanged from the value quoted earlier in \eqref{jotat} with a small change in the 
$d\psi$ fibration structure:
\bg\label{cbet} 
{\cal F}_3 = \left(k_1~ e_{\theta_1} \wedge e_{\phi_1} + k_2 ~dx_8 \wedge dx_9\right) \wedge (d\psi + \cos~\theta~\cos~\theta_1 d\phi_1), \nd
where ($k_1, k_2$) are the same as in \eqref{k1k2d}. However now
along with the three-form RR flux, we also have a source of NS three-form flux which is responsible for generating the NC deformation
in our system. This extra source of NS flux is given by:
\bg\label{h3dipuli}
{\cal H}_3 ~=~ {F_{2r}~{\rm sin}~2\theta\over 2 \left(\cos^2~\theta + F_2~\sin^2~\theta\right)^2}~e_r \wedge \hat{e}_\psi \wedge e_3  
- {F_2~\sin~\theta\over \cos^2~\theta + F_2~\sin^2~\theta} e_{\theta_1} \wedge e_{\phi_1}\wedge e_3, \nonumber\\ \nd
from where we see how $\theta$ creates the necessary NC deformation and $\hat{e}_\psi = d\psi + \cos~\theta~\cos~\theta_1 d\phi_1$ denotes the new $\psi$ fibration. 
Finally, the NC deformation also effects the type IIB dilaton, changing it from $e^{-\phi}$ to:
\bg\label{iibdiles}
e^{\phi_B} = {e^{-\phi}\over \sqrt{\cos^2\theta + F_2~\sin^2\theta}}. \nd

\subsubsection{Comparision with an alternative deformation \label{mothdaug}}

Here we pause a bit to ask the question whether the NC deformation that we study here is consistent with the procedure adopted in 
\cite{langland, wittenknots1} to localize the knots along the 
Euclidean $x_{0, 1, 2}$ directions. In the original construction of \cite{wittenknots1} an axionic background $C_0$ is switched on to provide a theta-angle to the gauge theory 
on the D3-branes (with the NS5-brane boundary). In our language this will dualize to a RR B-field switched on the wrapped D5-branes on the Taub-NUT two 
cycle. 
Note that this RR B-field
is in addition to the RR B-field generated by the D5-brane sources. 
The question now is how will this additional  
RR B-field change the background solution. To analyze this let us assume, for simplicity, that the RR B-field for the wrapped D5-brane sources is given by:
\bg\label{roro}
{\cal C}_2 = b_{\theta_1\phi_1}~d\theta_1 \wedge d\phi_1 + b_{89}~ dx_8 \wedge dx_9, \nd
with the metric as in \eqref{sugra1} and \eqref{chote} and ($b_{\theta_1 \phi_1}, b_{89}$) are functions of all the internal coordinates except ($\psi, \phi_1$) to maintain the
necessary isometries. Note that if $b_{\theta_1\phi_1} = b_{\theta_1\phi_1}(\psi)$ and $b_{89} = b_{89}(\psi, \phi_1)$, then:
\bg\label{dc2}
d{\cal C}_2 = \left(m_1~e_{\theta_1} \wedge e_{\phi_1} + m_2~dx_8 \wedge dx_9\right)\wedge \left(d\psi + m_3~\cos~\theta_1~d\phi_1\right), \nd
which resembles \eqref{jotat} but is closed and doesn't have the required isometries. We have defined the coefficients in the following way:
\bg\label{chepta}
m_1 = {\rm cosec}~\theta_1~{db_{\theta_1\phi_1}\over d\psi},~~~ m_2 = {\partial b_{89}\over \partial \psi}, 
~~~m_3 = \sec~\theta_1~\left({\partial b_{89}\over \partial \psi}\right)^{-1}\left({\partial b_{89}\over \partial \phi_1}\right). \nd
Therefore 
to be consistent with the RR field strength \eqref{jotat}, we can define:
\bg\label{jota}
{\cal F}_3 \equiv d{\cal C}_2 + {\rm sources}, \nd
with $d{\cal C}_2$ derivable from \eqref{roro} that preserves the ($\psi, \phi_1$) isometries.
What happens when a component like ${\cal C}_{\psi 3}$ is switched on? To be consistent with \cite{wittenknots1} this component 
should be a constant along the fractional D3-branes' direction but could be a function of the internal coordinates. 

The answer can be derived following certain well defined, but tedious, steps. The backreacted metric changes from \eqref{sugra1} and \eqref{chote} to the following:
\bg\label{hardface}
ds^2 & = & e^{\varphi_B}\left[-dt^2 + dx_1^2 + dx_2^2 + {dx_3^2 \over \cos^2~\theta + F_2 e^{2\phi}~\sin^2~\theta}\right] \\
&+ & e^{2\phi+\varphi_B}\left[F_1 dr^2 + {F_2 \left({d\psi\over \cos~\theta} + \cos~\theta_1~d\phi_1\right)^2\over 1 + F_2 e^{2\phi}~\tan^2~\theta} + 
F_3\left(d\theta_1^2 + \sin^2\theta_1 d\phi_1^2\right) + F_4\left(dx_8^2 + dx_9^2\right)\right] \nonumber \nd  
where $\theta$ will be related to the additional RR B-field component switched on. Comparing \eqref{metudip} and \eqref{hardface} we see they are formally equivalent: 
the Lorentz invariance  
along spacetime directions is broken in exactly the same way for both the cases; and the $\psi$-fibration structure match. The metric differs slightly 
along the ($\psi, x_3$) 
directions, 
and the warp factors are little different from \eqref{metudip}, but the essential features are reproduced in an identical way. The dilaton $e^{\varphi_B}$ is again a 
slight variant of \eqref{iibdiles} and takes the form:
\bg\label{kolabaz} e^{-\varphi_B} = {e^{\phi} \over \sqrt{\cos^2~\theta + F_2 e^{2\phi}~\sin^2~\theta}}. \nd
The RR B-field changes from what we started off in \eqref{roro} because of the backreactions from the additional RR B-field piece. The precise functional form can 
also be worked out 
with some efforts, and the result is:
\bg\label{chapmaro} 
{\cal C}_2 & = & \left({F_2~e^{2\phi}\tan~\theta \over \cos^2~\theta + F_2 e^{2\phi}~\sin^2~\theta}\right) ~(d\psi + \cos~\theta~\cos~\theta_1 d\phi_1) \wedge dx_3 \nonumber\\
 &+&  b_{\theta_1\phi_1}~d\theta_1 \wedge d\phi_1   
+ b_{89}~ dx_8 \wedge dx_9 \nd
where we see that the first term is precisely the additional RR B-field piece that is switched on to restrict the knots along the Euclidean $x_{0, 1, 2}$ directions. 
In the limit 
$\theta \to 0$ we get back \eqref{chote} and \eqref{roro}. 

Thus, comparing \eqref{metudip} and \eqref{hardface}, we see that NC (or dipole) deformation and the deformation from switching on RR component of the B-field 
essentially amount to the
same thing: they both restrict the knots along the $x_{0, 1, 2}$ directions, albeit in the Euclidean version, by breaking the Lorentz invariance 
along the $x_{0, 1, 2}$ and the $x_3$ 
directions\footnote{This is a bit sloppy as, we shall see later, restricting the knots along a particular subspace is more subtle.}. 
However the RR deformation is sometime hard to implement in the supergravity language as it relies on the precise values of the ${\cal C}_2$ 
components in the presence of 
sources. But now with our above-mentioned equivalence we can use the NC deformations to compare the results as the supergravity analysis that we
perform here will only be sensitive to the 
metric deformations! Henceforth we will mostly use the dipole (or NC) deformations 
to study the knots, unless mentioned otherwise, and compare with the RR deformations whenever possible as we go along.

\subsubsection{M-theory uplift and harmonic forms \label{jahangir}}

It is now instructive to analyze the M-theory uplift of the deformed background \eqref{metudip}. 
Before that however we can see how the intermediate type IIA background looks like by T-dualizing
along a compact orthogonal direction. There are no global one-cycle, but locally we have polar coordinates ($\theta_1, \phi_1$). There is no isometry along $\theta_1$ direction, so that 
leaves us only with the $\phi_1$ circle. Local T-duality along $\phi_1$ will give us D6-branes, originally 
wrapped along the two-sphere generated by the collapsing $\psi$ coordinate on the radial $r$ 
direction, and the $\phi_1$ circle. This configuration is stabilized against collapse by background fluxes, which we will determine below. The background metric for the wrapped D6-branes
is now given by:
\bg\label{d6brane}
ds^2  & = &   e^{-\phi}\left[-dt^2 + dx_1^2 + dx_2^2 + {dx_3^2\over \cos^2\theta + F_2~\sin^2\theta} + {(d\phi_1 + \widetilde{F}_2 ~\tan~\theta ~\sec~\theta~\cos~\theta_1~dx_3)^2 
\over \widetilde{F}_2 ~\cos^2\theta_1 + F_3~\sin^2\theta_1}\right] \nonumber\\
&&~~~~~~~~~~ + e^\phi\left[F_1~dr^2 + F_3~d\theta_1^2 + F_4~ds^2_{89} + \left({\widetilde{F}_2 F_3~\sin^2\theta_1~\sec^2\theta \over \widetilde{F}_2 ~\cos^2\theta_1 + F_3~\sin^2\theta_1}
\right)d\psi^2\right], \nd 
where we note that the Lorentz invariance along ($x_3, \phi_1$) directions is broken so that the knots are still localized along the $x_{0, 1, 2}$ directions, albeit in the Euclidean
version. Note also the non-trivial fibration of the $\phi_1$ circle, which in turn appears in the background NS two-form ${\cal B}_2$ as:
\bg\label{B2} 
{\cal B}_2 = && {\widetilde{F}_2~\cos~\theta_1~\sec~\theta \over \widetilde{F}_2 ~\cos^2\theta_1 + F_3~\sin^2\theta_1} 
\left(d\phi_1 + \widetilde{F}_2 ~\tan~\theta ~\sec~\theta~\cos~\theta_1~dx_3\right) \wedge d\psi \nonumber\\
&& ~ + \widetilde{F}_2 ~\tan~\theta~\sec^2\theta ~d\psi \wedge dx_3, \nd
from where the field strength ${\cal H}_3 = d{\cal B}_2$ can be determined. We have also defined $\widetilde{F}_2$ as:
\bg\label{f2f2}
\widetilde{F}_2 = {F_2\over 1 + F_2~\tan^2\theta}. \nd 
To complete the story we will need the type IIA dilaton and the RR fluxes. The dilaton is well defined and takes the form:
\bg\label{2adil}
e^{\phi_A} = {e^{-3\phi/2}\over \sqrt{\left(\cos^2\theta + F_2~\sin^2\theta\right)\left(\widetilde{F}_2 ~\cos^2\theta_1 + F_3~\sin^2\theta_1\right)}}, \nd
provided the warp factors ($F_2, F_3$) are well defined everywhere. Otherwise strong coupling will set in at the following two isolated points:
\bg\label{isopoint}
\left(\theta_1 = 0, F_2(r_1) = 0\right), ~~~~~~~~~~\left(\theta_1 = {\pi\over 2}, F_3(r_2) = 0\right), \nd
irrespective of whether there is any NC deformation on the type IIB side. In general however, for arbitrary choice of the warp factors, strong coupling will set in when 
$e^{-\phi} \to \infty$. This is the regime where the dynamics will be captured by M-theory. 

To study the RR fluxes we first note that in the type IIB framework, the RR three-form flux ${\cal F}_3$ is not closed and gives rise to the following source equation:
\bg\label{f3source}
d{\cal F}_3 = - k_2~\cos~\theta~ e_{\theta_1} \wedge e_{\phi_1} \wedge dx_8 \wedge dx_9 +
\left(k_{1a} ~e_{\theta_1} \wedge e_{\phi_1} + k_{2a}~dx_8 \wedge dx_9 \right)\wedge e_a \wedge {\hat e}_\psi, \nonumber\\ \nd
with $a \equiv$ ($\theta_1, r, 8, 9$) and $e_a \equiv$ ($d\theta_1, dr, dx_8, dx_9$).
The first term is the expected source term for the D5-branes located at a point in ($\theta_1, \phi_1, x_8, x_9$) space. The other two terms signify the fact that we have fractional 
D5-branes. This is also reflected on the type IIA two-form ${\cal F}_2$ as:
\bg\label{f2vperido}
d{\cal F}_2 =&& -k_2 ~\cos~\theta~\sin~\theta_1 ~d{\theta_1} \wedge dx_8 \wedge dx_9 \nonumber\\ 
&& -\left(\sin~\theta_1~k_{1a}~d{\theta_1} \wedge d\psi - \cos~\theta~\cos~\theta_1 ~k_{2a}~dx_8 \wedge dx_9\right) \wedge e_a, \nd
with the first line denoting the expected charge of the wrapped D6-branes. 

At strong type IIA coupling, we can analyze the dynamics using M-theory. The M-theory metric takes the following form:
\bg\label{curamo}
ds^2 & = & H_1\left[-dt^2 + dx_1^2 + dx_2^2 + H_2 ~dx_3^2 + H_3(d\phi_1 + f_3 dx_3)^2 + e^{2\phi}\left(F_1 dr^2 + H_4 d\psi^2\right)\right] \nonumber\\
& + & e^{2\phi} H_1\left[F_3~ d\theta_1^2 + F_4\left(dx_8^2 + dx_9^2\right)\right] + e^{-2\phi} H_1^{-2}\left(dx_{11} + {\bf A}_{1m} dx^m\right)^2, \nd
where we see that the second line reflects the {\it warped} Taub-NUT nature of the background using gauge field ${\bf A}_1$ from the source \eqref{f2vperido}. The warp factors 
$H_i$ and $f_3$ describing the background are defined as:
\bg\label{wardaf} 
&& H_1 = \left(\cos^2\theta + F_2~\sin^2 \theta \right)^{1/3 }\left(\widetilde{F}_2 ~\cos^2\theta_1 + F_3~\sin^2 \theta_1 \right)^{1/3} \nonumber\\
&& H_2 = {1\over \cos^2\theta + F_2~\sin^2 \theta}, ~~~ H_3 = {1\over \widetilde{F}_2 ~\cos^2\theta_1 + F_3~\sin^2 \theta_1} \nonumber\\
&& f_3 = \widetilde{F}_2 ~\tan~\theta ~\sec~\theta~\cos~\theta_1, ~~~ H_4 = {\widetilde{F}_2 F_3~\sin^2\theta_1~\sec^2\theta \over \widetilde{F}_2 ~\cos^2\theta_1 + F_3~\sin^2\theta_1}.\nd 
To proceed further we will have to define the type IIA gauge field from \eqref{f2vperido} as:
\bg\label{nrd}
{\cal F}_2 = d{\bf A}_1 \equiv \alpha_1~dx_8 \wedge dx_9 + \alpha_2~ dx_8 \wedge d\theta_1 + \alpha_3~dx_9 \wedge d\theta_1, \nd
with the background one-form ${\bf A}_1$ appears in the fibration structure of \eqref{curamo} giving the Taub-NUT form and
$\alpha_i \equiv \alpha_i(\theta_1, x_8, x_9)$ as some generic function of ($\theta_1, x_8, x_9$) at some fixed value of $r$ satisfying the constraint:
\bg\label{consleng}
{\partial \alpha_1 \over \partial \theta_1}  + {\partial \alpha_3 \over \partial x_8} - {\partial \alpha_2 \over \partial x_9}  = 0. \nd
Since most of the warp factors are functions of $r$, except $F_4$ and $e^\phi$ which are respectively
generic functions of ($x_8, x_9$) and ($x_8, x_9, \theta_1$) also, at a given point if $r$, i.e at $r = r_0$, we have a warped
Taub-NUT space specified by the following metric derivable from \eqref{curamo}:
\bg\label{warTN} 
ds^2_{\rm TN} = {G}_1~d\theta_1^2 + {G}_2~dx_8^2 + {G}_3~dx_9^2 + {G}_4 (dx_{11} + {\bf A}_1)^2, \nd
with ${G}_i$ given by the following expressions in terms of the warp factors $H_1$ given in \eqref{wardaf}, $F_i$ in \eqref{chote}, and the dilaton $e^{2\phi}$:
\bg\label{givalues}
{G}_1 = e^{2\phi}~H_1~F_3, ~~~~~~~~{G}_2 = {G}_3 = e^{2\phi}~H_1~F_4, ~~~~~~~~ {G}_4 = {1\over e^{2\phi}~H_1^{2}}. \nd
To proceed further we will assume, for simplicity, the warped Taub-NUT space described above in \eqref{warTN} is a {\it single centered} Taub-NUT space. This is clearly not an accurate 
description of the system as the warped Taub-NUT space is derived originally from $N$ wrapped D4-branes in type IIB theory. We will rectify the situation soon by resorting 
back to the original description, but for the time being a single-centered Taub-NUT space will suffice to illustrate the picture without going into too much technicalities. Having
said this, we now use the fact that a single-centered Taub-NUT space allows a {\it unique} normalizable harmonic form $\omega \equiv d\zeta$ which is self-dual or anti-self-dual i.e 
$\omega = \pm \ast_4~ \omega$. For our case, this is given by:
\bg\label{chokranto}
\zeta = g(\theta_1, x_8, x_9) \left(dx_{11} + {\bf A}_1\right), \nd
with $g(\theta_1, x_8, x_9)$ satisfying the following set of differential equations at $r$ fixed at $r = r_0$:
\bg\label{gsaisfy}
&& {1\over g} {\partial g \over \partial \theta_1} = \pm \alpha_1 \sqrt{{G}_1 {G}_4\over {G}_2 {G}_3} = \pm
{\alpha_1\over e^{2\phi} ~F_4}\sqrt{F_3\over \left(\cos^2\theta + F_2~\sin^2 \theta \right)\left(\widetilde{F}_2 ~\cos^2\theta_1 + F_3~\sin^2 \theta_1 \right)}\nonumber\\
&&{1\over g} {\partial g \over \partial x_8} = \pm \alpha_3 \sqrt{{G}_2 {G}_4\over {G}_1 {G}_3} 
= \pm {\alpha_3 ~e^{-2\phi} \over \sqrt{F_3\left(\cos^2\theta + F_2~\sin^2 \theta \right)\left(\widetilde{F}_2 ~\cos^2\theta_1 + F_3~\sin^2 \theta_1 \right)}}\\
&& {1\over g} {\partial g \over \partial x_9} = \mp \alpha_2 \sqrt{{G}_3 {G}_4\over {G}_1 {G}_2}
= \mp {\alpha_2 ~e^{-2\phi} \over \sqrt{F_3\left(\cos^2\theta + F_2~\sin^2 \theta \right)\left(\widetilde{F}_2 ~\cos^2\theta_1 + F_3~\sin^2 \theta_1 \right)}}. \nonumber \nd
The above set of partial differential equations are in general hard to solve if we don't know the precise functional forms of the warp factors and dilaton involved in the 
expressions above. However comparing \eqref{nrd} and \eqref{f2vperido} we see that $\alpha_1$ appearing above in \eqref{gsaisfy} should at least be proportional to $k_2$ 
defined in \eqref{k1k2d}. In other words, we can write $\alpha_1$ at $r = r_0$ as:
\bg\label{dilchoice}
\alpha_1(r_0, x_8, x_9, \theta_1) = e^{2\phi} ~F_4 \alpha_a(\theta_1), \nd
where $F_4 = F_4(r_0, x_8, x_9)$ and $\phi = \phi(r_0, x_8, x_9, \theta_1)$. Note that, with the choice of ${\cal F}_2$ in \eqref{nrd} and the wedge structure, we can allow 
the above functional form for $\alpha_1$ without spoiling the constraint equation \eqref{consleng}. This way the first equation in \eqref{gsaisfy} is
easily satisfied. However for the other two equations in \eqref{gsaisfy}, one simple way to solve it would be to allow 
the dilaton as well as ($\alpha_2, \alpha_3$) to be functions of ($r, x_8, x_9, \theta_1$), such that the following conditions are met:
\bg\label{dilu}
{\alpha_3 e^{-2\phi} \over \sqrt{\widetilde{F}_2~\cos^2\theta_1 + F_3~\sin^2\theta_1}} \equiv \beta_3(x_8), ~~~~~~
{\alpha_2 e^{-2\phi} \over \sqrt{\widetilde{F}_2~\cos^2\theta_1 + F_3~\sin^2\theta_1}} \equiv \beta_2(x_9). \nd
Let us also assume that $g$ appearing
in \eqref{chokranto} can be expressed as:
\bg\label{choko}
g(\theta_1, x_8, x_9) \equiv g_1(\theta_1) g_2(x_8) g_3(x_9). \nd
Thus plugging in \eqref{choko} into the differential equations \eqref{gsaisfy} and assuming, without loss of generality, $F_2(r_0) = b_0^{-1}$, we get the following functional form for $g$:
\bg\label{gchakka}
g(x_8, x_9, \theta_1) = g_0~{\rm exp}\left[\pm c_0\left(\int_0^{\theta_1} {\alpha_a \over \sqrt{\sin^2 \theta_1 + {\cos^2 \theta_1\over b_0 + \tan~\theta}}}~d\theta_1 
+ \int_0^{x_8} {\beta_3}~dx_8 - \int_0^{x_9} {\beta_2}~dx_9\right)\right], 
\nonumber\\ \nd
where for appropriate sign we should get a normalizable harmonic form $\omega$ and we have defined $c_0$ as $c^{-1}_0 = \sqrt{F_3(r_0)\left(\cos^2~\theta + b_0^{-1}\sin^2~\theta\right)}$. The 
normalizability is defined wrt ($x_8, x_9$) directions as $\theta_1$ is a compact angular coordinate. Thus the $\theta_1$ dependence of \eqref{gchakka} is redundant and we
can simplify \eqref{gchakka} by eliminating the $\theta_1$ dependence in the gauge field \eqref{nrd} i.e eliminating the $\alpha_1$ factor in \eqref{nrd}. Under this 
assumption the integrand in:
\bg\label{bladam}
\int_{TN} \omega \wedge \omega = \int 2g\left(\alpha_3 ~{\partial g \over \partial x_8} - \alpha_2 ~{\partial g \over \partial x_9}\right)~ d\theta_1 \wedge dx_8 \wedge dx_9 
\wedge dx_{11}, \nd
will be independent of $\theta_1$ provided ($\alpha_2, \alpha_3$) can be made independent of $\theta_1$ leading to 
a constant factor for the $\theta_1$ integral\footnote{In general however one should get an additional piece of the 
form $2g \alpha_1~ {\partial g\over \partial \theta_1}$ in \eqref{bladam}.} as $g$ in \eqref{gchakka} will now be a function of ($x_8, x_9$). The $\theta_1$ independency of 
($\alpha_2, \alpha_3$) is still consistent with \eqref{dilu}, but the question is whether this will be true for \eqref{nrd}. To see this, recall that ${\cal F}_2$ in 
\eqref{nrd} needs to satisfy:
\bg\label{lmeye}
{\cal F}_2 = d{\bf A}_1 + \Delta, ~~~~~~ d\Delta = {\rm sources}, \nd
where ${\bf A}_1$ would still be written as \eqref{nrd}, but now with only ($\alpha_2, \alpha_3$), and
appear in the M-theory fibration structure in the metric \eqref{curamo}; and the sources correspond to the D6-brane sources. We can distribute the sources appropriately 
such that \eqref{nrd} has $\alpha_2 = \alpha_2(x_9)$ and $\alpha_3 = \alpha_3(x_8)$ satisfying all the background constraints. The dilaton, which is a function of 
($r, x_8, x_9, \theta_1$), can be chosen from the start in \eqref{sugra1} to be of the form:
\bg\label{dileshwar}
e^{2\phi} = {e^{2\phi_0}Q(r, x_8, x_9)\over \sqrt{\widetilde{F}_2~ \cos^2\theta_1 + F_3~\sin^2\theta_1}}, \nd
which can then be used to determine the RR three-form flux ${\cal F}_3$ in \eqref{jotat} and \eqref{k1k2d} with the functional form for $Q(r, x_8, x_9)$ determined using 
supersymmetry via torsion classes\footnote{An example of supersymmetric compactification will be described in details later using torsion classes. For our case using torsion
classes may lead us to consider a more generic case with $F_4(r, x_8, x_9, \theta_1)$ instead of our present choice of $F_4(r, x_8, x_9)$.}.      
The $\theta_1$ independence of \eqref{bladam} will be useful later. Finally,
this harmonic form can be used to express the M-theory $G$-flux ${\cal G}_4$ as:
\bg\label{chorM} 
{\cal G}_4 = \langle {\cal G}_4 \rangle + {\cal F} \wedge \omega, \nd
where ${\cal F} = d{\cal A}$ is the field strength of the $U(1)$ gauge field ${\cal A}$ and $\langle {\cal G}_4\rangle$ is the background $G$-flux whose explicit form can be easily 
determined 
form the type IIB three-form fluxes ${\cal F}_3$ and ${\cal H}_3$. This can be worked out by the diligent reader, 
therefore we will not discuss this and instead we will concentrate on the M-theory uplift of the RR deformed 
background \eqref{hardface}, \eqref{chapmaro} and \eqref{kolabaz}. The M-theory metric is given as:
\bg\label{curamo2} 
ds^2 & = & {\widetilde H}_1\left[-dt^2 + dx_1^2 + dx_2^2 + {\widetilde H}_2 ~dx_3^2 + {\widetilde H}_3(d\phi_1 + f_3 dx_3)^2 
+ e^{2\phi}\left(F_1 dr^2 + {\widetilde H}_4 d\psi^2\right)\right] \nonumber\\
& + & e^{2\phi} {\widetilde H}_1\left[F_3 ~d\theta_1^2 + F_4\left(dx_8^2 + dx_9^2\right)\right] 
+ e^{-2\phi} {\widetilde H}_1^{-2} {\widetilde H}_2^{-1}\left(dx_{11} + {\bf A}_{1m} dx^m\right)^2,  \nd
where we see that the metric is almost similar to the one presented earlier with NC deformation in \eqref{curamo}.
In fact the coefficients are also identical to 
the ones in \eqref{wardaf}, namely:
\bg\label{wardaf2}
&& {\widetilde H}_1 = \left(\cos^2\theta + F_2~e^{2\phi}~\sin^2 \theta \right)^{1/3 }\left(\widetilde{F}_2 ~\cos^2\theta_1 + F_3~\sin^2 \theta_1 \right)^{1/3} \nonumber\\
&& {\widetilde H}_2 = {1\over \cos^2\theta + e^{2\phi}~F_2~\sin^2 \theta}, ~~~~ {\widetilde F}_2 = {F_2 \over 1 + e^{2\phi}~F_2~ \tan^2 \theta} \nonumber\\
&& {\widetilde H}_3 = {\widetilde H}_1^{-3}, ~~~~ f_3 = 0, ~~~~
{\widetilde H}_4 = {\widetilde{F}_2 F_3~\sin^2\theta_1~\sec^2\theta \over \widetilde{F}_2 ~\cos^2\theta_1 + F_3~\sin^2\theta_1}, \nd 
with the differences being the vanishing of $f_3$, and the existence of certain extra factors of $e^{2\phi}$. Finally,  
the gauge field appearing in the fibration structure of \eqref{curamo2} can be read from the $b_{\theta_1\phi_1}$ and $b_{3\phi_1}$ components of \eqref{chapmaro} as:
\bg\label{nrd2}
{\bf A}_1 = b_{\theta_1\phi_1} ~d\theta_1 + b_{3\phi_1}~dx_3. \nd
The next step would be to evaluate the field strength for ${\bf A}_1$ and bring it in the form \eqref{nrd} with 
the triplet ($\alpha_1, \alpha_2, \alpha_3$) such that we can eliminate $\alpha_1$ and make $\alpha_2 = \alpha_2(x_9), \alpha_3 = \alpha_3(x_8)$ 
at $r = r_0$ and fixed $x_3$. All these can be accomplished by a simple choice of the components in \eqref{chapmaro} and \eqref{nrd2}: 
\bg\label{dileshwar2}
db_{\theta_1\phi_1} = \alpha_2(x_8, x_9)~dx_8 + \alpha_3(x_8, x_9)~dx_9, ~~ b_{3\phi_1} = \alpha_1(\theta_1), ~~ 
{\partial\alpha_2 \over \partial x_9} - {\partial\alpha_3 \over \partial x_8} = 0. \nd
This way $\alpha_1$ piece in \eqref{nrd} will be absent at fixed $x_3$ 
and the harmonic function will be independent of $\theta_1$ in exactly the way we wanted. The dilaton can now be chosen 
as \eqref{dileshwar} with $\widetilde{F}_2$ defined as in \eqref{wardaf2} to satisfy the remaining constraints.   
Thus with the intial metric choice \eqref{sugra1} and \eqref{chote},  
alongwith the dilaton \eqref{dileshwar}, supersymmetric configuration can be constructed once the RR fluxes satisfy the second relation in \eqref{sugra1}. This can be 
verified by working out the torsion classes, but we will not do so here. 
Instead, in the 
following section, we will determine the 
four-dimensional action that may appear from the 11-dimensional M-theory supergravity action.

\subsubsection{First step towards a gauge theory \label{fstep}} 
 
To derive a four-dimensional gauge theory from M-theory we will start by assuming Lorentz invariance along ($x_0, x_1, x_2, \psi$). Looking at \eqref{curamo}, we see that this
is possible only if the dilaton
and the warp factor $H_4$ combination $e^{2\phi} H_4$
is expressed as:
\bg\label{boundwc}
e^{2\phi} H_4 ~ = ~ 1 + {\cal U}_4, \nd
with small ${\cal U}_4$ at all points in ($r, x_8, x_9, \theta_1$). In this limit,
comparing this with \eqref{dilu} and \eqref{wardaf}, it means ($\alpha_2, \alpha_3$) are chosen as
\bg\label{virginia}
&&\alpha_2(r, x_9, \theta_1) = {\beta_2(x_9) \left(\widetilde{F}_2~\cos^2\theta_1 + F_3~\sin^2\theta_1\right)^{3/2} \over \widetilde{F}_2 F_3 \sec^2\theta~\sin^2\theta_1}
+ {\cal O}({\cal U}_4) \nonumber\\
&& \alpha_3(r, x_8, \theta_1) = {\beta_3(x_8) \left(\widetilde{F}_2~\cos^2\theta_1 + F_3~\sin^2\theta_1\right)^{3/2} \over \widetilde{F}_2 F_3 \sec^2\theta~\sin^2\theta_1}
+ {\cal O}({\cal U}_4), \nd
for all points in ($r, x_8, x_9, \theta_1$) space except at $\theta_1 = 0$. At $\theta_1 = 0$ one has to resort back to the definition \eqref{dilu}.

Therefore for small ${\cal U}_4$, the metric along ($x_0, x_1, x_2, \psi$) is essentially $H_1$, and consequently the M-theory action 
with $l_p \equiv 1$ will have the following four-dimensional reduction:
\bg\label{mtheorya}
\int d^{11} x~{\cal G}_4 \wedge \ast_{\tiny{11}} {\cal G}_4 + \int {\cal C}_3 \wedge {\cal G}_4 \wedge {\cal G}_4 
= c_1 \int d^4 x~{\cal F} \wedge \ast_{\small 4} {\cal F} + c_2 \int {\cal F} \wedge {\cal F}, \nd
where we have ignored for the time being the seven-dimensional nature of the $U(1)$ theory by compactifying down to four-dimensions over the three-cycle $\Sigma_3$ parametrized by 
$\phi_1$ in \eqref{curamo} and the two-sphere determined by the degenerating $x_3$ fibration over the radial coordinate $r$. The coefficients 
$c_i$ appearing in \eqref{mtheorya} are given as:
\bg\label{c1c2c1c2}
c_1 = \int_{\Sigma_3} d^3\sigma \sqrt{g_3} \int_{\rm TN} \omega \wedge \ast_{\tiny{\rm TN}} \omega,
 ~~~~~~ c_2 = \int_{\Sigma_3} \langle {\cal C}_3 \rangle ~ \int_{\rm TN} \omega \wedge \omega, \nd
with $c_1$ giving us the $U(1)$ YM coupling whose value can be read off from $\omega$, using \eqref{gchakka},
and the internal metric along ($\phi_1, r, \psi$), using \eqref{curamo}; and $c_2$ giving us the $\Theta$ angle. Note also that $c_1$ and $c_2$ are related by:
\bg\label{c1c2rel}
c_2 = {\int_{\Sigma_3} \langle {\cal C}_3 \rangle\over \int_{\Sigma_3} d^3\sigma \sqrt{g_3}} ~c_1, \nd
which should be reminiscent of the relation between $\Theta$ and ${1\over g^2_{\rm YM}}$ discussed in \cite{wittenknots1}. To see the precise connection, let us go back to the original 
orientation of the D3-branes on the NS5-brane in {\bf Table \ref{wittenbranes}}. The D3-branes are oriented along $x_0, x_1, x_2$ 
and $\psi$ directions, and therefore since the M-theory 
Taub-NUT is oriented along ($x_8, x_9, \theta_1, x_{11}$), we are left with the three-cycle $\Sigma_3$ along ($x_3, r, \phi_1$) directions with metric:
\bg\label{metmuta}
g_3 = \begin{pmatrix} H_1H_2 + H_1H_3 f_3^2 & ~~~H_1 H_3 f_3  & 0 \\ {} & {} & {} \\ H_1 H_3 f_3 & H_1 H_3 & 0 \\ {} & {} & {} \\ 0 & 0 & H_1 e^{2\phi} F_1 \end{pmatrix}, \nd
which could be read from the metric \eqref{curamo}, and $H_i, f_3$ are defined in \eqref{wardaf} above. The above metric leads to the following value of the integral:
\bg\label{intig}
{v}_3 \equiv \int_{\Sigma_3} d^3\sigma \sqrt{g_3} = 2\pi R_3 \int_0^\infty dr~ e^\phi\sqrt{F_1}, \nd
at a fixed value for ($\theta_1, x_8, x_9$). In deriving \eqref{intig}, we have assumed $R_3$ to be the radius of the $x_3$ circle. The above integral is 
a well defined function because the dilaton is well defined at the two boundaries of $r$ and $F_1$ vanishes at the origin and goes to identity at $r \to \infty$. Thus \eqref{intig} 
will lead to some constant value at any given point of ($\theta_1, x_8, x_9$) space.

Coming to the M-theory three-form ${\cal C}_3$, we now require the component (${\cal C}_3$)$_{3r\phi_1}$ to compute
$c_2$ in \eqref{c1c2c1c2}. A naive computation from T-duality will yield zero value for this component\footnote{A more accurate statement is the following. Existence of the 
RR two-form ${\cal C}_2$ in \eqref{chapmaro} implies the three-form field strength components $({\cal F}_3)_{3\psi r}$ and $({\cal F}_3)_{3\phi_1 r}$, both of which under 
specific gauge transformations may yield a two-form field $({\cal C}_2)_{3r}$. However consistency would require this to be functions of ($\psi, \phi_1$) which, in our T-dual 
framework, would be impossible as we require the field components to be independent of the T-dual coordinates ($\psi, x_3, \phi_1$).}. 
However the scenario is subtle because of the fractional brane nature of the 
type IIB three-branes. The D5-${\overline{\rm D5}}$ nature of the fractional D3-branes imply that we need a small value of
NS B-field switched on along ($x_3, r$) directions to take care of the tachyons
\cite{DM2}. Consistency then requires us to have at least a RR two-form along ($x_3, r$) directions in type IIB side. This will dualize to the required ${\cal C}_3$ component
(${\cal C}_3$)$_{3r\phi_1}$ which, without loss of generalities, will be assumed to take the following form: 
\bg\label{chotas}
{\cal C}_3 ~\equiv ~ {N_{r}~\sin~2\theta~\cos~\theta~p(\theta_1, \theta)~q(\theta) \over 2(\cos^2~\theta + N~\sin^2~\theta)^2} 
~dr \wedge dx_3 \wedge d\phi_1, \nd
where $N \equiv N(r, \theta)$ such that $N$ remains arbitrary small for all $r$ and only at $r\to \infty$, $N \to 1$; and 
$p(\theta_1, \theta)$ and $q(\theta)$ are well-defined periodic functions of $\theta_1$ and $\theta$ respectively. This way EOMs will not be affected by the introduction of these
field components. 
Using this, the value of the integral in \eqref{c1c2c1c2} for $c_2$
is given by:
\bg\label{valuec2} 
\int_{\Sigma_3} \langle {\cal C}_3 \rangle = {2\over \pi}\int_0^{\pi/2} d\theta_1~ p(\theta_1, \theta) \int_{\Sigma_3}dr \wedge dx_3 \wedge d\phi_1  
{N_{r}~\sin~2\theta~\cos~\theta~q(\theta) \over (\cos^2~\theta + N~\sin^2~\theta)^2} = 2R_3 q(\theta)~\sin~\theta, \nonumber\\ \nd 
where we have absorbed the value of the $\theta_1$ integral in the definition of $R_3$ and $q$. Now
combining \eqref{intig} and \eqref{valuec2}, and making $q(\theta) = 1$ for simplicity, we find that $c_1$ and $c_2$ are related by:
\bg\label{chotsard}
c_2 = \sin~\theta~c_1 = \left({2\tan~{\theta\over 2}\over 1 + \tan^2{\theta\over 2}}\right) c_1 = \left({2a\over 1 + a^2}\right)c_1, \nd
where we have normalized the integral in \eqref{intig} to $2R_3$ to avoid some clutter. Furthermore, in \eqref{chotsard}, we have defined $a \equiv \tan~{\theta\over 2}$. It is interesting
that if we identify this $a$ with the same $a$ used in eq. (2.7) of \cite{wittenknots1}, we can compare \eqref{chotsard} 
with eq. (2.14) of \cite{wittenknots1} provided we define ($c_1, c_2$) as\footnote{The results don't match exactly as the 
above comparison is naive. The precise connection between $a$ of 
\cite{wittenknots1} and the supergravity parameters will be outlined later.}:
\bg\label{c12def}
c_1 \equiv {4\pi\over g^2_{\rm YM}}, ~~~~~~~~~ c_2 \equiv {\Theta\over 2\pi}. \nd 
What happens for the M-theory uplift \eqref{curamo2} for the type IIB background \eqref{hardface}, \eqref{chapmaro} and \eqref{kolabaz}? It is easy to see that the  
component of the ${\cal C}_3$ \eqref{chotas} remains unchanged, but $v_3$ defined in \eqref{intig} changes to the following:
\bg\label{intig2} 
v_3 &=& 2\pi R_3 ~\sec~\theta\int_0^\infty dr~e^\phi\sqrt{F_1\over 1+ e^{2\phi}~F_2~\tan^2\theta} \nonumber\\
&=& 2\pi R_3 ~\sec~\theta\int_0^\infty dr~e^\phi \sqrt{F_1} \left(1- {1\over 2} e^{2\phi}~F_2~\tan^2\theta + ...\right), \nd
at a fixed value of ($\theta_1, x_8, x_9$) space. The last equality is assuming small RR deformation parameter $\theta$, otherwise one will need the explicit form for the warp factor
$F_i$ and the dilaton $e^\phi$ to evaluate the three-volume $v_3$. Now, because of the change in the volume $v_3$, the relation between $c_2$ and $c_1$ becomes:
\bg\label{chotsard2}
c_2~ = ~{1\over 2}~ \sin~2\theta~c_1 ~ = ~ \left({\tan~{\theta} \over 1 + \tan^2 {\theta}}\right)c_1 ~ \equiv ~ {2a\over 1 + a^2}\left({1-a^2\over 1+a^2}\right)c_1, \nd
with corrections coming from the ${\cal O}(\theta^2)$ terms in \eqref{intig2}. This relation can be compared with \eqref{chotsard} and also with \cite{wittenknots1} where somewhat 
similar discussion appears from gauge theory point of view.  

\subsubsection{Including the effects of ${\cal U}_4$ \label{inclu}}

The above identification \eqref{chotsard} or \eqref{chotsard2}
is encouraging and points to the consistency of the picture from M-theory point of view. However generically ${\cal U}_4$ is never small everywhere, and therefore Lorentz invariance
cannot always be restored along the $\psi$ direction. In such a scenario we expect the gauge theory to have the following form:
\bg\label{gaugut}
{c_{11}. c_1\over v_3}\int d^4x \sum_{a, b} {\cal F}_{ab}{\cal F}^{ab} + {c_{12}. c_1\over v_3}\int d^4x \sum_{a} {\cal F}_{a\psi}{\cal F}^{a\psi}, \nd
where $a, b = 0, 1, 2$ and ($c_{11}, c_{12}$) will eventually be related to the YM coupling \eqref{c12def} after proper redefinitions of the gauge fields. We will do this later. 
However subtlety arises when we try to define these coefficients in terms of the background data because the components of the metric along directions orthogonal to the Taub-NUT
space as well as the dilaton do depend on the Taub-NUT coordinates ($\theta_1, x_8, x_9$). For example the first coefficient in \eqref{gaugut} can be expressed as:
\bg\label{c11c1v3}
{c_{11} c_1\over v_3} \equiv  4R_3 e^{2\phi_0}\sec~\theta \int d^4 \zeta~
\sqrt{F_1\widetilde{F}_2\over \widetilde{F}_2 - F_3} ~{\rm tan}^{-1}\left(\sqrt{\widetilde{F}_2 - F_3 \over F_3}\right) 
gQ \left(\alpha_3 ~{\partial g \over \partial x_8} - \alpha_2 ~{\partial g \over \partial x_9}\right), \nonumber\\ \nd
where $g = g(r, x_8, x_9)$ instead of $g(r_0, x_8, x_9, \theta_1)$ as in \eqref{gchakka}
and $Q = Q(r, x_8, x_9)$. We have defined $d^4\zeta$ as the integral over:
\bg\label{4int}
\int d^4\zeta \equiv \int_0^\infty dr \int_0^{R_8} dx_8 \int_0^{R_9} dx_9 \int_0^{R_{11}}dx_{11} \nd
with 
$R_n$ being the radius of the $n$-th direction, which could be compact or non-compact depending on the configuration. For example we expect $R_8$ or $R_9$ to be non-compact.   

Looking at \eqref{c11c1v3}, we see that there is a {\it mixing} between the Taub-NUT and the non Taub-NUT coordinates. However we can simplify the resulting formula 
by making two small assumptions: (a) we can take the constant leading term for the dilaton, namely $e^{2\phi_0}$, and (b) fix the Taub-NUT space at $r = r_0$. The latter
would mean that the $dr$ integral could be restricted {\it only} to the space orthogonal to our Taub-NUT configuration, whereas the former would imply that we do not have
to worry about the $dx_8$ and $dx_9$ integrals\footnote{If we define $Q$ appearing in \eqref{dileshwar} as $Q \equiv Q_1 \sqrt{F_3}$, then we see that the dilaton
varies between $e^{2\phi_0}{Q_1}$ and $e^{2\phi_0}{Q_1} \sqrt{2F_3\over \widetilde{F}_2 + F_3}$. For regimes where $\widetilde{F}_2 \to F_3$, the 
latter is simply ${e^{2\phi_0}{Q_1}}$. Therefore the choice of constant dilaton means that ${Q_1}$ do not vary significantly 
over the ($r, x_8, x_9$) space. This way issues related to strong coupling could be avoided.}. 
Note also that the average over $\theta_1$ coordinate that we perform here is consistent with \eqref{bladam} because one 
may assume as though the $d\theta_1$ integral is being {\it transferred} to the integrand over the space orthogonal to the Taub-NUT space. This is where our work on making the 
integrand in \eqref{bladam} independent of $\theta_1$ will pay off. 
Of course as we saw, a general analysis is not too hard to perform, but this is not necessary to elucidate the underlying physics.  

Therefore taking the two assumptions into account, the first coefficient $c_{11}$ is easy to work out, and is given by the following integral:
\bg\label{c11theta}
c_{11}(\theta) = {R_3~\sec~\theta}\int_0^\infty dr~e^{2\phi_0}\sqrt{F_1 \widetilde{F}_2 F_3\over \widetilde{F}_2 - F_3}
~{\rm ln}\left|{\sqrt{\widetilde{F}_2} +\sqrt{\widetilde{F}_2 - F_3}\over \sqrt{\widetilde{F}_2} -\sqrt{\widetilde{F}_2 - F_3}}\right|, \nd
where we have only taken the constant leading term for the dilaton. 
Additionally, the combination $\widetilde{F}_2 - F_3$ should be viewed as 
$\left|\widetilde{F}_2 - F_3\right|$ 
so that this will always be real. This means for our purpose we will always be choosing the metric ansatze \eqref{chote} with $\widetilde{F}_2 \ge F_3$ at all points in 
$r$, the radial coordinate\footnote{Note that the $r$ behavior of the warp-factors $F_i$ typically goes as $F_k = \sum_n \alpha_{kn} (r/r_o)^n$ where $r_o$ is the scale and 
the sum over $n$ can be from all positive and negative numbers depending on the model. This means, to maintain $\widetilde{F}_2 \ge F_3$ at {\it all} points in $r$, 
we will have to choose the 
functional behavior differently for $r < r_o$ and for $r > r_o$. Again, this subtlety is only because we restricted ourselves to a concrete example with 
$\widetilde{F}_2 \ge F_3$. We could take {\it generic} ($\widetilde{F}_2, F_3$) for our case, but then the analysis becomes a bit cumbersome although could nevertheless be 
performed. However since in the latter case we don't gain any new physics, we restrict ourselves with the former choice.}.  
This choice, although not generic, should suffice at the level of a concrete example. An alternative choice with $F_3 \ge \widetilde{F}_2$, at
all $r$, leads to:
\bg\label{meyefac}
c_{11}(\theta) = {2 R_3~\sec~\theta}\int_0^\infty dr
~e^{2\phi_0} \sqrt{F_1 \widetilde{F}_2 F_3\over F_3 - \widetilde{F}_2}~
{\rm tan}^{-1}\left(\sqrt{F_3 - \widetilde{F}_2 \over \widetilde{F}_2}\right), \nd
and could be considered instead of \eqref{c11theta} but we will only consider the former case namely $\widetilde{F}_2 \ge F_3$. 

The above integral \eqref{c11theta} is just a number and is well defined for all values of the warp factors even in the limits 
$F_3 = 0 = \widetilde{F}_2$ and $\widetilde{F}_2 = F_3$. On the other 
hand $c_{12}$ is more non-trivial to represent in integral form because $c_{12}$ depends on $H_4^{-1}$ given in \eqref{wardaf}, which unfortunately is not well defined at 
$\theta_1 = 0$. To deal with this we will express the integral form for $c_{12}$ in the following way:
\bg\label{krisis}
c_{12}(\theta) &=& 2R_3~\cos~\theta \int_0^\infty dr\sqrt{F_1(\widetilde{F}_2 - F_3)\over \widetilde{F}_2 F_3} 
\int_{-1}^1 dz~{\sqrt{z^2 + a^2}\over b^2-z^2}\\
& = & 2R_3~\cos~\theta \int_0^\infty dr~b_2
\left[b_3
~{\rm tanh}^{-1}\left({1\over b}\sqrt{F_3 + b^2(\widetilde{F}_2 - F_3)\over \widetilde{F}_2}\right)
+ {\rm ln}\left|{\sqrt{\widetilde{F}_2} - \sqrt{\widetilde{F}_2 - F_3}\over \sqrt{\widetilde{F}_2} + \sqrt{\widetilde{F}_2 - F_3}}\right|\right]\nonumber \nd
such that $b$ is the regularization factor introduced to avoid the $z = \pm 1$ singularities. We have also defined ($a, b_2, b_3$) in the following way:
\bg\label{aa2a3}
a = \sqrt{F_3 \over \widetilde{F}_2 - F_3}, ~~~~ b_2 = \sqrt{F_1(\widetilde{F}_2 - F_3)\over 4\widetilde{F}_2 F_3}, ~~~~
b_3 = {2\over b}\sqrt{F_3 + b^2(\widetilde{F}_2 - F_3)\over \widetilde{F}_2 - F_3}. \nd
Let us now study the limiting behavior of the integrand in \eqref{krisis}. In the limit $F_3$ vanishes for some point(s) in $r$, the integrand generically blows up
but we can arrange it such that this vanishes as:
\bg\label{integr1}
{}^{\rm lim}_{F_3 \to 0} ~{1\over \sqrt{F_3}}\left\{{\rm tanh}^{-1}\left[\sqrt{1+\left({1-b^2\over b^2}\right){F_3\over \widetilde{F}_2}}\right] 
+ {\rm ln}\left(\sqrt{\widetilde{F}_2} - \sqrt{\widetilde{F}_2-F_3}\right)\right\}~\to~ 0. \nonumber\\ \nd 
On the other hand, when $\widetilde{F}_2 \to F_3$ for certain value(s) of $r$, the integrand in \eqref{krisis} approaches the following limit:
\bg\label{inte2}
{\rm tanh}^{-1} \left({1\over b}\right), \nd
which blows up in the limit $b = 1$. But since $b$ is never identity $-$ the original integral \eqref{krisis} being not well-defined for $b = 1$ $-$ the value 
in \eqref{inte2} can be large but 
not infinite. However subtlety arises when $\widetilde{F}_2 \to 0$, because in this limit we expect $F_3$ to also vanish otherwise $\widetilde{F}_2 \ge F_3$ cannot be maintained. 
Furthermore, $F_3$ has to go to zero {\it faster} than $\widetilde{F}_2$. This then brings us to the case \eqref{integr1} 
studied above, and we can impose $\widetilde{F}_2 \to 0$ 
there. 
This means the integrand in \eqref{krisis} will be well defined at all points in ($r, x_8, x_9, \theta_1$) space 
even where both ($\widetilde{F}_2, F_3$) vanish, 
and the large value of \eqref{inte2}
can be absorbed in the definition of ${\cal A}_\psi$ in \eqref{gaugut}. 

Again, we should ask as to what happens once we consider the M-theory uplift \eqref{curamo2}. The coefficients in the metric \eqref{curamo2} are slightly different from the 
ones in \eqref{curamo} so we expect ($c_{11}, c_{12}$) to change a bit. Indeed that's what happens once we evaluate the precise forms for ($c_{11}, c_{12}$). The first coefficient 
$c_{11}$ is now given by:
\bg\label{c11theta2}
c_{11}(\theta) = {R_3~\sec~\theta}\int_0^\infty dr~e^{2\phi_0}\sqrt{F_1 \widetilde{F}_2 F_3\over b_4\left(\widetilde{F}_2 - F_3\right)}~
{\rm ln}\left|{\sqrt{\widetilde{F}_2} +\sqrt{\widetilde{F}_2 - F_3}\over \sqrt{\widetilde{F}_2} -\sqrt{\widetilde{F}_2 - F_3}}\right|, \nd
where $\widetilde{F}_2$ is now defined as in \eqref{wardaf2} with an extra factor of the dilaton $e^{2\phi}$. Unless mentioned otherwise, we will continue using the same 
notation for $\widetilde{F}_2$ as in \eqref{f2f2} to avoid clutter. It should be clear from the context which one is meant. As expected, \eqref{c11theta2}
is exactly as in \eqref{c11theta} except for the additional factor of $b_4$ defined as: 
\bg\label{bfour}
b_4 \equiv \cos^2\theta + e^{2\phi} ~F_2 ~\sin^2\theta \nd
in the $dr$ integral. Similarly, the $c_{12}$ coefficient 
is given by an expression of the form \eqref{krisis} except $b_2$ in \eqref{aa2a3} changes to ${b_2\over \sqrt{b_4}}$, i.e:
\bg\label{b2change} b_2 ~ \to ~ {b_2\over \sqrt{\cos^2\theta + e^{2\phi}~ F_2 ~\sin^2\theta}}. \nd
This concludes our discussion of the gauge theory from M-theory and we see that the components of the gauge fields, namely (${\cal A}_0, {\cal A}_1, {\cal A}_2$) can formally be 
distinguised from ${\cal A}_\psi$ because of their structure of the kinetic terms in \eqref{gaugut}. However
the picture that we developed so far is related to 
$U(1)$ theory, so the natural question is to ask whether we can extend the story to include non-abelian gauge theories. This is in general a hard question because the G-flux 
in the supergravity limit is always a $U(1)$ field. However if we are able to include M2-brane states then we should be able to study the non-abelian version of \eqref{mtheorya}. In the 
following we will analyze this picture in some details.

\subsubsection{Non-abelian enhancement and M2-branes \label{NAE}}

To proceed we will have to first find the two-cycles in the space given by the metric ansatze \eqref{warTN}, where we now take our background to be a  warped multi-centered 
Taub-NUT space. The idea is to wrap a M2-brane on each of the the two-cycles such that in the limit of vanishing size of the cycles, the M2-branes become tensionless giving rise
to enhanced gauge symmetry. This idea has been explored earlier in \cite{imamura} so we will be brief. Note that for this to happen, 
we will start by assuming that the circle parametrized by the coordinate $x_{11}$ {\it shrinks} to zero size at various points on the geodesic line in the 
($\theta_1, x_8, x_9$) space. This way we will have multiple two-cycles, giving rise to a warped multi-centered Taub-NUT space.
In other words, we can rewrite the warped Taub-NUT metric \eqref{warTN} in the following suggestive way:
\bg\label{kolabaaz}
ds^2_{\rm TN} = U^{-1} \left(dx_{11} + {\bf A}_1\right)^2 + U d {\vec{\bf x}}^2, \nd
where we have defined the variables appearing above, using the definitions in \eqref{givalues}, in the following way:
\bg\label{chotapot}
&& d {\vec{\bf x}}^2 = {1\over H_1(\theta_1)}\left[F_3 d\theta_1^2 + F_4(x_8, x_9)\left(dx_8^2 + dx_9^2\right)\right] \\
&&U \equiv e^{2\phi}H_1^2 = e^{2\phi} \left(\cos^2 \theta + {\sin^2 \theta\over b_0}\right)^{2/3}\left(F_3~\sin^2 \theta_1 
+ {\cos^2 \theta_1\over b_0 + \tan~\theta}\right)^{2/3}, 
\nonumber \nd
with $b_0 = F^{-1}_2(r_0)$ as before. Now since both $b_0$ and $F_3(r_0)$ are 
 ${\cal O}(1)$ numbers, and just for analytical simplicity if we take a small NC deformation $\theta$, then both $U$ and $H_1$ 
will be independent of ($\theta, \theta_1$) and $U$ can be expressed as:
\bg\label{uexp}
U(x_8, x_9) = 1 + \sum_{m = 1}^\infty \sum_{k = 1}^N {c_{mk}\over \vert l_{89} - l_k\vert^m}, \nd
stemming entirely from the dilaton $e^{2\phi}$, 
where $c_{mk}$ are certain constants associated with the $N$-centered warped Taub-NUT space and $l_{89}$ is the geodesic length in ($x_8, x_9$) space.   

We can simplify the subsequent analysis a bit more if we assume that the warp factor $F_4$ is only a function of $x_8$ at $r = r_0$ and is independent of $x_9$. Of course the generic case 
can also be done, but since this will not change any of the physics that we want to discuss here, we will resort to the simplest treatment here. Thus the mass of the wrapped M2-brane
between ($l_k, l_{k+1}$) two-cycle is then given by:
\bg\label{massM2}
m_{k, k+1} \equiv T_M S_{k, k+1} = \beta T_M R_{11} \int_{l_k}^{l_{k+1}} dx_8 \sqrt{F_4}, \nd
where $T_M$ is the tension of the membrane, $S_{k, k+1}$ is the area of the two-cycle between points ($l_k, l_{k+1}$) 
and $\beta$ is a constant that could be extracted from the coefficients $c_{mk}$ in \eqref{uexp} that is needed to avoid any conical singularities
in the system. The next step is easy and has been discussed in details in \cite{imamura}. The intersection matrices of the two cycle satisfy the following algebra:
\bg\label{intmat}
\left[S_{k, k+1}\right] ~o~ \left[S_{l, l+1}\right] = \Bigg\{ \begin{matrix} 2\delta_{kl}\\ -\delta_{l, k-1} \end{matrix} \nd 
which is exactly the Cartan matrix of $A_{N-1}$ algebra! Thus the enhanced gauge symmetry of the system leads to an $SU(N)$ group with the Cartan coming from the decomposition of the 
localised G-flux as \eqref{chorM} but now with:
\bg\label{cutota}
{\cal G}_4 = \langle {\cal G}_4 \rangle + \sum_{i = 1}^N {\cal F}_i \wedge \omega_i, \nd
with orthonormal harmonic forms $\omega_i$ associated the $i$-th two-cycle. All these harmonic forms can be easily derived from \eqref{gchakka} by restricting the ($x_8, x_9$) integrals 
over the two-cycles appropriately. Thus after the dust settles, and ignoring the seven-dimensional origin of the system for the time being, we expect the following non-abelian 
enhancement of the $U(1)$ theory discussed earlier in \eqref{mtheorya} for the D3-branes oriented as in {\bf Table \ref{wittenbranes}}:
\bg\label{chelara}
S_{YM} = c_1 \int d^4 x \left(\gamma_1 \sum_{a<b}~{\rm Tr}~{\cal F}_{ab}{\cal F}^{ab} + \gamma_2 \sum_{a}~{\rm Tr}~{\cal F}_{a\psi}{\cal F}^{a\psi}\right) + c_2 \int {\rm Tr}~ {\cal F} \wedge {\cal F}, \nonumber\\ \nd
with the trace in the adjoint representation of $SU(N)$ and ($c_1, c_2$) defined as in \eqref{c12def} and related by \eqref{c1c2rel} and \eqref{chotsard} (the correct relation will be provided
later). 
Note that we have inserted 
($\gamma_1, \gamma_2$) for the coefficients of the ${\cal F}_{ab}$ and ${\cal F}_{a\psi}$ terms respectively. We expect
$\gamma_1$ to be related to \eqref{c11theta} and \eqref{intig}; and $\gamma_2$ to be related to \eqref{krisis} and \eqref{intig} as in the $U(1)$ case described in \eqref{gaugut}. A
proof of this is hard, and in the following we will try to give some justification of this. 

So far we saw that the localized G-fluxes at the Taub-NUT singularities provide the Cartan of the gauge group and the wrapped M2-brane states provide the necessary charged states
to allow for the non-abelian enhancement. In fact the M2-brane states provide a two-dimensional sigma model description at weak string coupling that takes the following form:
\bg\label{sigmam}
S_2& =& \int d^2\sigma \sqrt{h}~h^{\alpha\beta}\left[f_1(\Phi_1, \Phi_2, \Phi_3) \partial_\alpha \lambda_1^\top \partial_\beta \lambda_1 + f_2(\Phi_1, \Phi_2, \Phi_3)  
\partial_\alpha \lambda_2 \partial_\beta \lambda_2\right] \\
 &+& \int d^2\sigma \sqrt{h}~ h^{\alpha\beta}\left[\sum_{k=3}^5 f_k(\Phi_1, \Phi_2, \Phi_3) \partial_\alpha \Phi_k^\top \partial_\beta \Phi_k + 
f_6(\Phi_1, \Phi_2, \Phi_3) \partial_\alpha \lambda_3^\top \partial_\beta \lambda_3 + ... \right], \nonumber \nd
where $h_{\alpha\beta}$ is the world-sheet metric, $f_l(\Phi_k)$ are the couplings, the dotted terms denote couplings to NS and RR fields 
including the fermions, and the various sigma model fields are defined as:
\bg\label{sigfield}
&& \lambda_1 = \begin{pmatrix} x_0 \\ x_1 \\ x_2 \end{pmatrix}, ~~~~~ \lambda_2 = \psi, ~~~~~ \lambda_3 = \begin{pmatrix} x_3 \\ \phi_1 \end{pmatrix}, \nonumber\\
&& \Phi_1 = \begin{pmatrix} x_8 \\ x_9 \end{pmatrix}, ~~~~~~ \Phi_2 = r, ~~~~~~ \Phi_3 = \theta_1. \nd
Due to the non-trivial interaction terms in \eqref{sigmam}, a detailed study of the spectra is hard. However we make a few observations. First, the couplings are not arbitary and 
can be worked out from \eqref{d6brane}. We will specifically concentrate on the first two interactions in \eqref{sigmam} as their fluctuations will be related to the 
four-dimensional gauge interactions. The ($f_1, f_2$) terms are given by:
\bg\label{camesee}
f_1(\Phi_1, \Phi_2, \Phi_3) \equiv f_1(r, x_8, x_9, \theta_1) = e^\phi, ~~~ 
f_2(\Phi_1, \Phi_2, \Phi_3) \equiv f_2(r, x_8, x_9, \theta_1) = {e^\phi \over 1 - {\cal U}_4}, \nonumber\\ \nd
where ${\cal U}_4$ is the same function that appeared in \eqref{boundwc} and entered in the derivation of the couplings ($c_{11}, c_{12}$) in \eqref{c11theta} and \eqref{krisis} 
respectively. Thus plugging \eqref{camesee} in the first two terms of \eqref{sigmam} 
leads us to the following 2d interacting lagrangian for the fields ($\lambda_1, \lambda_2$):
\bg\label{interlag}
{\cal L} = \partial_\alpha \lambda_1^\top \partial^\alpha \lambda_1 + \partial_\alpha \lambda_2 \partial^\alpha \lambda_2 
+ \phi \partial_\alpha \lambda_1^\top \partial^\alpha \lambda_1 
+ \left(\phi - {\cal U}_4 - \phi {\cal U}_4\right)\partial_\alpha \lambda_2 \partial^\alpha \lambda_2 + ...\nd
Secondly, the dilaton field $\phi$ interacts {\it equally} with all the components of the sigma model field $\lambda_1$, but has a different interaction with the sigma model field 
$\lambda_2$. This at least suggests that the three gauge fields (${\cal A}_0, {\cal A}_1, {\cal A}_2$) appearing from the corresponding vertex operator with $\lambda_1$ will have
identical gauge couplings, which would differ from the gauge coupling of the gauge field ${\cal A}_\psi$ appearing from $\lambda_2$. Thirdly, the appearance of ${\cal U}_4$, or 
more appropriately $e^{2\phi} H_4$ from \eqref{wardaf} with the same relative weight as in \eqref{curamo} points to the emergence of the coefficents ($c_{11}, c_{12}$) describing the 
gauge fields in four-dimensions. Therefore putting these together, and including the Chan-Paton factors, we expect
the possibility of the
emergent action \eqref{chelara}, with 
\bg\label{ansas}
\gamma_1 ~\propto ~ c_{11}, ~~~~~~~ \gamma_2 ~\propto ~ c_{12}. \nd
In addition to the emerging gauge theory description \eqref{chelara}, the M-theory gravitational coupling also leads to interesting four-dimensional gravitational coupling. For example
we can have the following correspondence:
\bg\label{grovcou}
\int {\cal C}_3 \wedge X_8 ~ \to ~ c_3 \int {\rm tr}~{\cal R} \wedge {\cal R}, \nd
which will become useful in studying gravitational and framing anomalies associated with the knots in a curved background as mentioned in \cite{wittenknots1}. We will discuss this later. 
In writing \eqref{grovcou} we have defined ${\cal R}$ as the four-dimensional curvature two-form, the trace over the Lorentz group, and the coefficient $c_3$ given via:
\bg\label{sinsag}
c_3 = \int_{\Sigma_3} \langle {\cal C}_3 \rangle \int_{\rm TN} p_1, \nd 
with $p_1$ being the first Pontryagin class defined over the warped Taub-NUT space (and as such should be an integer). 

\subsubsection{Dynamics on the three-dimensional boundary \label{dynamo}}

In writing \eqref{chelara} and \eqref{grovcou} we have inadvertently described the theory in four-dimensional spacetime without resorting to any boundary. The boundary description 
is important and as such lies in the heart of the problem. This description featured prominently in \cite{wittenknots1} and therefore we should see if our M-theory picture leads us to
the right boundary description.     

To infer about any boundary, we note that we have two possible four-dimensional description in the dual type IIB side. In one description, mentioned in the brane construction
{\bf Table \ref{wittenbranes}}, the D3-branes are oriented along ($x_0, x_1, x_2, \psi$) directions. In the other description, also in type IIB, the fractional D3-branes are oriented along
($x_0, x_1, x_2, x_3$). Thus we should look at the M-theory metric along ($x_0, x_1, x_2$) as well as along ($x_3, \psi$). This can be extracted from \eqref{curamo} and is given by:
\bg\label{012met}
ds^2 =  H_1\left(-dt^2 + dx_1^2 + dx_2^2\right) + H_1 H_2 ~dx_3^2 + e^{2\phi} H_1H_4 d\psi^2, \nd 
where we note that the Lorentz invariance along $x_3$ is broken by our choice of $H_2$ that depends on the NC deformation $\theta$ as depicted in \eqref{wardaf}; and the Lorentz invariance
along $\psi$ is broken both by our brane construction as well as the NC deformation, as depicted also in \eqref{wardaf}. 
This is at least one reason for localizing the knots along ($x_0, x_1, x_2$) directions albeit in the 
Euclidean version. The other reason, which also stems from the Lorentz invariance, is related to supersymmetry as described in \cite{wittenknots1}. Therefore from both viewpoints, 
namely the brane construction of {\bf Table \ref{wittenbranes}} and the fractional branes on warped Taub-NUT space, there is a reason to localize the knots along the Euclidean 
three dimensions.  

Having got the space along which knots could be described, we should now investigate the topological theory describing the knots from M-theory. Of course some parts of the 
{\it four-dimensional} theory is already at hand, this is given by \eqref{chelara} and \eqref{grovcou}. We will have to restrict them to the three-dimensional {\it boundary}. This
is the same boundary ${\bf W}$ that featured in \cite{wittenknots1}:
\bg\label{vwr+}
{\bf V} \equiv {\bf W} \times {\bf R}_+, \nd
with ${\bf R}_+$ related to $\psi$ described earlier. Note that in the language of fractional branes wrapped on two-cycle of our warped Taub-NUT space, ${\bf W}$ will be the same 
Euclidean three-dimensional space, although the four-dimensional space ($x_0, x_1, x_2, x_3$) doesn't have a representation like \eqref{vwr+}.

\subsubsection{Action for the three scalar fields in four-dimensions \label{actionf}}

Before moving to the three-dimensional description on the boundary ${\bf W}$, we should complete our four-dimensional description. This would require us to go back to the original 
seven-dimensional description that appears naturally from M-theory. The non-abelian seven-dimensional gauge field will have an action similar to \eqref{chelara}, but now the 
integral will be restricted to $d^7x$. The number of scalars in this description appears from various sources. A set of three non-abelian
scalar fields should appear from the dimensional reduction of our seven-dimensional non-abelian gauge
fields on $\Sigma_3$, and as such also appears from the wrapped M2-branes
fluctuating orthogonally to both the Taub-NUT and the four-dimensional space-time directions. 

It is instructive to work this out in some details as this will help us to unravel the BPS structure of the system. In this 
section we will concentrate on the scalars that come from the 
non-abelian gauge fields on $\Sigma_3$.
To start we will define our non-abelian gauge field as:
\bg\label{nagf}
{\cal A} &= & \alpha_1~e_3 + \alpha_2~e_r + \alpha_3~e_{\phi_1} \equiv {\cal A}_3 ~dx_3 + {\cal A}_r~dr + {\cal A}_{\phi_1}~d\phi_1 \nonumber\\
& = & {\left({\cal A}_3 - f_3 {\cal A}_{\phi_1}\right)\over \sqrt{H_1 H_2}}~e_3 + {e^{-\phi} {\cal A}_r \over \sqrt{F_1 H_1}}~e_r 
+ {{\cal A}_{\phi_1}\over \sqrt{H_1 H_3}}~e_{\phi_1}, 
\nd 
where these three components of the gauge field (${\cal A}_3, {\cal A}_r, {\cal A}_{\phi_1}$), that are now functions of ($x_0, x_1, x_2, \psi$),
 would appear as scalar fields in four-dimensional space (note $\alpha_i$ are also functions of ($r, \theta_1$)). 
These three scalar fields form a part of 
the ${\cal N} = 4$ vector multiplet, and we will discuss the remaining three scalar fields in the next subsection.
The functional forms for $H_i, F_1$ and $f_3$ have been defined in \eqref{wardaf}, and $e_i$ are given by:
\bg\label{e1giv}
e_3 = \sqrt{H_1 H_2} ~dx_3, ~~~ e_r = e^\phi \sqrt{H_1 F_1}~dr, ~~~~ e_{\phi_1} = \sqrt{H_1 H_3} \left(d\phi_1 + f_3 dx_3\right). \nd
Now using the gauge field ${\cal A}$ in \eqref{nagf}, and using the vielbeins $e_i$ in \eqref{e1giv} we can evaluate the following four-dimensional piece stemming from 
the interaction term of \eqref{chelara}:
\bg\label{chalbeta} 
S_{int} & = & \int {\rm Tr} \left({\cal A} \wedge {\cal A}\right) \wedge \ast  \left({\cal A} \wedge {\cal A}\right) \\
& = &  {R_3} \int d^4 x ~dr~ d\theta_1~H_1^2 \sqrt{H_4 F_1} e^{2\phi} 
~{\rm Tr}\Big([\alpha_1, \alpha_2]^2 + [\alpha_1, \alpha_3]^2 + [\alpha_2, \alpha_3]^2\Big), \nonumber 
\nd
where $d^4 x \equiv dt dx_1 dx_2 d\psi$ and the commutator brackets take the following form in terms of the gauge field components:
\bg\label{commbra}
&& [\alpha_1, \alpha_3] = {[{\cal A}_3, {\cal A}_{\phi_1}]\over H_1 \sqrt{H_2 H_3}}, ~~~ 
[\alpha_2, \alpha_3] = {e^{-\phi}[{\cal A}_r, {\cal A}_{\phi_1}]\over H_1 \sqrt{F_1 H_3}}\nonumber\\
&& [\alpha_1, \alpha_2] = {e^{-\phi}[{\cal A}_3, {\cal A}_r] + e^{-\phi} f_3 [{\cal A}_r, {\cal A}_{\phi_1}]\over H_1 \sqrt{H_2 F_1}}. \nd
To evaluate the functional form of the scalar action we need to plug in the values of the warp factors from \eqref{wardaf} in \eqref{commbra} and \eqref{chalbeta}. 
Doing this we get the following terms for the scalar field action in four-dimensional space:
\bg\label{chormar} 
S^{(1)}_{int} = \int d^4x~{\rm Tr}\Bigg{\{}a_1 \Bigg[{\cal A}_r, {\cal A}_{\phi_1} - {a_3 {\cal A}_3\over 2a_1} \Bigg]^2 
+ \left({4a_1 a_2 - a_3^2\over 4a_1}\right) \left[{\cal A}_3, {\cal A}_r\right]^2 +
a_4 \left[{\cal A}_3, {\cal A}_{\phi_1}\right]^2\Bigg{\}}, \nonumber\\ \nd
where $a_i \equiv a_i(\theta)$ are all functions of the constant NC parameter $\theta$ which are got by integrating out all the internal coordinates as well as averaging over $\theta_1$ 
coordinate. For example $a_1(\theta)$ will be defined as\footnote{All coefficients, including the ones for ${\cal A}_{\phi_1}$, henceforth
will be taken to be positive definite, unless mentioned otherwise. Any overall 
negative signs can be absorbed in the definition of the fields.}:
\bg\label{a1the} 
a_1(\theta) & = &  R_3 \int_0^\infty dr \int_0^\pi d\theta_1~\sqrt{H_4\over F_1}\left({1\over H_3} + {f_3^2\over H_2}\right)\\
& = & R_3~\sec~\theta \int_0^\infty dr ~\widetilde{a}_1\left[\sqrt{\widetilde{F}_2^2 F_3\over F_1} +
{\widetilde{a}_2F_3^{3/2}\over 2\widetilde{a}_1} \sqrt{\widetilde{F}_2\over F_1(\widetilde{F}_2 - F_3)} {\rm ln}\left|{\sqrt{\widetilde{F}_2} + \sqrt{\widetilde{F}_2-F_3}\over 
\sqrt{\widetilde{F}_2} - \sqrt{\widetilde{F}_2-F_3}} \right|\right] \nonumber \nd
where we have assumed that $\widetilde{F}_2$, defined in \eqref{f2f2}, satisfy 
$\widetilde{F}_2 > F_3$ at all points in $r$, otherwise we will need to replace this combination by $\vert \widetilde{F}_2 - F_3\vert$ to allow for real 
values of the above integral. We have also defined $\widetilde{a}_i$ as:
\bg\label{widaid}
\widetilde{a}_1 = 1 + {\tan^2\theta(1 + F_2~\tan^2\theta) \widetilde{F}_2^2 \over \widetilde{F}_2 - F_3}, ~~~
\widetilde{a}_2 = 1 - {\tan^2\theta(1 + F_2~\tan^2\theta)\widetilde{F}_2^2 \over \widetilde{F}_2 - F_3}. \nd 
The above integrand in \eqref{a1the} is well defined everywhere in $r$ and therefore integrates to a constant, i.e only a function of the constant NC parameter $\theta$ as predicted 
earlier. The other constants $a_i(\theta)$ are slightly simpler than \eqref{a1the}, and we will define them in the following. It is interesting to note that:
\bg\label{a3the}
a_3(\theta) = 2R_3 \int_0^\infty dr \int_0^\pi d\theta_1 ~{f_3\over H_2}\sqrt{H_4\over F_1} = 0, \nd
which mean that there are no unnecessary cross-terms in the scalar-field interactions \eqref{chormar}, as one might have expected from supersymmetric considerations. The other two 
coefficients are given as follows:
\bg\label{a2the}
a_2(\theta) & = & R_3 \int_0^\infty dr \int_0^\pi d\theta_1 ~{1\over H_2} \sqrt{H_4\over F_1} \\
& = & R_3~\sec~\theta \int_0^\infty dr \left(\cos^2\theta + F_2~\sin^2\theta\right) \sqrt{\widetilde{F}_2 F_3 \over F_1(\widetilde{F}_2- F_3)} {\rm ln} 
\left|{\sqrt{\widetilde{F}_2} + \sqrt{\widetilde{F}_2-F_3}\over \sqrt{\widetilde{F}_2} - \sqrt{\widetilde{F}_2-F_3}}\right|, \nonumber \nd
where the integrand is again a well defined function for all values of $r$, and therefore $a_2$ is just a function of the constant NC parameter $\theta$. On the other hand the coefficient
$a_4(\theta)$ is given by:
\bg\label{a4the1}
a_4(\theta) & = & {R_3}\int_0^\infty dr \int_0^\pi d\theta_1 {e^{2\phi_0}\sqrt{H_4 F_1}\over H_2 H_3} \\
& = & {R_3~\sec~\theta} \int_0^\infty dr~\widetilde{a}_4
\left[\sqrt{\widetilde{F}_2} + 
{F_3\over 2\sqrt{\widetilde{F}_2-F_3}}~{\rm ln}\left|{\sqrt{\widetilde{F}_2} + \sqrt{\widetilde{F}_2-F_3} \over \sqrt{\widetilde{F}_2} - \sqrt{\widetilde{F}_2-F_3}}\right|\right], 
\nonumber \nd
assuming as before the dilaton $e^{2\phi}$ to be given by the leading order constant piece $e^{2\phi_0}$. In that case 
$\widetilde{a}_4(r, \theta)$ is given by the following expression:
\bg\label{atil4}
\widetilde{a}_4(r, \theta) = e^{2\phi_0} \left(\cos^2\theta + F_2~\sin^2\theta\right)\sqrt{\widetilde{F}_2 F_3 F_1}. \nd
We can similarly work out the coefficients for the M-theory uplift \eqref{curamo2}. Interestingly, the functional forms for the $a_1$ and the $a_4$ coefficients for the new background 
are similar to the $a_4$ functional functional form \eqref{a4the1} except with $\widetilde{a}_4$ in \eqref{atil4} replaced by:
\bg\label{ati42}
\widetilde{a}_4 ~\to ~ \sqrt{b_4 F_1^{-1} \widetilde{F}_2 F_3}, ~~ {\rm and}~~ \widetilde{a}_4 ~\to ~ e^{2\phi_0}\sqrt{b^3_4 F_1 \widetilde{F}_2 F_3}, \nd
respectively with $b_4$ as in \eqref{bfour}. 
On the other hand, the functional form for the new $a_2$ is similar to the functional form for $a_2$ in \eqref{a2the}. The only difference being that the following 
replacement in \eqref{a2the}:   
\bg\label{repla} b_4(\phi = 0) ~ \to~ \sqrt{b_4}. \nd
We now have all the functional forms for $a_i$ given in terms of the constant NC parameter $\theta$. All the $a_i$ are finite numbers, and although one might worry about the case when
$F_3$ vanishes for some $r$ in \eqref{a1the}, \eqref{a2the}, \eqref{a4the1} 
because the logarithmic functions therein are not well defined, this is not a problem. The reason is that 
all the logarithmic functions in \eqref{a1the}, \eqref{a2the}, \eqref{a4the1} always come with a factor of $F_3$ attached to them, so when $F_3$ vanishes, the logarithmic functions 
also vanish. Thus after the dust settles, the interaction terms for the three scalars in \eqref{chalbeta} and \eqref{chormar} can now be expressed as:
\bg\label{chwasha}
S^{(1)}_{int} = \int d^4x \Big\{a_1(\theta){\rm Tr}\left[{\cal A}_r, {\cal A}_{\phi_1}\right]^2 + a_2(\theta){\rm Tr}\left[{\cal A}_3, {\cal A}_r\right]^2 
+ a_4(\theta){\rm Tr}\left[{\cal A}_3, {\cal A}_{\phi_1}\right]^2\Big\}. \nd
Having got the interaction terms, it is now instructive to work out the kinetic terms of the three scalars (${\cal A}_3, {\cal A}_r, {\cal A}_{\phi_1}$). As one might have expected, 
M-theory does reproduce the expected form of the kinetic terms, namely:
\bg\label{kintu} 
S^{(1)}_{kin} & = & \int d^4 x \Big\{ c_{\psi 3} {\rm Tr}\left({\cal D}_\psi{\cal A}_3\right)^2 + c_{\psi r} {\rm Tr}\left({\cal D}_\psi{\cal A}_r\right)^2 
+ c_{\psi\phi_1}{\rm Tr} \left({\cal D}_\psi{\cal A}_{\phi_1}\right)^2\nonumber\\ 
& + & \sum_{a= 0}^{2} \left[c_{a3} {\rm Tr}\left({\cal D}_a{\cal A}_3\right)^2 + c_{ar} {\rm Tr}\left({\cal D}_a{\cal A}_r\right)^2 
+ c_{a\phi_1} {\rm Tr}\left({\cal D}_a{\cal A}_{\phi_1}\right)^2\right]\Big\}, \nd
where ${\cal D}_a$ and ${\cal D}_\psi$ are defined using the four-dimensional gauge fields ${\cal A}_a \equiv ({\cal A}_0, {\cal A}_1, {\cal A}_2)$ and ${\cal A}_\psi$ in the usual way:
\bg\label{fintu}
{\cal D}_a\varphi \equiv \partial_a \varphi + i\left[{\cal A}_a, \varphi\right], ~~~~~~~ {\cal D}_\psi\varphi \equiv \partial_\psi \varphi + i\left[{\cal A}_\psi, \varphi\right]. \nd
The coefficients ($c_{am}, c_{\psi m}$), where $m = (3, r, \phi_1)$, are straightforward (albeit tedious) to work out from the background data. We will first tackle the easier 
ones. 
The coefficients $c_{a3}$ for all $a$'s take the following form:
\bg\label{tentu}
c_{a3}(\theta) = {R_3~\sec~\theta}\int_0^\infty dr~{e^{2\phi_0}\over H_2}\sqrt{F_1 \widetilde{F}_2 F_3\over \widetilde{F}_2 - F_3} 
~{\rm ln}\left|{\sqrt{\widetilde{F}_2} + \sqrt{\widetilde{F}_2 - F_3}\over \sqrt{\widetilde{F}_2} - \sqrt{\widetilde{F}_2 - F_3}}\right|,\nd
where $H_2$ is defined in \eqref{wardaf}, and the integrand is well defined when ($\widetilde{F}_2, F_3$) $\to 0$ as well as when $\widetilde{F}_2 \to F_3$. This means $c_{a3}$ 
is just a constant defined in terms of $\theta$, the NC parameter. Similarly the other three coefficients $c_{ar}$ are similar to \eqref{tentu} and take the following form:
\bg\label{bentu}
c_{ar}(\theta) = R_3~\sec~\theta \int_0^\infty dr \sqrt{\widetilde{F}_2 F_3 \over F_1(\widetilde{F}_2 - F_3)} 
~{\rm ln}\left|{\sqrt{\widetilde{F}_2} + \sqrt{\widetilde{F}_2 - F_3}\over \sqrt{\widetilde{F}_2} - \sqrt{\widetilde{F}_2 - F_3}}\right|, \nd
and is well defined at all the limits described above. 

The remaining three coefficients $c_{a\phi_1}$ are more complicated than \eqref{tentu} and \eqref{bentu} as they involve certain manipulations involving $c_{a3}$ in \eqref{tentu}. 
After the dust settles, the result is:
\bg\label{jhogra}
c_{a\phi_1}(\theta) & = & {R_3~\sec~\theta}\int_0^\infty dr~e^{2\phi_0}\sqrt{F_1 \widetilde{F}_2 F_3}
\left(1 + {\widetilde{F}_2^2 \over H_2(\widetilde{F}_2 - F_3)}~\tan^2\theta ~\sec^2\theta\right) \nonumber\\
&& ~~~~~~~~~~~~ \otimes \left(2\sqrt{\widetilde{F}_2} + {\widetilde{a}_2F_3\over \widetilde{a}_1\sqrt{\widetilde{F}_2 - F_3}}
~{\rm ln}\left|{\sqrt{\widetilde{F}_2} + \sqrt{\widetilde{F}_2 - F_3}\over \sqrt{\widetilde{F}_2} - \sqrt{\widetilde{F}_2 - F_3}}\right|\right). \nd
The integrand is well defined in the limit $\widetilde{F}_2 = F_3 = 0$, but seems to diverge in the limit $\widetilde{F}_ 2 \to F_3$. However as before, we should look at the limit
more carefully. If we assume $\widetilde{F}_2 - F_3 = \epsilon^2$, where $\epsilon \to 0$, then the relevant part of the integrand in \eqref{jhogra} takes the following form:
\bg\label{tunilight}  
{}^{\rm lim}_{\epsilon \to 0} ~{1\over \epsilon}\left(2\sqrt{1+{F_3\over \epsilon^2}} - {F_3\over \epsilon^2} ~{\rm ln}\left|{\sqrt{\widetilde{F}_2} +\epsilon \over 
\sqrt{\widetilde{F}_2} -\epsilon}\right|\right)~ \to ~ {4 \over 3 \sqrt{\widetilde{F}_2}} + {\cal O}(\epsilon), \nd
which 
implies that the integrand in \eqref{jhogra} is well-defined everywhere, and thus the corresponding integral leads to a constant function of the NC parameter $\theta$.

The integral form of the other two coefficients, namely $c_{\psi 3}$ and $c_{\psi r}$, have certain resemblance to \eqref{krisis} as for all three cases the integrand are somewhat 
similar. For example:
\bg\label{fintubabu}
c_{\psi 3}(\theta) =  2R_3~\cos~\theta \int_0^\infty dr~{b_2 J_3\over H_2}, ~~
c_{\psi r}(\theta) & = & 2R_3~e^{-2\phi_0}~\cos~\theta \int_0^\infty dr~{b_2 J_3\over F_1}, \nd
where the functional form for $J_3$ can be expressed from \eqref{krisis} as:
\bg\label{finbab}
J_3(r) \equiv b_3~{\rm tanh}^{-1}\left({1\over b}\sqrt{F_3 + b^2(\widetilde{F}_2 - F_3)\over \widetilde{F}_2}\right)+ {\rm ln}\left|{\sqrt{\widetilde{F}_2} 
- \sqrt{\widetilde{F}_2 - F_3}\over \sqrt{\widetilde{F}_2} + \sqrt{\widetilde{F}_2 - F_3}}\right|, \nd
with ($b_2, b_3$) as defined earlier in \eqref{aa2a3} and $H_2$ as in \eqref{wardaf}. Since the integrand in \eqref{krisis} is well-defined for the limits 
$\widetilde{F}_2 = F_3 = 0$ and $\widetilde{F}_2 \to F_3$, we expect the integrands in \eqref{fintubabu} to be well-defined as well. Note that only in the limit $H_2 = F_1 = 1$ we get:
\bg\label{kolaju}
c_{\psi 3} = c_{\psi r} = c_{12}, \nd
which is in general not true as $H_1$ is a function of ($r, \theta_1$) whereas $F_1$ is a function of $r$ only. However if $\widetilde{F}_2 \to F_3$, then $H_1$ becomes a function of 
$r$ only, and we can choose our starting metric \eqref{chote} with $F_2 = F_1^3$ in the absence of NC deformation. This choice is very special, so in general we don't expect 
\eqref{kolaju} to hold. 

The final coefficient $c_{\psi\phi_1}$ is a little harder to compute as it involves some mixing with $c_{\psi 3}$ in \eqref{fintubabu}, similar to \eqref{jhogra} derived earlier. The 
analysis nevertheless is straightforward, and is given by:
\bg\label{lebu}
c_{\psi\phi_1}(\theta) = \int_0^\infty dr\left[a_{01}~{\rm tanh}^{-1}\left(a_{02}\right) - b_{01}~{\rm ln}\left|b_{02}\right| - c_{01}\right], \nd
where the various coefficients appearing above are defined in the following way. The first three coefficients ($a_{01}, b_{01}, c_{01}$) receive 
contributions from $c_{\psi 3}$ of \eqref{fintubabu}. The other two ($a_{02}, b_{02}$) are more straightforward. We start with $a_{01}$:
\bg\label{coeffdeff}
a_{01} &=& {2 R_3~\cos~\theta \left[(1-b^2)F_3 + b^2 \widetilde{F}_2\right]^{3/2} \sqrt{F_1}\over b\sqrt{\widetilde{F}_2 F_3}} \nonumber\\
&& +~ {2bR_3 \tan^2\theta~\sec~\theta~\widetilde{F}^2_2\over H_2}\sqrt{(1-b^2)F_1 F_3 + b^2 F_1\widetilde{F}_2 \over \widetilde{F}_2 F_3}, \nd 
where the first line is the expected output directly from M-theory analysis, and the second line involves contribution from $c_{\psi 3}$ in \eqref{fintubabu}. The second 
coefficient $b_{01}$ also takes a somewhat similar form:
\bg\label{fclesb}
b_{01} & = & {1\over 2}R_3~\cos~\theta \left[{(3-2b^2)F_3 + 2b^2 \widetilde{F}_2\over \widetilde{F}_2 - F_3}\right]\sqrt{F_1(\widetilde{F}_2 - F_3)^3 \over \widetilde{F}_2 F_3}\nonumber\\
&& +~ {R_3 ~\tan^2\theta~\sec~\theta~\widetilde{F}^2_2\over 2 H_2}\left[{(1-2 b^2)F_3 + 2b^2 \widetilde{F}_2 \over \sqrt{\widetilde{F}_2- F_3}}\right] 
\sqrt{F_1\over \widetilde{F}_2 F_3}, \nd 
where again the second line appears from the $c_{\psi 3}$ coefficient of \eqref{fintubabu}. Finally the coefficient $c_{01}$ is given by:
\bg\label{brazle}
c_{01} = {R_3~\cos~\theta\left(\widetilde{F}_2-F_3 \right)\sqrt{F_1} \over \sqrt{{F}_3}} + {R_3 \widetilde{F}_2^2 ~\tan^2\theta~\sec~\theta\over H_2}\sqrt{F_1 \over F_3}, \nd
with the second term now appearing from $c_{\psi 3}$ piece. The other two factors, namely ($a_{02}, b_{02}$), are straightforward to work out and take the familiar forms:
\bg\label{duggar}
a_{02} = {1\over b}\sqrt{(1-b^2)F_3 + b^2 \widetilde{F}_2\over \widetilde{F}_2}, ~~~~
b_{02} = {\sqrt{\widetilde{F}_2} + \sqrt{\widetilde{F}_2-F_3} \over \sqrt{\widetilde{F}_2} - \sqrt{\widetilde{F}_2-F_3}}. \nd
Once again, it is time to look at the limiting behavior of the integrand when $F_3 \to 0$ and $F_3 \to \widetilde{F}_2$. The other limit of $\widetilde{F}_2 \to 0$ is contained
in the other two limits if we assume that $F_3$ goes to zero faster than $\widetilde{F}_2$. Thus in the limit $F_3 \to 0$, the integrand in \eqref{lebu} behaves as:
\bg\label{cmontot}
{}^{\rm lim}_{F_3\to 0}~{1\over\sqrt{F_3}}\left\{{\rm tanh}^{-1}\left[\sqrt{1+\left({1-b^2\over b^2}\right){F_3\over \widetilde{F}_2}}\right] 
+ {\rm ln}\left|\sqrt{\widetilde{F}_2} - \sqrt{\widetilde{F}_2-F_3}\right| 
- {\rm constant}\right\}, \nonumber\\ \nd 
which could be arranged to vanish as before. For the other limit $F_3 \to \widetilde{F}_2$, or alternatively as $\widetilde{F}_2 - F_3 = \epsilon^2 \to 0$, the integrand in 
\eqref{lebu} behaves as:
\bg\label{nerdip}
\sqrt{F_3}(1+ F_3)~{\rm tanh}^{-1}\left({1\over b}\right) - {F_3^2\over \epsilon} ~{\rm ln}\left|{1+\epsilon\over 1-\epsilon}\right| - {\epsilon^2 \over \sqrt{F_3}}, \nd
which vanishes in the limit $F_3$ goes to zero {\it slower} than $\epsilon^2$. However this limit, although would contradict with $\widetilde{F}_2 > F_3$ $-$ 
where we expect $F_3$ to vanish
{\it faster} than $\widetilde{F}_2$ $-$ would still be fine if we impose $\widetilde{F}_2 = F_3$ at the vanishing point. 

We are almost done, but before ending this section let us work out the gauge theory coefficients for the kinetic terms in \eqref{kintu} using the M-theory uplift \eqref{curamo2}
of the RR deformed background \eqref{hardface}. The coefficients are again easy to work out, and its no suprise that they don't change appreciably from what we computed above. 
For example the expressions for ($c_{\psi 3}, c_{\psi r}$) remain similar to \eqref{fintubabu} with the same $J_3$ as in \eqref{finbabu} except for the following changes:
\bg\label{fbaajatu}
H_2 ~ \to ~ \sqrt{\widetilde{H}_2}, ~~ {\rm and}~~ b_2 ~ \to ~ b_2 \sqrt{\widetilde{H}_2}, \nd
respectively, where $\widetilde{H}_2 \equiv b_4^{-1}$ is defined earlier in \eqref{wardaf2} and \eqref{bfour}. Similarly for the coefficient $c_{ar}$
the new expression is exactly as in \eqref{bentu} given above, except with the following replacement in the integrand of \eqref{bentu}:
\bg\label{fcblone}
\sqrt{F_1} ~ \to ~ \sqrt{F_1 \over \widetilde{H}_2}. \nd
For the other two coefficients $c_{a\phi_1}$ and $c_{\psi \phi_1}$ in \eqref{jhogra} and \eqref{lebu} respectively, 
the above replacement \eqref{fcblone} alongwith the vanishing of the $\widetilde{F}_2^2$ terms in \eqref{jhogra} and \eqref{coeffdeff}, \eqref{fclesb}, \eqref{brazle} 
respectively capture the new coefficients succinctly. Finally for the $c_{a3}$ coefficient \eqref{tentu}, all we need is to replace $H_2$ therein by $\sqrt{\widetilde{H}_2}$ of
\eqref{wardaf2} to get the correct expression.

Thus, after the dust settles, 
the three scalars coming from the seven-dimensional gauge fields, all combine together to reproduce the action \eqref{kintu} with the coefficients $c_{\psi m}$ and 
$c_{am}$ as well defined functions of the NC parameter $\theta$ or the RR deformed parameter $\theta$. 
In the following section we will discuss the remaining three scalars that come from the explicit form of the 
warped Taub-NUT geometry. 

\subsubsection{Action for the remaining three scalar fields \label{actionR}}

The remaining scalar fields, that fill the rest of the ${\cal N} = 4$ vector multiplet in four-dimensions, come precisely from the seven-dimensional vector multiplet. In M-theory
they should appear from our warped Taub-NUT configuration.  
The zero-mode fluctuations of the $N$-centered Taub-NUT space, namely:
\bg\label{nos}
N(2h_{11} + 1) = 3N, \nd
which would appear in our four-dimensional description on ${\bf V}$, provide the Cartan of the $A_{N-1}$ algebra for the seven-dimensional theory. The fluctuations of the wrapped
M2-branes along the Taub-NUT directions provide the necessary roots and weights of the $A_{N-1}$ algebra leading to the non-abelian enhancement of the three scalars in the
vector multiplet of the seven-dimensional theory.

To analyze these scalars, let us first discuss the abelian version of the model that would come from the zero mode fluctuations of our warped Taub-NUT space. These fluctuations are not hard
to work out from the M-theory Einstein term, and have the following action derivable from the supergravity lagrangian:
\bg\label{himan}
\int d^{11}x~\delta\left(\sqrt{g_{11}}R_{11}\right) ~\propto~ \int d^4x \sum_{k=1}^3 \left[\sum_{a=0}^2 b_{ak} \left(\partial_a \varphi_k\right)^2 
+ b_{\psi k} \left(\partial_\psi \varphi_k\right)^2\right], \nd
where ($\varphi_1, \varphi_2, \varphi_3$) are the three abelian scalars, and $\delta$ denote the combination of the three fluctuations of the internal Taub-NUT space. In writing \eqref{himan} 
we have assumed that the fluctuations are only functions of the spacetime coordinates ($x_0, x_1, x_2, \psi$). The coefficient $b_{ak}$ for a given ($a, k$) is a function 
of the NC parameter $\theta$ and can be expressed in terms of the warp factors as:
\bg\label{boka}
b_{ak}(\theta) = {2 R_3~\sec~\theta}\int_0^\infty dr ~e^{2\phi_0}\left(\cos^2\theta+F_2\sin^2\theta\right)^{1/3} F_3^{1/3} \sqrt{F_1\widetilde{F}_2} 
~\Theta_{12}, \nd
where we see that all the nine coeffcients have identical functional form because of the isometry along the ($x_0, x_1, x_2$) directions. We have also defined $\Theta_{12}$ 
using Hypergeometric function in the 
following way:
\bg\label{theta12}
\Theta_{12} = {}_2{\bf F}_1\left({1\over 6}, {1\over 2}; {3\over 2}; {F_3-\widetilde{F}_2\over F_3}\right). \nd
Let us now check the limits. When $F_3 = 0$, the integrand in \eqref{boka} vanishes, and so it is well defined. On the other hand, when $F_3 \to \widetilde{F}_2$, the Hypergeometric 
function $\Theta_{12} = 1$, and the integral is again well defined provided none of the warp factors blow up at $r \to \infty$. However subtlety arises once we use the warp factors to 
define the other coefficient $b_{\psi k}$. The form of $b_{\psi k}$ for any $k$ is more non-trivial compared to \eqref{boka}, and takes the following form:
\bg\label{bokcho}
b_{\psi k}(\theta) = {2 R_3~\cos~\theta \over b^2}\int_0^\infty dr \left(\cos^2\theta + F_2~\sin^2\theta\right)^{1/3} F_3^{1/3} \sqrt{F_1 \over \widetilde{F}_2}~\Theta_{34}, \nd
where as before all the three coefficients have identical functional forms, and they differ from \eqref{boka} because the Lorentz invariance along $\psi$ direction is broken. The functional 
form for $\Theta_{34}$ is now defined in terms of a certain Appell function in the following way: 
\bg\label{appell}
\Theta_{34} = ~ {\bf F}_1\left({1\over 2}; -{5\over 6}; 1; {3\over 2}; {F_3 - \widetilde{F}_2 \over F_3}; {1\over b^2}\right), \nd
where $b^2$ is the same regularization parameter used earlier in \eqref{krisis} to avoid certain singularities. Note that when
$F_3 \to \widetilde{F}_2$, the Appell function blows up in the limit $b\to 1$, but $b$ is not necessarily identity. 
This way the integrand will be well defined everywhere. Alternatively, the 
field $\varphi_k$ could be made independent of $\psi$ altogether. We will discuss a variant of the latter idea soon when we study the boundary dynamics in more details. 

Our discussions so far have mostly concentrated on the abelian scalar fields. To study the non-abelian scalars we will, without loss of generalities, define the scalar fields 
again as $\varphi_k$ where $\varphi_k \equiv \varphi_k^a T^a$ with $T^a$ being the generator of $SU(N)$ in the adjoint representation. The extension of \eqref{himan} to the 
non-abelian version is now straightforward:
\bg\label{himthi} 
S^{(2)}_{kin} = \int d^4x \sum_{k=1}^3 \left[\sum_{a=0}^2 b_{ak}~ {\rm Tr}\left({\cal D}_a \varphi_k\right)^2 + b_{\psi k} ~{\rm Tr}\left({\cal D}_\psi \varphi_k\right)^2\right], \nd   
where the trace is in the adjoint representation and ${\cal D}_{a, \psi}$ are the covariant derivatives with respect to the four-dimensional bulk gauge fields 
(${\cal A}_a, {\cal A}_\psi$) as described in \eqref{fintu}.

To proceed further we shall use various arguments to justify the remaining interaction terms. Maximal supersymmetry tells us that the remaining scalars should at least have the following
form of the lagrangian:
\bg\label{lengor} 
{\cal L}_{\varphi} = \beta_1 {\rm Tr}\left({\cal D}_m \varphi_k\right)^2 + \beta_2 {\rm Tr}\left[\varphi_k, \varphi_l\right]^2 
+ \beta_3{\rm Tr}\left[{\cal A}_{\{3, r, \phi_1\}}, \varphi_k\right]^2, \nd
where we determined the form of $\beta_1$ in \eqref{himthi} above. Additionally, multiple D6-branes wrapped on a 3-cycle of a manifold will have the world-volume dynamics given 
by a non-abelian Born-Infeld action in a curved space. What curvatures are we interested in 
from M-theory point of view? Looking at the analysis done in the earlier subsections, we see
that the emergent dynamics of the seven-dimensional gauge theory from M-theory is simply an interacting non-abelian vector multiplet in a {\it curved} space with a metric given by the 
first line of \eqref{curamo}. In fact this is consistent with the matrix formalism of M-theory also. Multiple D6-branes in a curved background can be studied as a M(atrix) theory on warped 
multi-centered Taub-NUT space \cite{rajaraman, hanany} where the seven-dimensional gauge theory appears on a curved ambient space {\it orthogonal} to the warped Taub-NUT background.
  
With this in mind, the rest of the discussions is now straightforward and will follow the pattern developed in \eqref{kintu}. The interaction terms of the three scalars will not only 
involve self interactions, but also interactions with the other three scalars (${\cal A}_3, {\cal A}_r, {\cal A}_\psi$)
that we studied in the previous subsection. The interaction terms then take the following form:
\bg\label{finbabu}
{\cal L}^{(2)}_{int} = \sum_{k, l} d_{kl} ~{\rm Tr}\left[\varphi_k, \varphi_l\right]^2 + \sum_{k=1}^3 \Big\{c_{rk} ~{\rm Tr}\left[{\cal A}_r, \varphi_k\right]^2 
+ c_{3k} ~{\rm Tr}\left[{\cal A}_3, \varphi_k\right]^2 + c_{\phi_1 k} ~{\rm Tr}\left[{\cal A}_{\phi_1}, \varphi_k\right]^2\Big\}. \nonumber\\ \nd 
Let us first study the self-interaction terms. These terms have coefficients $d_{kl}$ as depicted above, and since all these scalars appear in a democratic way, we expect the 
coefficients $d_{kl}$ to be the same for all choices of $k$ and $l$. 
This is indeed what is bourne out from our analysis, and the coefficient $d_{kl}$ for {\it any} ($k, l$) is given by:
\bg\label{venu}
d_{kl}(\theta) = {1\over 2} {R_3~\sec~\theta}\int_0^\infty dr~e^{2\phi_0}\sqrt{F_1\widetilde{F}_2 F_3}
\left(\cos^2\theta+F_2\sin^2\theta\right)^{2/3} \Theta_{56}, \nd 
where $\Theta_{56}$ now involves another Hypergeometric function that can be expressed, in combination with other warp factors, in the following way:
\bg\label{chottal} 
\Theta_{56} = F_3^{1/6} {}_2{\bf F}_1\left({1\over 2}, {5\over 6}; {3\over 2}; {F_3-\widetilde{F}_2\over F_3}\right) + 3 \widetilde{F}^{1/6}_2,  \nd 
that approaches 1 in the limit $F_3 \to \widetilde{F}_2$. This means the integrand in \eqref{venu} is well defined when $F_3 \to 0$ and when $F_3 \to \widetilde{F}_2$. 

The interaction of the scalars $\varphi_k$ with the other three scalars (${\cal A}_3, {\cal A}_r, {\cal A}_{\phi_1}$) can now be determined using similar Hypergeometric functions. For example
the coefficient $c_{rk}$ can be expressed as:
\bg\label{himleng}
c_{rk}(\theta) = 2 R_3~\sec~\theta \int_0^\infty dr ~F_3^{1/3} \left(\cos^2\theta+F_2\sin^2\theta\right)^{1/3} \sqrt{\widetilde{F}_2 \over F_1}~\Theta_{12}, \nd 
in terms of the Hypergeometric function $\Theta_{12}$ given in \eqref{theta12}, 
which implies that the limiting behaviors of the integrand \eqref{himleng} for $F_3 \to 0$ and $F_3 \to \widetilde{F}_2$ remain well-defined. 
The other coefficient $c_{3k}$ now has a form given by:
\bg\label{keyjar}
c_{3k}(\theta) = {2 R_3~\sec~\theta} \int_0^\infty dr ~e^{2\phi_0} 
~F_3^{1/3} \sqrt{\widetilde{F}_2 F_1}\left(\cos^2\theta+F_2\sin^2\theta\right)^{4/3} 
\Theta_{12}, \nd   
using the same Hypergeometric function $\Theta_{12}$ as in \eqref{theta12}. 
The above integrand is also well-defined in the limits $F_3 \to 0$ and $F_3 \to \widetilde{F}_2$ as before because $\Theta_{12}$ is well behaved in the latter limit. 

Finally, the last three coefficients $c_{\phi_1 k}$ for any $k$ are more complicated than the other coefficients that we derived earlier. However as before we do expect all the three coefficients 
to be identical because of the isometry of the three scalars. Thus for any given $k$, we get:
\bg\label{dhorshala}
c_{\phi_1 k}(\theta) = {R_3~\sec~\theta} \int_0^\infty dr ~e^{2\phi_0} \left(\cos^2\theta+F_2\sin^2\theta\right)^{1/3} 
\sqrt{F_1 \widetilde{F}_2 F_3}~\Pi_{78}, \nd 
which is well defined in the limit $F_3 \to 0$. For the other limit $F_3 \to \widetilde{F}_2$ we need to know the behavior of $\Pi_{78}$. Our analysis shows that
$\Pi_{78}$ can be expressed in the following way:
\bg\label{pi78}
\Pi_{78} \equiv \hat{\Pi}_{78} + 3~\tan^2\theta~\sec^2\theta ~\widetilde{F}_2^2 \left(\cos^2\theta+F_2\sin^2\theta\right) \widetilde{\Pi}_{78}, \nd
where, compared to our earlier analysis, this is a more complicated form because of the fibration structure of $\phi_1$ in the metric \eqref{curamo}. The variables 
 $\hat{\Pi}_{78}$ and $\widetilde{\Pi}_{78}$ are both expressed in terms of the Hypergeometric function $\Theta_{12}$, given earlier in \eqref{theta12}, and the warp factors as:
\bg\label{chotsar}
\hat{\Pi}_{78} = {3\over 4}~\widetilde{F}_2^{5/6} + {5\over 4}~F_3^{5/6} \Theta_{12}, ~~~~~~
\widetilde{\Pi}_{78} = {\widetilde{F}_2^{5/6} - F_3^{5/6} \Theta_{12} \over 4({\widetilde F}_2 - F_3)}. \nd
Now the limiting behavior of $F_3 \to \widetilde{F}_2$ is easy to determine. Since the Hypergeometric function $\Theta_{12}$ approaches identity in this limit, $\widetilde{\Pi}_{78}$ 
vanishes and $\hat{\Pi}_{78} \to 8 \widetilde{F}_2^{5/6}$. This way the integrand in \eqref{dhorshala} is well defined everywhere. 

For the M-theory background \eqref{curamo2}, one may similarly work out the coefficients as we had done earlier. We expect, as before, the results to not change significantly and 
indeed this is what appears from concrete computations. For example, for the coefficients ($b_{ak}, b_{\psi k}, c_{rk}$) in \eqref{boka}, \eqref{bokcho} and \eqref{himleng} 
respectively, 
the integral expressions remain unchanged upto the 
following replacements in each of the above integrands: 
\bg\label{lemurgi}
b_4^{1/3}(\phi = 0) ~ \to ~ {1\over b_4^{1/6}}, \nd
where $b_4 \equiv \widetilde{H}_2^{-1}$ has been defined earlier in \eqref{wardaf2} and 
\eqref{bfour}. In a similar vein, the integral expressions for $d_{kl}$ in \eqref{venu} and $c_{3k}$ in 
\eqref{keyjar} remain unchanged for the new background \eqref{curamo2}, except, with the following replacements:
\bg\label{mcakrisis}
b_4^{2/3}(\phi = 0) ~ \to ~ b_4^{1/6}, ~~~ {\rm and} ~~~ b_4^{4/3}(\phi = 0) ~ \to ~ b_4^{5/6}, \nd
respectively. This means all the Hypergeometric and the Appell functions preserve their forms for the RR deformed background \eqref{hardface}. Finally, the only expression that 
changes significantly is the expression for $c_{\phi_1 k}$ in \eqref{dhorshala}. The new expression for $c_{\phi_1 k}$ doesn't have the second $\widetilde{F}_2^2$ term of 
\eqref{dhorshala}. This is of course expected. However the first term of \eqref{dhorshala} is reproduced in a similar fashion except with the following replacement:  
\bg\label{jotadhar} b_4^{1/3}(\phi = 0) ~ \to ~ b_4^{5/6}. \nd
We have now completed the discussions of the full gauge theory action in four-dimensions using a warped multi-centered Taub-NUT space in M-theory. In the following subsection we will derive
the Bogomolnyi-Hitchin-Nahm (BHN) type of equation from our gauge theory data which will help us to search for, among other things, the Nahm poles.  

\subsubsection{A derivation of the BHN type of equation \label{bhuc}}

Before proceeding further, let us summarize our results so far. The full non-abelian $SU(N)$
gauge theory action that we get from our M-theory construction, from a warped seven-dimensional non-compact manifold that is topologically of the 
form:
\bg\label{pwhite}
{\bf TN}_N \times \Sigma_3, \nd
with {\it compact} $\Sigma_3$ and a $N$-centered warped Taub-NUT space ${\bf TN}_N$, 
can now be assimilated together from \eqref{chelara}, \eqref{kintu}, \eqref{himthi}, \eqref{chwasha} and \eqref{finbabu} (or with the corresponding RR deformed ones), 
to give us the following total 
action: 
\bg\label{stotal}
S_{total} &=& {c_1\over v_3} \int d^4 x \left(c_{11} \sum_{a<b}~{\rm Tr}~{\cal F}_{ab}{\cal F}^{ab} + c_{12} \sum_{a}~{\rm Tr}~{\cal F}_{a\psi}{\cal F}^{a\psi}\right) 
+ c_2 \int {\rm Tr}~ {\cal F} \wedge {\cal F} \nonumber\\
&+& {c_1\over v_3}\int d^4x\Big\{ c_{\psi 3}~ {\rm Tr}\left({\cal D}_\psi{\cal A}_3\right)^2 + c_{\psi r} ~{\rm Tr}\left({\cal D}_\psi{\cal A}_r\right)^2
+ c_{\psi\phi_1}~ {\rm Tr} \left({\cal D}_\psi{\cal A}_{\phi_1}\right)^2\nonumber\\
& + & \sum_{a= 0}^{2} \left[c_{a3}~ {\rm Tr}\left({\cal D}_a{\cal A}_3\right)^2 + c_{ar} ~{\rm Tr}\left({\cal D}_a{\cal A}_r\right)^2
+ c_{a\phi_1}~ {\rm Tr}\left({\cal D}_a{\cal A}_{\phi_1}\right)^2\right] \Big\}\nonumber\\
& + & \int d^4x \sum_{k=1}^3 \left[\sum_{a=0}^2 b_{ak}~ {\rm Tr}\left({\cal D}_a \varphi_k\right)^2 + b_{\psi k} ~{\rm Tr}\left({\cal D}_\psi \varphi_k\right)^2\right] \\
&+& \int d^4x \bigg\{ {c_1\over v_3}\Big(a_1~ {\rm Tr}\left[{\cal A}_r, {\cal A}_{\phi_1}\right]^2 + a_2 ~{\rm Tr}\left[{\cal A}_3, {\cal A}_r\right]^2
+ a_4~ {\rm Tr}\left[{\cal A}_3, {\cal A}_{\phi_1}\right]^2\Big) \nonumber\\
&+& \sum_{k, l} d_{kl} ~{\rm Tr}\left[\varphi_k, \varphi_l\right]^2 + \sum_{k=1}^3 \Big(c_{rk} ~{\rm Tr}\left[{\cal A}_r, \varphi_k\right]^2
+ c_{3k} ~{\rm Tr}\left[{\cal A}_3, \varphi_k\right]^2 + c_{\phi_1 k} ~{\rm Tr}\left[{\cal A}_{\phi_1}, \varphi_k\right]^2\Big)\bigg\}, \nonumber \nd
where the coefficients ($a_m, c_{mn}, b_{mn}, d_{mn}$) for all values of ($m, n$) specified above 
are functions of the constant NC or RR
parameter $\theta$. Since we have maintained supersymmetry in the M-theory construction, we expect the action to have, at least for certain
choices of the warp-factors, the maximal
${\cal N} = 4$ supersymmetry. In 
fact the choice of supersymmetry depends on the supersymmetry of the original type IIB background \eqref{sugra1} and \eqref{chote}. For specific choices of 
$F_i$ in \eqref{chote}, one of the NS5-brane in {\bf Table \ref{wittenbranes}} can be moved away from the other to allow for the maximal ${\cal N} = 4$ supersymmetry. Generically however 
\eqref{chote} has a ${\cal N} = 2$ or ${\cal N} = 1$ supersymmetry, implying atmost a $G_2$ structure for the M-theory seven-manifold \eqref{pwhite}.   

Looking at \eqref{stotal}, one may note that all the NC or RR
deformations appear {\it only} as constant coefficients for various terms in \eqref{stotal}. The presence or absence 
of the NC or RR deformations will not change the form of the effective action, except alter the coefficients ($c_{mn}, b_{mn}, d_{mn}, a_m$) a bit.  
An interesting question at this stage is to see what additional constraints on these coefficients appear from minimizing the energy of the system. These would of course
be the BPS conditions, 
and once the BPS conditions are satisfied the EOMs will be automatically satisfied. Our original configuration \eqref{sugra1} with the choice of dilaton \eqref{dileshwar} 
and the internal space \eqref{chote} satisfy EOMs in the absence of any BPS states on the type IIB fractional D3-branes. To satisfy the EOMs in the {\it presence} of
the BPS states would require us to find static configurations on the branes that minimize the total energy of the system. This in turn would require us to compute the 
Hamiltonian and search for the static BPS configurations by minimizing this.   

To determine the constraints on the warp-factors, i.e the constant coefficients ($c_{mn}, b_{mn}, d_{mn}, a_m$) appearing in \eqref{stotal}, we first proceed to determine the 
the BPS configurations. For consistency,
these configurations should satisfy the Gauss' constraint. We isolate the scalar ${\cal A}_3$, and express the Gauss' law constraint in the following
way:
\bg\label{gauss}
c_{11} {\cal D}_\alpha {\cal F}_{\alpha 0} + c_{12} {\cal D}_\psi {\cal F}_{\psi 0} &=& ic_{03}\left[{\cal A}_3, {\cal D}_0 {\cal A}_3\right]  + 
ic_{0r}\left[{\cal A}_r, {\cal D}_0 {\cal A}_r\right] + ic_{0\phi_1}\left[{\cal A}_{\phi_1}, {\cal D}_0 {\cal A}_{\phi_1}\right] \nonumber\\ 
&&~~~~~~~~~ + \sum_{k = 1}^3~{iv_3 b_{0k}\over c_1} \left[\varphi_k, {\cal D}_0\varphi_k\right], \nd
where ($c_{mn}, b_{mn}$) are exactly the coefficients that appear in \eqref{stotal}. We have also divided $a =$ ($ 0, 1, 2$) $\equiv$ ($0, \alpha$) where 
$\alpha = 1, 2$. 

Secondly, looking at {\bf Table \ref{brenulia}} we can identify the scalar fields $\overrightarrow{X}$ and $\overrightarrow{Y}$ used in \cite{wittenknots1}. This will be useful 
when we want to express the BHN equations in terms of the scalar field components used here. The scalar fields $\overrightarrow{X}$ and $\overrightarrow{Y}$ can be identified as:
\bg\label{xandy}
\overrightarrow{X} ~ \equiv ~ \left({\cal A}_3, \varphi_1, \varphi_2\right), ~~~~~~~
\overrightarrow{Y} ~ \equiv ~ \left({\cal A}_r, {\cal A}_{\phi_1}, \varphi_3\right), \nd
which appears from the fact that a part of 
the Coulomb branch for the NS5-D3 system as shown in {\bf Table \ref{brenulia}}, is along the ($x_3, x_8, x_9$) directions . 
This also means, 
associated with the components of the gauge fields ${\cal A}_\mu =$ (${\cal A}_0, {\cal A}_1, {\cal A}_2, {\cal A}_\psi$) in four-dimensions, we can now identify 
approximately the four scalars 
used in \cite{wittenknots1} as\footnote{Note that the identification \eqref{4scalars} differs slightly from \cite{wittenknots1}. For example, using \eqref{4scalars}, 
$\overrightarrow{X}$ would be ($\phi_1, \phi_2, \phi_3$), whereas in \cite{wittenknots1} it is ($\phi_0, \phi_1, \phi_2$). We will consider a different mapping of the scalars in \eqref{foursca} later.
Furthermore to avoid cluttering of symbols we will use 
the same symbol
to denote the {\it twisted} and the {\it untwisted} scalars of \cite{wittenknots1}, unless mentioned otherwise. 
It should hopefully
be clear from the context which one is meant. \label{shollow}}:
\bg\label{4scalars}
\left(\phi_0, \phi_1, \phi_2, \phi_3\right) ~ \propto ~ \left(\varphi_3, \varphi_1, \varphi_2, {\cal A}_3\right), \nd
which, as described in \cite{wittenknots1}, can be made by picking the three scalar fields in $\overrightarrow{X}$ and one scalar field from $\overrightarrow{Y}$ (which we take 
here as $\varphi_3$). This means the complex $\sigma$ field of \cite{wittenknots1}, for our case will become:
\bg\label{sigma}
\sigma ~\equiv ~ {\cal A}_r + i {\cal A}_{\phi_1}. \nd 
The Gauss law constraint and the identification of the scalar fields 
will lead us to compute the Hamiltonian from the total effective action \eqref{stotal}. Isolating the same scalar ${\cal A}_3$, 
the expression for the Hamiltonian, for the case when $c_2 = 0$ in \eqref{stotal}, 
can be expressed as sum of squares of various terms in the following way:
\bg\label{hamilbeta} 
{\cal H} & = & \int d^3 x ~{\rm Tr}\Bigg\{\sum_{\alpha =1}^2 {c_1\over v_3} \left(\sqrt{c_{11}} {\cal F}_{\alpha 0} - \sqrt{c_{\alpha 3}} {\cal D}_\alpha {\cal A}_3\right)^2 
+ {c_1\over v_3} \left(\sqrt{c_{12}} {\cal F}_{\psi 0} - \sqrt{c_{\psi 3}} {\cal D}_\psi {\cal A}_3\right)^2 \nonumber\\
&+& {c_1\over v_3} \left(\sqrt{c_{0r}}{\cal D}_0 {\cal A}_r - i\sqrt{a_2}[{\cal A}_3, {\cal A}_r]\right)^2    
+ {c_1\over v_3} \left(\sqrt{c_{0\phi_1}}{\cal D}_0 {\cal A}_{\phi_1} - i\sqrt{a_4}[{\cal A}_3, {\cal A}_{\phi_1}]\right)^2 \nonumber\\
&+& {c_1\over v_3}\left(s^{(1)} c_{\psi r}({\cal D}_\psi {\cal A}_r)^2+ s^{(2)} c_{\psi \phi_1}({\cal D}_\psi {\cal A}_{\phi_1})^2 + t^{(1)} c_{\beta r}({\cal D}_\beta {\cal A}_r)^2
+ t^{(2)} c_{\beta \phi_1}({\cal D}_\beta {\cal A}_{\phi_1})^2\right)\nonumber\\
& + & \sum_{k = 1}^3 \left(\sqrt{b_{0k}}{\cal D}_0 \varphi_k - i\sqrt{c_{3k}}[{\cal A}_3, {\varphi}_k]\right)^2 + {c_1 c_{03}\over v_3}\left({\cal D}_0 {\cal A}_3\right)^2 
+ \sum_{\alpha, \beta = 1}^2 \bigg(\sqrt{c_1c_{11}\over 2 v_3} {\cal F}_{\alpha\beta} \nonumber \\
&+&\sqrt{c_1c_{\psi r}\over v_3}s^{(1)}_{\alpha\beta}\epsilon_{\alpha \beta \psi r} {\cal D}_\psi {\cal A}_r     
+\sqrt{c_1c_{\psi \phi_1}\over v_3}s^{(2)}_{\alpha\beta}\epsilon_{\alpha \beta \psi \phi_1} {\cal D}_\psi {\cal A}_{\phi_1} 
+\sum_{\delta = 1}^3\sum_{k =1}^3 \sqrt{b_{\delta k}}\epsilon_{\alpha \beta} \cdot m^{(1)}_{\delta k} {\cal D}_\delta \varphi_k \nonumber\\
&-& \sum_{k, l} ig^{(1)}_{\alpha\beta kl}\sqrt{d_{kl}}\left[\varphi_k, \varphi_l\right] 
- \sum_{k=1}^3 i\Big(g^{(2)}_{\alpha\beta k}\sqrt{c_{rk}}\left[{\cal A}_r, \varphi_k\right]          
+ g^{(3)}_{\alpha\beta k}\sqrt{c_{\phi_1 k}} \left[{\cal A}_{\phi_1}, \varphi_k\right]\Big) \nonumber\\ 
&-& ig^{(4)}_{\alpha\beta} \sqrt{c_1 a_1 \over  v_3}\left[{\cal A}_r, {\cal A}_{\phi_1}\right]\bigg)^2 
+ {\left({\bf Q_E} + {\bf Q_M}\right)\delta^3 x\over {\rm dim}~G}
+ \sum_{\alpha = 1}^2 \bigg(\sqrt{c_1c_{12}\over 2 v_3} {\cal F}_{\alpha\psi} +\sqrt{c_1c_{\beta r}\over v_3}t^{(1)}_{\alpha}
\epsilon_{\alpha \psi \beta r} {\cal D}_\beta {\cal A}_r \nonumber\\    
&+&\sqrt{c_1c_{\beta \phi_1}\over v_3}t^{(2)}_{\alpha}\epsilon_{\alpha \psi \beta \phi_1} {\cal D}_\beta {\cal A}_{\phi_1} 
+ \sum_{\delta = 1}^3\sum_{k =1}^3 \sqrt{b_{\delta k}}\epsilon_{\alpha \psi}\cdot m^{(2)}_{\delta k} {\cal D}_\delta \varphi_k 
- \sum_{k, l} ih^{(1)}_{\alpha\psi kl}\sqrt{d_{kl}}\left[\varphi_k, \varphi_l\right] \nonumber\\
&-& \sum_{k=1}^3 i\Big(h^{(2)}_{\alpha\psi k}\sqrt{c_{rk}}\left[{\cal A}_r, \varphi_k\right]
+h^{(3)}_{\alpha \psi k}\sqrt{c_{\phi_1 k}} \left[{\cal A}_{\phi_1}, \varphi_k\right]\Big) 
- i h^{(4)}_{\alpha \psi}\sqrt{c_1 a_1 \over  v_3}\left[{\cal A}_r, {\cal A}_{\phi_1}\right]\bigg)^2 \nonumber\\
& + & \sum_{k, l} q^{(1)}_{kl} d_{kl} ~\left[\varphi_k, \varphi_l\right]^2 + \sum_{k=1}^3 \sum_{\gamma=2}^3 q^{(\gamma)}_k c_{y_\gamma k} 
~\left[{\cal A}_{y_\gamma}, \varphi_k\right]^2
+ {q^{(4)}c_1 a_1 \over  v_3}\left[{\cal A}_r, {\cal A}_{\phi_1}\right]^2\Bigg\}, \nd 
where ${\bf Q_E}$ and ${\bf Q_M}$ are the electric and the magnetic charges respectively, which will be determined later; dim $G$ is the dimension of the group;  
and $\delta \equiv$ ($\alpha, \psi$), $(y_2, y_3) \equiv (r, \phi_1)$. 
Most of coefficients appearing in \eqref{hamilbeta} have been 
determined earlier, which the readers may want to look up. The other coefficients appearing above are defined in the following way:
\bg\label{codelulu}
g^{(1)}_{\alpha\beta kl} ~ \equiv ~ g^{(1)}_{[\alpha\beta] [kl]}, ~~~~ g^{(m)}_{\alpha\beta k} ~ \equiv ~ g^{(m)}_{[\alpha\beta] k}, ~~~~ 
g^{(m)}_{\alpha\beta} ~ \equiv ~ g^{(m)}_{[\alpha\beta]}, \nd
\begin{table}[h!]
 \begin{center}
\begin{tabular}{|c||c||c|c|c|c|c|c|c|c|c|c|c|}\hline Theory & Configurations & $x_0$ & $x_1$ & $x_2$
& $x_3$ & $\theta_1$ & $\phi_1$ & $\psi$ & $r$ & $x_8$ & $x_9$ & $x_{11}$ \\ \hline
IIB & NS5 & $\surd$  & $\surd$   & $\surd$   & $\surd$ & $\ast$  & $\ast$ & $\ast$ & $\ast$ & $\surd$
& $\surd$ & $\ast$ \\  \hline
IIB & D3 & $\surd$  & $\surd$   & $\surd$   & $\ast$ & $\ast$ & $\ast$ & $\surd$ & $\ast$ & $\ast$
& $\ast$ & $\ast$ \\  \hline
IIA & D4 & $\surd$  & $\surd$   & $\surd$   & $\surd$ & $\ast$  & $\ast$ & $\surd$ & $\ast$ & $\ast$  & $\ast$ & $\ast$ \\  \hline
IIB & D5/$\overline{\rm D5}$ & $\surd$  & $\surd$   & $\surd$   & $\surd$ & $\ast$  & $\ast$ & $\surd$ & $\surd$ & $\ast$  & $\ast$ & $\ast$ \\  \hline
IIA & D6 & $\surd$  & $\surd$   & $\surd$   & $\surd$ & $\ast$  & $\surd$ & $\surd$ & $\surd$ & $\ast$  & $\ast$ & $\ast$ \\  \hline
M & ${\rm TN}_N$ & $\ast$  & $\ast$   & $\ast$   & $\ast$ & $\surd$  & $\ast$ & $\ast$ & $\ast$ & $\surd$  & $\surd$ & $\surd$ \\  \hline
M & $\Sigma_3$ & $\ast$  & $\ast$   & $\ast$   & $\surd$ & $\ast$  & $\ast$ & $\surd$ & $\surd$ & $\ast$  & $\ast$ & $\ast$ \\  \hline
\end{tabular}
\end{center}
  \caption{The orientations of branes and manifolds at various stages of dualities in our set-up.}
  \label{brenulia}
\end{table}
\noindent and similarly for ($h^{(j)}_{...}, s^{(j)}_{...}, t^{(j)}_{...}$).
In other words they are all generically taken 
to be anti-symmetric\footnote{For $\varphi_k$ it will be instructive to resort to the identification \eqref{4scalars} to discuss anti-symmetry.} 
with respect to ($\alpha, \beta$), ($\alpha, \psi$), and ($k, l$), except for $m^{(j)}_{\delta k}$ where the symmetric part will play some role later. 
Assuming this, the relation between them are now easy to work out from the definition of the Hamiltonian in \eqref{hamilbeta} as:    
\bg\label{coeffrelo}
&&2\left|g^{(4)}_{12}\right|^2 + \left|h^{(4)}_{1\psi}\right|^2 + \left|h^{(4)}_{2\psi}\right|^2 - q^{(4)} ~ = ~ 1\nonumber\\ 
&&2\left|g^{(n)}_{12k}\right|^2 + \left|h^{(n)}_{1\psi k}\right|^2 + \left|h^{(n)}_{2\psi k}\right|^2 - q^{(n)}_{k} ~ = ~ 1\nonumber\\ 
&&2\left|g^{(1)}_{12kl}\right|^2 + \left|h^{(1)}_{1\psi kl}\right|^2 + \left|h^{(1)}_{2\psi kl}\right|^2 - q^{(1)}_{kl} ~ = ~ 1\nonumber\\ 
&& 2\left|s^{(l)}_{12}\right|^2 + s^{(l)}  = 1, \sum_{\alpha = 1}^2\left|t^{(l)}_{\alpha}\right|^2 + t^{(l)}  = 1, \sum_{j=1}^2\left|m^{(j)}_{\delta k}\right|^2 = {1\over 2}, \nd
where $n = 2, 3$ and $l = 1, 2$. Note that the last relation for coefficients
$m^{(j)}_{\delta k}$ can have additional pieces depending on how the 
kinetic piece $({\cal D}_\delta \varphi_k)^2$ is defined in the action \eqref{stotal}. We will discuss this later. In general however all 
the coefficients appearing above are generic (they should of course satisfy \eqref{coeffrelo})
and we will determine them for a special configuration that resonates with \cite{wittenknots1}. 
For the time being we want to identify
generic BPS configurations by minimizing the energy of the system. We start by taking {\it static} configurations with the following gauge choice:
\bg\label{gaugec}
{\cal A}_0 ~ = ~ {\cal A}_3, \nd
which is motivated, in retrospect, from our choice of isolating the scalar field ${\cal A}_3$ from the very begining in the expression for the Hamiltonian \eqref{hamilbeta}. 
The gauge choice \eqref{gaugec} implies the following constraints on ${\cal A}_3$ field from \eqref{hamilbeta}:
\bg\label{sbtaaja}
&& {\cal D}_0 {\cal A}_3 = 0, ~~~~~ \left(\sqrt{b_{0k}} - \sqrt{c_{3k}}\right)^2 \left[{\cal A}_3, {\varphi}_k\right]^2 = 0 \nonumber\\
&& \left(\sqrt{c_{11}} - \sqrt{c_{\alpha 3}}\right)^2 \left({\cal D}_\alpha {\cal A}_3\right)^2 = 0, ~~~~
\left(\sqrt{c_{12}} - \sqrt{c_{\psi 3}}\right)^2 \left({\cal D}_\psi {\cal A}_3\right)^2 = 0 \nonumber\\
&& \left(\sqrt{c_{0r}} - \sqrt{a_2}\right)^2 \left[{\cal A}_3, {\cal A}_r\right]^2 = 0, ~~~~
\left(\sqrt{c_{0\phi_1}} - \sqrt{a_4}\right)^2 \left[{\cal A}_3, {\cal A}_{\phi_1}\right]^2 = 0. \nd
The first equation is automatically satisfied once we demand static configurations. The other covariant derivatives, or the commutator brackets cannot vanish unless we take trivial 
solutions. This observation leads to two possible set of solutions to the system of equations in \eqref{sbtaaja}. The first set of solutions is when ${\cal A}_3 = 0$. The second set
of solutions is for 
the coefficients, associated to the various configurations of the ${\cal A}_3$ fields, to vanish. In the following, we will first discuss the second set of solutions wherein the 
coefficients vanish. 
To check whether this is possible, let us study the coefficient associated with ${\cal D}_\alpha {\cal A}_3$.
Comparing \eqref{c11theta} and \eqref{tentu} and for the benefit of discussion we can re-express the two coefficients appearing in \eqref{sbtaaja} as: 
\bg\label{pacchor} 
&& c_{11}(\theta) = {R_3~\sec~\theta}\int_0^\infty dr~e^{2\phi_0}\sqrt{F_1 \widetilde{F}_2 F_3\over \widetilde{F}_2 - F_3}
~{\rm ln}\left|{\sqrt{\widetilde{F}_2} +\sqrt{\widetilde{F}_2 - F_3}\over \sqrt{\widetilde{F}_2} -\sqrt{\widetilde{F}_2 - F_3}}\right| \nonumber\\
&& c_{\alpha 3}(\theta) = {R_3~\sec~\theta}\int_0^\infty dr~{e^{2\phi_0}\over H_2} \sqrt{F_1 \widetilde{F}_2 F_3\over \widetilde{F}_2 - F_3}
~{\rm ln}\left|{\sqrt{\widetilde{F}_2} +\sqrt{\widetilde{F}_2 - F_3}\over \sqrt{\widetilde{F}_2} -\sqrt{\widetilde{F}_2 - F_3}}\right|. \nd
We see that they are exactly identical except for the appearance of the $H_2$ term in the second integral. In fact this observation repeats for all the doublet 
coeffcients appearing in \eqref{sbtaaja}, 
namely, ($c_{12}, c_{\psi 3}$) in \eqref{krisis} and \eqref{fintubabu} respectively; ($c_{0r}, a_2$) in \eqref{bentu} and \eqref{a2the} respectively; 
($c_{0\phi_1}, a_4$) in \eqref{jhogra} and \eqref{a4the1} respectively; 
and ($b_{0k}, c_{3k}$) in \eqref{boka} and \eqref{keyjar} respectively, in exactly the same way: they all differ by the
presence of the $H_2$ term in the integral! This conclusion will not change if we take the RR deformation instead, or if we consider the full expression for the dilaton 
\eqref{dileshwar}. All the differences of the coefficients in \eqref{sbtaaja} take the following form:
\bg\label{ckjwala}
c_{(a)} - c_{(b)} \equiv \int_0^\infty dr~G_{(ab)}(r) (1 - b_4), \nd
where $c_{(a)} \equiv$ ($c_{mn}, b_{mn}, d_{mn}, a_m$), $b_4$ as defined in \eqref{bfour},  
and the explicit forms of the $G_{(ab)}$ functions can be read up from \eqref{c11theta}, \eqref{tentu}, \eqref{krisis}, \eqref{fintubabu}
etc., as mentioned above. The result for RR deformation can be expressed as \eqref{ckjwala} with $b_4(\phi)$, whereas with $b_4(\phi = 0)$ we get the results for the NC deformation. 
Therefore the vanishing of the integral in \eqref{ckjwala} implies the vanishing of the NC or the RR deformation parameter $\theta$, or in the language of \eqref{c12def}, the 
vanishing of $\Theta$ implying further that in our four-dimensional gauge theory:
\bg\label{taut}
\tau ~ \equiv ~{4\pi i\over g^2_{\rm YM}}.\nd  
This is of course consistent with our simplifying choice of $c_2 = 0$ in \eqref{stotal} and \eqref{hamilbeta} and also with  
the observations of \cite{langland}, \cite{wittenknots1} and \cite{wittenknots2}, namely that the four-dimensional supersymmetry 
in the presence of BPS configurations\footnote{For example like Wilson loops etc., that we will discuss soon.} is 
only preserved when $\theta$ vanishes. However when ${\cal A}_3$ vanishes, which is our second set of solutions, we are basically restricted to the three-dimensional boundary 
${\bf W}$ of \eqref{vwr+} where $\theta$ in general could be non-zero\footnote{Here $c_2$ may be made to vanish by taking $q(\theta) = 0$ for non-zero $\theta$. Thus switching on $q(\theta)$ would 
imply switching on $c_2$.}. 
Therefore to summarize, we have the following two sets of solutions:
\bg\label{twosets}
&& {\rm Set}~1: ~({\cal A}_3 ~\ne~ 0, ~~ \theta ~ = ~ 0)  \nonumber\\
&& {\rm Set}~2: ~({\cal A}_3 ~=~ 0, ~~ \theta ~ \ne ~ 0). \nd  
Our next series of conditions, which in principle should be valid for either of the above two sets of solutions \eqref{twosets} but will only consider Set 2 henceforth,  
appear from looking at the third and the last lines of \eqref{hamilbeta}. Since the coefficients ($c_{\psi r}, c_{\psi \phi_1}, c_{\beta r}, c_{\beta \phi_1}$)
in \eqref{fintubabu}, \eqref{lebu}, \eqref{bentu} and \eqref{jhogra} respectively are all non-zero, and we will assume ($s^{(n)}, t^{(n)}, q^{(4)}$) 
also to be generically 
non-zero, minimization of the Hamiltonian \eqref{hamilbeta} implies the following conditions on the two scalar fields ${\cal A}_{r}$ and ${\cal A}_{\phi_1}$:
\bg\label{scalfy} 
{\cal D}_\eta {\cal A}_{r} ~ = ~ {\cal D}_\eta {\cal A}_{\phi_1} ~ = ~ \left[{\cal A}_{r}, {\cal A}_{\phi_1}\right] ~ = ~ 0, \nd
with $\eta \equiv$ ($\alpha, \psi$). Thus these scalar fields, appearing in $\overrightarrow{Y}$ in \eqref{xandy}, are covariantly constants and have a vanishing commutator bracket. In the 
language of the complex field $\sigma$ in \eqref{sigma}, the relations in \eqref{scalfy} imply the following conditions on ($\sigma, \bar{\sigma}$):
\bg\label{rater}
{\cal D}_\eta \sigma  ~ = ~ {\cal D}_\eta \bar{\sigma} ~ = ~ \left[\sigma, \bar{\sigma}\right] ~ = ~ 0, \nd
which is also the conditions imposed on ($\sigma, \bar{\sigma}$) fields in \cite{wittenknots1}. Additionally,
it is interesting to note that, since we took ($s^{(n)}, t^{(n)}, q^{(4)}$) to be
non-zero, the first and the last set of equations in \eqref{coeffrelo} can be easily satisfied. Thus they do not impose further constraints on the BPS equations \eqref{scalfy}. Finally,
we can completely decouple the scalars (${\cal A}_r, {\cal A}_{\phi_1}$) if we demand:
\bg\label{kyabol}
\left[{\cal A}_r, \varphi_k\right] ~ = ~ \left[{\cal A}_{\phi_1}, \varphi_k\right] ~ = ~ 0, \nd
for any values of $q_k^{(\gamma)}$ in \eqref{hamilbeta}. This way the second set of equations for $n = 2, 3$ in \eqref{coeffrelo}
can also be easily satisfied without introducing any additional constraints.

We are finally left with two sets of equations in \eqref{coeffrelo} that need to be  satisfied. These are important equations as they deal with the commutator brackets 
$\left[\varphi_k, \varphi_l\right]$ and covariant derivatives ${\cal D}_\delta \varphi_k$. 
We first demand that the commutator brackets do not vanish $-$ at least not all the brackets $-$
to avoid the system from becoming completely trivial. This immediately implies $q_{kl}^{(1)} =0$ 
for some choices of ($k, l$) to 
satisfy the BPS conditions from the Hamiltonian \eqref{hamilbeta} (see the last line of \eqref{hamilbeta}). The equations for the other coefficients from \eqref{coeffrelo} then 
become:
\bg\label{hotat}
2\left|g^{(1)}_{12kl}\right|^2 + \left|h^{(1)}_{1\psi kl}\right|^2 + \left|h^{(1)}_{2\psi kl}\right|^2 ~ = ~ 1, ~~~ \left|m^{(1)}_{\delta k}\right|^2 
+ \left|m^{(2)}_{\delta k}\right|^2 = {1\over 2}, \nd
again for the specific choices of ($k, l$). 
To see what values of the coefficients could solve the above set of equations \eqref{hotat}, let us write down the corresponding BPS equations that
use these coefficients. 
The  simplest case is when only {\it one} commutator bracket doesn't vanish, i.e when $q^{(1)}_{12} = 0$. This means  
the field $\varphi_3$ will commute with the other two scalar fields $\varphi_1$ and $\varphi_2$. In other words, we 
take\footnote{One might worry that \eqref{phi3c} could be too strong a constraint that would eventually trivialize some of the boundary terms in \eqref{QoM}, \eqref{nest} or in \eqref{aletex}. 
This is however not true because the boundary theory will be developed {\it without} resorting to any constaints so that the boundary degrees of freedom may capture the fluctuations over any classical 
configurations. As an aside, note that we can allow all but one of $q^{(1)}_{kl}$ to vanish so that we are not obliged to impose the full set of \eqref{phi3c}. The remaining decouplings may be achieved 
by choosing appropriate values for $g_{12kl}^{(1)}, h_{a\psi kl}^{(1)}$.}: 
\bg\label{phi3c}\left[\varphi_3, \varphi_1\right] ~ = ~ \left[\varphi_3, \varphi_2\right] ~ = ~ 0.\nd
The first equation of \eqref{hotat} then
connects the gauge-field ${\cal F}_{12}$ with the scalar fields in $\overrightarrow{X}$ defined earlier as \eqref{xandy} in the following way\footnote{Expectedly, because of our
gauge choice \eqref{gaugec}, the Nahm equation will have ${\cal D}_\psi \varphi_3$ and $[\varphi_1, \varphi_2]$ which is slightly different from what one would have expected from the 
orientations of the branes 
in {\bf Table \ref{brenulia}}. This generic formalism is more useful for later development so we will mostly concentrate on this. Again, a more {\it standard} formalism is
also possible and we will discuss it briefly for the gauge choice \eqref{gaugec2} later in this section.}:
\bg\label{f12bhn}
{\cal F}_{12} + \sqrt{b_{\psi 3} v_3\over c_1 c_{11}}~{\cal D}_\psi \varphi_3 
+ \sqrt{b_{12} v_3\over c_1 c_{11}}\left({\cal D}_1 \varphi_2 - {\cal D}_2 {\varphi}_1\right)
- 2 i\sqrt{v_3 d_{12}\over c_1 c_{11}}~\left[\varphi_1, \varphi_2\right] = 0, \nd
where ($b_{\psi 3}, b_{12}, c_1, c_{11}, v_3, d_{12}$) 
are given in \eqref{bokcho}, \eqref{boka}, \eqref{c1c2c1c2}, \eqref{c11theta}, \eqref{intig} and \eqref{venu} respectively. The above equation 
is one of the Bogomolnyi-Hitchin-Nahm (BHN) equation that appears from our analysis. In fact the generic equation that we get from \eqref{hamilbeta} is more complicated than 
\eqref{f12bhn}, but we have simplified the system by assuming the following values of the coefficients: 
\bg\label{g12kl}
g^{(1)}_{1212} ~ = ~ m^{(1)}_{\psi 3} ~ = ~ m^{(1)}_{12} ~ = ~ {1\over \sqrt{2}}, \nd
with other coefficients, except $m^{(j)}_{11}$ and $m^{(j)}_{22}$, 
vanishing. This in turn
is motivated in part to bring the BHN equation in a more {\it standard} form like \eqref{f12bhn} with
\bg\label{bhn1} 
m^{(j)}_{11}\sqrt{b_{11}}~{\cal D}_1\varphi_1 + m^{(j)}_{22}\sqrt{b_{22}}~{\cal D}_2\varphi_2 ~ = ~0 ~ \equiv ~ {\cal D}_1\varphi_1 + {\cal D}_2\varphi_2, \nd
which involves the symmetric coefficients $m^{(j)}_{11}$ and $m^{(j)}_{22}$ with, as we'll see below, $j = 2$ to avoid 
contradictions\footnote{We could also get \eqref{bhn1} by adding a term 
$\left({\sum}_{a} m_{aa} {\cal D}_a\varphi_a\right)^2$ to the Hamiltonian \eqref{hamilbeta}. This will 
only change the last equation in \eqref{coeffrelo}. \label{constraint}}. 
Without loss of generalities, they are taken to be equal; 
and $b_{11} = b_{22}$ as can be inferred from \eqref{boka}.

The choice \eqref{g12kl}, when plugged in \eqref{hotat}, would imply that both $h^{(1)}_{1\psi 12}$ as well as $h^{(2)}_{2\psi 12}$ vanish. However other coefficients can be 
non-zero, and as before we will make the following choice of the coefficients:
\bg\label{consco}
-h^{(1)}_{1\psi 1 \psi} ~ = ~ -h^{(2)}_{2\psi 2\psi} ~ = ~ m^{(2)}_{\beta 3} ~ = ~ m^{(2)}_{\psi \beta} ~ = ~ {1\over \sqrt{2}}, \nd
with the rest taken to be zero. For the time,
the above choice should be viewed as being motivated by consistency, and we will go beyond these special choices of coefficients 
\eqref{g12kl} and \eqref{consco} later on. With this in mind,
the BPS conditions lead to the following additional equation:
\bg\label{meyepita}
{\cal F}_{\alpha\psi} -6 \sum_{\delta, k} \sqrt{2 b_{\delta k} v_3\over c_1 c_{12}}~\epsilon_{[\alpha\psi} m^{(2)}_{\delta k]}{\cal D}_\delta \varphi_k 
+ \sqrt{2 b_{\psi \alpha} v_3\over c_1 c_{12}}~\epsilon_{\alpha\psi} m^{(2)}_{\psi \alpha}{\cal D}_\psi \varphi_\alpha 
~ = ~ 0, \nd 
where $\alpha = 1, 2$; $b_{\psi \alpha}$ and $b_{\alpha 3}$ as given in \eqref{bokcho} and \eqref{boka} respectively, 
and ($v_3, c_1, c_{11}, c_{12}$) are given in \eqref{intig}, 
\eqref{c1c2c1c2}, \eqref{c11theta} and \eqref{krisis} respectively. Note the way we arranged the anti-symmetric pieces together. 
This could be taken as the definition of the term $\epsilon_{ab}\cdot m^{(k)}_{cd}$ in \eqref{hamilbeta}.
We could do the same for \eqref{f12bhn}, but that is not necessary 
because of our choice of coefficients \eqref{g12kl}. 
The above equation is valid for Set 1 in \eqref{twosets}, but we can always use Set 2 by switching on the NC 
or the RR
parameter $\theta$ and interpret the coefficients apearing in \eqref{f12bhn} accordingly. For this case, \eqref{meyepita} will give rise to the following two equations:
\bg\label{bhishon}
&&{\cal F}_{1\psi} + \sqrt{b_{23} v_3\over c_1 c_{12}}~{\cal D}_2 \varphi_3 + \sqrt{b_{\psi 1} v_3\over c_1 c_{12}}~{\cal D}_\psi \varphi_1 ~ = ~ 0 \nonumber\\
&&{\cal F}_{2\psi} + \sqrt{b_{13} v_3\over c_1 c_{12}}~{\cal D}_1 \varphi_3 + \sqrt{b_{\psi 2} v_3\over c_1 c_{12}}~{\cal D}_\psi \varphi_2 ~ = ~ 0, \nd
without involving any commutator brackets. Thus combining \eqref{f12bhn} with the two equations in \eqref{bhishon}, for Set 2 in \eqref{twosets}, we have our three BHN equations 
for the system.

Before ending this section, let us what would happen if our gauge choice were different from \eqref{gaugec}. One example would be to choose the 
following gauge where:
\bg\label{gaugec2} {\cal A}_0 = {\cal A}_r. \nd
Looking at the action \eqref{stotal} we see that there is a symmetry between $x_3$ and $r$, implying that we can re-write the Hamiltonian \eqref{hamilbeta} in the gauge 
\eqref{gaugec2} simply by exchanging the two coordinates! The BPS condition then changes from \eqref{sbtaaja} to the following new conditions that are easy to derive:
\bg\label{sbtaaja2}
&& {\cal D}_0 {\cal A}_r = 0, ~~~~~ \left(\sqrt{b_{0k}} - \sqrt{c_{rk}}\right)^2 \left[{\cal A}_r, {\varphi}_k\right]^2 = 0 \nonumber\\
&& \left(\sqrt{c_{11}} - \sqrt{c_{\alpha r}}\right)^2 \left({\cal D}_\alpha {\cal A}_r\right)^2 = 0, ~~~~
\left(\sqrt{c_{12}} - \sqrt{c_{\psi r}}\right)^2 \left({\cal D}_\psi {\cal A}_r\right)^2 = 0 \nonumber\\
&& \left(\sqrt{c_{03}} - \sqrt{a_2}\right)^2 \left[{\cal A}_3, {\cal A}_r\right]^2 = 0, ~~~~
\left(\sqrt{c_{0\phi_1}} - \sqrt{a_1}\right)^2 \left[{\cal A}_r, {\cal A}_{\phi_1}\right]^2 = 0. \nd
The non-trivial issue is to verify that the coefficients do vanish in the limit ${\cal A}_r \ne 0$, just as it were for the case when ${\cal A}_3 \ne 0$ in \eqref{sbtaaja}. To see whether 
this is still the case, let us consider two coefficients $c_{0\phi_1}$ in \eqref{jhogra} and $a_1$ in \eqref{a1the}. For the benefit of the discussion, we reproduce them once again as:
\bg\label{cibelma}
&& a_1(\theta)  =  R_3~\sec~\theta \int_0^\infty dr ~{\widetilde{a_1}\sqrt{F_1 \widetilde{F}_2 F_3} \over F_1} \left(2 \sqrt{\widetilde{F}_2} 
+ {\widetilde{a}_2F_3 \over \widetilde{a}_1 \sqrt{\widetilde{F}_2 - F_3}} {\rm ln}\left|{\sqrt{\widetilde{F}_2} + \sqrt{\widetilde{F}_2-F_3}\over 
\sqrt{\widetilde{F}_2} - \sqrt{\widetilde{F}_2-F_3}} \right|\right) \nonumber\\
&& c_{0\phi_1}(\theta)  =  {R_3~\sec~\theta}\int_0^\infty dr~{\widetilde{a_1}\sqrt{F_1 \widetilde{F}_2 F_3}\over e^{-2\phi_0}}
\left(2\sqrt{\widetilde{F}_2} + {\widetilde{a}_2F_3\over \widetilde{a}_1\sqrt{\widetilde{F}_2 - F_3}}
~{\rm ln}\left|{\sqrt{\widetilde{F}_2} + \sqrt{\widetilde{F}_2 - F_3}\over \sqrt{\widetilde{F}_2} - \sqrt{\widetilde{F}_2 - F_3}}\right|\right), \nonumber\\ \nd
where $\widetilde{a_1}$ and $\widetilde{a_2}$ are defined in \eqref{widaid}. The above two expressions for the coefficients are well defined for any choices of the 
warp-factors $F_1$ as we discussed earlier. We now see that the two coefficients in \eqref{cibelma} would be the same when:
\bg\label{chor} e^{2\phi_0} F_1 = 1. \nd
This condition on $F_1$ remains the same if we compare the other coeffcients appearing in \eqref{sbtaaja2} namely ($b_{0k}, c_{rk}$) from \eqref{boka} and \eqref{himleng}; 
($c_{11}, c_{\alpha r}$) from \eqref{c11theta} and \eqref{bentu}; ($c_{12}, c_{\psi r}$) from \eqref{krisis} and \eqref{finbab}; and ($c_{03}, a_2$) from \eqref{tentu} and 
\eqref{a2the} respectively. This is illustrated in {\bf Table \ref{gajarh}}. However since $F_1$ is taken to be a non-trivial function in general, it may not always be possible to impose 
\eqref{chor}. Thus in this gauge we can take ${\cal A}_r = 0$ and $\theta \ne 0$. Interestingly however demanding ${\cal A}_r \ne 0$ doesn't imply vanishing $\theta$. This is therefore
different from \eqref{twosets} that we had for the ${\cal A}_3$ gauge.  

Most of the other details, regarding the Hamiltonian, Hitchin equations etc should be similar to what we 
discussed earlier once we replace $x_3$ with $r$. This also means that the complex $\sigma$ field \eqref{sigma} will now be $\sigma = {\cal A}_3 + i {\cal A}_{\phi_1}$ satisfying relations
similar to \eqref{rater}. The decoupling of the ${\cal A}_3$ and ${\cal A}_{\phi_1}$ scalars would follow relations similar to \eqref{kyabol}.  

\begin{table}[h!]
 \begin{center}
\begin{tabular}{|c||c||c|}\hline ${\cal A}_0 = {\cal A}_3$ & ${\cal A}_0 = {\cal A}_r$ & Relevant Equations \\ \hline
$c_{11}, c_{\alpha 3}$ & $c_{11}, c_{\alpha r}$ & \eqref{c11theta}, \eqref{tentu}, \eqref{bentu} \\ \hline
$b_{0k}, c_{3 k}$ & $b_{0k}, c_{r k}$ & \eqref{boka}, \eqref{keyjar}, \eqref{himleng} \\ \hline
$c_{12}, c_{\psi 3}$ & $c_{12}, c_{\psi r}$ & \eqref{krisis}, \eqref{fintubabu}, \eqref{finbab} \\ \hline
$c_{0r}, a_2$ & $c_{03}, a_2$ & \eqref{bentu}, \eqref{tentu}, \eqref{a2the}\\ \hline
$c_{0 \phi_1}, a_{4}$ & $c_{0 \phi_1}, a_{1}$ & \eqref{jhogra}, \eqref{a4the1}, \eqref{a1the} \\ \hline
$H_2 = 1$ & $e^{2\phi_0} F_1 = 1$ & \eqref{wardaf}, \eqref{chote} \\ \hline
\end{tabular}
\end{center}
\caption{Comparing various pairs of coefficients in the action for two different gauge choices
${\cal A}_0 = {\cal A}_3$ and
${\cal A}_0 = {\cal A}_r$. The last entries
give us the BPS conditions which can be got by
demanding equality between the individual pair of coefficients for the two gauge choices.}
\label{gajarh}
\end{table}

We could also discuss a slightly different formalism with the gauge choice \eqref{gaugec2} where the Nahm equation from the corresponding BHN equation may take a more standard 
form\footnote{Alternatively we could take the same gauge choice \eqref{gaugec} but use a different mapping \eqref{foursca} of the scalars instead of the original mapping \eqref{4scalars}. In fact
the mapping \eqref{foursca} will be useful later to elucidate the physics in the presence of a surface operator.}. 
For example
with a different choice of the Hamiltonian we may get our BHN equation to take the following form that is a slight variant of \eqref{f12bhn}:
\bg\label{pharmacy}
{\cal F}_{12} + \sqrt{c_{\psi 3}\over c_{11}}~{\cal D}_\psi {\cal A}_3 - 2 i\sqrt{v_3 d_{12}\over c_1 c_{11}}~[\varphi_1, \varphi_2]~ = ~ 0, \nd
and similarly for the equations for ${\cal F}_{\alpha \psi}$. We can see that the Nahm reduction of the above equation implies that the scalar fluctuations (${\cal A}_3, \varphi_1, \varphi_2$)
are all restricted to the Coulomb branch of the original D3-brane picture as depicted in {\bf Table \ref{wittenbranes}}. This also means that the decoupled complex scalar $\sigma$ is now 
completely the Higgs branch scalar field combination $\sigma = \varphi_3 + i {\cal A}_{\phi_1}$. The story could be developed further, more or less along the line of our earlier discussions,
but we will not do it here and instead leave it as an exercise for our diligent reader.

\subsubsection{First look at the $t$ parameter and the BHN equations \label{flook}}

The analysis that we performed in the above section assumed $c_2 = 0$ for simplicity. It is now time to switch on the $c_2$ parameter and
see how the results changes. In the process we can 
analyze the three BHN equations \eqref{f12bhn} and \eqref{bhishon}. Our procedure would be to compare our results with the ones 
given in \cite{wittenknots1} and 
\cite{wittenknots2} and express them in a language suitable for later developments. First, we will write our complexified gauge 
coupling $\tau$ using supergravity variables. 
Switching on $c_2$ in \eqref{stotal} and \eqref{hamilbeta}, this is expressed as:
\bg\label{taude}
\tau ~\equiv ~ c_1\left(q~\sin~\theta + {ic_{11}\over v_3}\right), \nd
where the expression for ($c_1, c_{11}, v_3, q$) are given earlier as \eqref{c1c2c1c2}, \eqref{c11theta}, \eqref{intig} and \eqref{chotas}
respectively. The above expression \eqref{taude} is 
for NC deformation, and if we replace $\sin~\theta$ with ${1\over 2}~\sin~2\theta$ 
and assume that ($c_{11}, v_3$) are now given by \eqref{c11theta2} and \eqref{intig2} respectively, we
will get the functional form for $\tau$ with RR deformation $\theta$. In the following however we will continue using the NC deformation $\theta$, although the RR deformation is
equally easy to implement. 
To proceed, let us define another quantity called $t$, in the following way:
\bg\label{tkya}
t ~\equiv~  \pm {\left|\tau\right|\over \tau} ~= ~ \pm \left({v_3 q~\sin~\theta\over \sqrt{c_{11}^2 + v_3^2 q^2~\sin^2\theta}} 
- {ic_{11} \over \sqrt{c_{11}^2 + v_3^2 q^2~\sin^2\theta}}\right), \nd
which is in general a complex number, and becomes a purely imaginary number $t = \pm i$ when the $\theta$ parameter vanishes or when $c_{11}$ becomes very large compared to other 
parameters appearing in \eqref{tkya}. On the other hand when $v_3 q~\sin~\theta >> c_{11}$, $t$ approaches $t = \pm 1$.  
Once we replace $\sin~\theta$ by ${1\over 2} ~\sin~2\theta$, 
alongwith certain appropriate changes mentioned above, we will 
get the expression for the RR deformation. Note that similar arguments can be made for the limit $t = \pm i$, whereas for the other limit $t = \pm 1$, the condition becomes
$v_3 q~\sin~2\theta >> 2c_{11}$. 

What is the usefulness of the parameter $t$? As discussed in \cite{wittenknots1} and in \cite{wittenknots2} $t$ is useful in expressing the BHN equation in terms of topologically 
twisted variables\footnote{There are other and more deeper reasons for introducing $t$ in gauge theory, especially topological field theory, which will be elaborated later.}. 
In general however we don't have to incorporate topological twist to express the BHN equation in terms of $t$. For example the BHN equations, as they appear 
in \cite{wittenknots1} with topological twist, can be expressed as:
\bg\label{bhnwitten}
\left(F - \phi \wedge \phi + t d_A \phi\right)^+ = \left(F - \phi \wedge \phi - t^{-1} d_A \phi\right)^- = D_\mu \phi^\mu = 0, \nd
where $\phi_\mu$ are twisted scalar fields (see details in \cite{wittenknots1}), the $\pm$ appearing above denote self-dual and anti-self-dual expressions respectively. 
Without the topological twist, the last equation in \eqref{bhnwitten} is clearly our equation \eqref{bhn1}. 

Adding the self-dual and the anti-self-dual parts of \eqref{bhnwitten}, and removing the topological twist 
so as to express everything in the language of standard gauge theory\footnote{We are a bit hand-wavy in describing the details here, but before the readers despair we want to assure that
our sloppiness will be rectified in the following sections.}, 
the equation that we get for the $F_{12}$ component the gauge fields can be expressed as:
\bg\label{fkhan}
F_{12} + \left({t + t^{-1}\over 2}\right) D_\psi \phi_0 + \left({t-t^{-1}\over 2}\right) D_{[1}\phi_{2]} + 2 \left[\phi_1, \phi_2\right] = 0, \nd
where we have assumed the four-dimensional coordinates to be ($x_0, x_1, x_2, \psi$). Before comparing this equation with 
\eqref{f12bhn}, we should ask whether incorporating $c_2$ back in \eqref{hamilbeta} changes the form of \eqref{f12bhn}. The gauge theory
part of the action \eqref{stotal} now reproduces the following Hamiltonian\footnote{Needless to say, this is the special case with $c_{11} \propto c_{12}$, where $c_{11}$ and $c_{12}$ are defined 
in \eqref{c11theta} and \eqref{krisis} respectively. The picture is not hard to generalize, but we will not do so here.}:
\bg\label{hamilchondro}
{\cal H}_2 ~ = ~ {2i \over \tau - \bar{\tau}} {\rm Tr} \left({c_1 c_{11} {\cal F}^{0i}\over v_3} + \tau ~\epsilon^{ijk}{\cal F}_{jk}\right) 
\left({c_1 c_{11} {\cal F}_{0i}\over v_3} + \bar{\tau} ~\epsilon_{ilm}{\cal F}^{lm}\right), \nd 
where $\tau$ is given earlier in \eqref{taude}. In the presence of the scalar fields of \eqref{stotal}, the above Hamiltonian will 
reproduce the Hamiltonian \eqref{hamilbeta} apart from the additional pieces: 
\bg\label{toppiece}
c_1 q~\sin~\theta \int {\rm Tr} ~{\cal F} \wedge {\cal F} + {v_3 c_1 q^2\over c_{11}}~\sin^2\theta \int {\rm Tr}~{\cal F} \wedge \ast {\cal F}, \nd
depending on how all the terms are arranged as sum of squares. An alternative way of putting ${\cal F}$ and $\ast {\cal F}$ {\it inside} the
sum of squares could also be performed, but in the end the final results shouldn't differ. The former way of separating the topological 
piece from the non-topological pieces has one advantage: the BHN equations \eqref{f12bhn} etc., remain mostly unaltered. 

The definition of $t$ in \eqref{tkya} is motivated from \cite{wittenknots1}, and one may see that when $\theta = 0$, $t$ takes the value of $\pm i$. However what definition of 
$t$ we use is up to us: for every choice of $t$ there is a topological field theory although choosing a $t$ that may be an arbitrary complex number would break supersymmetry.
Furthermore the appearance of $q(\theta)$ in \eqref{tkya} will complicate the subsequent analysis as knowing the precise value of $q(\theta)$ from \eqref{chotas}
requires knowing the background fluxes in M-theory 
in full details. We can then use our freedom to choose $\theta$, using Set 2 in \eqref{twosets}, to make $q(\theta) = 1$ for $\theta = \beta$.
Therefore let us define $t$, when $\theta = \beta$, using the functional form similar to \eqref{tkya} but without any adjoining $q(\beta)$, namely\footnote{We could 
also define $\widetilde{v}_3(\theta) \equiv v_3(\theta) q(\theta)$ and replace all $v_3$ appearing below by $\widetilde{v}_3$. This will lead to identical conclusion.}:
\bg\label{tbol}
t ~\equiv~  \pm \left({v_3~\sin~\beta\over \sqrt{c_{11}^2 + v_3^2 ~\sin^2\beta}} - {ic_{11} \over \sqrt{c_{11}^2 + v_3^2 ~\sin^2\beta}}\right), \nd 
but now with $\beta$, a specific angle, instead of the generic NC parameter $\theta$, that can be used to parametrize the warp-factors $F_i$ in the following way:
\bg\label{pateraba}
F_{k} \equiv F_k(r; \beta), ~~~~~~~ F_4 \equiv F_4(r, x_8, x_9; \beta), \nd
in \eqref{chote}, where $k = 1, 2, 3$. The question that we want to ask is whether this 
could lead to a consistent description. 

Before answering this, we should also note that the scalar fields used here are
($\phi_0, \phi_1, \phi_2$), which should be compared to \eqref{4scalars}, and also note
the apparent
absence of $i$ in the equation compared to our set-up\footnote{We define $D_a\phi_c = \partial_a \phi_c + [A_a, \phi_c]$ compared to ${\cal D}_a\phi_c$ that has an $i$ in the 
definition (see \eqref{fintu}).}. 
However, with $\left| t\right|^2 = 1$ and $t$ given as \eqref{tbol}, $t + t^{-1}$ is real but $t - t^{-1}$ cannot be
real\footnote{Unless of course $t = \pm 1$, in which case $t - t^{-1} = 0$. We will discuss this case later.}. 
This means, and according to \eqref{4scalars}, we can now identify our relevant scalars and gauge-field components with the
ones in \cite{wittenknots1} in the following way:
\bg\label{namtha} 
&&{\cal A}_\mu = -iA_\mu, ~~~~ \varphi_3 = - i \phi_0, ~~~~ \varphi_1 =  \left({c_1 {\cal C}_{11}\over v_3 d_{12}}\right)^{1/4} \phi_1, ~~~~
\varphi_2 =  \left({c_1 {\cal C}_{11}\over v_3 d_{12}}\right)^{1/4} \phi_2 \\
&& {\cal F}_{\mu\nu} = - i F_{\mu\nu}, {\cal D}_\alpha \varphi_1 = \left({c_1 {\cal C}_{11}\over v_3 d_{12}}\right)^{1/4} D_\alpha \phi_1, 
{\cal D}_\alpha \varphi_2 = \left({c_1 {\cal C}_{11}\over v_3 d_{12}}\right)^{1/4} D_\alpha \phi_2, 
{\cal D}_\beta\varphi_3 = -i D_\beta\phi_0, \nonumber \nd  
where $D_\alpha \phi_k = \partial_\alpha \phi_k + [A_\alpha, \phi_k]$; 
($c_1, v_3, d_{12}$) are defined earlier in \eqref{c1c2c1c2},  \eqref{intig} and \eqref{venu} respectively; and the new parameter ${\cal C}_{11}$ can be expressed as:
\bg\label{racsolar} 
{\cal C}_{11} \equiv c_{11}\left(1 + {v_3^2 ~ \sin^2\beta\over c_{11}^2}\right), \nd
where $c_{11}$ is given in \eqref{c11theta}. For vanishing $q(\beta)$, ${\cal C}_{11}$ and $c_{11}$ coincide.  
Therefore using the identifications 
\eqref{namtha}, we can reexpress \eqref{f12bhn} in the following suggestive way:
\bg\label{blada}
F_{12} + \left({b_{\psi 3} v_3\over c_1 {\cal C}_{11}}\right)^{1/2} D_\psi \phi_0 + i\left({b^2_{12} v_3\over c_1 {\cal C}_{11} d_{12}}\right)^{1/4} D_{[1} \phi_{2]} 
+ 2 \left[\phi_1, \phi_2\right] = 0, \nd
where ($b_{\psi 3}, b_{12}$) are defined in \eqref{bokcho} and \eqref{boka} respectively. Comparing \eqref{blada} with \eqref{fkhan}, we can easily identify:
\bg\label{tequa}
t + t^{-1} = 2 \left({b_{\psi 3} v_3\over c_1 {\cal C}_{11}}\right)^{1/2} \equiv 2 \xi_1, ~~~
t - t^{-1} = 2 i\left({b^2_{12} v_3\over c_1 {\cal C}_{11} d_{12}}\right)^{1/4} \equiv 2 i\xi_2, \nd
where $\xi_i$ are defined accordingly. Note that there are two equations for $t$ and therefore we should expect some relation between $\xi_1$ and $\xi_2$. 
Solving the first equation in \eqref{tequa} gives us the following expression for $t$:
\bg\label{t1}
t = \xi_1 \pm i \sqrt{1-\xi^2_1}, \nd
which should now be compared to \eqref{tkya} that we found earlier. Equation \eqref{t1} implies two possible values for $t$ (which are the two solutions of the quadratic 
equation \eqref{tequa}), consistent with \eqref{tkya}. Therefore using \eqref{t1}, \eqref{tequa} in \eqref{tkya}, we get:
\bg\label{tantan} 
\sin^2 \beta ~ = ~ {c_{11}(\beta) b_{\psi 3}(\beta)\over c_1(\beta) v_3(\beta)}, \nd
where the $\beta$ dependence of $c_{11}(\beta)$ and $b_{\psi 3}(\beta)$ can be read from \eqref{c11theta} and \eqref{bokcho} respectively in the limit $\theta = \beta$ when we assume that the 
warp factors are parametrized by $\beta$. 

Observe that the above equation \eqref{tantan} has {\it two} free variables: the parameter $\beta$, and the asymptotic value of the gauge field $e^{2\phi_0}$.
Thus the above relation connects $\beta$ with $e^{2\phi_0}$. To determine them individually we will require another relation between them. In fact this appears from
the second equation for $t$ in \eqref{tequa} in the following way. Solving it, we get:
\bg\label{tt2}
t ~ = ~ i\xi_2 \pm \sqrt{1-\xi_2^2}. \nd
This should be related to \eqref{t1}, otherwise it will lead to certain inevitable contradictions. Equating \eqref{tt2} to \eqref{t1}, leads to:
\bg\label{xi12}
\xi_1^2 + \xi_2^2 = 1, \nd
which when expressed in terms of supergravity variables described above in \eqref{tequa}, leads to the following relation between the coefficients:
\bg\label{malabar}
b_{\psi 3}\sqrt{v_3 \over c_1 {\cal C}_{11}} + {b_{12} \over \sqrt{d_{12}}} = \sqrt{c_1 {\cal C}_{11}\over v_3}, \nd
which as expected should provide another relation between $\beta$ and $e^{2\phi_0}$. To see this let us go back to the definitions of the parameters appearing in \eqref{malabar} and
\eqref{tantan} all in the limit $\theta = \beta$: 
$b_{12}(\beta)$ in \eqref{boka}, $d_{12}(\beta)$ in \eqref{venu}, $v_3(\beta)$ in \eqref{intig}, $c_{11}(\beta)$ in \eqref{c11theta}, $b_{\psi 3}(\beta)$
in \eqref{bokcho} and $c_1(\beta)$ in \eqref{c1c2c1c2}, and isolate their $e^{2\phi_0}$ dependences in the following way:
\bg\label{rtalcan} 
&& c_1(\beta) \equiv e^{\phi_0}\langle c_1(\beta)\rangle, ~~~~~ v_3(\beta) \equiv e^{\phi_0} \langle v_3(\beta)\rangle, 
~~~~~ b_{\psi 3}(\beta) \equiv \langle b_{\psi 3}(\beta)\rangle \nonumber\\
&& b_{12}(\beta) \equiv e^{2\phi_0} \langle b_{12}(\beta)\rangle,~~  d_{12}(\beta) \equiv e^{2\phi_0} \langle d_{12}(\beta)\rangle,~~
c_{11}(\beta) \equiv e^{2\phi_0} \langle c_{11}(\beta)\rangle, \nd
here $\langle a_{mn}\rangle$ is simply used to denote the form for $a_{mn}$ sans the dilaton dependence $e^{\phi_0}$. Plugging \eqref{rtalcan} in \eqref{tantan} and \eqref{malabar}, we
get the following relations between the two free parameters $\beta$ and $e^{\phi_0}$:
\bg\label{rta} 
e^{2\phi_0} = {{\hat b}_1(\beta)\over {\hat b}_3(\beta) - {\hat b}_2(\beta)}, ~~~~~~~e^{2\phi_0} = {{\hat a}_1(\beta) \over {\hat a}_3(\beta) - {\hat a}_2(\beta)}, \nd
which when solved simultaneously should provide the values for $\beta$, the parameter used for defining $t$ at $\theta = \beta$, and $e^{\phi_0}$, the aymptotic value of the dilaton. 
The coefficients appearing in \eqref{rta} are defined, using \eqref{rtalcan}, in the following way:
\bg\label{alcan}
&& {\hat a}_1 = \sqrt{\langle b_{\psi 3} \rangle^2\langle v_3\rangle \over \langle c_1\rangle \langle {\cal C}_{11} \rangle}, ~~
{\hat a}_2 = {\langle b_{12} \rangle \over \sqrt{\langle d_{12} \rangle}}, ~~ {\hat a}_3 = \sqrt{\langle c_1\rangle \langle {\cal C}_{11} \rangle \over \langle v_3 \rangle}\nonumber\\
&& {\hat b}_2 = \langle {\cal C}_{11}\rangle^2 \langle b_{\psi 3}\rangle, ~~ {\hat b}_1 = \langle v_3\rangle^2 \langle b_{\psi 3}\rangle \sin^2\beta, 
~~{\hat b}_3 = \langle v_3\rangle \langle c_1\rangle \langle {\cal C}_{11}\rangle \sin^2\beta, \nd
where we have defined $\langle {\cal C}_{11}\rangle$ using the relation ${\cal C}_{11} = e^{2\phi_0}\langle {\cal C}_{11}\rangle$, which is similar to $c_{11}$ defined in 
\eqref{rtalcan} above. However the definition of ${\cal C}_{11}$ in \eqref{racsolar} will yield:
\bg\label{helsolari}
{\cal C}_{11} = e^{2\phi_0}\langle {\cal C}_{11} \rangle + {\cal O}(\phi_0), \nd
and therefore in the limit $\phi_0 << 1$, the above analysis can be trusted. Additionally, 
since $e^{\phi_0}$ is a positive definite quantity, the two equations in \eqref{rta} only makes sense if ${\hat b}_3 \ge {\hat b}_2$ and ${\hat a}_3 \ge {\hat a}_2$.
In the language of
the gauge theory coefficients, this would imply:
\bg\label{ssage}
{\langle v_3 \rangle \langle c_1 \rangle \over \langle {\cal C}_{11}\rangle \langle b_{\psi 3}\rangle} ~ \ge ~  {\rm cosec}^2\beta, 
~~~~ {\langle {\cal C}_{11} \rangle \langle c_1\rangle \over \langle v_3 \rangle} ~ \ge ~ {\langle b_{12} \rangle^2 \over \langle d_{12} \rangle}, \nd   
where ($c_1, c_{11}, v_3, b_{\psi 3}, b_{12}, d_{12}$) 
are defined in \eqref{c1c2c1c2}, \eqref{c11theta}, \eqref{intig}, \eqref{bokcho}, \eqref{boka} and \eqref{venu} respectively. We expect the condition \eqref{ssage} to be
compatible with the following equation, used to determine the parameter $\beta$:
\bg\label{theeq}
{{\hat b}_1(\beta)\over {\hat b}_3(\beta) - {\hat b}_2(\beta)} ~= ~ {{\hat a}_1(\beta) \over {\hat a}_3(\beta) - {\hat a}_2(\beta)}, \nd
which indeed is the case as \eqref{theeq} leads to the following relation between the gauge theory coefficients formed as a juxtaposition of the two inequalities, discussed above
in \eqref{ssage}, in the following way:
\bg\label{iggy}
{\langle v_3 \rangle \langle c_1 \rangle \sin^2\beta \over \langle {\cal C}_{11}\rangle \langle b_{\psi 3}\rangle} ~ = ~ 
\sqrt{\langle {\cal C}_{11} \rangle \langle c_1\rangle \langle d_{12} \rangle \over \langle v_3 \rangle \langle b_{12} \rangle^2}. \nd
So far the analysis have moved smoothly and we have results that are apparently self-consistent. There is however one issue that is not completely satisfactory, and it appears 
at the point where we identified the scalars, namely ($\varphi_1, \varphi_2, \varphi_3$) with the ones of \cite{wittenknots1}, namely 
($\phi_0, \phi_1, \phi_2$), in \eqref{namtha}. Using the identification \eqref{namtha}, the resulting action does not have the full canonical form. A way out of this would be
to insert $\sqrt{-1}$ in the definition of ($\varphi_1, \varphi_2$) in \eqref{namtha}. However this will imply $t - t^{-1}$ to be real once we identify \eqref{f12bhn} with 
\eqref{fkhan}, leading to a contradiction, {\it unless} we impose the following condition:
\bg\label{michel}
{D}_{[1}\phi_{2]} ~ \equiv ~ {D}_1 \phi_2 - {D}_2 \phi_1 ~ = ~ 0. \nd  
Now with appropriate identification of the scalars ($\varphi_1, \varphi_2$) with ($\phi_1, \phi_2$), the BHN equation for our case takes the following form:
\bg\label{binu}
F_{12} + \left({b_{\psi 3} v_3\over c_1 {\cal C}_{11}}\right)^{1/2} D_\psi \phi_0 + 2 \left[\phi_1, \phi_2\right] = 0, \nd
which one may now compare with the BHN equation discussed in \cite{wittenknots1} and \cite{wittenknots2} for $t \ne \pm 1$.  
The way we have defined things here, the BHN equation comes with relative plus signs, but
we can always redefine the variables so as to allow for the anti-symmetric condition \eqref{michel}. 

The discussion in the last couple of pages was intended to convince the reader that we have ample independence in defining the parameter $t$. Once the parameter $t$ is chosen,
we can define the other variables in the problem appropriately to give us consistent results as we saw above. 
For $\theta \ne 0$, $t$ is in general a complex number different from $\pm i$, and therefore a definition like \eqref{tkya}, used in \cite{wittenknots1}, could 
as well suffice without resorting to the fixed parameter $\beta$ to make $q(\beta) = 1$. 
However, now due to \eqref{hamilchondro}, the BHN equation will change a little from 
\eqref{blada} to the following more generic form:
\bg\label{bhorthi}
F_{12} + \left[{b_{\psi 3}(\tau - \bar\tau)\over 2i \vert\tau\vert^2}\right]^{1/2} D_\psi \phi_0 
+ i\left[{b^2_{12} (\tau - \bar\tau)\over 8i \vert\tau\vert^2 d_{12}}\right]^{1/4} D_{[1} \phi_{2]} +  2 \left[\phi_1, \phi_2\right] = 0, \nd
by appropriately defining $m^{(1)}_{\delta k}$ and $g^{(1)}_{\alpha \beta kl}$ 
in \eqref{hamilbeta} and using the scaling relations similar to \eqref{namtha}. Note that the form of 
\eqref{bhorthi} may not be unique if we allow for other components of the scalar fields. However once we choose the appropriate number of scalar fields, we may use the components 
$m^{(1)}_{\delta k}$ and $g^{(1)}_{\alpha \beta kl}$ to always bring the BHN equation into the form \eqref{bhorthi}. 

Comparing \eqref{bhorthi} with \eqref{fkhan}, and using the definition of $t$ as in \eqref{tkya}, it is easy to see that the NC parameter $\theta$ now satisfies a relation similar to 
\eqref{tantan}:
\bg\label{tanta2}
\sin^2\theta = {b_{\psi 3}(\theta) c_{11}(\theta)\over q^2(\theta) c_1(\theta) v_3(\theta)}. \nd
We should note a few details regarding the above relation. One, 
for the RR deformation, the LHS of the above relation \eqref{tanta2} will be replaced by ${1\over 2}~\sin~2\theta$ alongwith appropriate changes to $v_3$ as in \eqref{intig2},  $c_{11}$ as in \eqref{c11theta2} and $b_{\psi 3}$ as in \eqref{lemurgi} with the functional form for $c_1$ remaining similar to \eqref{c1c2c1c2} as before\footnote{As discussed earlier, this change is valid only for small RR deformation parameter $\theta$. For finite $\theta$ the relation \eqref{chotsard2} gets corrected, and therefore the LHS of \eqref{tantan} will change accordingly.}. 
Two, when $\theta$ vanishes, we expect the RHS of \eqref{tanta2} to vanish. This may not be too obvious from the form of $b_{\psi 3}$ in 
\eqref{lemurgi}, so we may use an alternative way to express this by redefining $b_{\psi 3}$ as:
\bg\label{bpsi3a}
b_{\psi 3} = {\sigma_0 c_1 c_{11} \over v_3}, \nd
where $\sigma_0(\theta)$ is a positive definite $\theta$-dependent constant. We can now use \eqref{bpsi3a} to rewrite \eqref{tanta2} in the following suggestive way:
\bg\label{lrus}
{c_1 q~\sin~\theta \over \left({c_1 c_{11}\over v_3}\right)} ~ \equiv ~ {\Theta/2\pi \over 4\pi/g^2_{YM}} ~ = ~ \sqrt{\sigma_0}, \nd
from where the vanishing of $b_{\psi 3}$ when $\theta$ vanishes amounts to the vanishing of $\sigma_0$. While the above step may not shed much tranparency to the vanishing issue, 
our rewrite of \eqref{tanta2} in terms of \eqref{lrus} will be useful later on.

On the other hand, we can use \eqref{tt2} to express the second term in the BHN equation \eqref{bhorthi} in terms of the known variables. This will give us:
\bg\label{safox}
{b^2_{12}\over d_{12} (1 + \sigma_0)} ~ = ~ {c_1 c_{11}\over v_3}. \nd
The above relation should be compatible with \eqref{xi12} and \eqref{malabar} even if we switch off $\theta$ in our equations. In general, equation like \eqref{malabar} follows 
provided $c_{11}$ is replaced by its $\theta$-dependent cousin:
\bg\label{vannah}
c_{11}(0) ~ \to ~ \sec \theta~c_{11}(0) - 2 R_3 ~\sin^2\theta~\sec^3\theta \int_0^\infty dr ~e^{2\phi}~F_2 ~{\partial G_0\over \partial F_2}~{\partial F_2\over \partial r}, \nd
in \eqref{malabar} for small $\theta$, where $G_0(F_1, F_2, F_3)$ is the integrand in \eqref{c11theta}. 
Other relations like the ones discussed above should follow, and one may easily check that the overall picture is still expectedly consistent. 
We will not 
elaborate further on this, instead however we will try to express \eqref{safox} in a way that may be a bit more transparent with the analysis of \cite{wittenknots1} by redefining $b_{12}$
and $d_{12}$ as:
\bg\label{nunchat} 
b_{12} = {\gamma_0 c_1 c_{11} \over v_3},~~~~~~~ d_{12} = {\kappa_0 c_1 c_{11} \over v_3}, \nd 
which is similar to the definition \eqref{bpsi3a} studied above. The coefficients ($\gamma_0, \kappa_0$) are constants, just like $\sigma_0$ in \eqref{bpsi3a} above. 
They can be related to each other via:
\bg\label{phomar}
\gamma_0^2 ~ = ~ \kappa_0(1 + \sigma_0), \nd
which is easily got by plugging \eqref{nunchat} in \eqref{safox}. We could also rewrite all the other coefficients appearing in our original lagrangian \eqref{stotal} as 
\eqref{nunchat} so that they are all proportional to ${c_1c_{11} \over v_3} \equiv {4\pi\over g^2_{YM}}$. This way the overall four-dimensional lagrangian will take the familiar
form given in \cite{wittenknots1} and a direct comparison to the results of \cite{wittenknots1} can then be performed succinctly. We will however leave this as an exercise for our 
attentive readers.  

Let us now come to the other two BHN equations for our case, namely the two equations in \eqref{bhishon}. We can rewrite them using $t$ and the definitions \eqref{namtha} in the 
following way:
\bg\label{avarse}
{F}_{a\psi} + \left({t + t^{-1}\over 2}\right){D}_b\phi_0 +  \left({t - t^{-1}\over 2}\right){D}_\psi\phi_a  = 0, \nd
where $a = 1, 2$ and we can allow a relative sign difference by 
allowing the sign choice for ($\varphi_1, \varphi_2$) identifications in \eqref{namtha}. As before, noticing that 
$t- t^{-1}$ cannot be real, and preserving the canonical form of the action, we 
conclude\footnote{There is an alternate way of expressing \eqref{tucci}, after twisting, that is sometime useful although the resulting constraint may be a bit weaker than \eqref{tucci}. To see this
combine the two relations in \eqref{tucci} as: 
$$D_\psi \phi_1 - i D_\psi\phi_2 = \partial_\psi \varphi_{12} + \left[A_\psi - i\phi_0, \varphi_{12}\right] ~ = ~ 0$$
where $\varphi_{12} \equiv \phi_1 - i \phi_2$ with $\phi_i$ being the twisted scalar (see footnote \ref{shollow}) and we have used a shifted gauge field using the twisted scalar $\phi_0$. Since $\phi_0$ 
decouples via \eqref{phi3c} (using the identification \eqref{4scalars}) both unshifted and the shifted fields will have the same effect here.}: 
\bg\label{tucci}
{D}_\psi \phi_a ~ = ~ 0 ~~~~ \Longrightarrow ~~~~ {D}_\psi \phi_1 ~ = ~ {D}_\psi \phi_2 ~ = ~ 0. \nd
At this stage there seems to be two possibilities: we can either identify $t + t^{-1}$ with the coefficients of the ${D}_b \phi_0$ terms, or we can assume that ${D}_b\phi_0$
terms themselves vanish. The former leads to two relations, but since $b_{23} = b_{13}$ we will only have one quadratic equation in $t$. However we will have to identify this to the 
one that we got earlier in \eqref{tequa} otherwise there will be contradictions. This means:
\bg\label{biratpa}
{c_{11}\over c_{12}} ~ = ~ {b_{\psi 3} \over b_{23}}. \nd
Looking at \eqref{c11theta} for $c_{11}$, \eqref{krisis} for $c_{12}$, \eqref{bokcho} for $b_{\psi 3}$ and \eqref{boka} for $b_{23} = b_{13}$, we can see that \eqref{biratpa} is 
definitely
{\it not} generic. Under special choices of the warp factors one might be able to recover \eqref{biratpa} but generically \eqref{biratpa} will be hard to satisfy. Thus the second option 
seems more viable. Interestingly, imposing the second condition:
\bg\label{permonir}
{D}_1\phi_0 ~ = ~ {D}_2\phi_0 ~ = ~ 0 ~~~~ \Longrightarrow ~~~~ F_{1\psi} ~ = ~ F_{2\psi} ~ = ~ 0, \nd
which is equivalent to putting a flat connection along $\psi$ direction. This further means, from \eqref{tucci}, the scalar fields ($\phi_1, \phi_2$) are covariantly constant along
$\psi$ direction, with $\phi_0$ being covariantly constant along ($x_1, x_2$) directions. Thus the non-trivial scalar fields $\phi_1 \equiv \phi_1(x_1, x_2, \psi)$ and 
 $\phi_2 \equiv \phi_2(x_1, x_2, \psi)$ 
satisfy:
\bg\label{prolpart}
&&{D}_1 \phi_1 ~ =  -{D}_2 \phi_2, ~~~~~~~ {D}_1 \phi_2  ~= ~ {D}_2 \phi_1 \nonumber\\
&& F_{12} + \left[{b_{\psi 3}(\tau - \bar{\tau}\over 2i \vert \tau\vert^2}\right]^{1/2} D_\psi \phi_0 + 2\left[\phi_1, \phi_2\right] = 0, \nd   
assuming $\phi_0$ to not be covariantly constant along $\psi$ direction. The system is therefore tightly constrained, but note that for $t = \pm 1$, the second constraint in 
\eqref{prolpart} is relaxed\footnote{The first constraint can be expressed as $D_0\phi_0 + D_1\phi_1 + D_2\phi_2 + D_\psi\phi_3 = D_1\phi_1 + D_2\phi_2 = 0$, where we have 
defined ${\cal A}_3 = - i \phi_3$. This is exactly $D_\mu\phi_\mu = 0$ in \eqref{bhnwitten}.}. 
The first and the third equation in \eqref{prolpart} are thus related to the equations \eqref{bhnwitten} (see also \cite{wittenknots1} and 
\cite{wittenknots2}). The Gauss law equation \eqref{gauss} puts no additional constraints on ($\phi_1, \phi_2$) in this gauge.    

We will soon solve these set of equations, but for the time being we will postpone this to concentrate on identifying the supergravity variables used here to the gauge-theory variables
described in \cite{wittenknots1} and \cite{wittenknots2}.  

\subsubsection{Identifying supergravity and gauge theory parameters \label{boundth}} 

In the previous section we have developed the full gauge theory data from our M-theory analysis. It is encouraging to see how the Bogomolnyi-Hitchin-Nahm (BHN) equation appears naturally
from out set-up. However we have been a bit sloppy in describing two things: the appearance of $t$ given in \eqref{tkya} and the appearance of $a$ to describe the boundary gauge theory
as in \cite{langland}, \cite{wittenknots1}, and \cite{wittenknots2}. Our initial identification of $a$ with the NC parameter $\theta$ in \eqref{c12def}, although matched with 
\cite{wittenknots1}, was actually accidental. Once the effect of ${\cal U}_4$ in \eqref{boundwc} is added, we no longer expect $a = \tan~{\theta\over 2}$ for both NC and RR deformations. 
The identification of $a$ with the sugra variables will have to be more non-trivial, and finding this will allow us to describe the other parameter, called $t$ here \eqref{tkya} and in 
\cite{wittenknots1} and \cite{wittenknots2} respectively, more succinctly.  

With all the development that we carried out in the previous section, it is not too hard to make an ans\"atze for $a$ using the background data. In the begining we used \eqref{chotsard} to define $a$ for the Yang-Mills data ($c_1, c_2$). However now the Yang-Mills data have changed by the inclusion of ${\cal U}_4$. Let us then define $a$ using the new 
data in the following way:
\bg\label{kurianm}
{{\Theta/ 2\pi}\over {4\pi/g^2_{YM}}} ~=~ {v_3 q~\sin~\theta\over c_{11}}~ \equiv ~ {2a\over 1 - a^2}, \nd
where $c_{11}$ is given in \eqref{c11theta} and $v_3$ is given in \eqref{intig}. This
would be the natural extension of \eqref{chotsard} and 
is motivated by the connection between the gauge theory ${\Theta\over 2\pi}$ parameter and the Yang-Mills coupling ${4\pi\over g^2_{YM}}$ described in \cite{langland} and 
\cite{wittenknots1}; and also in \eqref{lrus} earlier. The above relation to $a$ 
will continue to hold once we replace the $\sin~\theta$ appearing in \eqref{kurianm} by ${1\over 2}~\sin~2\theta$, where $\theta$ 
will now be the RR deformation. 
For our case and assuming $\theta$,
for simplicity, is providing the NC deformation, the definition of $a$ in terms of the sugra variables can then be expressed as\footnote{There is a relative sign ambiguity, 
but that can 
be absorbed by redefining $\theta$.}:
\bg\label{mithrak}
a ~ \equiv ~ \sqrt{1 + {c_{11}^2\over v_3^2 q^2~ \sin^2\theta}} - {c_{11}\over v_3 q~\sin~\theta}, \nd
that follow naturally from \eqref{kurianm}. 
Additionally it is easy to verify, for NC deformation, the definition of $t$ in \eqref{tkya} can be re-expressed in terms of $a$ as:
\bg\label{mkmaja}
t ~=~ {2a\over 1+ a^2} - i\left({1-a^2\over 1+a^2}\right) ~ \equiv -i\left({1+ia\over 1-ia}\right), \nd
precisely as in \cite{langland} and \cite{wittenknots1}. Once again, with appropriate modification, one may describe 
an exactly similar relation with the RR deformation parameter $\theta$. 

So far our discussions have been self-consistent, and the results could be compared to \cite{wittenknots1}. However note that the introduction of the $t$ parameter in our model is 
not unique. There are other ways to introduce this parameter which may also lead to consistent results. In the following we will elaborate this and in turn determine the electric and the 
magnetic charges ${\bf Q_E}$ and ${\bf Q_M}$ respectively in \eqref{hamilbeta}. To start, we will first rewrite the relevant parts of the Hamiltonian ${\cal H}$ using \eqref{hamilchondro}
once we switch on $c_2$ parameter, in the following way:
\bg\label{horsho}
{\cal H} & = & 
\sum_{\alpha, \beta = 1}^2 \int d^3x~{\rm Tr}\bigg(\sqrt{2i \vert\tau\vert^2\over \tau - \bar{\tau}} 
{\cal F}_{\alpha\beta} +\sum_{\delta, k = 1}^3 \sqrt{b_{\delta k}}\epsilon_{\alpha \beta} m^{(1)}_{\delta k} {\cal D}_\delta \varphi_k 
-\sum_{k, l} ig^{(1)}_{\alpha\beta kl}\sqrt{d_{kl}}\left[\varphi_k, \varphi_l\right]\bigg)^2 \nonumber\\
&&+ \sum_{\alpha = 1}^2 \int d^3x~{\rm Tr}\bigg(\sqrt{4i \vert\tau\vert^2\over \tau - \bar{\tau}} 
{\cal F}_{\alpha\psi} +\sum_{\delta, k = 1}^3 \sqrt{b_{\delta k}}\epsilon_{\alpha \psi} m^{(2)}_{\delta k} {\cal D}_\delta \varphi_k
-\sum_{k, l} ih^{(1)}_{\alpha\psi kl}\sqrt{d_{kl}}\left[\varphi_k, \varphi_l\right]\bigg)^2 \nonumber\\ 
&& + {1\over 2} \int d^3x~\epsilon_{0\alpha\beta\gamma}  \left(\tau + \bar{\tau}\right) 
{\rm Tr}~{\cal F}^{0\alpha} {\cal F}^{\beta\gamma} + {\bf Q_E}  + {\bf Q_M}, 
\nd
where $\tau$ is given by \eqref{taude}, and the other parameters have been defined earlier\footnote{The electric and magnetic charges ${\bf Q_E}$ and ${\bf Q_M}$ respectively are $c$-numbers as
should be evident from \eqref{hamilbeta} and the dim $G$ piece is removed by taking the adjoint trace.}. 
We expect ${\bf Q_E} = 0$ if the warp-factors satisfy \eqref{sbtaaja}. To determine ${\bf Q_M}$, we can take the following simplifying condition that we discussed earlier:
\bg\label{simcondd}
&& \sqrt{b_{\delta k}}m^{(1)}_{\delta k} ~ = ~ 
\sqrt{b_{\delta k}}m^{(2)}_{\delta k} ~= ~ - \epsilon_{\delta k} \sqrt{2i \vert\tau\vert^2\over \tau - \bar{\tau}} \nonumber\\
&& g^{(1)}_{\alpha\beta kl} =  -\eta_{k\alpha} \eta_{l\beta}\sqrt{2i \vert\tau\vert^2\over d_{\alpha\beta}(\tau - \bar{\tau})}, 
~~~ h^{(1)}_{\alpha\psi kl} =  -\eta_{k\alpha} \eta_{l\psi}\sqrt{2i \vert\tau\vert^2\over d_{\alpha\psi}(\tau - \bar{\tau})}, \nd  
which would still satisfy the consistency relations \eqref{coeffrelo} because the other coefficient, namely $q^{(1)}_{kl}$, 
that does not appear in \eqref{simcondd}, is undetermined and can be used to our 
advantage to solve \eqref{coeffrelo}. Note that \eqref{simcondd} is more generic than our earlier choices \eqref{g12kl} and
\eqref{consco}, and thus the BHN equations for ${\cal F}_{\alpha \psi}$ will differ 
from \eqref{meyepita} and \eqref{bhishon}\footnote{The decoupling of the two scalars $\sigma$ and $\bar{\sigma}$ as given in \eqref{scalfy}, \eqref{rater} and \eqref{kyabol}
still holds and therefore they 
do not appear in \eqref{horsho}. This situation will change in the presence of {\it surface operators} and other defects, which will be discussed in section \ref{surfu}.}. 
This is good because it simplifies the form for ${\bf Q_M}$, which in our case will be given by (see also \cite{leeyee}):
\bg\label{QoM}
{\bf Q_M} = {2i \vert\tau\vert^2\over \tau - \bar{\tau}} \int d^3x ~\partial_\psi \left\{\epsilon^{\alpha\beta k}{\rm Tr}\left(
\varphi_k {\cal F}_{\alpha\beta} + {i\over 3}\varphi_k[\varphi_\alpha, \varphi_\beta] +  \varphi_{\alpha} {\cal D}_\beta \varphi_{k}\right)\right\}, \nd
where the subscript on the scalar fields $\varphi_m$ are to be interpreted in the way described earlier. In the absence of any boundary, ${\bf Q_M} = 0$, as should be obvious from \eqref{QoM}. 
In the presence of the boundary ${\bf W}$ along ($x_0, x_1, x_2$), as described in sec. (\ref{dynamo}), one might combine the ${\bf Q_M}$ piece \eqref{QoM} with the topological term 
in \eqref{horsho}, to write the following boundary action\footnote{The existence of $dx_0$ implies that the action \eqref{bandu} is still in the Lorenzian frame, although an extension to the 
Euclidean frame is straightformward and will be discussed below.}:
\bg\label{bandu} 
S_{bnd} & = & \int_{{\bf V}} dx_0 ~{\bf Q_M} + {\tau + \bar{\tau}\over 2}\int_{{\bf V}} {\rm Tr}~{\cal F} \wedge {\cal F} 
=  {\tau + \bar{\tau}\over 2}\int_{\bf W} {\rm Tr}\left({\cal A} \wedge d{\cal A} + {2i\over 3} {\cal A} \wedge {\cal A} \wedge {\cal A}\right) 
\nonumber\\ 
&+& ~{2i \vert\tau\vert^2\over \tau - \bar{\tau}} \int_{\bf W} dx_0 dx_1dx_2 ~\epsilon^{\alpha\beta k}{\rm Tr}\left(
{\cal F}_{\alpha\beta} \varphi_k + {i\over 3}\varphi_k[\varphi_\alpha, \varphi_\beta] +  \varphi_{\alpha} {\cal D}_\beta \varphi_{k}\right), \nd
where ${\bf V} = {\bf W} \times {\bf R_+}$ as described in sec. (\ref{dynamo}). Under {\it twisting}, the three scalars ($\varphi_1, \varphi_2, \varphi_3$) become one-forms\footnote{Note that previously \eqref{namtha} was used to relate scalar fields $\varphi_k$ 
with scalar fields $\phi_m$. Here we relate scalar fields 
$\varphi_k$ with one-forms $\phi_\mu$. Since we are using the same notations for scalar fields and one-forms, we hope the readers will not be 
confused as which one is meant should be clear from the context.} 
$\phi = \sum_{\mu = 0}^2 \phi_\mu dx^\mu$,
and therefore one might be tempted to declare \eqref{bandu} as the required boundary topological action for the three-dimensional theory once we 
convert to Euclidean signature. In fact under twisting and Euclideanisation, \eqref{bandu} almost resembles eq. (2.54) and (2.55) of \cite{wittenknots1} provided:
\bg\label{phimu}
\phi_\mu ~\to ~ \left({t^2-1\over 2t}\right)\phi_\mu, \nd  
with $t$ as in \eqref{tkya}. Unfortunately however the coefficients appearing in the two terms of \eqref{bandu} do not match with the ones in eq. (2.54) and (2.55) of \cite{wittenknots1}. One
might think that a different scaling of all the fields could bring \eqref{bandu} in the required form where one could compare with \cite{wittenknots1}. While this might be possible, the 
physics leading to the correct boundary topological action is more subtle, and the action that we got in \eqref{bandu}, despite its encouraging similarity, is not the complete story. 

What have we missed? First note that in the absence of any boundary our analysis from \eqref{QoM} and \eqref{bandu} would have implied {\it zero} 
boundary action. However once we {\it twist} our scalar fields ($\varphi_1, \varphi_2, \varphi_3$) to ($\phi_0, \phi_1, \phi_2$) we expect, again 
in the absence of any boundary, the action $S_{total}$ \eqref{stotal} to be expressible as:
\bg\label{stwist}
S_{total} \to \widetilde{S}_{total} = \{{\bf Q}, ....\} + (b_2 + c_2)\int_{\bf V} {\rm Tr}~{\cal F} \wedge {\cal F}, \nd
where ${\bf Q}$ is the topological charge, $c_2$ is given earlier as in \eqref{c1c2c1c2} and $b_2$ is a new coefficient that is not visible in the 
untwisted theory (see also \cite{langland, wittenknots1}). When the theory has a boundary, we expect the second term in \eqref{stwist} to give us:
\bg\label{kensec}
S^{(1)}_{bnd} = (b_2 + c_2) \int_{\bf W} {\rm Tr}\left({\cal A} \wedge d{\cal A} + {2i\over 3} {\cal A} \wedge {\cal A} \wedge {\cal A}\right), \nd
which differs from the coefficient $\tau + \bar{\tau}$ of the Chern-Simons term that we got earlier in \eqref{bandu}. This difference is crucial and
will help us to get the correct boundary theory. 

However \eqref{kensec} is not the only boundary term that we get from our analysis. We expect some variants of the second term in \eqref{bandu} to 
also show up, albeit with twisted scalar fields. Infact this turns out to be the case, and once we ignore the scalings \eqref{phimu} and 
\eqref{simcondd}, the boundary terms that we get are now:
\bg\label{nest}
S^{(2)}_{bnd} = \int_{\bf W} {\rm Tr}\left(2 d_1 {\cal F} \wedge \phi + {i d_2\over 3} \phi \wedge \phi \wedge \phi 
+ d_3 ~\phi \wedge d_{\cal A}\phi\right), \nd
where $d_k$ coefficients depend on $m^{(1)}_{kl}$ and $g^{(1)}_{abkl}$ appearing in \eqref{hamilbeta} and \eqref{hamilchondro}, 
$d_{\cal A} = d + 2 i{\cal A}$ is the 
covariant derivative expressed in differential geometry language and $\phi$ is the one-form constructed from the 
twisted scalars $\phi_\mu$ as depicted above. The extra factors of 2 in \eqref{nest} as well as in the definition of $d_{\cal A}$ are meant to relate the wedge products with the 
commutator brackets. 

At this stage one might conclude that we have all the necessary couplings for our topologically twisted theory. However this is not the case. We have
ignored few other possible ingredients in our construction associated with couplings of the scalar fields. The first one being related to Myers effect \cite{myers},  
namely the fact that the fractional D3-branes could also be thought of as the {puffed up} version of a {\it single} spherical 
fractional D5-brane\footnote{Recall the {\it fractional} brane origin of the D3-brane, namely it being a D5 - ${\overline {\rm D5}}$ pair.  
In the presence of multiple fractional D3-branes, there will be multiple pairs of D5 - ${\overline {\rm D5}}$ branes 
wrapped on the 
Taub-NUT two-cycles. Once we move the ${\overline {\rm D5}}$ branes along the Coulomb branch in the IIB picture, we can describe the physics  
using a  multi-centered Taub-NUT configuration in the M-theory lift. Thus in the spherical D5-brane picture, the bound fractional D3-branes are secretly 
D5 - ${\overline {\rm D5}}$ pairs much like bound D0-branes on a spherical D2-brane.}.   

It is crucial to get the orientations of various branes right. The wrapped D5 - ${\overline {\rm D5}}$ 
pairs are oriented along 
($x_0, x_1, x_2, x_3, r, \psi$) such that the D3-branes that we are concerned with can be viewed as along ($x_0, x_1, x_2, \psi$). The effective theory on the D3-branes have
been worked out in details in earlier sections using M-theory multi Taub-NUT configuration oriented along ($\theta_1, x_8, x_9, x_{11}$). The spherical D5-brane (which has no net 
D5-brane charge) is along the space-time directions ($x_0, x_1, x_2, \psi$) with a two-dimensional projection along ($\theta_1, x_8, x_9$)  
directions for both the gauge choices ${\cal A}_0 = {\cal A}_3$ and ${\cal A}_0 = {\cal A}_r$ respectively. 

The second type of couplings could be associated with the interactions of the NS three-form field strengths with the non-abelian brane configuration. These couplings are different from the 
usual couplings of the NS three-form field strengths with the brane in the sense that the couplings originate from the {\it orthogonal} components of the three-form field strengths with 
the non-abelian scalars of the brane (thus they are absent in the abelian case). 

The final set of couplings appear when one goes from the non-abelian nature of the scalars to their {\it twisted} version. To see this consider the boundary coupling \eqref{nest}.  
If we do not resort to the simplifying conditions \eqref{simcondd}, we see that the $d_k$ coefficients satisfy: 
\bg\label{punjlad} d_1 \propto m^{(1)}_{\delta k} \sqrt{b_{\delta k}}, ~~~~~~ d_3 \propto g^{(1)}_{\alpha \beta k l}\sqrt{d_{kl}}, ~~~~~~ d_2 \propto d_1 d_3, \nd
which is direct descendent of the properties of $d_k$ {\it before} twisting. The constraint \eqref{punjlad} may not hold once we twist the scalars. However if we want to keep 
the constraint \eqref{punjlad}, we can insert an additional cubic coupling of the twisted scalars. All these can be achieved by allowing the following couplings:  
\bg\label{myersc}
S_{add} = {i\over 3}\int dx_0 dx_1 dx_2 d\psi ~{\rm Tr}\left(\Phi^i \Phi^j \Phi^k\right)\left[e_1 \left({\cal F}_7\right)_{012\psi ijk} + 
e_2 \left({\cal H}_3\right)_{ijk}\right], \nd
where we expect $e_1$ to be proportional to $\pm e_2$ with the sign determining whether it is a brane or an anti-brane, and $\Phi^i$ are the scalar fields 
$\varphi_k$ that we discussed above. The seven-form field strength accomodates both the Myers effect as well as the changes in the coupling when ones goes from one 
description to another\footnote{The seven-form field strength originates from dimensional reduction of a nine-form field strength of the form ${\cal F}_9 = \ast dC_0 + F_9$,
where $C_0$ is the axion and $F_9$ is a nine-form $d_5 \epsilon_{0123\psi r \theta_1 89}$ with constant coefficient $d_5$.
For the specific case that we study we have no axion switched on, and no three-form with components $({\cal H}_3)_{\theta_1 89}$. However this is not generic, as we can easily change
the identification of the scalars \eqref{4scalars} to allow for the required components of the three and the effective seven forms. To take care of this we express the couplings 
generically as \eqref{myersc}.}. 
This can be seen by twisting the non-abelian scalar in \eqref{myersc} to reproduce the 
following boundary action:
\bg\label{bnd3}
S^{(3)}_{bnd} = {i\over 3}(e_1 n_1 + e_2 n_2) \int_{\bf W} {\rm Tr} \left(\phi \wedge \phi \wedge \phi\right), \nd
where $n_1$ and $n_2$ are related to the expectation values of ${\cal F}_7$ and ${\cal H}_3$ respectively. In deriving \eqref{bnd3} 
we have assumed the integrand in \eqref{myersc} to be independent of $\psi$. 

We now have all the necessary boundary bosonic couplings. Combining \eqref{kensec}, \eqref{nest} and \eqref{bnd3}, we can get the full action on the 
boundary ${\bf W}$, parametrized by coordinates ($x_0, x_1, x_2$), as:
\bg\label{aletex}
S_{bnd} &=& (b_2 + c_2) \int_{\bf W} {\rm Tr}\left({\cal A} \wedge d{\cal A} + {2i\over 3} {\cal A} \wedge {\cal A} \wedge {\cal A}\right)\\
&+ & \int_{\bf W} {\rm Tr}\left[2 d_1 {\cal F} \wedge \phi + {i\over 3} (d_2 + n_1 e_1 + n_2 e_2)\phi \wedge \phi \wedge \phi + d_3 ~\phi \wedge d_{\cal A}\phi\right]. \nonumber \nd
Comparing the boundary action with \eqref{horsho}, we can make a few observations on the $d_k$ coefficients without actually computing them. First, and as we discussed above, we can 
continue using \eqref{punjlad} even when we have twisted scalars. 
Thus the second coefficient $d_2$ gets fixed once ($d_1, d_3$) are determined. Secondly, we can use the ambiguity of ($m^{(1)}_{\delta k}, g^{(1)}_{\alpha\beta k l}$) to fix 
the form of $d_3$ in terms of $d_1$. As we discussed, from \eqref{punjlad}, this way $d_2$ also gets fixed in the process once $d_3$ is fixed. Thus we can have:
\bg\label{d1d2d3}
d_3 ~ = ~ {d_1^2 \over b_2 + c_2}, ~~~~~~~ d_2 ~ = ~ {d_1^3\over (b_2 + c_2)^2}, \nd
where ($b_2, c_2$) are the coefficients that appear in \eqref{aletex}. The ($b_2 + c_2$) factors in the $d_k$ coefficients guarantee that the Chern-Simons coupling remain 
($b_2 + c_2$) instead of shifting to another value. The choice \eqref{d1d2d3} is motivated from the scaling argument that we performed earlier in \eqref{phimu}. 

The last bit of information that we need to complete the story is the value for the interaction term \eqref{myersc}. As we see in \eqref{myersc}, the values for ($n_1, n_2$) 
depend on the background fluxes ${\cal F}_7$ and ${\cal H}_3$. We can fix the background data from the start in \eqref{sugra1} in such a way that:
\bg\label{myerbeta}
n_1e_1 + n_2 e_2 ~ \equiv ~ d_2 ~ = ~ {d_1^3\over (b_2 + c_2)^2}, \nd  
which in fact governs the way the warp-factors $F_i$ in \eqref{chote} are chosen. This is good because so far we have left the warp-factors $F_i$ in \eqref{chote} undetermined. 
Thus after the dust settles, our boundary action takes the following form:
\bg\label{bonbund}
S_{bnd} &=& (b_2 + c_2) \int_{\bf W} {\rm Tr}\left({\cal A} \wedge d{\cal A} + {2i\over 3} {\cal A} \wedge {\cal A} \wedge {\cal A}\right)\\
&+ & \int_{\bf W} {\rm Tr}\left\{2 d_1 {\cal F} \wedge \phi + {2i\over 3} \left[{d_1^3 \over (b_2+c_2)^2}\right]\phi \wedge \phi \wedge \phi 
+ \left({d_1^2\over b_2 + c_2}\right)\phi \wedge d_{\cal A}\phi\right\} \nonumber \\
&=& (b_2 + c_2) \int_{\bf W} {\rm Tr}\bigg\{\left[{\cal A} + \left({d_1\over b_2 + c_2}\right)\phi\right] \wedge d\left[{\cal A} + \left({d_1\over b_2 + c_2}\right)\phi\right] \nonumber\\
&+ & {2i\over 3} \left[{\cal A} + \left({d_1\over b_2 + c_2}\right)\phi\right] \wedge 
\left[{\cal A} + \left({d_1\over b_2 + c_2}\right)\phi\right] \wedge \left[{\cal A} + \left({d_1\over b_2 + c_2}\right)\phi\right]\bigg\}, \nonumber \nd
where the coefficients $b_2$ and $d_1$ are yet to be determined from the background data. Interestingly however,
even though we do not have the precise functional form for the two 
coefficients $b_2$ and $d_1$,  
 the second equality combines the original gauge field ${\cal A}$ with the twisted scalar field $\phi$ to give us a 
new gauge field:
\bg\label{newgf}
{\cal A}_d \equiv {\cal A} + \left({d_1\over b_2 + c_2}\right)\phi, \nd
using which we have defined 
another Chern-Simons theory with a coupling constant ($b_2 + c_2$) in the following way:
\bg\label{plaza}
S_{bnd} = (b_2 + c_2) \int_{\bf W} {\rm Tr}\left({\cal A}_d \wedge d{\cal A}_d + {2i\over 3} {\cal A}_d \wedge {\cal A}_d \wedge {\cal A}_d \right), \nd
which is the topological field theory that we have for our boundary manifold ${\bf W}$. One may check that our considerations have led to the same topological theory envisioned by 
Witten in \cite{wittenknots1} but using completely different techniques.

\subsubsection{More on the Chern-Simons theory and S-duality \label{sdualt}}

There are a few details regarding the Chern-Simons theory written above in \eqref{plaza} that needs clarifications. First, 
the Chern-Simons theory is expressed in terms of the modified gauge field ${\cal A}_d$ which in turn can be expressed in terms of the original gauge field
${\cal A}$ and the twisted scalar $\phi$ via \eqref{newgf}. The factor $d_1$ appearing above is not arbitrary and can be determined using supersymmetry condition:
\bg\label{sosie}
\delta{\cal A}_\mu + {d_1\over b_2 + c_2} \delta{\phi_\mu} ~ = ~ - i \bar{\lambda} \left(\Gamma_\mu + {d_1\over b_2 + c_2} \Gamma_{4+\mu}\right) \epsilon ~ = 0, \nd
where $\lambda$ is the fermion of the supersymmetrc multiplet and $\epsilon$ is the supersymmetric transformation parameter. We have used the similar notations to express the 
$\Gamma$-matrices as in \cite{wittenknots1} and therefore the RHS of \eqref{sosie} follow same algebra as in \cite{wittenknots1}. 

The $\Gamma$-matrices chosen here are the {\it flat} space $\Gamma$-matrices as they are related to the effective theory \eqref{stotal} defined on four-dimensional spacetime 
parametrized by ($x_0, x_1, x_2, \psi$). Although our model is inherently supersymmetric from the start, it may be interesting to revisit the issue of supersymmetry so we could 
directly compare our analysis with that of \cite{wittenknots1}. The original orientations of the branes are given in {\bf Table {\ref{brenulia}}} and therefore it is easy to see that we
have the required Lorentz symmetry of:
\bg\label{sorus}
SO(1, 2) ~\times ~SO(3) ~\times ~SO(3), \nd
where $SO(1, 2)$ correspondings to Lorentz rotation along ($x_0, x_1, x_2$) directions; the first $SO(3)$ corresponds to rotation along ($x_3, x_8, x_9$) directions associated with the 
Coulomb branch of the theory on the D3-branes; and the second $SO(3)$ corresponds to rotation along ($r, \theta_1, \phi_1$) directions. 
In the dual type IIB theory where we have wrapped D5/${\overline{\rm D5}}$ branes on two-cycle of a Taub-NUT space we can easily allow the
 symmetry \eqref{sorus} to persist by putting some mild constraints on the warp factors $F_i$. Note that this is not a {\it necessary} constraint, 
so at this stage we can see that for certain 
choices of the warp-factors we can reproduce precisely the results of \cite{wittenknots1}. Similar arguments can be given for our M-theory construction where we only have a Taub-NUT space
with background fluxes. 

Finding a symmetry like \eqref{sorus} in our construction means that we can channel the results of \cite{wittenknots1} more directly. For example one persistent questions has been the 
identity of the parameter $t$ in our set-up. In the last couple of sections we have mentioned how $t$ could appear in our set-up, and in fact this parameter played important roles in 
\cite{langland}, \cite{wittenknots1} and \cite{wittenknots2}, so the natural question is to ask where a parameter like $t$ could fit in our analysis. 

To answer this question, it may be intructive to search for the source of $t$ in, for example, \cite{wittenknots1}. The ${\bf 16}$ dimensional fermionic component in our model decomposes 
as two copies of (${\bf 2}, {\bf 2}, {\bf 2}$) of the symmetry group \eqref{sorus} which, following \cite{wittenknots1}, we write it as a vector space ${\bf V_8} \otimes {\bf V_2}$. 
Thus a supersymmety parameter $\epsilon$ appearing in \eqref{sosie} above can be expressed as $\epsilon = \eta \otimes \epsilon_0$, where $\eta$ is an element of ${\bf V_8}$ and 
$\epsilon_0$ is an element of ${\bf V_2}$. Supersymmetry therefore requires us to find two functions ($Q_2, Q_3$) that may be used to express the susy relation:
\bg\label{sosie2} \left[{\bf 1} + {1\over 2}\left(Q_2 - Q_3\right){\bf B_0} + {1\over 2}\left(Q_2 + Q_3\right){\bf B_1}\right]\epsilon_0 ~=~ 
\left(\begin{matrix} 1 & ~ & ~Q_2 \\ ~Q_3 &~ & 1 \end{matrix}\right)\epsilon_0 ~ = ~ 0, \nd
where ${\bf B_0}$ and ${\bf B_1}$ are two two-dimensional matrices given in eq (2.4) of \cite{wittenknots1}; and 
$\epsilon_0$ is normalised as $\epsilon_0 = \left(\begin{matrix} -a \\ ~1 \end{matrix}\right)$ similar to \cite{wittenknots1}.
This is the same $a$ that appears in 
\eqref{kurianm} above and is related to the $\theta$-angle via \eqref{mithrak}. The two functions ($Q_2, Q_3$) are then functions of the parameter $a$ and it is easy to see that 
to solve \eqref{sosie2} we need: 
\bg\label{Q23} Q_2 ~ \equiv ~ a, ~~~~~~ Q_3 \equiv ~ {1\over a}.\nd
The picture developed above is before {\it twisting}, and so the natural question is to ask about the susy condition after twisting. Again following the notation of \cite{wittenknots1},
we can define the susy parameter $\epsilon$ to be $\epsilon = \epsilon_L + t \epsilon_R$. This is where the parameter $t$ appears in our picture, and one can easily see that 
$t$ has to be a function of $a$ so that a relation like \eqref{sosie2} may be constructed for $\epsilon$ after twisting. What value of $t(a)$ is allowed so that supersymmetry is 
preserved both before and after twisting? The answer, as worked out in \cite{wittenknots1}, is:
\bg\label{tanda}
t = -i\left({1+ia\over 1-ia}\right), \nd
which matches precisely with \eqref{mkmaja}. This is not surprising because we have tailored our definition of $t$ in \eqref{tkya} so as to reproduce the correct answer \eqref{tanda}, 
although we should note that the definition of $t$ as $\pm {\vert\tau\vert\over\tau}$ is {\it not} with an arbritary $\tau$ \eqref{taude}, but with a $\tau$ constrained via 
\eqref{kurianm}. 
  
The parameter $t$, as mentioned above is expressed in terms of $a$ which, in the original construction of Witten \cite{wittenknots1} is related to the axionic background. For us, looking 
at the RR deformation \eqref{chapmaro}, the axion in our original NS5-D3 brane construction {\bf Table \ref{wittenbranes}} 
will be given by the following expression:
\bg\label{axeon} 
{\cal C}_0 ~ = ~ {F_2 e^{2\phi} ~\tan~\theta \over \cos^2\theta + F_2 e^{2\phi}~\sin^2 \theta}\bigg\vert_{r = r_0}, \nd 
where the parameters have been described earlier. Note that the D3-branes in {\bf Table \ref{wittenbranes}} are located at some fixed value of $r = r_0$ as they are oriented along 
($x_0, x_1, x_2, \psi$).
This should be contrasted with the 
dual D5-${\overline{\rm D5}}$ picture where the branes wrap the two-sphere along the ($\psi, r$) directions. This is of course the reason for the $r$ integrals in all the 
coefficients appearing in \eqref{stotal}. 

Dualizing ${\cal C}_0$ gives us RR two-form $({\cal C}_2)_{3\psi}$ as we would have expected from \eqref{chapmaro}, and from the background \eqref{hardface}. This is not quite the two-form we 
require from M-theory point of view to reproduce the topological coupling in \eqref{stotal}, but as discussed earlier, the existence of a small amount of NS B-field on the two-sphere
oriented along ($\psi, r$) directions tells us that we can also allow a RR two-form $({\cal C}_2)_{3r}$. Lifting this to M-theory yields a three-form $({\cal C}_3)_{3r\phi_1}$ as given in 
\eqref{chotas} which we can re-express in the following form:
\bg\label{chotabeta}
{\cal C}_3 = p(\theta_1, \theta)~q(\theta)~\sin~\theta~d\zeta_\theta \wedge dx_3 \wedge d\phi_1, \nd
where $p(\theta_1, \theta)$ and $q(\theta)$
are arbitrary periodic functions of ($\theta_1, \theta$) respectively as described in \eqref{chotas}, and $\zeta_\theta$ is given in terms of a slowly varying function
$N(r, \theta)$ as:
\bg\label{zetthet}
\zeta_\theta ~ = ~ {N(r, \theta)\over \cos^2\theta + N(r, \theta)~\sin^2\theta}. \nd
The smallness of $N(r, \theta)$ in fact tells us that switching on \eqref{chotabeta} will change the background very slightly in M-theory. 
The function $\zeta_\theta$ is of the form \eqref{axeon}, so that the 
three-form does give us the required topological term or, in other words, the coefficient $c_2$ of the topological term. 

On the other hand if we normalize our warp-factor and the dilaton to satisfy $F_2 e^{2\phi} = 1$ at $r = r_0$, then from \eqref{axeon} we see that $C_0 = \tan~\theta$. 
We can go back to our definition of $a$ in \eqref{mithrak} and ask for what values of $q(\theta)$, $a$ becomes $\tan~{\theta\over 2}$. The answer is the following $\theta$-dependence
for $q(\theta)$:
\bg\label{qutta} 
q(\theta) ~ = ~ {c_{11} ~{\rm sec}~\theta\over v_3}, \nd 
which may be easily derived from \eqref{chotas} and \eqref{kurianm}. It is interesting that if we plug in \eqref{qutta} in \eqref{chotsard}, the coefficient $c_2$ becomes: 
\bg\label{c2nowb} 
c_2 ~ = ~ {c_1 c_{11}\over v_3}~\tan~\theta ~ = ~ {4\pi\over g^2_{YM}}\left({2a\over 1 - a^2}\right), \nd 
where we have normalized $v_3$ as $v_3 = 2 R_3$. The above relation is precisely the coefficient of the $\Theta$-parameter in \cite{wittenknots1}.  

All the above discussions point to the consistency of our model, both in terms of reproducing the correct boundary theory as well as comparing our results to that of 
\cite{wittenknots1}. One issue that we haven't discussed so far is the issue of S-duality that forms an integral part of the discussion in \cite{wittenknots1}. Can we analyze the
S-dual picture completely in terms of a supergravity background with fluxes and without branes, as we did for the case before S-duality? 

The answer turns out to be in affirmative although the computations are a bit more subtle now. 
Our aim is to address the analysis completely in terms of supergravity fields with no branes, so the first choice of S-dualizing the brane constructions in 
{\bf Table \ref{wittenbranes}} doesn't seem to give us the required answer as an S-duality leads to D3-branes perpendicular to the D5-brane. A further T-duality may lead to 
D4-D6 system which when lifted to M-theory will have M5-branes in a Taub-NUT geometry. This is not what we are aiming for, so we have to look for alternative scenario to 
study the S-dual background. Interestingly the D4-D6 system has been used in \cite{wittenknots1} to study the S-dual model. 

The alternative scenario appears from the wrapped D5-brane construction that we developed earlier. The D5-${\overline{\rm D5}}$ branes wrap the two-cycle of a Taub-NUT geometry and
we move the ${\overline{\rm D5}}$-branes along the Coulomb branch to study the wrapped D5-branes on the Taub-NUT two-cycle. 
This
picture, as we discussed earlier is not only {\it equivalent} to the brane construction but has a distinct advantage over the brane model when expressing the 
explicit supergravity solution.

S-dualizing the wrapped D5-branes, give us wrapped NS5-branes on the Taub-NUT two-cycle. The directions are important: the NS5-branes are oriented along ($x_0, x_1, x_2, x_3$) and
wrap two-cycle of the Taub-NUT oriented along ($\psi, r$) directions. 
The remaining two directions of the Taub-NUT are along ($\theta_1, \phi_1$) directions. A T-duality orthogonal to the
wrapped NS5-branes, i.e along $\phi_1$ direction, converts it to a multi-centered Taub-NUT space in type IIA theory warping 
the original Taub-NUT geometry suitably. Thus we have the 
following scenario.

\vskip.1in

\noindent $\bullet$ A muti-centered deformed Taub-NUT geometry in type IIA theory where the four-dimensional gauge theory can be studied from dimensional reduction of type IIA 
fields over the multi Taub-NUT space in the way we described earlier. 

\vskip.1in

\noindent $\bullet$ A M-theory uplift of the type IIA geometry where the multi Taub-NUT space develops further warping yet retaining the essential topological properties of the 
underlying space. The four-dimensional gauge theory can now be recovered from the dimensional reduction over the Taub-NUT space and over the M-theory circle.     

\vskip.1in

\noindent Both the above techniques will give us the required four-dimensional gauge theory, but the latter method might be suitable to {\it compare} with the results 
that we had 
earlier from M-theory. To start therefore let us write the metric in type IIA theory:
\bg\label{oldport}
ds^2 & = & -dt^2 + dx_1^2 + dx_2^2 + e^{2\phi} F_1 dr^2 + {dx_3^2\over \cos^2\theta + F_2 e^{2\phi}~\sin^2\theta} + 
\left({e^{2\phi} \widetilde{F}_2 F_3 \sec^2\theta ~\sin^2\theta_1 \over \widetilde{F}_2 ~\cos^2\theta_1 + F_3 ~\sin^2\theta_1}\right) d\psi^2 \nonumber\\
&& ~~~~~~ +  e^{2\phi}\left[F_3 d\theta_1^2 + F_4(dx_8^2 + dx_9^2)\right] +
{\left(d\phi_1 + b_{\phi_1 3} dx_3 + b_{\phi_1 \theta_1} d\theta_1\right)^2 \over e^{2\phi}\left(\widetilde{F}_2 ~\cos^2\theta_1 + F_3 ~\sin^2\theta_1\right)} \nd
where the second line is the warped Taub-NUT space that appears from the wrapped NS5-branes, $b_{\theta_1\phi_1}$ is the component of the RR B-field appearing in 
\eqref{chapmaro} and $b_{3\phi_1}$ is the RR deformation in \eqref{chapmaro} and is given by the following expression:
\bg\label{jutamar}
b_{\phi_1 3} = \widetilde{F}_2 e^{2\phi}~\tan~\theta~\sec~\theta~\cos~\theta_1. \nd
It is interesting that the Taub-NUT fibration structure depends on the $x_3$ direction, and the $F_1$ warp-factors are at least functions of the radial coordinate $r$. Thus 
the Taub-NUT space is non-trivially fibered over the six-dimensional base and at a given point ($r, x_3$) we can have a well-defined warped Taub-NUT manifold. 

The fluxes on the other hand are mostly NS fluxes as the only non-trivial RR flux component is the three-form $({\cal C}_3)_{\psi r \phi_1}$ appearing from the NS B-field 
switched on the two cycle in the type IIB side to cancel the D5-${\overline{\rm D5}}$ tachyons. This is a small amount of flux, which in turn allows us to have the 
NS B-field component $b_{3r}$ appearing from the RR two-form potential $({\cal C}_2)_{3r}$ responsible for \eqref{chotas}. The NS B-field in type IIA is then the following:
\bg\label{nesti}
{\cal B}_2 & = & {\widetilde{F}_2~ \cos~\theta_1 ~\sec~\theta \over \widetilde{F}_2~ \cos^2\theta_1 + F_3 ~\sin^2 \theta_1}~ 
\left(d\phi_1 + b_{\phi_1 3}dx_3 + b_{\phi_1 \theta_1} d\theta_1\right) \wedge d\psi \nonumber\\
&+ &  \widetilde{F}_2 ~e^{2\phi}~\tan~\theta ~\sec^2\theta ~dx_3 \wedge d\psi + b_{89}~dx_8 \wedge dx_9 + b_{3r}~dx_3 \wedge dr, \nd
with $b_{89}$ as it appears in \eqref{chapmaro}, and the functional form of the $b_{3r}$ component will be similar to \eqref{chotas} i.e we expect $b_{3r}$ to take the 
following form:
\bg\label{wonwom}
b_{3r} ~ = ~ {N_r~\sin~2\theta~\cos~\theta~p(\theta_1, \theta)~q(\theta) \over 2(\cos^2\theta + N~\sin^2\theta)^2}. \nd
On the other hand the behavior of the type IIA dilaton is interesting. Unlike its type IIB counterpart \eqref{kolabaz}, the parameter $e^\phi$ only appears in the 
subleading term, and the functional form is given by: 
\bg\label{dildeke}
e^{\varphi_A} ~ = ~ {\sec~\theta\over \sqrt{F_2~\cos^2 \theta_1 + F_3(1 + F_2~e^{2\phi}~\tan^2 \theta)\sin^2 \theta_1}}, \nd
which means that the type IIA background is in general not weakly coupled. One may compare this to the type IIA dilaton that we get from the background \eqref{kolabaz} by 
T-dualizing along direction $\phi_1$ as: 
\bg\label{vargas}
e^{\varphi_A} = {e^{-3\phi/2}\sqrt{\cos~\theta}\left(1 + F_2 e^{2\phi}~\tan^2\theta\right)^{3/4}\over \sqrt{F_2~\cos^2 \theta_1 + F_3(1 + F_2~e^{2\phi}~\tan^2 \theta)\sin^2 \theta_1}}. \nd
We see that there exists a tunable parameter $e^{-3\phi/2}$ that helps us to realize the M-theory uplift. Such a tunable parameter is absent in \eqref{dildeke}. In fact 
in the limit $\phi \to \pm\infty$, \eqref{dildeke} yields 
\bg\label{flowetu}
e^{\varphi_A} = {\sec~\theta\over \sqrt{F_2~\cos^2\theta_1 + F_3~\sin^2\theta_1}}\bigg\vert_{\phi \to -\infty}, ~~~
e^{\varphi_A} = \left({{\rm cosec}~\theta~{\rm cosec}~\theta_1 \over \sqrt{F_2~F_3}}\right)e^{-\phi}\bigg\vert_{\phi \to +\infty}. \nd
The former being an ${\cal O}(1)$ number; whereas the latter vanishes implying that strong type IIA coupling may be reached although infinite coupling will not be.  
Thus studying the background using M-theory might be more appropriate which, as we had 
anticipated earlier, puts an emphasis on the eleven-dimensional uplift. The story herein should then be somewhat similar to the one that we developed earlier, and therefore the
first step would be the derivation of the harmonic forms. As before, we will first attempt the single-centered case and then extend this to the multi Taub-NUT picture. 

At a given point in ($r, x_3$), the taub-NUT space takes a simple form if we, without loss of generalities, put $F_2(r_0) = F_3(r_0) \equiv a$. The other warp-factor 
$F_4$ remains a function of ($x_8, x_9$) as before. Thus the warped Taub-NUT space at a given point on ($r, x_3$) takes the following form:
\bg\label{outside}
ds^2 = e^{2\phi}\left(a~d\theta_1^2 + F_4~ds^2_{89}\right) + {e^{-2\phi} + a~\tan^2\theta\over a + a^2 ~e^{2\phi}~\tan^2\theta~\sin^2\theta_1}~
\left(d\phi_1 +  b_{\phi_1\theta_1} d\theta_1\right)^2. \nd 
The harmonic form will again be written as $\widetilde\omega = d\widetilde\zeta$ 
with the property that $\widetilde\omega = \pm \ast_4 \widetilde\omega$, where the Hodge-star is over the Taub-NUT space \eqref{outside}. 
The one-form $\widetilde\zeta$ is expressed as:
\bg\label{perez}
\widetilde\zeta ~ \equiv ~ g(\theta_1, x_8, x_9) \left(d\phi_1 + b_{\phi_1\theta_1}~d\theta_1\right), \nd
where we have used the same notation $g$ that we had used earlier in \eqref{chokranto}. 
The functional form of $g$ remains unchanged if we go to M-theory (despite the fact that in M-theory the 
warping of our Taub-NUT \eqref{outside} is different). Again, as before we expect $g$ in \eqref{perez} to satisfy the following set of equations:
\bg\label{dilbeta5}
&&{1\over g}{\partial g\over \partial \theta_1} ~ = ~ \pm {\alpha_1\over e^{2\phi} F_4} \sqrt{1 + a~e^{2\phi}~\tan^2\theta \over 1 + a~e^{2\phi}~\tan^2\theta~\sin^2 \theta_1} \nonumber\\
&& {1\over g}{\partial g\over \partial x_8} ~ = ~ \pm {\alpha_3\over a~e^{2\phi}} \sqrt{1 + a~e^{2\phi}~\tan^2\theta \over 1 + a~e^{2\phi}~\tan^2\theta~\sin^2 \theta_1} \nonumber\\ 
&& {1\over g}{\partial g\over \partial x_9} ~ = ~ \pm {\alpha_2\over a~e^{2\phi}} \sqrt{1 + a~e^{2\phi}~\tan^2\theta \over 1 + a~e^{2\phi}~\tan^2\theta~\sin^2 \theta_1} \nd 
where $\alpha_2$ and $\alpha_3$ are used to express the type IIB B-field component $b_{\theta_1\phi_1}$ as \eqref{dileshwar2}; and the vanishing of $\alpha_1$ would imply the 
$\theta_1$ independence of the $g$ function in \eqref{perez}. If we now assume that the dilaton satisfies:
\bg\label{dilchut}
e^{2\phi} ~ = ~ {e^{2\phi_0}\over \sqrt{F_3}}\left({\widetilde{Q}(r, x_8, x_9)\over \sqrt{\widetilde{F}_2~\cos^2 \theta_1 + F_3~\sin^2 \theta_1}}\right), \nd
which when compared to \eqref{dileshwar} would imply $Q(r, x_8, x_9) = {\widetilde{Q}\over \sqrt{F_3}}$, we maintain the expected consistency in every duality frames. On the other
hand the type IIA dilaton $e^{\varphi_A}$ at the given point $r = r_0$, in the limit with small $\theta$, is given by:
\bg\label{dkald}
e^{\varphi_A} ~ = ~ {\sec~\theta\over \sqrt{a}} + {\cal O}(\theta^2). \nd
Since $a$ is a finite non-zero number, the type IIA coupling is finite and an ${\cal O}(1)$ number at least for small $\theta$. Thus eleven-dimensional supergravity analysis may not
be able to capture the full details of the theory. This is clear when we try to compute the four-dimensional axionic coupling from dimensional reduction over the Taub-NUT space using an
analysis similar to \eqref{mtheorya}. The functional form of the three-form entering the topological coupling of M-theory \eqref{mtheorya} is similar to \eqref{chotas} although the 
components are $({\cal C}_3)_{3 r, 11}$ appearing in turn from the uplift of \eqref{wonwom}. The precise form is given via:
\bg\label{uruktu}
\widetilde{c}_2 ~ & = &~ \int_{\widetilde{\Sigma}_3} {\cal C}_3 \int_{\rm TN} \widetilde\omega \wedge \widetilde\omega \\
 &= & ~- \int 2\langle {\cal C}_3 \rangle ~g^2 \left(\alpha_3^2 - \alpha_2^2\right)~
dr \wedge dx_3 \wedge d\phi_1 \wedge d\theta_1 \wedge dx_8 \wedge dx_9 \wedge dx_{11}, \nonumber \nd
where $\langle {\cal C}_3\rangle$ is the {\it value} of the three-form that we got in \eqref{chotas} and $\widetilde{\Sigma}_3$ 
is the three-cycle along ($r, x_3, x_{11}$). Expectedly 
the orientation of  $\widetilde{\Sigma}_3$ differs from the three-cycle $\Sigma_3$ used earlier in \eqref{c1c2c1c2}. This is consistent with the fact that the Taub-NUT spaces in both cases
are oriented slightly 
differently as we saw above. Thus once we re-arrange the integral properly, we see that $\tilde{c}_2$ differs from $c_2$ in \eqref{c1c2c1c2} by at least an overall minus sign, although
the full behavior of $\tilde{c}_2$ would require us to get higher order terms in M-theory. The sign difference indicates S-duality at play, so this is consistent with expectation. 

The question however is why we should expect higher order corrections here. The answer lies in \eqref{flowetu}. The type IIA couplings is of ${\cal O}(1)$, and so the 11-dimensional circle 
has a {\it finite} radius. Thus there is an infinite tower of KK states that would contribute to the M-theory spectra which in turn would enter the supergravity loops to change the 
background solution. Of course very massive KK states can be integrated out in the Wilsonian action, but light states would affect the background. When the radius of the 11-dimensional circle
is infinite, the type IIA coupling is infinite and the theory is governed by eleven-dimensional supergravity only.  

The above discussion implies that the values of ($\alpha_2, \alpha_3$) from \eqref{dileshwar2} that appears in the S-dual picture should receive correction so that 
$\int \widetilde\omega \wedge \widetilde\omega$ 
computed above in \eqref{uruktu} from \eqref{dilbeta5} will differ from the one given earlier in \eqref{bladam}. Thus we expect:
\bg\label{dhechu}
\widetilde{c}_2 ~ = ~ - c_2 \Bigg[{R_{11}\over 2\pi}
\left({\int_{\rm TN_2} \widetilde\omega \wedge \widetilde\omega  \over \int_{\rm TN_1} \omega \wedge \omega}\right) + {\cal O}(\delta F_i)\Bigg], \nd
where we should remember that the two Taub-NUT spaces discussed above (respectively as ${\rm TN}_1$ in \eqref{curamo2} and ${\rm TN}_2$ as \eqref{outside}) 
not only have different orientations but also slightly different warp-factors; $R_{11}$ is the eleven-dimensional radius; and 
the corrections $\delta F_i$ to the warp-factors $F_i$ 
are the corrections 
to $\langle {\cal C}_3\rangle$. 

The Yang-Mills coupling should also change accordingly. To see this we should compute $\widetilde{c}_{11}$, the equivalence of $c_{11}$ given earlier. We proceed by first
defining $\hat{F}_i = F_i + \delta F_i$ for $i = 1, 3, 4$ and $\hat{F}_2 = \widetilde{F}_2 + \delta \widetilde{F}_2$, where the variations represent possible quantum
corrections to the warp factors. To the first approximation we will assume that there are no {\it extra} 
cross-terms in the type IIA metric \eqref{oldport} coming from the 
quantum corrections. A full generalization is technically challenging because eliminating the cross-terms by redefining the coordinates can make the resultant 
warp-factors to be functions of all the internal coordinates. However since $\widetilde{c}_{11}$ involves finding the determinant of the metric along the 
directions orthogonal to the Taub-NUT space, the cross-terms (which are of the same order as $\delta F_i$) would mostly contribute to ${\cal O}[(\delta F_k)^2]$. Thus 
the ${\cal O}(\delta F_i)$ contributions to the determinant can be viewed coming entirely from the warp-factor fluctuations of the metric \eqref{oldport}.    

This then gives us the explicit form for $\widetilde{c}_{11}$ in terms of the warp-factors $\hat{F}_i$, which have been defined above. The form is similar to what we 
had earlier because, 
as one may verify, the deformations to the type IIA metric \eqref{oldport} coming from M-theory uplift simply gets cancelled in the final 
expression:
\bg\label{pachmukh} 
\widetilde{c}_{11} &= & {R_3 R_{11}\over 2\pi}~\sec~\theta \int_0^\infty dr~e^{2\phi_0} \sqrt{{F}_1 \widetilde{F}_2 {F}_3\over b_4(\widetilde{F}_2 - {F}_3)} 
~{\rm ln}\left|{\sqrt{\widetilde{F}_2} + \sqrt{\widetilde{F}_2 - {F}_3}\over \sqrt{\widetilde{F}_2} - \sqrt{\widetilde{F}_2 - {F}_3}}\right| \nonumber\\
 && ~~~~~ + \int_0^\infty dr \left({\mathbb{B}_1~\delta F_1 \over F_1} + {\mathbb{B}_2~\delta{\widetilde{F}_2}\over \widetilde{F}_2} + {\mathbb{B}_3~\delta F_3 \over F_3} 
+ {\mathbb{B}_4~\delta \phi \over \phi}\right),\nd  
where $b_4$ is given in \eqref{bfour}, and the first term above is similar to \eqref{c11theta2} except for the additional factor of $R_{11}$, the eleven-dimensional 
radius. The correction terms given in terms of $\mathbb{B}_i$ are all functions of the warp-factors $F_i$, as one may easily derive. This means that the four-dimensional 
Yang-Mills coupling can now be expressed as:
\bg\label{brazpac}
{\widetilde{c}_1 \widetilde{c}_{11}\over \widetilde{v}_3} ~ = ~ {c_1 c_{11}\over v_3} 
\Bigg[{R_{11}\over 2\pi}
\left({\int_{\rm TN_2} \widetilde\omega \wedge \widetilde\omega  \over \int_{\rm TN_1} \omega \wedge \omega}\right) + {\cal O}(\delta F_i)\Bigg], \nd
where it should be clear from the context that the volumes of the three-cycles $\widetilde{v}_3$ and $v_3$ have different orientations, the former being along 
($x_3, r, x_{11}$) and the latter being along ($x_3, r, \phi_1$). However since $\widetilde{c}_1$ and $c_1$ are also oriented differently, the ratios 
${\widetilde{c}_1\over \widetilde{v}_3}$ and ${c_1\over v_3}$ match precisely with the orientations of the Taub-NUT spaces ${\rm TN}_2$ and ${\rm TN}_1$ respectively. 

The ${\cal O}(\delta F_i)$ corrections appearing in \eqref{brazpac} and \eqref{dhechu} are, at this stage, arbitrary but we expect them to be proportional to each 
other\footnote{Both the ${\cal O}(\delta F_i)$ corrections are integrated over all the coordinates, and especially $r$ and $\theta_1$, so they are only functions of 
the NC (or RR parameter) $\theta$.}.
In general they are not equal, so it will be instructive to see how they are related to each other. To analyze this let us express the ${\cal O}(\delta F_i)$ 
corrections to \eqref{brazpac} and \eqref{dhechu} to be ${\cal O}(\delta F^{(a)}_i)$ and ${\cal O}(\delta F^{(b)}_i)$  respectively. This means we can rewrite 
\eqref{dhechu} with the same coefficient of ${c_1 c_{11}\over v_3}$ as in \eqref{brazpac} but with an extra factor of:  
\bg\label{thebox}
{q(\theta)\over \widetilde{q}(\theta)} ~ \equiv ~ 1 ~+ ~ {{\cal O}(\delta F^{(b)}_i) - {\cal O}(\delta F^{(a)}_i) 
\over {R_{11}\over 2\pi}\left({\int_{\rm TN_2} \widetilde\omega \wedge \widetilde\omega  \over \int_{\rm TN_1} \omega \wedge \omega}\right) + {\cal O}(\delta F^{(a)}_i)}, \nd
where $\widetilde{q}(\theta)$ is similar to the arbitrary small parameter $q(\theta)$ that appeared in \eqref{chotas} in the definition of $\langle {\cal C}_3\rangle$. 
The above manipulation is useful because we can now express the complex coupling $\widetilde{\tau}$ for the S-dual theory to be:
\bg\label{asffacdes} 
\widetilde{\tau} ~ = ~ \widetilde{c}_2 + {i \widetilde{c_1} \widetilde{c}_{11}\over \widetilde{v_3}}
~= ~ {\left(-c_2 + {ic_1 c_{11}\over v_3}\right) \over
\Big[{R_{11}\over 2\pi}\left({\int_{\rm TN_2} \widetilde\omega \wedge \widetilde\omega  \over \int_{\rm TN_1} \omega \wedge \omega}\right) + {\cal O}(\delta F^{(a)}_i)\Big]^{-1}}, \nd
where all the parameters appearing above are functions of the RR (or NC) parameter $\theta$ as we discussed earlier. Furthermore,
the form of the denominator in \eqref{asffacdes} is written in a suggestive way so that one may connect this to the expected S-dual result:
\bg\label{sdualres} 
\widetilde{\tau} ~ = ~ -{1\over \tau} ~ = ~ - {\overline{\tau} \over \vert\tau\vert^2} ~ =  ~ {-c_2 + {ic_1 c_{11}\over v_3}\over ~c_2^2 + {c_1^2 c_{11}^2\over v_3^2}} ~ = ~ 
{v_3 \over c_1 c_{11}}\Bigg[{i - {2a\over 1 - a^2} \over \left({1+a^2\over 1 - a^2}\right)^2}\Bigg], \nd
provided of course that the denominator in \eqref{asffacdes} is equal to $\vert \tau \vert^2$. In the last equality above, we have invoked \eqref{c2nowb} which relates 
$c_2$ and ${c_1 c_{11}\over v_3}$ so that the ratio is completely expressed in terms of the parameter $a$. In this form, it may be easier to relate the denominator of 
\eqref{asffacdes} to the denominator in \eqref{sdualres}.  

Having described the S-duality in some details from supergravity, the next question is 
how should we go about defining a parameter like ${t}$, now to be renamed $\widetilde{t}$, in the S-dual theory. A naive description, following \eqref{tkya}:
\bg\label{lakla}
\widetilde{t} ~ \equiv ~ {\bar{\widetilde\tau} \over \vert\widetilde{\tau}\vert}, \nd
cannot quite be the right description for $\widetilde{t}$ simply because the {\it definition} of $t$ as in \eqref{tkya} only works when the 
four-dimensional Yang-Mills coupling and the $\Theta$-parameter are related via \eqref{kurianm}. Since the relation between Yang-Mills coupling and the $\Theta$-parameter changes under 
S-duality, \eqref{lakla} cannot be the right definition. We need to look for an alternative definition for $\widetilde{t}$ that may capture the right behavior in the S-dual theory. 

The clue comes from the connection between $\epsilon_0$, the susy transformation parameter before twisting, and $\epsilon$, the susy transformation parameter after twisting via the 
relation $\epsilon = \eta \otimes \epsilon_0$ where $\eta \in {\bf V}_8$. There exist an operator, defined in terms of $Q_2$ and $Q_3$ in \eqref{sosie2}, that may act on both 
$\epsilon$ and $\epsilon_0$ to annihilate them. The value of $t$ for which this could happen is of course \eqref{mkmaja} or \eqref{tanda}. Under a S-duality we should now ask how
$\epsilon$ and $\epsilon_0$ transform. We expect:
\bg\label{sdualexp}
\epsilon ~ \to ~ \widetilde{\epsilon} ~ \equiv ~ \widetilde{\epsilon}_L + \widetilde{t} ~\widetilde{\epsilon}_R ~ = ~ {\rm exp}\left({{\bf Q}_a}\right) \epsilon, \nd
where ${\bf Q}_a$ is an element of the S-duality group. On the other hand, a transformation like \eqref{sdualexp}, allows us to construct the following transformation laws for the 
individual components of $\epsilon$, namely $\epsilon_L$ and $\epsilon_R$, as:
\bg\label{LRtra}
\epsilon_L ~ \to ~ \widetilde{\epsilon}_L ~ \equiv ~ {\rm exp}\left({{\bf Q}_a}\right) \epsilon_L, 
~~~~~~~~ \epsilon_R ~ \to ~ \widetilde{\epsilon}_R ~ \equiv ~ {\rm exp}\left({{\bf Q}_b}\right)  \epsilon_R, \nd
where ${\bf Q}_a$ and ${\bf Q}_b$ are in general {\it not} equal to each other, although could be commuting. However a transformation like \eqref{LRtra} with unequal ${\bf Q}_a$  and 
${\bf Q}_b$ will not be consistent with \eqref{sdualexp}, unless we demand $t$ to also transform in the following way:
\bg\label{ttt}
t ~ \to ~ \widetilde{t} ~ \equiv ~ {\rm exp}\left({{\bf Q}_a - {\bf Q}_b}\right) t, \nd
under S-duality. Note that, with \eqref{ttt}, the transformations of $\epsilon$ as well as $\epsilon_0$ under S-duality are consistent to each other. This means, while we needed to use
a relation like \eqref{sosie2} to express $t$ in terms of the parameter of $\epsilon_0$ in \eqref{tanda}, the form for $\widetilde{t}$ can be inferred from \eqref{ttt} directly provided
we know the forms of (${\bf Q}_a, {\bf Q}_b$). 

Our simple consideration has yielded the transformation rule for $t$, but not the forms for (${\bf Q}_a, {\bf Q}_b$). At this stage, and as we 
mentioned above, we can say that they are commuting but unequal. The functional forms for (${\bf Q}_a, {\bf Q}_b$) require a more detailed analysis along the lines of \cite{langland}, 
wherein it is shown that ${\bf Q}_b = \overline{\bf Q}_a$, and the following transformation rule:
\bg\label{mipag}
\widetilde{t} ~ = ~ {\rm exp}\left(2i{\rm Im}~{\bf Q}_a\right) t ~ = ~ {c\tau + d\over \vert c\tau + d\vert}~t, \nd
where the last equality uses elements of the $SL(2, {\bf Z})$ 
group\footnote{Note that when $\tau = {4\pi i \over g^2_{YM}}$ or $\tau = {i\over g_s}$, then $\widetilde{t} = t$ in the limit $d = 0$. This makes sense because the ten-dimensional 
fermionic action in type IIB supergravity in the string frame has the form $\int d^{10}x ~e^{-2\varphi_B}\sqrt{g_{10}}~ \overline{\Psi}\Gamma^N D_N \Psi$ (plus interactions) which 
does not require any additional scaling of the fermions when $\varphi_B \to - \varphi_B$. However when the axion ${C}_0$ is present, the story is more involved. This is similar to
what we see from four-dimensional point of view too as depicted in \eqref{mipag}.}.  
As expected the definition of $\widetilde{t}$ is different from \eqref{lakla}. A little work, following 
\eqref{mipag} and \cite{wittenknots1}, will give us $\widetilde{t} = 1$. 

The choice of $\widetilde{t} = 1$ in the S-dual side may be a bit puzzling from the corresponding supergravity point of view. Before S-duality, the parameter $t$ can be related 
to the supergravity variables via the two relations in \eqref{tequa} or via \eqref{t1} and \eqref{tt2}. If we assume similar relations now between $\widetilde{t}$ and the sugra 
variables for the S-dual metric, we face a contradiction because the vanishing of $\widetilde{t} - \widetilde{t}^{-1}$ would imply the vanishing of corresponding $b_{12}$ coefficient, 
but this coefficient in the S-dual metric clearly doesn't vanish. The reason why we see an apparent contradiction is because we have assumed that the S-dual constraint equations 
would follow similar pattern of derivation as elaborated for the pre S-dual scenario. That this may not happen is already been anticipated in footnote {\ref{constraint}}: we may get 
{\it same} set of constraints via adding two additional terms to the Hamiltonian, instead of mapping the picture to the one involving $t$. From this point of view, there is no 
need to make any extra connection to the $t$ variable because supergravity by itself knows all about the fermionic structure from the start. As such, the S-dual picture is also 
self contained.  

However the mapping to $t$ in \eqref{tequa} is not without its own merit. It showed us how to connect our set of solutions to the localization equations of \cite{wittenknots1} and 
\cite{langland}. Interestingly, adding the aforementioned two set of terms to the Hamiltonian would not have changed our conclusions, or the path of derivations, regarding the 
background constraints! The mapping to $t$ in the pre S-dual picture showed us another layer of hidden structures in our construction. In the S-dual picture no contradictions 
will now arise even if don't make any mention of the $\widetilde{t}$ parameter from supergravity point of view. The BHN equations would continue to resemble the ones in 
\eqref{prolpart}, but now expressed in terms of the S-dual fields.

\subsection{Types of solutions: surface operators and opers \label{surfu}}

In the above sections we have managed to discuss the appearance of the BHN equations, including the boundary Chern-Simons theory \eqref{plaza} using the twisted 
gauge field \eqref{newgf}, from M-theory. 
The key question to ask now is the locations of the knots. In other words, what additional ingredients do we need to construct knots in this theory? 
In the following we will discuss this and
other related issues. Our aim would also be to build a bridge between {\bf Model A} and {\bf Model B} using our set-up that we developed above. As we shall see, the key player 
for both the models would be the surface operators.  

\subsubsection{M2-brane states, surface operators and the BHN equations \label{bathle}}

Lets us start with M2-brane wrapping the two-cycle of our Taub-NUT space. The Taub-NUT space is oriented along directions ($\theta_1, x_8, x_9, x_{11}$) with 
$x_{11}$ being the Taub-NUT circle. This means the M2-brane will be a source of a point charge in the remaining $6+1$ dimensional orthogonal space in the
following way:
\bg\label{m2source} 
\int {\cal C} ~ = ~ \int {\cal A} \wedge \omega ~ = ~ \int {\cal A}_0 ~dx_0 \int_{\rm TN} \omega ~ \equiv ~ q \int {\cal A}_0 ~dx_0, \nd
where the value of the charge $q$ appears from the integral of the harmonic two-form $\omega$ over the Taub-NUT space. Reducing down to the $3+1$ dimensional space, this would lead to the 
non-abelian enhancement in the presence of multiple wrapped M2-branes on the two-cycle, as discussed in section \ref{NAE} and in \eqref{chelara}. 
 
For our case this is {\it not} what we need to study the knots: The wrapped M2-branes on Taub-NUT two-cycles could only enhance the gauge symmetry but will not give us the required Wilson 
loops necessary to study knots. What other M2-brane states can we study here? This then brings us to few other possible configurations of M2-branes that can be realized in the 
Taub-NUT background. As we shall see, the most relevant ones will be related to the {\it surface operators} in our $3+1$ dimensional gauge theory.  

Our first configuration that we want to entertain can be realized directly in the original brane construction in {\bf Table \ref{wittenbranes}}, or more appropriately the T-dual one 
given in {\bf Table \ref{modbranes}} with the second NS5-brane removed. This way we can simply keep two parallel NS5-branes oriented along ($x_0, x_1, x_2, x_3, x_8, x_9$) with 
D4-branes and a D2-brane oriented as in {\bf Table \ref{surfaceoper}}. 
\begin{table}[h!]
 \begin{center}
\begin{tabular}{|c||c|c|c|c|c|c|c|c|c|c|}\hline Directions & $x_0$ & $x_1$ & $x_2$
& $x_3$ & $\theta_1$ & $\phi_1$ & $\psi$ & $r$ & $x_8$ & $x_9$ \\ \hline
NS5 & $\surd$  & $\surd$   & $\surd$   & $\surd$ & $\ast$  & $\ast$ & $\ast$ & $\ast$ & $\surd$ & $\surd$ \\  \hline
D4 & $\surd$  & $\surd$   & $\surd$   & $\surd$ & $\ast$  & $\ast$ & $\surd$ & $\ast$ & $\ast$  & $\ast$ \\  \hline
D2 & $\surd$  & $\surd$   & $\ast$   & $\ast$ & $\ast$  & $\surd$ & $\ast$ & $\ast$ & $\ast$  & $\ast$ \\  \hline
  \end{tabular}
\end{center}
  \caption{The orientation of a D2-brane as a surface operator in $3+1$ dimensional non-compact directions.}
\label{surfaceoper}
\end{table}
\noindent The D2-brane state, which is a co-dimension two defect, acts as a surface operator in $3+1$ dimensional gauge theory. This has been described in many recent works 
(see \cite{gukovwitten2}, \cite{gukovwitten}, \cite{gukovsurf} for discussions
on the subject and for references) 
which the readers may 
refer to for details. For us, we want to lift this configuration to M-theory by first dualizing this to type IIB theory (see details in earlier sections), followed by
shrinking the $\phi_1$ circle to zero size and then opening up the eleventh-direction. The M2-brane state in M-theory 
is now depicted in {\bf Table \ref{m2m2}}. In type IIA theory this will simply be a D2-brane embedded inside D6-branes. It is also easy to make the system non-abelian by taking multiple 
M2-branes, or in type IIA theory, multiple embedded D2-branes inside D6-branes. 
\begin{table}[h!]
\begin{center}
\begin{tabular}{|c||c|c|c|c|c|c|c|c|c|c|c|}\hline Directions & $x_0$ & $x_1$ & $x_2$
& $x_3$ & $\theta_1$ & $\phi_1$ & $\psi$ & $r$ & $x_8$ & $x_9$ & $x_{11}$\\ \hline
Geometry & $\ast$  & $\ast$   & $\ast$   & $\ast$ & $\surd$  & $\surd$ & $\surd$ & $\surd$ & $\ast$ & $\ast$ & $\ast$ \\  \hline
Taub-NUT & $\ast$  & $\ast$   & $\ast$   & $\ast$ & $\surd$  & $\ast$ & $\ast$ & $\ast$ & $\surd$  & $\surd$ & $\surd$ \\  \hline
M2 & $\surd$  & $\surd$   & $\ast$   & $\ast$ & $\ast$  & $\ast$ & $\surd$ & $\ast$ & $\ast$  & $\ast$ & $\ast$ \\  \hline
  \end{tabular}
\end{center}
  \caption{M2-brane state in the warped Taub-NUT background. The warping appears from non-trivial geometry, shown above, and G-fluxes, discussed earlier.}
\label{m2m2}
\end{table}

\noindent Our goal now is to find how the M2-branes modify the BHN equations that we discussed earlier. In particular we would like to see how, for example, the background constraint 
equations \eqref{prolpart} (or \eqref{pharmacy}), 
change in the presence of the M2-brane states. A direct study of multiple M2-brane states in M-theory following \cite{bagger} would make our analysis harder. 
However the fact that, in the dual type IIA side, the D2-brane states are bound states with the D6-branes make this analysis a bit easier because the bound D2-brane states could be considered 
as instantons on the D6-branes. In M-theory therefore the M2-brane states would simply be provided by {\it localized} G-fluxes, and the M2-branes' charge $Q_2$ would appear from:
\bg\label{m2charge} 
Q_2 ~= ~ \int_{\Sigma_8} {\cal G}_4 \wedge {\cal G}_4 ~ = ~ \int_{\Sigma_4} \langle {\cal F}\rangle  \wedge \langle {\cal F} \rangle \int_{\rm TN} \omega \wedge \omega, \nd
where $\Sigma_8 = \Sigma_4 \times {\rm TN}$, with $\Sigma_4$ being a four-dimensional surface oriented along ($x_2, x_3, r, \phi_1$) and the orientation of the Taub-NUT space as before. 
This means, on one hand, switching on the above-mentioned instanton implies switching on the following components of the 
seven-dimensional gauge fields: ${\cal A}_2, {\cal A}_3, {\cal A}_r, {\cal A}_{\phi_1}$. On the other hand, from our four-dimensional point of view with the action \eqref{stotal}, 
having an instanton \eqref{m2charge} implies switching on the four-dimensional gauge field component ${\cal A}_2$ and the three scalar fields
(${\cal A}_3, \sigma, \bar{\sigma}$) where $\sigma$ is defined in \eqref{sigma}.  

The above discussion implies that, in the presence of M2-branes, we can entertain a more elaborate decomposition than envisioned in \eqref{chorM} by taking into account {\it localized}
G-fluxes of \eqref{m2charge} alongwith the usual G-fluxes in the following way:
\bg\label{nayak}
{\cal G}_4 ~ = ~ \langle {\cal G}_4 \rangle + \langle {\cal F} \rangle \wedge \omega + \left({\cal F} + {\cal B}_2 \right) \wedge \omega +  {\cal G} \varphi_o 
+ {\cal H}_3 \wedge \zeta, \nd
where $\omega = d\zeta$ has been defined earlier, ${\cal H}_3  = d{\cal B}_2$ is the three-form, $\varphi_o$ is the harmonic zero-form defined on the warped Taub-NUT space, and 
${\cal G}$ is the fluctuation of the four-form in the seven-dimensional spacetime orthogonal to the warped Taub-NUT space. The four-form piece ${\cal H}_3 \wedge \zeta$ only 
contributes to the ten-dimensional type IIA action, and so we can ignore this for our case. This means we can also absorb ${\cal B}_2$ in the definition of ${\cal F}$ without any
loss of generalities. 
 
Plugging \eqref{nayak} in the M-theory action along the lines of \eqref{mtheorya} will not only reproduce back the total four-dimensional action \eqref{stotal} from the zero mode fluctuations
of the fluxes and fields over the warped Taub-NUT space, but will also give us the additional M2-brane piece $Q_2\int {\cal C}_{01\psi}~dx_0 \wedge dx_1 \wedge d\psi$. This means the BHN 
equation \eqref{f12bhn} will remain unchanged if the internal instanton contibutions to the 
charge piece \eqref{m2charge} come only from the background scalar fields (${\cal A}_3, \sigma, \bar\sigma$). The precise conditions, to first approximations, 
are modifications of \eqref{scalfy} and \eqref{rater}
in the following way:
\bg\label{metromeye}
&&{\cal D}_\eta \delta\sigma ~ = ~ {\cal D}_\eta \delta{\bar\sigma} ~ = ~ 0 \nonumber\\
&& [\delta\sigma, \bar\sigma] + [\sigma, \delta\bar\sigma] ~ = ~ [\delta\sigma, \varphi_k] ~ = ~ [\delta\bar\sigma, \varphi_k] ~ = ~ 0, \nd
where $\delta\sigma$ and $\delta\bar\sigma$ are the fluctuations of the scalar fields ($\sigma, \bar\sigma$) in the presence of 
the instanton \eqref{nayak}. The other two fluctuations of the components of the gauge fields $\delta{\cal A}_2$ and $\delta{\cal A}_3$ would in principle only redefine the BHN equation
\eqref{prolpart} and the gauge condition \eqref{gaugec} respectively without changing the content of the equations. We will however 
retain the gauge condition \eqref{gaugec} by resorting to ${\cal A}_3 = 0$ case\footnote{This way ${\cal F}_{\alpha 3} \equiv -\partial_3 {\cal A}_\alpha$ for both abelian and non-abelian cases.}. 

However subtlety comes when we look at the other set of the BHN equations, namely \eqref{meyepita} or \eqref{bhishon}. Considering the $c_2 = 0$ case for simplicity, the BHN equation for
the ${\cal F}_{\alpha\psi}$ components of the gauge fields can be rewritten in a more complete form, in the absence of M2-branes, as:
\bg\label{falppsi}
{\cal F}_{\alpha\psi} + \sum_{\beta, k = 1}^3 \sqrt{2 b_{\beta k} v_3\over c_1 c_{12}}~
\epsilon_{\alpha \psi} m^{(2)}_{\beta k} ~{\cal D}_\beta \varphi_k - i \sum_{k, l = 1}^3 \sqrt{2 d_{kl} v_3\over c_1 c_{12}}~ h^{(1)}_{\alpha \psi kl}~ [\varphi_k, \varphi_l] ~ = ~ 0, \nd
where the coefficients appearing above have been defined earlier. In \eqref{meyepita} and \eqref{bhishon} we had taken the simplifying assumption where only $q^{(1)}_{12}$ vanishes. 
Generically however $q^{(1)}_{kl} = 0$ for all choices of ($k, l$). Additionally we can demand non-zero values for the coeffcients $h^{(1)}_{\alpha \psi kl}$. This way we no longer have 
to decouple $\varphi_3$ as in \eqref{phi3c}. On the other hand, if we don't want to change \eqref{prolpart}, we can easily take appropriate values for the coefficients 
$g^{(1)}_{\alpha\beta k l}$ satisfying the third constraint in \eqref{coeffrelo}. 
    
The discussion in the above paragraph was intended to establish a link between the BHN equation \eqref{falppsi} and the surface operators that we discussed at the begining of this section.  
In the type IIA side, as depicted in {\bf Table \ref{surfaceoper}}, the D2-branes intersect the D4-branes along ($x_0, x_1$) directions and therefore the support $D$ of the surface 
operator should be along $x_2 = \psi = 0$ (recall that $x_3$ direction is a compact circle for us). When one of the parallel NS5 is sufficiently far away the 
supersymmetry on the D4-branes is ${\cal N} = 4$ and therefore, as discussed in \cite{gukovwitten}, the supersymmetry preserved by the surface operator is ($4, 4$) supersymmetry 
from two-dimensional point of view. Using the language of M-theory construction discussed in {\bf Table \ref{m2m2}}, the ($4, 4$) vector multiplet contains vector fields with 
components (${\cal A}_0, {\cal A}_1$) and four scalars (${\cal A}_3, \sigma, \bar{\sigma}, \varphi_1$) all in the adjoint representations of the gauge group. The ($4, 4$) 
hypermultiplet is constructed from the remaining two gauge field components (${\cal A}_2, {\cal A}_\psi$) and the two scalars
($\varphi_2, \varphi_3$)\footnote{Following \eqref{4scalars} one might have expected the two scalars to be ($\varphi_2, {\cal A}_3$). This unfortunately will not work with the 
gauge choice \eqref{gaugec}. However since $h^{(1)}_{\alpha\psi kl} = h^{(1)}_{[\alpha\psi][kl]}$ this is not an issue for us, and we can as well choose the two scalars to be
($\varphi_2, \varphi_3$). Additionally note that while the components of the gauge fields that enter
the vector multiplet and the hypermultiplet are fixed, we have some independence in distributing the scalars in the two multiplets. This independence stems from two sources, one, 
our choice of the 
gauge \eqref{gaugec} or \eqref{gaugec2} and, two, the definition of the decoupled scalars ($\sigma, \bar\sigma$).}.  

Looking at the components of the hypermultiplets, we see that the BHN equation \eqref{falppsi} can be used to capture the behavior of the hypermultiplets of the two-dimensional theory. In fact
we are interested in $\alpha = 2$ BHN equation in \eqref{falppsi}. In other words, we have the following BHN equation, again in the absence of any M2-branes, 
associated to the ${\cal F}_{2\psi}$ component of the gauge field:
\bg\label{rashi}
{\cal F}_{2\psi} + \sqrt{2 v_3\over c_1 c_{12}} \left(m^{(2)}_{23}\sqrt{b_{23}} ~{\cal D}_2\varphi_3 + m^{(2)}_{\psi 2}\sqrt{b_{\psi 2}} ~{\cal D}_\psi\varphi_2\right) 
-2 i \sqrt{2 d_{23}v_3\over c_1 c_{12}} ~h^{(1)}_{2\psi 2 3}~[\varphi_2, \varphi_3] ~ = ~ 0, \nonumber\\ \nd 
where the coefficients $b_{23}, b_{\psi 2}$ and $d_{23}$ are given in \eqref{boka}, \eqref{bokcho} and \eqref{venu} respectively; with $m^{(2)}_{\alpha \beta}$ satisfying the 
constraint given by the last equation in \eqref{coeffrelo}. 
Note that keeping \eqref{prolpart} unchanged means that $m^{(2)}_{\psi 2} = \pm m^{(2)}_{23}$, where the sign ambiguity will be 
fixed soon. In addition, we will make a small change in the identification of the scalars given earlier in \eqref{4scalars} to the following:
\bg\label{foursca}
\left(\phi_0, \phi_1, \phi_2, \phi_3\right) ~\propto~ \left({\cal A}_3, \varphi_1, \varphi_2, \varphi_3\right), \nd
which will be more useful for us than the earlier identification. Interestingly \eqref{foursca} implies that the Coulomb branch scalar $\overrightarrow{X}$ will be ($\phi_0, \phi_1, \phi_2$) 
exactly as in \cite{wittenknots1} (see also footnote \ref{shollow} and {\bf Table \ref{pirates}}). 
Now defining:
\bg\label{devkanta}
\Phi_2 ~ \equiv ~ -i\sqrt{2 v_3 b_{\psi 2} \over c_1 c_{12}}~m^{(2)}_{23}~\varphi_2, ~~~~~~~~ \Phi_3 ~ \equiv ~ -i\sqrt{2 v_3 b_{23} \over c_1 c_{12}}~m^{(2)}_{23}~\varphi_3, \nd
where $c_1, c_{12}$ and $v_3$ have been defined earlier in \eqref{c1c2c1c2}, \eqref{krisis} and \eqref{intig} respectively, we can plug this in \eqref{rashi} to rewrite it as:
\bg\label{swapan}
{\cal F}_{2\psi} - i\left({\cal D}_2 \Phi_3 \pm {\cal D}_\psi \Phi_2\right) + i \sqrt{2 d_{23} c_1 c_{12}\over v_3 b_{23} b_{\psi 2}} \left({h^{(1)}_{2\psi 23}\over 
\vert m^{(2)}_{\psi 2} m^{(2)}_{2 3}\vert}\right) [\Phi_2, \Phi_3] ~ = ~ 0. \nd 
The sign ambiguity appearing above can be fixed by looking at the constraints on the scalar fields in \eqref{prolpart}. If we want similar conditions for our present case too, then 
we expect the {\it full} set of BHN equations to be an appropriate modification of \eqref{swapan} in the following way:
\bg\label{vedpra}
&& F_{2\psi} + c_0 ~D_1\Phi_0 - [\Phi_2, \Phi_3] ~ = 0 \nonumber\\
&& D_2\Phi_2 + D_\psi \Phi_3 ~ = ~ 0, ~~~D_\psi\Phi_2 - D_2\Phi_3 ~ = ~ 0, \nd
where $c_0$ is a constant that we will derive below. Note that there is no relative constant in the second equation in \eqref{vedpra}. This is only in the simplifying case where
$b_{\psi 3} = b_{23}$, with $b_{\psi k}$ as given in \eqref{bokcho} and $b_{ak}$ as given in \eqref{boka}, 
otherwise we expect a relative ratio of ${b_{\psi 3} \over b_{23}}$.  
The two scalar fields ($\Phi_2, \Phi_3$) have already been identified in \eqref{devkanta}, so $\Phi_0$ appearing in \eqref{vedpra} 
can only be proportional to $\varphi_1$ or ${\cal A}_3$. However it cannot be proportional to $\varphi_1$ because of the derivative structure in the first equation of \eqref{vedpra}. Thus
$\Phi_0$ should be proportional to ${\cal A}_3$, but since the value of ${\cal A}_3$ is fixed via the gauge choice \eqref{gaugec} at least to the first 
approximation\footnote{Looking at the Hamiltonian \eqref{hamilbeta}, which is written as sum of squares, we can easily infer that ${\cal A}_3$ do not appear in the squared piece with 
${\cal F}_{\alpha \psi}$. This of course is because of our gauge choice \eqref{gaugec} hence it is no surprise that $\Phi_0$ vanishes in \eqref{vedpra}.},   
we conclude that $\Phi_0 = 0$ here. This not only fixes the
sign ambiguity in \eqref{swapan}, but also gives rise to the Hitchin's equation which are precisely the conditions for supersymmetry with the hypermultiplets!
 
\begin{table}[h!]
\begin{center}
\begin{tabular}{|c||c||c|c|}\hline Epsilon factor & BHN decomposition & Map 1: \eqref{4scalars} & Map 2: \eqref{foursca}\\ \hline 
$12~ \otimes$ ($0\psi \oplus\psi 0$) & $D_0\phi_3 - D_\psi \phi_0$ & ${\cal D}_0{\cal A}_3 - {\cal D}_\psi\varphi_3$  & ${\cal D}_0\varphi_3 - {\cal D}_\psi {\cal A}_3$ \\
$12~ \otimes$ ($12 \oplus 21$) & $D_1\phi_2 - D_2 \phi_1$ & ${\cal D}_1{\varphi}_2 - {\cal D}_2\varphi_1$  & ${\cal D}_1{\varphi}_2 - {\cal D}_2\varphi_1$  \\ \hline
$1\psi~ \otimes$ ($02 \oplus 20$) & $D_0\phi_2 - D_2 \phi_0$ & ${\cal D}_0{\varphi}_2 - {\cal D}_2\varphi_3$  & ${\cal D}_0\varphi_2 - {\cal D}_2 {\cal A}_3$ \\
$1\psi~ \otimes$ ($1\psi \oplus \psi 1$) & $D_1\phi_3 - D_\psi \phi_1$ & ${\cal D}_1{\cal A}_3 - {\cal D}_\psi\varphi_1$  & ${\cal D}_1\varphi_3 - {\cal D}_\psi {\varphi}_1$ \\ \hline
$2\psi~ \otimes$ ($01 \oplus 10$) & $D_0\phi_1 - D_1 \phi_0$ & ${\cal D}_0{\varphi}_1 - {\cal D}_1\varphi_3$  & ${\cal D}_0\varphi_1 - {\cal D}_1 {\cal A}_3$ \\
$2\psi~ \otimes$ ($2\psi \oplus \psi 2$) & $D_2\phi_3 - D_\psi \phi_2$ & ${\cal D}_2{\cal A}_3 - {\cal D}_\psi\varphi_2$  & ${\cal D}_2\varphi_3 - {\cal D}_\psi {\varphi}_2$ \\ \hline
  \end{tabular}
\end{center}
  \caption{Various terms in the BHN equations coming from the two scalar fields mapping choices 1 and 2 respectively. The first
column is the epsilon tensor decomposition along the lines of our earlier discussion, where only the relevant pieces are shown. The second column correspond to the parts of the BHN equations associated to the 
epsilon decomposition. Finally columns 3 and 4 are related to the pieces of the BHN equations once we use the mappings 1 and 2 respectively.}
\label{pirates}
\end{table} 

The coefficient $c_0$ is not zero, and fixing this will also tell us how $F_{2\psi}$ appearing in \eqref{vedpra} is related to ${\cal F}_{2\psi}$ appearing in \eqref{swapan}. To 
see how the latter transformation occurs, we define:
\bg\label{moyukh} 
&& {\cal A}_2 = -{i A_2\over \sqrt{c_0}}, ~~~~~~ {\cal A}_\psi = -{i A_\psi \over \sqrt{c_0}}, ~~~~~~~ x_2 = \bar{x}_2 \sqrt{c_0}, ~~~~~~~ \psi = \bar{\psi}_2 \sqrt{c_0} \nonumber\\ 
&& A = A_2 ~d\bar{x}_2 + A_{\psi} ~d\bar{\psi}, ~~~~~ \Phi = \Phi_2 ~d\bar{x}_2 + \Phi_3 ~d\bar{\psi}, ~~~~~ d_A = d + [A, ~].  \nd
The first line of the above set of equations when plugged in \eqref{swapan} gives us \eqref{vedpra} with vanishing $\Phi_0$. Once we plug in the second line of \eqref{moyukh} in 
\eqref{vedpra}, we can rewrite \eqref{vedpra} as:
\bg\label{may6baz}
F - \Phi \wedge \Phi ~ = ~ 0, ~~~~~ d_A \Phi ~ = ~ 0 ~ = ~ d_A \ast \Phi, \nd
which, as discussed above, are precisely the set of Hitchin's equations that appeared in \cite{gukovwitten2}, \cite{gukovwitten}, \cite{gukovsurf}
describing the scenario when we do not consider the singularity associated with the 
surface operators. The hodge star\footnote{Our choice of hodge star is slightly different from the ones taken in \cite{gukovwitten2}, \cite{gukovwitten}, \cite{gukovsurf} and 
in \cite{wittenknots1}, but the essential content is captured in \eqref{may6baz}. \label{diali}} is 
defined in the two-dimensional space parametrized by ($\bar{x}_2, \bar{\psi}$), 
and $c_0$ appearing in \eqref{vedpra} as well as \eqref{moyukh} is at least proportional to inverse of the 
coefficient of the commutator piece in \eqref{rashi}, i.e:
\bg\label{casin}
c_0 ~ \propto ~ {\left(m^{(2)}_{23}\right)^2\over h^{(1)}_{2\psi 2 3}} \sqrt{v_3 b_{23} b_{\psi 2}\over c_1 d_{23} c_{12}}. \nd   
The above derivations are encouraging and allow us to make the first step in deriving the behavior of the surface operator from M2-branes embedded in 
non-trivial geometry and fluxes in M-theory. The question now is: how is the 
{\it singularity} of the support $D$ of the surface operator manifested in the Hitchin's equations \eqref{may6baz}? 

To analyze this we will have to go beyond \eqref{metromeye} and look at \eqref{hamilbeta} more carefully. There is no reason for the two scalars ($\sigma, \bar{\sigma}$) to 
completely decouple $-$ like \eqref{scalfy} and \eqref{rater} $-$ now. The original constraints that governed the decoupling conditions appeared in \eqref{coeffrelo}, which we can 
rewrite in the following way: 
\bg\label{coeff45}
2\left|s^{(l)}_{12}\right|^2 + s^{(l)}  = 1, ~~~~~~~~ \sum_{\alpha = 1}^2\left|t^{(l)}_{\alpha}\right|^2 + t^{(l)}  = 1, \nd 
where all the parameters appearing above are described in \eqref{hamilbeta}, and we can choose $l = 1, 2$ for our case. Additionally, 
we have assumed $s^{(l)}$  and $t^{(l)}$ to be positive definite 
integers, and therefore  
the decoupling conditions in \eqref{scalfy} and \eqref{rater} were simply the non-vanishing of them, i.e:
\bg\label{madi}
s^{(l)} ~ > ~ 0, ~~~~~~~ t^{(l)} ~ > ~ 0. \nd
The constraint \eqref{scalfy} and \eqref{rater} imposed via \eqref{madi} in \eqref{hamilbeta} now would be harder to implement completely in the presence of the localized G-fluxes along 
($x_2, x_3, r, \phi_1$). However,
we might still be able to argue for $\psi$ independence of the scalar fields $\sigma$ and $\bar{\sigma}$, but $\beta$ independence cannot hold now. Thus 
the first constraint in \eqref{madi} above may still hold, but $t^{(l)}$ has to vanish in the Hamiltonian \eqref{hamilbeta}. Similarly $q^{(4)}$ appearing in the first equation, 
as well as $q^{(1)}_{kl}$ in the third equation, of \eqref{coeffrelo} will also have to vanish. This way we will only have:
\bg\label{sinlee}
{\cal D}_\psi \sigma ~ = ~ {\cal D}_\psi \bar{\sigma} ~ = ~ 0, \nd
and not the full constraints \eqref{scalfy} and \eqref{rater}. What about \eqref{kyabol}? Recall that this was imposed via switching on $q_k^{(\gamma)}$ in \eqref{hamilbeta} and 
appears in the second constraint relation \eqref{coeffrelo}. There is no reason why this could be non-zero now so, as a most generic condition, we will assume that this coefficient 
also vanishes. This way \eqref{kyabol} may not hold in the presence of the localized G-fluxes. 

There are two ways to proceed now. One, we can assume that all the BHN equations, namely \eqref{prolpart} and \eqref{falppsi}, get contributions from the scalar fields ($\sigma, \bar{\sigma}$);
and two, only \eqref{falppsi} gets contributions from the ($\sigma, \bar{\sigma}$) scalar fields with \eqref{prolpart} remaining unchanged. The latter would imply that we impose:
\bg\label{moson}
g^{(2)}_{\alpha \beta k}  ~ = ~ g^{(3)}_{\alpha \beta k}  ~ = ~ g^{(4)}_{\alpha \beta} ~ = ~ 0, \nd
in \eqref{coeffrelo} along with \eqref{sinlee}. Additionally the instantonic configuration, that results in the M2-brane states via \eqref{m2charge} and in the G-flux decomposition 
\eqref{nayak}, can be generated for our case from the following gauge field configurations\footnote{Note that we haven't made a distinction between 
($\sigma, \bar\sigma$) and ($\langle\sigma\rangle, \langle\bar\sigma\rangle$) to avoid clutter. Since ($\sigma, \bar\sigma$) only appear for our instanton configuration, switching on them 
means we have switched on their expectation values. This should be clear from the context.}:
\bg\label{oneill}
\langle {\cal A}_r\rangle (r, x_2) ~ = ~ {\sigma + \bar\sigma\over 2},
 ~~~~~~~ \langle {\cal A}_{\phi_1}\rangle (\phi_1, x_2, x_3) ~ = ~ {\sigma - \bar\sigma\over 2i}, \nd
from where we can have $\langle {\cal F}_{2r}\rangle$ and $\langle {\cal F}_{3\phi_1}\rangle$ as the source for the M2-brane charges \eqref{m2charge}. This choice of components is fairly 
generic and helps us avoid 
switching on components like $\langle {\cal F}_{2\phi_1}\rangle$, $\langle {\cal F}_{23}\rangle$, $\langle {\cal F}_{3r}\rangle$
and $\langle {\cal F}_{r\phi_1}\rangle$ at least in the abelian case (which we will finally resort to). 
Again, we can always go to more elaborate scenario but since many of the extra components can be eliminated by gauge transformations, 
with no additional physics insights, we can narrow our choice to the simple case of \eqref{oneill}. Of course
the above discussion does not in any way imply that {\it fluctuations} 
${\cal A}_1$ and ${\cal A}_2$ are defined as \eqref{oneill}. The fluctuations remain functions of the space coordinates ($x_1, x_2, \psi$) so that the components 
${\cal F}_{\alpha \beta}$ and ${\cal F}_{\alpha \psi}$ defined appropriately are related by the BHN equations. 

This then brings us to the BHN equation, in the presence of the instanton source \eqref{oneill}, for the component ${\cal F}_{\alpha\psi}$. As mentioned earlier, we are interested 
in the component ${\cal F}_{2\psi}$. The BHN equation for this is given by\footnote{Note that $\overline{{\cal D}_2\sigma}$ 
is defined with respect to the gauge field $\bar{{\cal A}_2}$.
However if we use $D_2\sigma$ instead of ${\cal D}_2\sigma$, these two definitions of covariant derivative being 
connected via ${\cal A}_2 = - i A_2$ as in \eqref{moyukh} assuming $c_0 = 1$, then $\overline{{\cal D}_2\sigma} = D_2\bar{\sigma}$ assuming $A_2$ to be purely real. 
Thus, unless mentioned otherwise, we will continue using the field strength ${\cal F}_{\alpha \beta}$ 
defined with respect to the gauge fields ${\cal A}_\alpha$ and ${\cal A}_\beta$ instead of the field strength $F_{\alpha\beta}$. Note that they are related via: 
${\cal F}_{\alpha\beta} = -i F_{\alpha\beta}$.}:
\bg\label{zoey}
{\cal F}_{2\psi} -i \gamma_4 [\varphi_2, \varphi_3] + 2\epsilon_{2\psi}~ {\bf Re}\left(\gamma_5 {\cal D}_2 \sigma \right) = 
2i~{\bf Re}\left(\gamma_1 [\bar\sigma, \varphi_2] + \gamma_2 [\bar\sigma, \varphi_3]\right) + \gamma_3 [\bar\sigma, \sigma], \nonumber\\ \nd
along with the two additional conditions on $\varphi_2$ and $\varphi_3$ as given in \eqref{vedpra} with suitable modifications. The other coefficients appearing in \eqref{zoey} are 
defined in the following way:
\bg\label{house}
&&\gamma_1 ~ = ~ {1\over 2} \sqrt{2 v_3\over c_1 c_{12}}\left[h^{(2)}_{2\psi 2} \sqrt{c_{r2}}  + i h^{(3)}_{2\psi 2} \sqrt{c_{\phi_1 2}}\right], 
~~\gamma_2 ~ = ~ {1\over 2} \sqrt{2 v_3\over c_1 c_{12}}\left[h^{(2)}_{2\psi 3} \sqrt{c_{r3}}  + i h^{(3)}_{2\psi 3} \sqrt{c_{\phi_1 3}}\right] \nonumber\\ 
&& \gamma_3 ~ = ~ {h^{(4)}_{2\psi} \sqrt{2 a_1}\over 2\sqrt{c_{12}}}, 
~~ \gamma_4 ~ = ~ {h^{(1)}_{2\psi 2 3}\sqrt{8 d_{23} v_3}\over \sqrt{c_1 c_{12}}}, ~~
\gamma_5 ~ = ~ {1\over \sqrt{2 c_{12}}}\left[t^{(1)}_{2r} \sqrt{c_{r2}}  - i t^{(2)}_{2\phi_1} \sqrt{c_{\phi_1 2}}\right],  
 \nd   
where we have defined the coeffcients $c_{ar}$ in \eqref{bentu}, $c_{a\phi_1}$ in \eqref{jhogra}, $d_{23}$ in \eqref{venu}, $c_{12}$ in \eqref{krisis}, $c_1$ in \eqref{c1c2c1c2} and 
$v_3$ in \eqref{intig}. The other coefficents appearing in \eqref{house} are defined in \eqref{coeffrelo} except the two {\it new} coefficients $t^{(1)}_{2r}$ and 
$t^{(2)}_{2\phi_1}$. These two coefficients replace the previous two coefficients $t^{(1)}_2$ and $t^{(2)}_2$ respectively, appearing in the Hamiltonian \eqref{hamilbeta} and the 
constraint equations \eqref{coeffrelo}, via:
\bg\label{20tigoenda}
t^{(k)}_2 \epsilon_{2\psi a s_k} ~ \to ~ \epsilon_{2\psi} ~t^{(k)}_{as_k}, \nd
where $k =$ (1, 2) and $s_k$ are coordinates defined as $s_1 = r, s_2 = \phi_1$. 
One immediate advantage of this replacement in \eqref{hamilbeta} is that $a$ in \eqref{20tigoenda} can take values $a = 1$ or $a = 2$ and is thus not restricted by the total antisymmetry
constraint. The constraint relation for $t^{(k)}_{ar}$ is similar to what we had for $t^{(k)}_\alpha$ in \eqref{coeffrelo}, namely:
\bg\label{sksamagra}
\sum_{a = 1}^2 \vert t^{(k)}_{as_k}\vert^2 ~ + ~ t^{(k)} ~ = ~ 1. \nd
Clearly for $t^{(k)} = 0$, this change doesn't alter any of our earlier results because of the decoupling of the ($\sigma, \bar\sigma$) fields. However now that ($\sigma, \bar\sigma$) 
are relevant, introducing $t^{(k)}_{as_k}$ can make our analysis more generic. Note that we are {\it not} required to make similar changes to $s^{(k)}_{\alpha\beta}$ in \eqref{hamilbeta} 
and \eqref{coeffrelo} because of \eqref{sinlee}. 

The ${\cal F}_{2\psi}$ BHN equation \eqref{zoey} seems more involved and therefore it will be instructive to rewrite it in a slightly different way so as to simplify the 
appearance of the equation. To proceed, let us define two new fields $\hat{\varphi}_2$ and $\hat{\varphi}_3$ using our old fields ${\varphi}_2$ and ${\varphi}_3$    
in the following way:
\bg\label{londonk}
\hat{\varphi}_2 ~ = ~ \varphi_2 + 2 {\bf Re}\left({\bar{\gamma}_2 \sigma \over \gamma_4}\right), ~~~~~~
\hat{\varphi}_3 ~ = ~ \varphi_3 - 2 {\bf Re}\left({\bar{\gamma}_1 \sigma \over \gamma_4}\right), \nd
where $\gamma_1, \gamma_2$ and $\gamma_4$ are defined in \eqref{house}. The fields are defined in such a way so that the commutator between them takes the following form:
\bg\label{sirK}
[\hat{\varphi}_2, \hat{\varphi}_3] ~ = ~ [{\varphi}_2, {\varphi}_3] + 2{\bf Re} \left({\gamma_1\over \gamma_4}[\bar\sigma, \varphi_2] 
+ {\gamma_2\over \gamma_4}[\bar\sigma, \varphi_3]\right) + 2i {\bf Im}\left({\gamma_1\bar\gamma_2\over \gamma_4^2}\right) [\bar\sigma, \sigma], \nonumber\\ \nd 
where $\gamma_4$ is real but $\gamma_1$ and $\gamma_2$ are complex numbers.  Interestingly, when we compare \eqref{sirK}
to the terms involving commutator brackets in the BHN equation \eqref{zoey}, we see that they are identical provided we identify $\gamma_3$ to $\gamma_1, \gamma_2$ and $\gamma_4$ 
in the following way:
\bg\label{verlag}
\gamma_3 ~ \equiv ~ -2{\bf Im}\left({\gamma_1 \bar\gamma_2\over \gamma_4}\right). \nd
Looking at the $\gamma_i$ defined in \eqref{house} and comparing the terms appearing in the definition of $\gamma_i$ with the ones in \eqref{coeffrelo}, we see that the above 
identification \eqref{verlag} implies the following relations between the coefficients:
\bg\label{piripiri}
{2 h^{(4)}_{2\psi} h^{(1)}_{2\psi 23} \over h^{(3)}_{2\psi 3} h^{(2)}_{2\psi 2} - h^{(2)}_{2\psi 3} h^{(3)}_{2\psi 2}} ~ = ~ 
\sqrt{v_3 c_{3r} c_{2\phi_1}\over a_1 c_1 d_{23}}. \nd 
The RHS of the above relation is defined with respect to the background warp-factors and $\theta$-parameter, whereas the LHS is only defined via 
\eqref{coeffrelo}. Thus satisfying \eqref{piripiri} doesn't seem hard. In fact we can make arbitrary choices for $h^{(2)}_{2\psi k}$ and $h^{(3)}_{2\psi k}$ satisfying
\eqref{coeffrelo}, and then arrange $h^{(4)}_{2\psi}$ to satisfy \eqref{piripiri}. This immediately implies that we can rewrite the BHN equation \eqref{zoey} in the 
following way:
\bg\label{alujen} 
{\cal F}_{2\psi} - i \gamma_4 \left[\hat{\varphi}_2, \hat{\varphi}_3\right] ~ = ~ - 2\epsilon_{2\psi} ~{\bf Re}\left(\gamma_5 {\cal D}_2\sigma\right). \nd
To bring the above equation in a more suggestive format, we can start by defining the fields $\hat{\Phi}_k$ for $k =$ ($2, 3$) 
as in \eqref{devkanta} and then construct one-forms out of them in a way 
similar to the definition we gave earlier in \eqref{moyukh}. More precisely:
\bg\label{coben}
\hat{\Phi}_k ~\equiv ~ -i\sqrt{\gamma_4} ~\hat{\varphi}_k, ~~~~~ {\Phi} ~ \equiv ~ \hat{\Phi}_2 ~dx_2 + \hat{\Phi}_3 ~d\psi, \nd
along with the gauge field components combined together to construct another one-form $A$ exactly as in \eqref{moyukh}, but now without any $c_0$ factor. To avoid clutter we 
removed the hat on $\Phi$. These redefinitions now convert the
BHN equation \eqref{alujen} to the following form:
\bg\label{motson}
F  - {\Phi} \wedge {\Phi}   =  - 2 ~{\bf Re}\left(\gamma_5 d_A \sigma\right), \nd
which is surprisingly similar to the first equation in \eqref{may6baz}, except that the RHS is no longer zero but is proportional to $d_A\sigma$. 
Note however the absence of the $i$ factor
in the RHS of \eqref{motson}. This is because we have absorbed the $i$ in the definition of $\sigma$ (this makes sense because $\sigma$, as constructed from ${\cal A}_r$ and 
${\cal A}_{\phi_1}$, go to $-i\sigma$ when we define ${\cal A}_\alpha = -i A_\alpha$). On the other hand, 
if we also redefine ${\cal A}_\psi$ in the following way:
\bg\label{virgo3}
{\cal A}_\psi ~ \to ~ \hat{\cal A}_\psi ~ \equiv ~ {\cal A}_\psi + 2 ~{\bf Re}\left(\gamma_5 \sigma\right), \nd
keeping the other gauge field components, i.e (${\cal A}_0, {\cal A}_1, {\cal A}_2$) same as before, then the  
BHN equation doesn't change and takes the form as the first equation in \eqref{may6baz}.  Thus there seems to be two ways of expressing the BHN equation for this case: one, if we assume 
that the gauge field components remain as before\footnote{With the assumption that, 
due to the instantonic background, ${\cal A}_2$ will be defined as $\langle {\cal A}_2 \rangle$ plus fluctuation.}, then the RHS of the BHN equation receives correction from the 
($\sigma, \bar\sigma$) fields 
as \eqref{motson}; and two, if we assume that ${\cal A}_\psi$ is defined using the ($\sigma, \bar\sigma$) fields then the RHS of the BHN equation vanishes. For the time being we 
will continue with first case, and consider the second case later.

Let us now turn our attention to the other parts of the BHN equations, namely the ones constraining $\varphi_2$ and $\varphi_3$ as in \eqref{vedpra}. To analyze them now, and as before,
we will consider the simplifying assumption of $b_{\psi k} = b_{ak}$ where the functional forms of $b_{\psi k}$ and $b_{ak}$ appear in \eqref{bokcho} and \eqref{boka} 
respectively\footnote{As mentioned earlier, there is no need for making this assumption other than for the sole reason of simplifying the form of the equations. Thus if we do away with this
assumption, the equations in \eqref{londonkk} will have relative coefficients but no new physics.}. 
The constraining equations now take the following form:
\bg\label{londonkk}
{\cal D}_2 \hat{\varphi}_2 + {\cal D}_\psi\hat{\varphi}_3 ~ = ~ {2\over \gamma_4}~{\bf Re}\left(\bar{\gamma}_2 {\cal D}_2 \sigma\right), ~~
{\cal D}_\psi \hat{\varphi}_2 - {\cal D}_2 \hat{\varphi}_3 ~ = ~ {2\over \gamma_4}~{\bf Re}\left(\bar{\gamma}_1 {\cal D}_2 \sigma\right), \nd
where the hatted fields are defined as in \eqref{londonk}. Alternatively we could also use the one-form $\Phi$, defined in \eqref{coben}, to rewrite the full set of BHN equations 
for our case. Combining \eqref{motson} with \eqref{londonkk}, we collect all the BHN equations together as:
\bg\label{niceg}
&& F  - {\Phi} \wedge {\Phi}   =  - 2 ~{\bf Re}\left(\gamma_5 d_A \sigma\right) \nonumber\\  
&& d_A \Phi ~ = ~ {2\over \sqrt{\gamma_4}}~{\bf Re}\left(\bar{\gamma}_1 d_A \sigma\right),~~~ 
d_A \ast \Phi ~ = - {2\over \sqrt{\gamma_4}}~{\bf Re}\left(\bar{\gamma}_2 d_A \sigma\right), \nd
where the hodge star is in two-dimensions, the gauge field components are ($A_2, A_\psi$) and $\gamma_i$ are defined in \eqref{house}. 
One may now compare our set of equations \eqref{niceg} for the surface operator to the ones appearing in 
\cite{gukovwitten2}, \cite{gukovwitten}, \cite{gukovsurf} and \cite{wittenknots1}:
\bg\label{sandstime} 
F - \phi \wedge \phi ~ = ~ 2\pi \alpha ~\delta_{\bf K}, ~~~ d_A \phi ~ = ~ 2\pi \gamma~\delta_{\bf K}, ~~~d_A\ast \phi ~ = ~ 2\pi \beta~\delta_{\bf K}, \nd
where $\delta_{\bf K}$ is a delta function that is Poincare dual to the knot ${\bf K}$. We have modified the hodge star so that now it is in two-dimensions (see footnote \ref{diali}). 
Comparing \eqref{sandstime} 
with \eqref{niceg} it is clear that $\phi$ in \eqref{sandstime} can be identified with $\Phi$ in \eqref{niceg}: they represent similar fields. On the other hand, 
the RHS of the equations 
have three different constants $\left(\gamma_5, {\gamma_2\over \sqrt{\gamma_4}}, {\gamma_1\over \sqrt{\gamma_4}}\right)$ and two functions $d_A \sigma$ and $d_A \bar\sigma$. 
These two functions are clearly composed of $\langle {\cal A}_2 \rangle, \langle {\cal A}_r \rangle$ and $\langle {\cal A}_{\phi_1} \rangle$ which form our instanton
configuration giving rise to localized G-fluxes and M2-brane charges in \eqref{m2charge} and \eqref{nayak} respectively. In the small instanton limit \cite{smallins}, 
where they indeed become M2-brane states, the two functions become highly localized so that they are like delta functions in the ($x_2, \psi$) plane i.e the plane orthogonal to 
our M2-brane states along ($x_0, x_1$) directions\footnote{It is not essential to go to the small instanton limit. All we need is finite localizations of the two functions.}. 
This is where we can make the following identifications between ($\alpha, \beta, \gamma$) appearing in \eqref{sandstime} and 
($\gamma_i, \sigma, \bar\sigma$) appearing in \eqref{niceg} and \eqref{house}:
\bg\label{lateni}
&& \alpha~\delta_{\bf K} ~ \equiv ~ {1\over \pi} \left[{\bf Im}(\gamma_5){\bf Im}(d_A \sigma) - {\bf Re}(\gamma_5){\bf Re}(d_A \sigma)\right]\nonumber\\                  
&& \beta~\delta_{\bf K} ~ \equiv ~ {1\over \pi\sqrt{\gamma_4}} \left[{\bf Im}({\overline\gamma}_2){\bf Im}(d_A \sigma)  - {\bf Re}({\overline\gamma}_2){\bf Re}(d_A \sigma)\right]\nonumber\\                  
&& \gamma~\delta_{\bf K} ~ \equiv ~ {1\over \pi\sqrt{\gamma_4}} \left[{\bf Re}(\gamma_1){\bf Re}(d_A \sigma) + {\bf Im}(\gamma_1){\bf Im}(d_A \sigma) \right]. \nd
The overall sign is irrelevant for us, as this can be absorbed by simultaneously shifting $\Phi \to -\Phi$ and $\sigma \to -\sigma$. Thus in the limit when $d_A\sigma$ approaches
$(1+i) \delta_{\bf K}$, at least when $K$ is a straight line along $x_1$ direction, the ($\alpha, \beta, \gamma$) coefficients in \eqref{sandstime} and \eqref{lateni}
can be mapped to the parameters in the Hamiltonian 
\eqref{hamilbeta} in the following way:
\bg\label{f685}
&& \alpha = -{1\over \pi\sqrt{2 c_{12}}}\left[t^{(1)}_{2r}\sqrt{c_{2r}} + t^{(2)}_{2\phi_1}\sqrt{c_{2\phi_1}}\right]\\
&& \beta = {1\over 2\pi}\left[{h^{(2)}_{2\psi 3} \sqrt{c_{3r}} + h^{(3)}_{2\psi 3} \sqrt{c_{3\phi_1}} \over \sqrt{h^{(1)}_{2\psi 23} \sqrt{2 c_1 c_{12}d_{23}v_3^{-1}}}}\right], ~~
\gamma = {1\over 2\pi}\left[{h^{(2)}_{2\psi 2} \sqrt{c_{2r}} + h^{(3)}_{2\psi 2} \sqrt{c_{2\phi_1}} \over \sqrt{h^{(1)}_{2\psi 23} \sqrt{2 c_1 c_{12} d_{23}v_3^{-1}}}}\right], \nonumber \nd
where all the parameters appearing above have been defined earlier, for example $c_{ar}$ in \eqref{bentu}, $c_{a\phi_1}$ in \eqref{jhogra}, $d_{23}$ in \eqref{venu}, $c_{12}$ in 
\eqref{krisis} and the other parameters in \eqref{coeffrelo} and in \eqref{sksamagra}. 

The above identification \eqref{f685} is highly suggestive of type IIA small instantons on D6-branes modelling as surface operators in the boundary three dimensional theory. However to 
complete the picture we will not only have to derive the BHN equations for the other components of the gauge fields but also find the boundary theory along similar lines 
to the technique developed in section \ref{boundth}. To proceed, let us first derive the BHN equations for the field strength ${\cal F}_{1\psi}$, which means we are looking 
at the gauge fields ${\cal A}_1$ and ${\cal A}_\psi$ and scalar fields $\varphi_1$ and $\varphi_3$ (see \eqref{foursca}). 
The $\sigma$ and $\bar\sigma$ fields will appear again, but since they are 
independent of $x_1$ direction, we are not compelled to make a redefinition like \eqref{20tigoenda}, or even go to \eqref{sksamagra}. In fact the {\it same} parameters $t^{(1)}_{2r}$ 
and $t^{(2)}_{2\phi_1}$ that appeared earlier in defining the BHN equations for ${\cal F}_{2\psi}$ will show up again here because the coefficients of $t^{(1)}_{1r}$ and 
$t^{(2)}_{1\phi_1}$ vanish in the Hamiltonian \eqref{hamilbeta}. Combining everything together, the ${\cal F}_{1\psi}$ BHN equation takes the following form:    
\bg\label{putloc} 
{\cal F}_{1\psi} -i \widetilde{\gamma}_4 [\varphi_1, \varphi_3] + 2\epsilon_{1\psi}~ {\bf Re}\left(\widetilde{\gamma}_5 {\cal D}_2 \sigma \right) = 
2i~{\bf Re}\left(\widetilde{\gamma}_1 [\bar\sigma, \varphi_1] + \widetilde{\gamma}_2 [\bar\sigma, \varphi_3]\right) + \widetilde{\gamma}_3 [\bar\sigma, \sigma], \nonumber\\ \nd
which is in fact a variant of the BHN equation \eqref{zoey} for ${\cal F}_{2\psi}$. As expected \eqref{putloc} relates the scalar fields $\varphi_1$ and $\varphi_3$, however the 
third term appears as ${\cal D}_2 \sigma$ instead of ${\cal D}_1 \sigma$. This is because of the comments that we made above. The other coefficients i.e 
$\widetilde{\gamma}_k$ are defined, also as a variation of \eqref{house}, in the following way: 
\bg\label{choshma} 
&&\widetilde{\gamma}_1 ~ = ~ {1\over 2} \sqrt{2 v_3\over c_1 c_{12}}\left[h^{(2)}_{1\psi 1} \sqrt{c_{r1}}  + i h^{(3)}_{1\psi 1} \sqrt{c_{\phi_1 1}}\right], 
~~\widetilde{\gamma}_2 ~ = ~ {1\over 2} \sqrt{2 v_3\over c_1 c_{12}}\left[h^{(2)}_{1\psi 3} \sqrt{c_{r3}}  + i h^{(3)}_{1\psi 3} \sqrt{c_{\phi_1 3}}\right] \nonumber\\ 
&& \widetilde{\gamma}_3 ~ = ~ {h^{(4)}_{1\psi} \sqrt{2 a_1}\over 2\sqrt{c_{12}}}, ~~ \widetilde{\gamma}_4 ~ = ~ {h^{(1)}_{1\psi 1 3}\sqrt{8 d_{13} v_3}\over \sqrt{c_1 c_{12}}}, 
~~\widetilde{\gamma}_5 ~ = ~ {1\over 2\sqrt{c_{12}}}\left[t^{(1)}_{2r} \sqrt{c_{r2}}  - i t^{(2)}_{2\phi_1} \sqrt{c_{\phi_1 2}}\right]. \nd 
The above set of coefficients can be related to the coefficients \eqref{house} in the following way. It is easy to see that $\gamma_5 = \widetilde{\gamma}_5$. Furthermore, looking
at the coefficients $c_{ar}, c_{a\phi_1}$ and $d_{kl}$ in \eqref{bentu}, \eqref{jhogra} and \eqref{venu} we can easily infer:
\bg\label{mothotel}
c_{1r} ~ = ~ c_{2r}, ~~~~~c_{1\phi_1} ~ = ~ c_{2\phi_1}, ~~~~~ d_{13} ~ = ~ d_{23}, \nd 
so that the only distinguishing factors between $\gamma_k$ and $\widetilde{\gamma}_k$ are the coefficients $h^{(\alpha)}_{a\psi a}$, $h^{(\alpha)}_{a\psi 3}$, 
$h^{(1)}_{a\psi a 3}$ and $h^{(4)}_{a\psi}$ where $a = $ (1, 2) and $\alpha = $ (2, 3). Other than these factors, the BHN equations for ${\cal F}_{1\psi}$ and ${\cal F}_{2\psi}$ given 
in \eqref{putloc} and \eqref{zoey} respectively are perfectly symmetrical. These factors, on the other hand, are controlled by \eqref{coeffrelo} which are in fact the {\it only} 
defining equations for them. Thus one assumption would be to take the individual pieces to be equal to each other. In other words, we can demand:
\bg\label{fannyp}
h^{(\alpha)}_{1\psi 1} ~ = ~ h^{(\alpha)}_{2\psi 2}, ~~~~ h^{(\alpha)}_{1\psi 3} ~ = ~ h^{(\alpha)}_{2\psi 3}, ~~~~ h^{(1)}_{1\psi 1 3} ~ = ~ h^{(1)}_{2\psi 2 3}, ~~~~
h^{(4)}_{1\psi} ~ = ~ h^{(4)}_{2\psi}, \nd
so that $\gamma_k = \widetilde{\gamma}_k$ in the BHN equation \eqref{putloc}. Note that with the identification \eqref{fannyp} it almost implies 
that the BHN equations, given in \eqref{zoey} and \eqref{putloc}, are identical via
the exchange of 1 and 2 in the subscripts of the gauge and the scalar fields. The only difference is that the ``symmetry'' between the two equations is
broken by the existence of ${\cal D}_2\sigma$ and ${\cal D}_2\bar\sigma$. 

Unfortunately the above assumption is too restrictive and could potentially lead to additional constraints when all the background equations are laid out. Therefore 
we will start by defining a field $\hat{\varphi}_1$ exactly as $\hat{\varphi}_2$ in \eqref{londonk} using $\widetilde{\gamma}_2$ and $\widetilde{\gamma}_4$. 
This way of defining $\hat{\varphi}_1$ has an 
immediate advantage: the commutator bracket of $\hat{\varphi}_1$ and $\hat{\varphi}_3$ will take similar form as \eqref{sirK}, i.e   
\bg\label{tarza}
[\hat{\varphi}_1, \hat{\varphi}_3] ~ = ~ [{\varphi}_1, {\varphi}_3] + 2{\bf Re} \left({{\gamma}_1\over {\gamma}_4}[\bar\sigma, \varphi_1] 
+ {\widetilde{\gamma}_2\over \widetilde{\gamma}_4}[\bar\sigma, \varphi_3]\right) + 2i {\bf Im}\left({{\gamma}_1\bar{\widetilde{\gamma}}_2\over \gamma_4 \widetilde{\gamma}_4}\right) 
[\bar\sigma, \sigma], \nonumber\\ \nd 
with $\widetilde{\gamma}_3$ identified as \eqref{verlag} except the $\bar\gamma_2$ therein is replaced by $\bar{\widetilde\gamma}_2$; and $\widetilde{\gamma}_1$ is proportional to $\gamma_1$ with 
the proportionality constant being the ratio ${\widetilde{\gamma}_4\over \gamma_4}$. 
The next set of manipulations are important. We can use \eqref{tarza} to express the BHN equation \eqref{putloc} as \eqref{alujen}. However since the scalar fields $\sigma$ and 
$\bar\sigma$ are independent of $x_1$ coordinate, and using the gauge field definition $\hat{\cal A}_\psi$ as given in \eqref{virgo3}, we see that the ${\cal F}_{1\psi}$ and 
the ${\cal F}_{2\psi}$ BHN equations take the following form:
\bg\label{karfish}
&&\hat{\cal F}_{2\psi} - i \gamma_4 \left[\hat{\varphi}_2, \hat{\varphi}_3\right]  =  0 \nonumber\\
&& \hat{\cal F}_{1\psi} - i \widetilde{\gamma}_4 \left[\hat{\varphi}_1, \hat{\varphi}_3\right] = - 2\epsilon_{1\psi} ~{\bf Re}\left[\gamma_5 {\cal D}_{(2, 1)}\sigma\right], \nd 
where $\hat{\cal F}_{a\psi}$ is the field strength for the gauge fields ${\cal A}_a$ and $\hat{\cal A}_\psi$ with $a = $ (1, 2) in the standard way; and the covariant 
derivative ${\cal D}_{(a, b)}$ is defined in the following way:
\bg\label{arbitrage}
{\cal D}_{(a, b)}\sigma ~ \equiv ~ \partial_a \sigma + i\left[{\cal A}_a - {\cal A}_b, \sigma\right], \nd 
using the difference of two gauge fields ${\cal A}_a$ and ${\cal A}_b$, instead of just ${\cal A}_a$ as we had before. The other equations, for example the constraining equations for 
the scalar fields ($\hat{\varphi}_2, \hat{\varphi}_3$) given earlier in \eqref{londonkk}, and the equations for the other pair of scalar fields ($\hat{\varphi}_1, \hat{\varphi}_3$) now take the
following form:
\bg\label{rodmoore}
&& {\cal D}_2 \hat{\varphi}_2 + {\hat{\cal D}}_\psi \hat{\varphi}_3 ~ = ~ 
2i \left[{\bf Re}\left(\gamma_5 \sigma\right), \hat{\varphi}_3\right] + 2{\bf Re}\left({\bar{\gamma}_2 {\cal D}_2 \sigma \over \gamma_4}\right) \nonumber\\
&& {\hat{\cal D}}_\psi \hat{\varphi}_2 - {\cal D}_2 \hat{\varphi}_3 ~ = ~ 
2i \left[{\bf Re}\left(\gamma_5 \sigma\right), \hat{\varphi}_2\right] + 2{\bf Re}\left({\bar{\gamma}_1 {\cal D}_2 \sigma \over \gamma_4}\right) \nonumber\\
&& {\cal D}_1 \hat{\varphi}_1 + {\hat{\cal D}}_\psi \hat{\varphi}_3 ~ = ~ 2i \left[{\bf Re}\left(\gamma_5 \sigma\right), \hat{\varphi}_3\right] 
+ {2i\over \widetilde{\gamma}_4}\left[{\cal A}_1, {\bf Re}\left(\bar{\widetilde\gamma}_2 \sigma\right)\right] \nonumber\\
&& {\hat{\cal D}}_\psi \hat{\varphi}_1 - {\cal D}_1 \hat{\varphi}_3  ~ = ~ 2i \left[{\bf Re}\left(\gamma_5 \sigma\right), \hat{\varphi}_1\right]
+ {2i\over {\gamma}_4}\left[{\cal A}_1, {\bf Re}\left(\bar{\gamma}_1 \sigma\right)\right],  \nd
where $\hat{\cal D}_\psi \sigma$ is the covariant derivative defined with respect to the gauge field $\hat{\cal A}_\psi$ \eqref{virgo3}. 
In terms of the unshifted field ${\cal A}_\psi$, the RHS of the above 
set of equations \eqref{rodmoore} will not have the commutator brackets. It is also instructive to work out the commutator bracket for $\hat{\varphi}_1$ and $\hat{\varphi}_2$:
\bg\label{kfish}
\left[\hat{\varphi}_1, \hat{\varphi}_2\right] ~ = ~ \left[{\varphi}_1, {\varphi}_2\right] + {2\over \gamma_4}\left[\varphi_1, {\bf Re}\left(\bar{\gamma}_2\sigma\right)\right]
- {2\over \widetilde{\gamma}_4}\left[\varphi_2, {\bf Re}\left(\bar{\widetilde\gamma}_2\sigma\right)\right] 
+ {\bar{\widetilde\gamma}_2 {\gamma}_2 - \bar{\gamma}_2 {\widetilde\gamma}_2 \over \gamma_4 {\widetilde\gamma}_4} \left[\sigma, \bar\sigma\right], \nonumber\\ \nd
where $\gamma_k$ and $\widetilde{\gamma}_k$ have been defined earlier in \eqref{house} and \eqref{choshma} respectively. Note that if we had applied the identifications \eqref{fannyp}, the 
commutator piece $[\sigma, \bar\sigma]$ in \eqref{kfish} would be absent. However as mentioned earlier, the identifications \eqref{fannyp} are 
not only over-constraining but also inconsistent.
We will therefore refrain from using them and stick with the commutator brackets in \eqref{kfish}. Additionally now:  
\bg\label{nadali}
\widetilde{\gamma}_3 = -2 {\rm Im}\left({\gamma_1 {\overline{\widetilde{\gamma}}}_2 \over \gamma_4}\right). \nd
We will use the above informations, including \eqref{moson}, 
to determine the BHN equation corresponding to the gauge field strength ${\cal F}_{12}$ in the presence of the instanton background. To start, let us define 
few things that will help us express the background more succinctly:
\bg\label{antique}
&& j_1  \equiv  m_{11}^{(1)}\sqrt{b_{11}}, ~~~~~ j_2 ~ \equiv ~ m_{12}^{(1)}\sqrt{b_{12}} \nonumber\\
&& \Gamma_1  \equiv -2 g_{1212}^{(1)} \sqrt{d_{12}}~{\bf Re}\left({\bar\gamma_2 \sigma\over \gamma_4}\right), ~~~
\Gamma_2 \equiv  2 g_{1212}^{(1)} \sqrt{d_{12}}~{\bf Re}\left({\bar{\widetilde\gamma}_2 \sigma\over \widetilde\gamma_4}\right), \nd  
where $b_{11}$ and $b_{12}$ coefficients are defined in \eqref{boka}, $d_{12}$ coefficient is defined in \eqref{venu}, and ($m_{11}^{(1)}, m_{12}^{(1)}, g_{1212}^{(1)}$) coefficients are defined in 
\eqref{coeffrelo} where we have assumed $m_{11}^{(1)} = m_{22}^{(1)}$ for simplicity. 
Note that ($j_1, j_2$) are numbers whereas ($\Gamma_1, \Gamma_2$) are scalar fields expressed using $\sigma$ and $\bar\sigma$. Using these we define three fields:
\bg\label{suncasa}
{\cal A}_x \equiv -\left({j_1\Gamma_1 + j_2 \Gamma_2 \over j_1^2 + j_2^2}\right), ~~~ {\cal A}_y \equiv {j_2\Gamma_1 - j_1 \Gamma_2 \over j_1^2 + j_2^2}, ~~~
{\cal A}_z \equiv -{\bar\sigma\over 4}\left({\bar\gamma_2 \widetilde\gamma_2 - \gamma_2 \bar{\widetilde\gamma}_2 \over j_2 \bar{\widetilde\gamma}_2 \gamma_4 - 
j_1 \bar\gamma_2 \widetilde\gamma_4}\right). \nonumber\\ \nd
These fields are written in a suggestive way so that they could be used as components of a vector field although ($x, y, z$) 
are not related to spacetime coordinates (they are simply parameters here). We can now use \eqref{antique} and \eqref{suncasa} to express the BHN equation for the gauge field strength 
${\cal F}_{12}$ in the following way (see also {\bf Table \ref{pirates}}):
\bg\label{mktan94}
&& {\cal F}_{12}  -i \left({m_{\psi 3}^{(1)} \sqrt{2v_3 b_{\psi 3}}\over \sqrt{c_1c_{11}}}\right) {\cal D}_\psi \phi_0 +
{m_{11}^{(1)} \sqrt{2 v_3 b_{11}}\over \sqrt{c_1c_{11}}}\left[{\cal D}_{(1, x)} \hat{\varphi}_1 + {\cal D}_{(2, y)} \hat{\varphi}_2 
- 2{\bf Re}\left({\bar\gamma_2 {\cal D}_{(2, z)} \sigma \over \gamma_4}\right)\right] \nonumber\\
&& + {m_{12}^{(1)} \sqrt{2 v_3 b_{12}}\over \sqrt{c_1c_{11}}}\left[{\cal D}_{(1, x)} \hat{\varphi}_2 - {\cal D}_{(2, y)} \hat{\varphi}_1 
+ 2{\bf Re}\left({\bar{\widetilde\gamma}_2 {\cal D}_{(2, z)} \sigma \over \widetilde\gamma_4}\right)\right] 
- i\left({2 g_{1212}^{(1)} \sqrt{2 v_3 d_{12}}\over \sqrt{c_1c_{11}}}\right) \left[\hat\varphi_1, \hat\varphi_2\right] = 0, \nonumber\\ \nd
where the new covariant derivative ${\cal D}_{(a, b)}$ is defined as in \eqref{arbitrage} now using the fields \eqref{suncasa}; the hatted scalar fields $\hat\varphi_k$
appear in \eqref{londonk}; $\gamma_k$ 
and $\widetilde{\gamma}_k$ are parameters given in \eqref{house} and \eqref{choshma} respectively; and $v_3$ is defined in \eqref{intig}. All other parameters have been defined earlier which the 
readers may refer to for details.

We now make a few observations. It is easy to see that when $\sigma = 0$, the above BHN equation \eqref{mktan94} goes back to the BHN equation derived earlier in \eqref{f12bhn} when we use the 
map \eqref{4scalars} alongwith the 
values of the parameters given in \eqref{g12kl}. The $\phi_0$ field appearing in \eqref{mktan94} is the same field that appeared in \eqref{vedpra} before. Using the scalar field map \eqref{foursca}, 
$\phi_0 \propto {\cal A}_3$, whereas using the map \eqref{4scalars}, $\phi_0 \propto \varphi_3$ as can also be inferred from column 4 in {\bf Table \ref{pirates}}.   
The additional constraint \eqref{bhn1} that we impose on the scalar fields $\varphi_1$ and $\varphi_2$ should continue to hold even for the 
case where we have nonzero $\sigma$. This immediately gives us our first constraint equation, in the same vein as \eqref{bhn1}, to be:
\bg\label{mklee}
{\cal D}_{(1, x)} \hat{\varphi}_1 + {\cal D}_{(2, y)} \hat{\varphi}_2 ~ = ~ 2{\bf Re}\left({\bar\gamma_2 {\cal D}_{(2, z)} \sigma \over \gamma_4}\right). \nd  
In some sense this could be taken as the {\it defining} equation for hatted scalar fields $\hat\varphi_1$ and $\hat\varphi_2$. Comparing \eqref{mklee} with the first and the third equations in 
\eqref{rodmoore}, we see that the constraints appear differently because of the structure of the covariant derivative \eqref{arbitrage}. In fact if we did not impose the constraint \eqref{moson}, 
we could have easily absorbed this in the definition of the fields \eqref{suncasa}. Thus the form of \eqref{mktan94} is generic enough even in the absence of \eqref{moson}.

Once \eqref{mklee} is applied on \eqref{mktan94}, the form of the ${\cal F}_{12}$ BHN equation is now almost identical to \eqref{f12bhn} except with extra ($\sigma, \bar\sigma$) dependences as 
we discussed above. Thus we could express it as \eqref{fkhan} using the $t$ parameter given in \eqref{tbol}. Following similar criteria as developed in section \ref{flook}, and without going
into details, we can again demand the coefficient of $t - t^{-1}$ piece to vanish. For the present case, this takes the following form:
\bg\label{kyalee}
{\cal D}_{(1, x)} \hat{\varphi}_2 - {\cal D}_{(2, y)} \hat{\varphi}_1 = - 2{\bf Re}\left({\bar{\widetilde\gamma}_2 {\cal D}_{(2, z)} \sigma \over \widetilde\gamma_4}\right), \nd
which becomes \eqref{michel} when $\sigma = \bar\sigma = 0$ once we appropriately redefine the scalar fields. Now putting everything together, the ${\cal F}_{12}$ BHN equation is identical (at least 
in form) 
to the one that we had earlier for $c_2 = 0$ in \eqref{binu}, namely:
\bg\label{asnles} 
{\cal F}_{12} -i \left({m_{\psi 3}^{(1)} \sqrt{2 v_3 b_{\psi 3}}\over \sqrt{c_1c_{11}}}\right) {\cal D}_\psi \phi_0 
- i\left({2 g_{1212}^{(1)} \sqrt{2 v_3 d_{12}}\over \sqrt{c_1c_{11}}}\right) \left[\hat\varphi_1, \hat\varphi_2\right] = 0. \nd
Comparing the set of equations, \eqref{asnles}, \eqref{mklee} and \eqref{kyalee} to \eqref{karfish} and \eqref{rodmoore}, we observe that \eqref{asnles} is expressed in terms of 
${\cal A}_\psi$ instead of $\hat{\cal A}_\psi$ as \eqref{virgo3}. The difference between the covariant derivatives may be expressed in terms of commutator brackets in the following way:
\bg\label{ashori}
\left({\cal D}_\psi - \hat{\cal D}_\psi\right) \phi_0 ~ \equiv ~  2i \left[\phi_0, {\bf Re}\left(\gamma_5 \sigma\right)\right]. \nd
This would change the form of our BHN equation \eqref{asnles} by putting extra commutator brackets. This is {\it not} what we want so alternatively we could retain the form of the BHN equation as
in \eqref{asnles} with $\hat{\cal D}_\psi \phi_0$  instead of ${\cal D}_\psi \phi_0$ 
and {\it no} extra commutator terms, but change the RHS of the two constraint equations for the scalar fields $\hat\varphi_1$ and $\hat\varphi_2$ by replacing the covariant derivative 
${\cal D}_{(2.z)}\sigma$ by:
\bg\label{chukmukh}
{\cal D}_{(2, z, w)}\sigma ~ \equiv ~ \partial_2\sigma + i\left[{\cal A}_2 - {\cal A}_z - {\cal A}_w, \sigma\right], \nd
in {\it both} \eqref{kyalee} and \eqref{mklee}. The above definition of the covariant derivative, in the similar vein as \eqref{arbitrage}, is arranged in such a way as to absorb the commutator 
brackets appearing in \eqref{ashori} by defining a field ${\cal A}_w$ as:
\bg\label{awdef}
{\cal A}_w ~\equiv ~ \left({m^{(1)}_{\psi 3} \gamma_5 \gamma_4 \widetilde\gamma_4 \over j_1 \widetilde\gamma_4 \bar\gamma_2 - j_2 \bar{\widetilde\gamma}_2 \gamma_4}\right) \phi_0, \nd
where $j_i$ are defined in \eqref{antique}, $\gamma_k$ in \eqref{house} and $\widetilde\gamma_k$ in \eqref{choshma}. The other coefficient $m^{(1)}_{\psi 3}$ appears in \eqref{coeffrelo}. The
above definition of ${\cal A}_w$ differs crucially from the three fields ${\cal A}_x$, ${\cal A}_y$ and ${\cal A}_z$  appearing in \eqref{suncasa} in the sense that it is not given in terms 
of the instanton fields ($\sigma, \bar\sigma$). Instead it is expressed in terms of the scalar field $\phi_0$ whose value 
in general is only known by solving the BHN equation \eqref{asnles}, although for the present case this vanishes.

The above observation of cyclicity
is not new, and in fact did show up already in \eqref{karfish} when we had used ${\cal D}_{(2, 1)}\sigma$ to express the BHN equation for $\hat{\cal F}_{1\psi}$. The field 
${\cal A}_1$ appears on both sides of the equation \eqref{karfish}. Thus it can only be solved order by order in terms of any small parameter used to express the field ${\cal A}_1$. Similar issue
also showed up for the constraint equations \eqref{rodmoore}: the fields $\hat\varphi_k$ appear on both sides of the equations rendering exact solutions harder to determine. The Hamiltonian, on the
other hand, retains its form \eqref{hamilbeta} as:
\bg\label{fracture}
{\cal H} & = & {c_1c_{11}\over v_3}\int d^3 x {\rm Tr} \Bigg\{{c_{12}\over c_{11}}\left(\hat{\cal F}_{1\psi} - i \widetilde{\gamma}_4 \left[\hat{\varphi}_1, \hat{\varphi}_3\right] 
+ 2\epsilon_{1\psi} ~{\bf Re}\left[\gamma_5 {\cal D}_{(2, 1)}\sigma\right]\right)^2
+ {c_{12}\over c_{11}}\left(\hat{\cal F}_{2\psi} - i \gamma_4 \left[\hat{\varphi}_2, \hat{\varphi}_3\right]\right)^2 \nonumber\\
&&~~~~~~ + \Bigg[{\cal F}_{12}  -i \left({m_{\psi 3}^{(1)} \sqrt{2v_3 b_{\psi 3}}\over \sqrt{c_1c_{11}}}\right) \hat{\cal D}_\psi \phi_0   
- i\left({2 g_{1212}^{(1)} \sqrt{2 v_3 d_{12}}\over \sqrt{c_1c_{11}}}\right) \left[\hat\varphi_1, \hat\varphi_2\right]\Bigg]^2\Bigg\} + {\bf Q_M},\nonumber\\  \nd
except with hatted scalar fields that originate from the extra ($\sigma, \bar\sigma$) fields. Due to the $\sigma$ and $\bar\sigma$ dependences, the magnetic charge
${\bf Q_M}$ will now be different from what we had before in \eqref{QoM}\footnote{To compare the magnetic charge to \eqref{QoM}, we need to put $c_2 = 0$ in \eqref{QoM}.}, although the electric 
charge would still vanish with a suitable gauge choice as before. 

Before determining the magnetic charge ${\bf Q_M}$, let us try to simplify the first set of BHN equations \eqref{karfish} and \eqref{rodmoore}. One simple way to keep the right hand sides of the 
equations simple is to go to the {\it abelian} case. In the abelian case, all commutator terms vanish and the rest of the BHN equations \eqref{mklee}, \eqref{kyalee} and \eqref{asnles} alongwith
\eqref{karfish} and \eqref{rodmoore} take the following simple form:
\bg\label{kings}
&& \hat{\cal F}_{2\psi} = \hat{\cal F}_{1\psi} + 2\epsilon_{1\psi} {\bf Re}\left(\gamma_5 \partial_2\sigma\right) = {\cal F}_{12} + \gamma_6 \partial_\psi \phi_0 = 0 \\
&& \partial_\psi \hat\varphi_2  - \partial_2 \hat\varphi_3  = 2 {\bf Re}\left({\bar\gamma_1 \partial_2 \sigma\over \gamma_4}\right), ~~
\partial_1 \hat\varphi_2 - \partial_2 \hat\varphi_1  = - 2 {\bf Re}\left({\bar{\widetilde\gamma}_2 \partial_2 \sigma\over \widetilde{\gamma}_4}\right) \nonumber\\
&& \partial_1 \hat\varphi_1 + \partial_\psi \hat\varphi_3 = \partial_\psi \hat\varphi_1 - \partial_1 \hat\varphi_3 = 0,~~
\partial_2 \hat\varphi_2  + \partial_\psi \hat\varphi_3  = \partial_1 \hat\varphi_1 + \partial_2 \hat\varphi_2  = 2 {\bf Re}\left({\bar\gamma_2 \partial_2 \sigma\over \gamma_4}\right), \nonumber 
\nd       
where $\gamma_6$ is the coefficient of ${\cal D}_\psi \phi_0$ term in \eqref{asnles}. The above set of equations immediately implies that the un-hatted scalar fields $\varphi_1$, $\varphi_2$
and $\varphi_3$ are independent of $x_1$, $x_2$ and $\psi$ directions respectively\footnote{In other words: $\varphi_1 \equiv \varphi_1(x_2, \psi), \varphi_2 \equiv \varphi_2(x_1, \psi)$ and 
$\varphi_3 \equiv \varphi_3(x_1, x_2)$. Being static solutions they are of course independent of $x_0$ direction. A very simple solution, and definitely not the most generic one, of the set of 
equations in \eqref{ekraat} is to take $\varphi_1 \equiv A \psi + B x_2, \varphi_2 \equiv B x_1 + C \psi$ and $\varphi_3 \equiv A x_1 + C x_2$ where ($A, B, C$) are constants.}. 
In addition, they are related to each other via:
\bg\label{ekraat}
\partial_\psi \varphi_1 = \partial_1 \varphi_3, ~~~~ \partial_2 \varphi_1 = \partial_1 \varphi_2, ~~~~ \partial_\psi \varphi_2 = \partial_2 \varphi_3. \nd  
With all these we are almost ready to derive the boundary theory. Our starting point would be to switch on the $c_2$ parameter. The changes in the Hamiltonian \eqref{fracture} would be 
similar to what we had earlier in \eqref{horsho}, and therefore choosing the coefficients in the Hamiltonian \eqref{fracture} as in \eqref{simcondd}, the magnetic charge will take the 
following form:
\bg\label{abandon}
{\bf Q_M} = {4i \vert\tau\vert^2\over \tau - \bar{\tau}} \int d^3x ~\partial_\psi \left(\sum_{\alpha, \beta = 1}^2 \sum_{k = 1}^3\epsilon^{\alpha\beta k}{\cal F}_{\alpha\beta}\hat{\varphi}_k 
+ \sum_{k, l, m = 1}^3\hat{\varphi}_k \partial_l \hat{\varphi}_m +{i(\tau-\bar\tau) \over 2 \vert\tau\vert} {\cal A}_1 {\bf Re}\left(\gamma_5 \partial_2 \sigma\right)\right), \nonumber\\ \nd
which differs from \eqref{QoM} in two ways: first, due to the abelian nature we no longer have the commutator brackets, therefore no cubic terms in fields; and secondly, we have an {\it extra} 
term proportional to ${\cal A}_1$. The proportionality factor is some combination of $\partial_2\sigma$ and $\partial_2\bar\sigma$ that would vanish in the absence of the surface operators. 

The physics that we developed here is all in the absence of any twisting, and therefore the picture will change once we introduce twisting exactly as we had in section \ref{boundth}.
Following similar procedure as before, we {\it twist} the scalar fields ($\hat\varphi_1, \hat\varphi_2, \hat\varphi_3$)
to one forms ($\hat\phi_1, \hat\phi_2, \hat\phi_3$),
along the lines of \eqref{foursca}, but now for the hatted fields\footnote{The procedure is similar to what we had in \eqref{namtha}, but now appropriately modified by the mapping \eqref{foursca}.}. 
In the absence of the linear term in ${\cal A}_1$ the procedure of getting the boundary theory is similar to \eqref{bonbund}, namely:
\bg\label{copycat}
S^{(1)}_{bnd} & = & (b_2 + c_2) \int_{\bf W} {\cal A} \wedge d{\cal A} + \int_{\bf W} \Bigg\{2d_1~{\cal F} \wedge \hat{\phi} + \left({d_1^2\over b_2 + c_2}\right) \hat\phi \wedge d\hat\phi\Bigg\} \\
&=& (b_2 + c_2) \int_{\bf W} \left[{\cal A} + \left({d_1\over b_2 + c_2}\right)\hat\phi\right] \wedge d\left[{\cal A} + \left({d_1\over b_2 + c_2}\right)\hat\phi\right] 
\equiv {k\over 4\pi}\int_{\bf W} {\cal A}_d \wedge d{\cal A}_d, \nonumber \nd
where $b_2$ appears in exactly the same way as in \eqref{stwist} before, albeit now in the abelian case, alongwith similar definition for ${\cal A}_d$ as in \eqref{newgf} but now with 
$\hat\phi_\mu$ instead of $\phi_\mu$. The parameters $c_2$ and  
 $d_1$ are determined from \eqref{c1c2rel} and the supersymmetry condition \eqref{sosie} respectively, as before. The linear term in ${\cal A}_1$ then adds a new term to the 
boundary action \eqref{copycat}:
\bg\label{jugnu}
S^{(2)}_{bnd} = d_4 \int_{\bf W} dx_0 dx_1 dx_2 ~{\cal A}_1 {\bf Re}\left(\gamma_5 \partial_2\sigma\right) \equiv Q_2 \int dx_1 {\cal A}_1, \nd
where $d_4$ is a constant that may be read off from \eqref{abandon} after twisting and $Q_2$ appears in the same limit that converted \eqref{lateni} to \eqref{f685} namely when 
$\partial_2\sigma = \left({1+i\over 2\gamma_5}\right)\delta_K$ where $K$ is a straight line along $x_1$ direction (in a more generic situation, $K$ will be a closed loop in the $x_1-x_2$ plane). 
Note that the integrand 
in \eqref{jugnu} is independent of $x_0$, so the $dx_0$ integral can be localized by our choice of $\delta_K$. Combining \eqref{copycat} and \eqref{jugnu}, we now get our 
complete boundary theory to be:
\bg\label{rambof}
S_{bnd} = (b_2 + c_2) \int_{\bf W} {\cal A}_d \wedge d{\cal A}_d + Q_2 \oint_K {\cal A}, \nd
where the second integral is now over a closed loop $K$, in the ($x_1, x_2$) plane, instead of a straight line along $x_1$ in \eqref{jugnu} above.   
At this stage one might compare \eqref{rambof} with the boundary theory that appears in \cite{gukovwitten2}, \cite{gukovwitten}, \cite{gukovsurf} and \cite{wittenknots1}. Note the appearance 
of ${\cal A}_d$ instead of ${\cal A}$ for the abelian Chern-Simons term. Interestingly the equation of motion from \eqref{rambof} becomes:
\bg\label{darkness}
{\cal F}_d ~ = ~ - {Q_2\over 2(b_2 + c_2)}~ \delta_K, \nd
where $\delta_K$, the Poincare dual of $K$, is the same singularity that appeared earlier. In this form \eqref{darkness} resembles closer to the analysis presented in section 6 of 
\cite{wittenknots1} in the sense that we can assume ${\cal A}_d$ to have a singularity along $K$ with the monodromy around $K$ to be: 
\bg\label{pocket}
{\cal M} \equiv {\rm exp}\left[-{iQ_2\over 2(b_2 + c_2)}\right]. \nd   
Note that the denominator in the monodromy formula \eqref{pocket} has the factor $b_2 + c_2$, which is $\Psi$ in the notation of \cite{wittenknots1}. This of course appears because of twisting
in the supergravity formalism, as we saw above. What is interesting however is that the denominator will not change if we go from the abelian to the non-abelian case as can be inferred from 
our earlier derivations although the boundary theory will change from it simple form \eqref{rambof} to its, more non-trivial, non-abelian generalization.

\subsubsection{Surface operators and knot configurations \label{knotty}}

All our above discussions are consistent with the series of papers \cite{gukovwitten2}, \cite{gukovwitten}, and \cite{gukovsurf} modulo couple of subtleties that 
we have kept under the rug so far, and they have to do with the precise structures of our D2-brane surface operator. The first subtlety arises when we look carefully at the orientations of the 
D2-brane in our problem. 
The orientation of the D2-brane is given in {\bf Table \ref{surfaceoper}}, and
we discussed how this appears in the BHN equations using the M-theory uplift given in {\bf Table \ref{m2m2}}. The analysis that we presented above works when the D2-brane circles the $\phi_1$ 
direction completely. In type IIB dual, this is a D3-brane stretched between the D5-${\overline{\rm D5}}$ pairs wrapped on the Taub-NUT two-cycles oriented along ($r, \psi$) directions as depicted 
in {\bf Table \ref{d3chappal}}. From here the result of {\bf Table \ref{m2m2}} can be easily inferred by T-dualizing along the compact $\phi_1$ direction and lifting the resulting configuration 
to M-theory.

\begin{table}[h!] 
 \begin{center}
\begin{tabular}{|c||c|c|c|c|c|c|c|c|c|c|}\hline Directions & $x_0$ & $x_1$ & $x_2$
& $x_3$ & $\theta_1$ & $\phi_1$ & $\psi$ & $r$ & $x_8$ & $x_9$ \\ \hline
Taub-NUT & $\ast$  & $\ast$   & $\ast$   & $\ast$ & $\surd$  & $\surd$ & $\surd$ & $\surd$ & $\ast$ & $\ast$ \\  \hline
D5-${\overline{\rm D5}}$ & $\surd$  & $\surd$   & $\surd$   & $\surd$ & $\ast$  & $\ast$ & $\surd$ & $\surd$ & $\ast$  & $\ast$ \\  \hline
D3 & $\surd$  & $\surd$   & $\ast$   & $\ast$ & $\ast$  & $\surd$ & $\surd$ & $\ast$ & $\ast$  & $\ast$ \\  \hline
  \end{tabular}
\end{center}
  \caption{The orientation of a D3-brane between the wrapped five-branes.}
\label{d3chappal}
\end{table}

The story however gets more complicated if the D3-brane is stretched, not completely along the $\phi_1$ circle, but only between the five-branes. A T-duality along $\phi_1$ direction now will 
only lead to a {\it fractional} D2-brane, which is a D4-${\overline{\rm D4}}$ pair wrapped on certain two-cycle in the internal space. The internal space, before T-duality, was a Taub-NUT 
manifold. However after T-duality, we expect the internal geometry to take the form as given in \eqref{d6brane}, namely: 
\bg\label{reckoning}
ds_6^2 = e^\phi\left(F_1~dr^2 + F_3~d\theta_1^2 + F_4~ds^2_{89}\right) + {\cal C}_1 (d\phi_1 + \chi~\cos~\theta_1 dx_3)^2 + {\cal C}_2 d\psi^2, \nd
where we see that the $\phi_1$ circle is non-trivially fibered over the $x_3$ circle. The reason for this is because of certain B-field components 
in the type IIB side as we saw in sections \ref{aluraj} and \ref{mothdaug}. The $\psi$ direction now no longer has the Taub-NUT fibration structure but still allows the six-branes to wrap around 
($\psi, r, \phi_1$) directions in the way described in section \ref{jahangir}. The other coefficients appearing in \eqref{reckoning} are defined using the $\theta$ parameter 
and the warp factors $F_i$ as (see also \eqref{d6brane}): 
\bg\label{kyloren} 
{\cal C}_1 \equiv {e^{-\phi} \over \widetilde{F}_2 ~\cos^2\theta_1 + F_3~\sin^2\theta_1}, ~~~
{\cal C}_2 \equiv {e^\phi~\widetilde{F}_2 F_3~\sin^2\theta_1~\sec^2\theta \over \widetilde{F}_2 ~\cos^2\theta_1 + F_3~\sin^2\theta_1}, ~~~
\chi \equiv \widetilde{F}_2 ~\tan~\theta ~\sec~\theta. \nonumber\\ \nd
The type IIA metric \eqref{reckoning} is in general a non-K\"ahler manifold and therefore the fractional two-brane may be thought of as D4-${\overline{\rm D4}}$ wrapped on a two-cycle 
${\bf \Sigma}_2$ 
in the non-K\"ahler space \eqref{reckoning}. The M-theory uplift will then be a $G_2$ structure manifold oriented\footnote{At a given point in the $x_3$ circle.} 
along ($\theta_1, \phi_1, r, \psi, x_8, x_9, x_{11}$) and a fractional M2-brane
state oriented along ($x_0, x_1, \psi$) that could be viewed as wrapped M5-brane on ${\bf \Sigma}_2 \times {\bf S}^1_{11}$ where  ${\bf S}^1_{11}$ is the eleven-dimensional circle. At energies smaller 
than the size of the internal cycle, the analysis that we performed above will suffice. 

The second subtlety also has to do with the precise orientation of our D2-brane surface operator. The surface operator that we discussed here is a co-dimension two singularity in four-dimensions, 
and is a co-dimension {\it one} singularity in the boundary three-dimensions. However what we need is a co-dimension two singularity in both three and four-dimensions 
\cite{wittenknots1}, \cite{surfknot}. One way out will be to change the orientations of our D2-brane in {\bf Table \ref{surfaceoper}} so that the D2-brane is now oriented along ($x_0, \psi, \phi_1$)
directions. This way, not only in our four-dimensional space ($x_0, x_1, x_2, \psi$) it is a co-dimension two singularity but is also a co-dimension two singularity in the three-dimensional boundary 
oriented along ($x_0, x_1, x_2$) directions. However, since the D2-brane has only temporal direction along the boundary, the line integral 
would vanish because of our gauge choice \eqref{gaugec} or \eqref{gaugec2}. Thus what we need here instead is a one-dimensional {\it curve} in the ($x_0, x_1, x_2$) plane. 
Lifting this configuration to M-theory will now have D0-brane whose precise contributions to our BHN equations should mimic what we had earlier.
Note that 
changing the orientation of the D2-brane from $\phi_1$ to any other orthogonal compact direction will uplift to a M2-brane but the orientation of the resulting brane is such that
it cannot always be percieved as an instanton contributing to the BHN equations\footnote{Unless one of the direction is along $r$. We will discuss this case later.}. 
As such the analysis will be harder to perform.

\begin{figure}[t]\centering \includegraphics[width=0.8\textwidth]{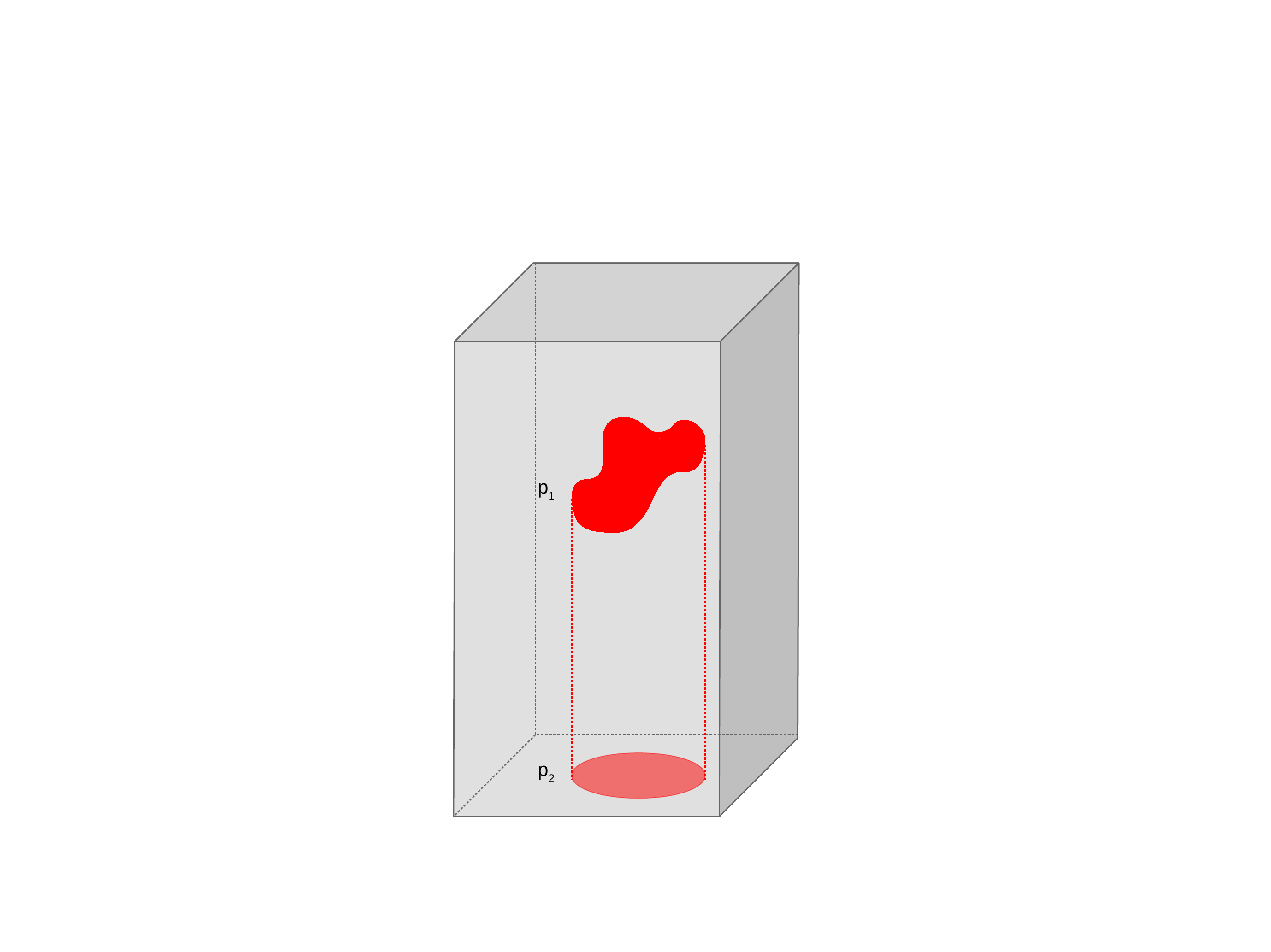}\caption{A loop $K$, denoted by $p_2$, in the ($x_1, x_2$) plane can be lifted up to form a knot ${\bf K}$, denoted by $p_1$, once we go to the Euclidean space. Non-trivial Wilson loop can now be constructed by integrating the twisted gauge field ${\cal A}_d$ along the knot $p_1$.}\label{fig1}\end{figure}

Alternatively
we can go to Euclidean space where the co-dimension two singularity is a curve in a three-manifold with  
non-trivial topology. This will be our knot configuration.
This means a co-dimension two singularity in four-dimensional space ${\bf V}$ as in \eqref{vwr+} will now be of the form: 
\bg\label{cnot}
{\bf C} \equiv {\bf K} \times {\bf R}_+, \nd
where ${\bf K}$ is a knot in three-dimensional Euclidean space (not to be confused with the loop $K$ discussed earlier in \eqref{rambof}) and ${\bf R}_+$ is our $\psi$ direction. In the equivalent 
Minkowski space, ${\bf K}$ would be a one-dimensional curve in ($x_0, x_1, x_2$) plane. 
In the above
discussion of putting a co-dimension two singularity along ($x_0, \psi, \phi_1$) directions the charge of the dual D0-brane bound state (with D6-branes) appears from:
\bg\label{dragonfly}
\int_{\Sigma_{11}} {\cal C}_3 \wedge {\cal G}_4 \wedge {\cal G}_4 = \int {\cal A}_0 dx_0 \int_{\Sigma_6} \langle {\cal F} \rangle \wedge \langle {\cal F} \rangle \wedge \langle {\cal F} \rangle 
\int_{\rm TN} \omega \wedge \omega, \nd
as such this amounts to switching on two extra components of gauge fields $\langle {\cal A}_1\rangle$ and $\langle {\cal A}_\psi \rangle$ in addition to what we had earlier. The caveat however is that, 
as discussed above, the temporal loop would vanish if we want to maintain our gauge choice \eqref{gaugec} or \eqref{gaugec2}. On the other hand, once we take a curve in the 
($x_0, x_1, x_2$) space, this issue doesn't arise and knots can arise naturally (see also {\bf Fig \ref{fig1}}). 

In the same vein if we allow the co-dimension two singularity to be along ($x_0, \psi, r$) directions, then the dual M2-brane state will be along ($x_0, \phi_1, r$) directions. Going to the 
Euclidean space we can allow the co-dimension two singularity to be along ${\bf C} \times {\bf R}$, where ${\bf C}$ is the surface given in \eqref{cnot} and ${\bf R}$ is the radial direction $r$. 
The dual M2-brane state then would be along ${\bf K} \times {\bf R} \times \phi_1$, where ${\bf K}$ is the knot.  
In the IIA framework this
is again an instanton in a four-dimensional space, whose two coordinates are ($x_3, \psi$) and the other two coordinates are orthogonal to the knot ${\bf K}$.    

\begin{figure}[t]\centering \includegraphics[width=0.7\textwidth]{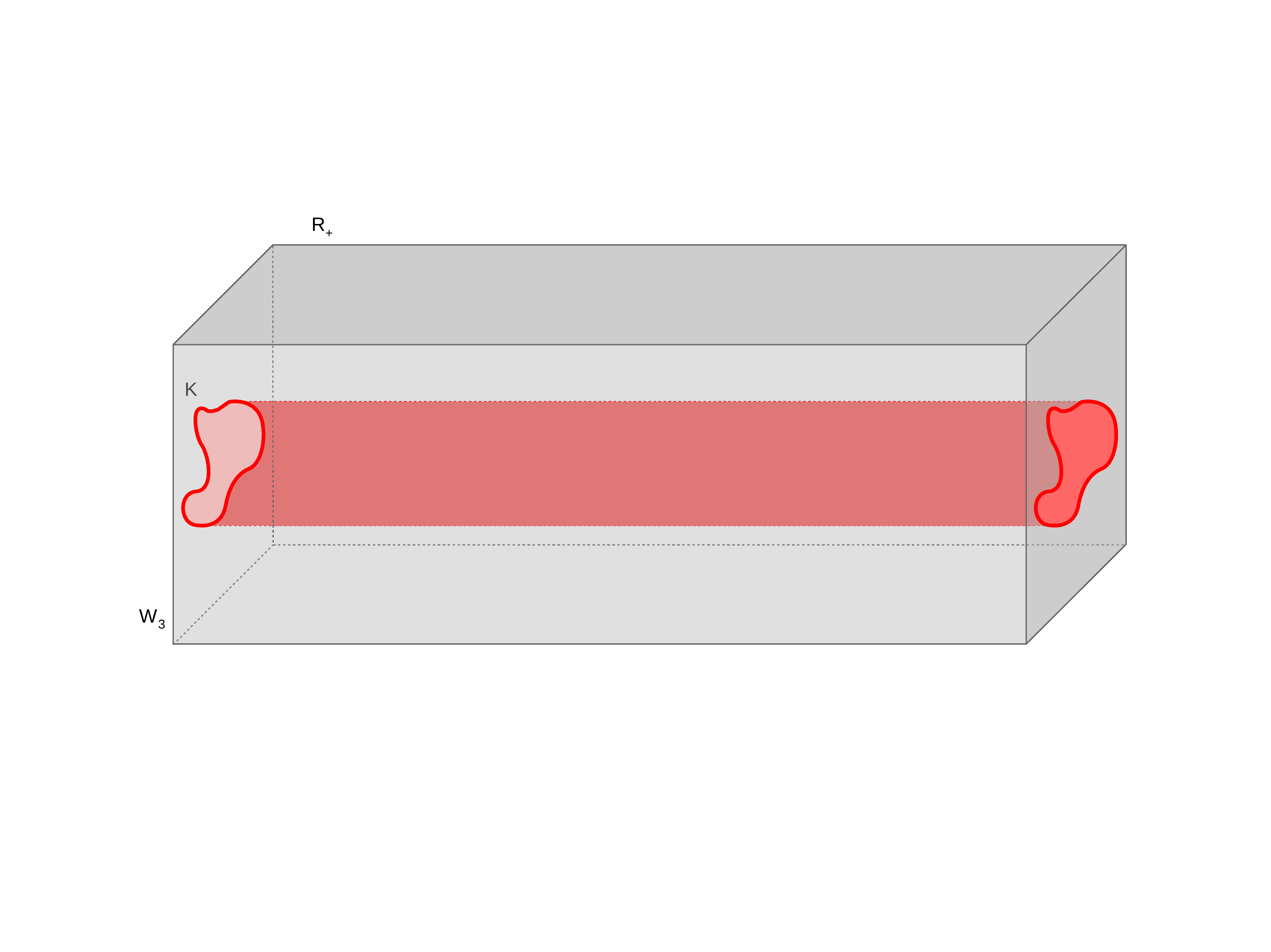}
\vskip-0.6in
\caption{A surface operator constructed out of a M2-brane intersects the three-dimensional Euclidean boundary ${\bf W}$ (or ${\bf W}_3$ in the language of \cite{wittenknots1}) along a knot ${\bf K}$ and is stretched along the remaining $\psi$ direction, which we denote here as ${\bf R}_+$. As such it is a co-dimension two singularity both on the three-dimensional boundary 
${\bf W}_3$ as well as the four-dimensional space ${\bf V} \equiv {\bf W}_3 \times {\bf R}_+$.}
\label{fig2} \end{figure}

\begin{figure}[t]\centering \includegraphics[width=0.8\textwidth]{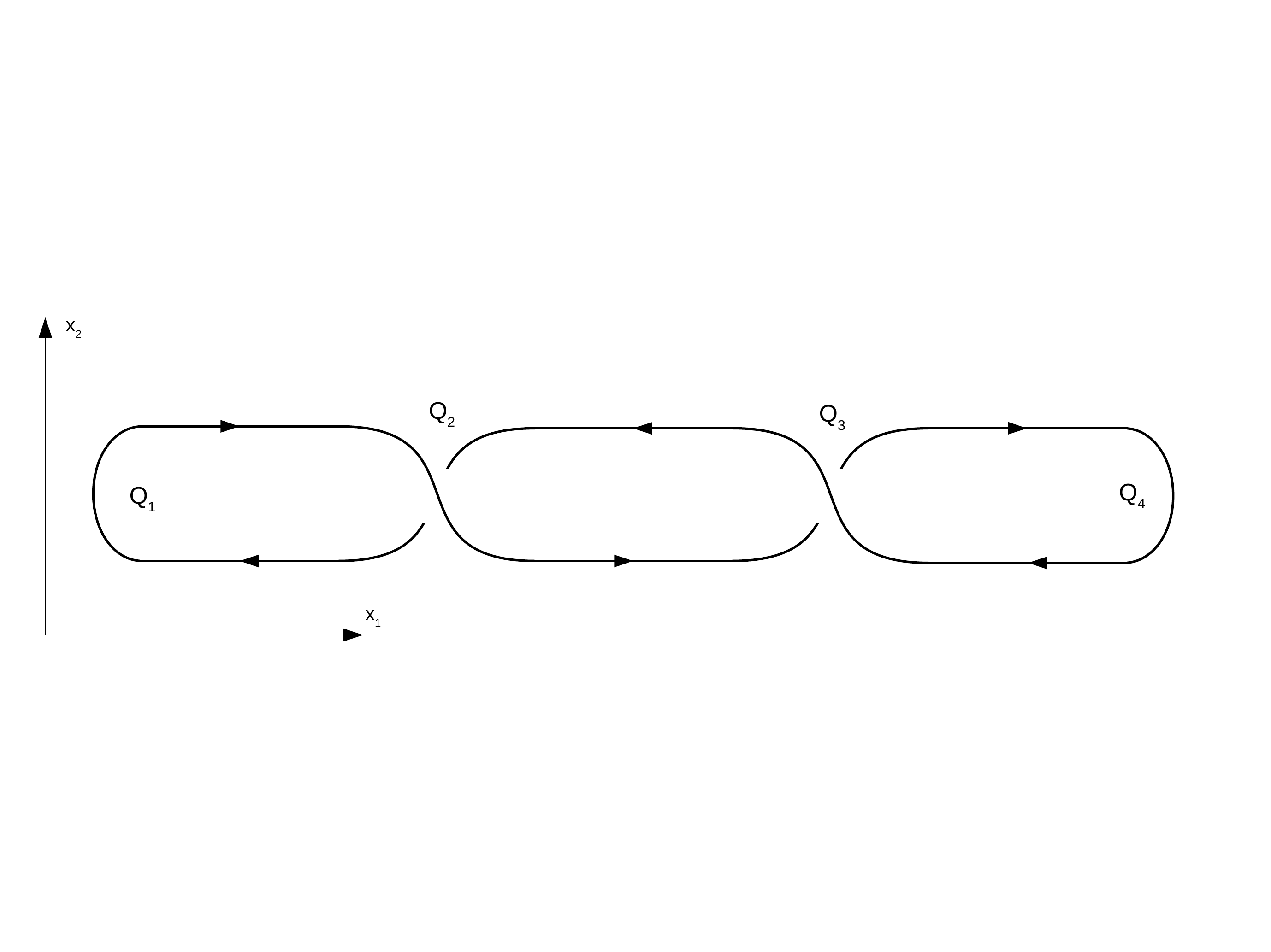} 
\vskip-1in
\caption{An unknot configuration drawn almost parallel to the $x_1$ axis to simplify the computation of the Wilson loop. Thus away from the regions 
denoted by $Q_i$, we can restrict the Wilson line integral to be only along $x_1$.}
\label{fig3} \end{figure}

Thus for either of the case discussed above, the co-dimensional two singularity in the Euclidean space is identical and is given in {\bf Fig \ref{fig2}}, although the M-theory uplifts differ. 
Previously when the co-dimension two singularity was along ($x_0, x_1$) the equations governed by the hypermultiplet scalars (${\cal A}_2, {\cal A}_\psi, \varphi_2, \varphi_3$) were the Hitchin's
equations \eqref{may6baz} from the BHN equation \eqref{swapan} in the absence of the surface operator; and \eqref{niceg} from the BHN equation \eqref{zoey} when the surface operator is present. 
Now our hypermultiplet scalars would appear from directions {\it orthogonal} to the knot ${\bf K}$ therefore the analysis will be different. However if we consider a knot 
configuration given in {\bf Fig \ref{fig3}}, away from the neighborhood points $Q_i$,  we have:
\bg\label{krisisjun25}
\oint_{\bf K} {\cal A} ~ \to ~ \int_{x_1} {\cal A}_1 dx_1, \nd
then again we expect the local picture to be similar, namely, the Hitchin's equations \eqref{may6baz} get suitably modified like \eqref{zoey} (although $\sigma$ and $\bar\sigma$
in \eqref{zoey} need to be interpreted carefully now).

We are now getting closer to the approach initiated in the series of papers \cite{gukovwitten2}, \cite{gukovwitten}, \cite{surfknot} and \cite{gukovsurf} and also in \cite{wittenknots1}. The
co-dimension two singularity in Euclidean space that we discuss here is clearly related to the {\it monodromy defect} studied in \cite{wittenknots1} and \cite{surfknot}. Moreover, since we 
study static configurations (using the Hamiltonian \eqref{hamilbeta}), the temporal direction $x_0$ remains suppressed and the co-dimension one singularity in the three-dimensional boundary of our earlier 
discussions continues to provide accurate description of the singularity structure of the (4, 4) hypermultiplets {\it locally}, although the global picture may be different.
This shift of our view point from 
global to local is not just a mere rephrasing of \eqref{krisisjun25} but more of a helpful calculational tool where analysis pertaining to specific knots could at least be addressed. In particular, for
the present context, this helps us to channel our earlier computations to analyze non-trivial knot configurations instead of just closed loops discussed in \eqref{rambof}. 

We can make our analysis a bit more precise. In the presence of the knot ${\bf K}$, the part of the boundary three-dimensional action \eqref{copycat} for the abelian case remains unchanged in form with
${\cal A}_d$ defined appropriately with $\hat{\phi}$. The additional piece of the action will be more non-trivial than \eqref{jugnu} as now we expect the 
integral to be over a knot ${\bf K}$. The total action will then take the form similar to \eqref{rambof} with the loop $K$ replaced by the knot ${\bf K}$ and ${\cal A}$ by ${\cal A}_d$. For 
completeness we reproduce this again as:
\bg\label{winter}
S_{bnd} = (b_2 + c_2) \int_{\bf W} {\cal A}_d \wedge d{\cal A}_d + Q_2 \oint_{\bf K} {\cal A}_d, \nd
where $Q_2$ can be calculated from M-theory using either the dual D0-brane charge \eqref{dragonfly}, or the dual M2-brane charge depending on our choice of orientation. For the latter case 
$\hat\phi$ in the definition of ${\cal A}_d$ will take a different form (that can be determined with some effort, but we will not do so here). Various other details like the field strength
${\cal F}_d$ as well as the monodromy around ${\bf K}$ remain similar to \eqref{darkness} and \eqref{pocket} respectively. Furthermore, the presence of ${\cal A}_d$ in the integral over the 
knot ${\bf K}$ can now be directly hinted from \eqref{zoey} and \eqref{putloc} by the following replacement:
\bg\label{deadof} 
{\cal F}_{\alpha\psi} ~ \to ~ {\cal F}_{\alpha\psi} + \langle {\cal F}_{\alpha\psi}\rangle \equiv {\cal F}_{\alpha\psi} + g_\alpha {\cal D}_\psi\hat{\varphi}_\alpha, \nd
where $g_\alpha$ is an appropriate constant and there is no sum over alpha. Indeed the above defines the gauge field ${\cal A}_{d, \alpha} \equiv {\cal A}_\alpha + g_\alpha \hat{\varphi}_\alpha$ that
eventually appears through the boundary magnetic charge ${\bf Q}_M$ into the boundary coupling \eqref{winter}. One may easily see that in our earlier derivation this replacement for ${\cal A}_2$ was 
not necessary despite the existence of $\langle {\cal A}_2 \rangle$ because the instanton configuration therein was defined in the space parametrized by ($x_2, x_3, r, \phi_1$) 
and thus independent of the $\psi$ coordinate\footnote{We define $\langle{\cal F}_{2\psi}\rangle = \partial_2\langle{\cal A}_{\psi}\rangle - \partial_\psi\langle{\cal A}_{2}\rangle + 
i\left[\langle{\cal A}_{2}\rangle, \langle{\cal A}_{\psi}\rangle\right]$ which is proportional to $\partial_\psi\langle{\cal A}_{2}\rangle$  
for the case studied earlier because $\langle{\cal A}_\psi\rangle$ vanishes, but now, for the present case, has all the terms.}.  
However now the dual D0-brane charge \eqref{dragonfly} does depend on all coordinates orthogonal to the Taub-NUT space and as such \eqref{deadof} 
becomes necessary.  

\begin{figure}[t]\centering \includegraphics[width=0.8\textwidth]{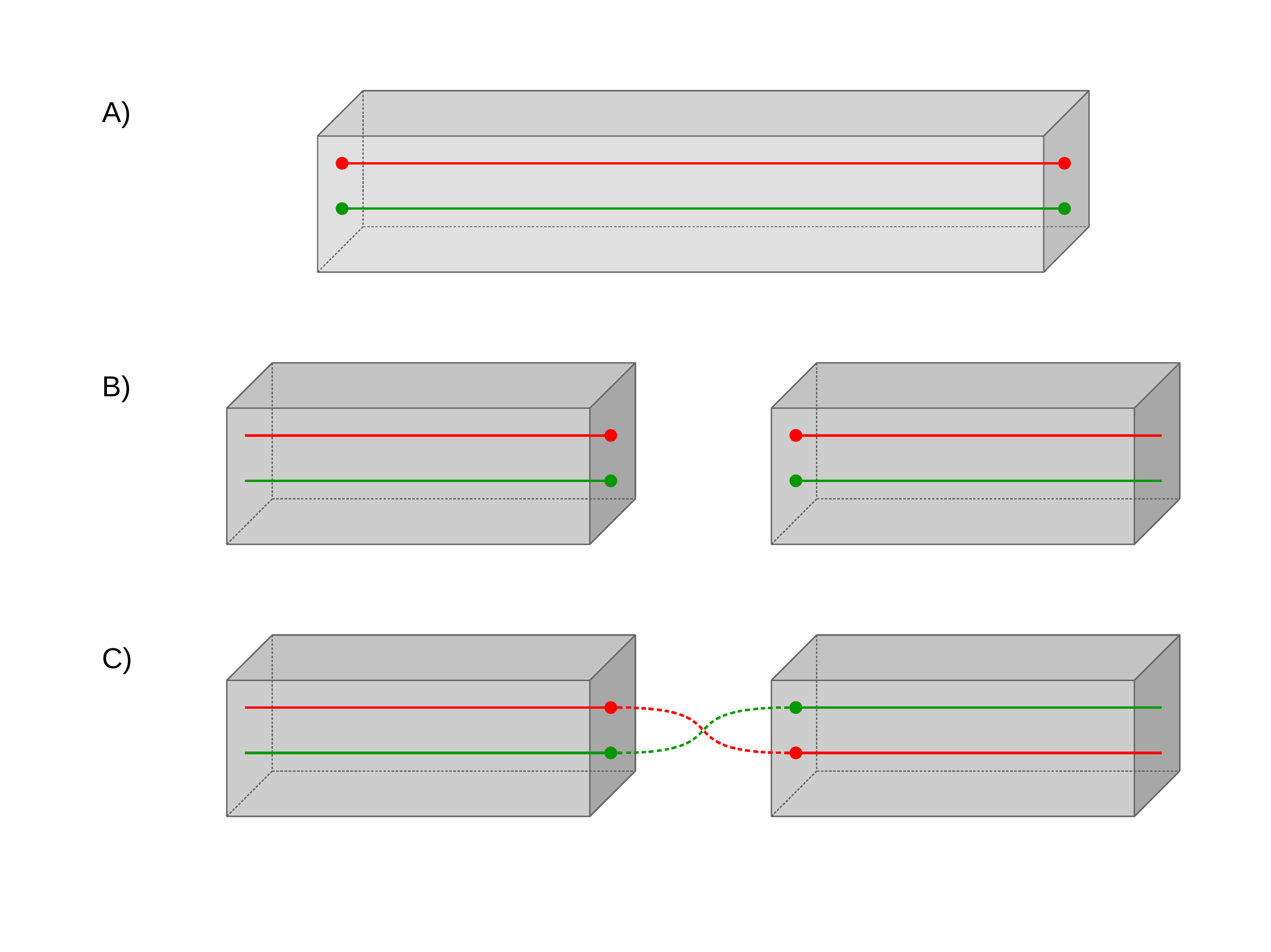} \caption{Two Wilson lines in the three-dimensional boundary denoted by (A) is arranged so that they are parallel to the $x_1$ axis. 
In (B) we split them via Heegaard splitting and they are rejoined in (C) via a braid group action. This procedure allows us to introduce non-trivial structures to the Wilson lines.}
\label{fig4} \end{figure}

Our short discussion above shows that, at least for the abelian case, the boundary theory appearing from the magnetic charge ${\bf Q}_M$ in the presence of a surface operator does have all the 
essential properties to study knot configurations. The brief mismatch that we had earlier in the boundary theory \eqref{rambof} goes away once the background is correctly defined as we see in 
\eqref{winter}. All this is satisfactory and one might, at this stage, even speculate how the {\it non-abelian} extension may look like. The non-abelian
boundary Chern-Simons theory will have the form 
\eqref{plaza}, but now ${\cal A}_{d, \alpha}$ will have to be defined with respect to $\hat{\phi}_\alpha$. The knot will then be added as a linear term in ${\cal A}_d$, just as in 
\eqref{winter}, but now to \eqref{plaza}. The above statements are easy to state but a direct 
derivation of the boundary action along the lines of our earlier discussion is unfortunately 
harder because of the issues pointed out above. We will therefore relegate a detailed discussion to the sequel of this 
paper and instead make some generic statements here.

There is one puzzle however that we need to clarify. 
The non-abelian Chern-Simons theory \eqref{plaza} with the coupling ${k \over 4\pi} \equiv b_2 + c_2$, appearing in \eqref{stwist} and \eqref{c1c2c1c2},  
is well defined in a path integral only when $k$ is an integer. With a gauge group $G$, the path 
integral representation is given by:
\bg\label{lonkeys}
Z({\bf K}, k, G) = \int {\cal D} {\cal A}_d ~{\rm exp}\Big[iS_{bnd}(k, {\cal A}_d)\Big] {\rm Tr}_R {P}~{\rm exp}\left(Q_2\oint_{\bf K} {\cal A}_d\right), \nd
where the integral
is over all gauge connections ${\cal A}_d$ modulo gauge transformations. What happens when $k$ is not an integer? 
This could in general be the case because both $b_2$ and $c_2$, given in \eqref{stwist} and \eqref{c1c2c1c2} respectively, appear from supergravity analysis and are 
as such not restricted to be integers.
It turns out, when $k$ is {\it not} an integer, we can still perform  
the path integral by {\it complexifying} the gauge field 
${\cal A}_d$. The story becomes more interesting now, and has been discussed in much details in \cite{jimjam}. This analytical continuation of Chern-Simons theory at the boundary proceeds 
in few steps: one, to change the measure of the path integral; two, to incorporate the complex conjugate piece in the path integral and then three, to assume the complex conjugate piece, 
constructed from $\overline{\cal A}_d$, to be 
independent of the one constructed from ${\cal A}_d$ \cite{jimjam}. In other words, we change \eqref{lonkeys} to:
\bg\label{fireworks} 
Z({\bf K}, k, \widetilde{k}, G) = \int_{\cal C} {\cal D} {\cal A}_d {\cal D}\widetilde{\cal A}_d
~{\rm exp}\Big[iS_{bnd}\left(k, {\cal A}_d\right) + iS_{bnd}\left(\widetilde{k}, \widetilde{\cal A}_d\right)\Big] 
{\rm Tr}_R {P}~{\rm exp}\left(Q_2\oint_{\bf K} {\cal A}_d\right), \nonumber\\ \nd 
where both $\widetilde{\cal A}_d$ and $\widetilde{k}$ are in general different from $\overline{\cal A}_d$ and $k$ respectively. The choice of the integration cycle ${\cal C}$ is subtle and is 
captured by finding critical points of the modified Chern-Simons action appearing in \eqref{fireworks} and then expressing ${\cal C}$ in terms of the so-called Lefshetz thimbles \cite{jimjam}.
The integrals over these Lefshetz thimbles should always converge, and this way finite values could be determined for the path integral \eqref{fireworks}\footnote{Clearly this is a playground
for using Morse theory and the theory of steepest descent as have been exemplified by \cite{jimjam}.}. 

\begin{figure}[t]\centering \includegraphics[width=0.8\textwidth]{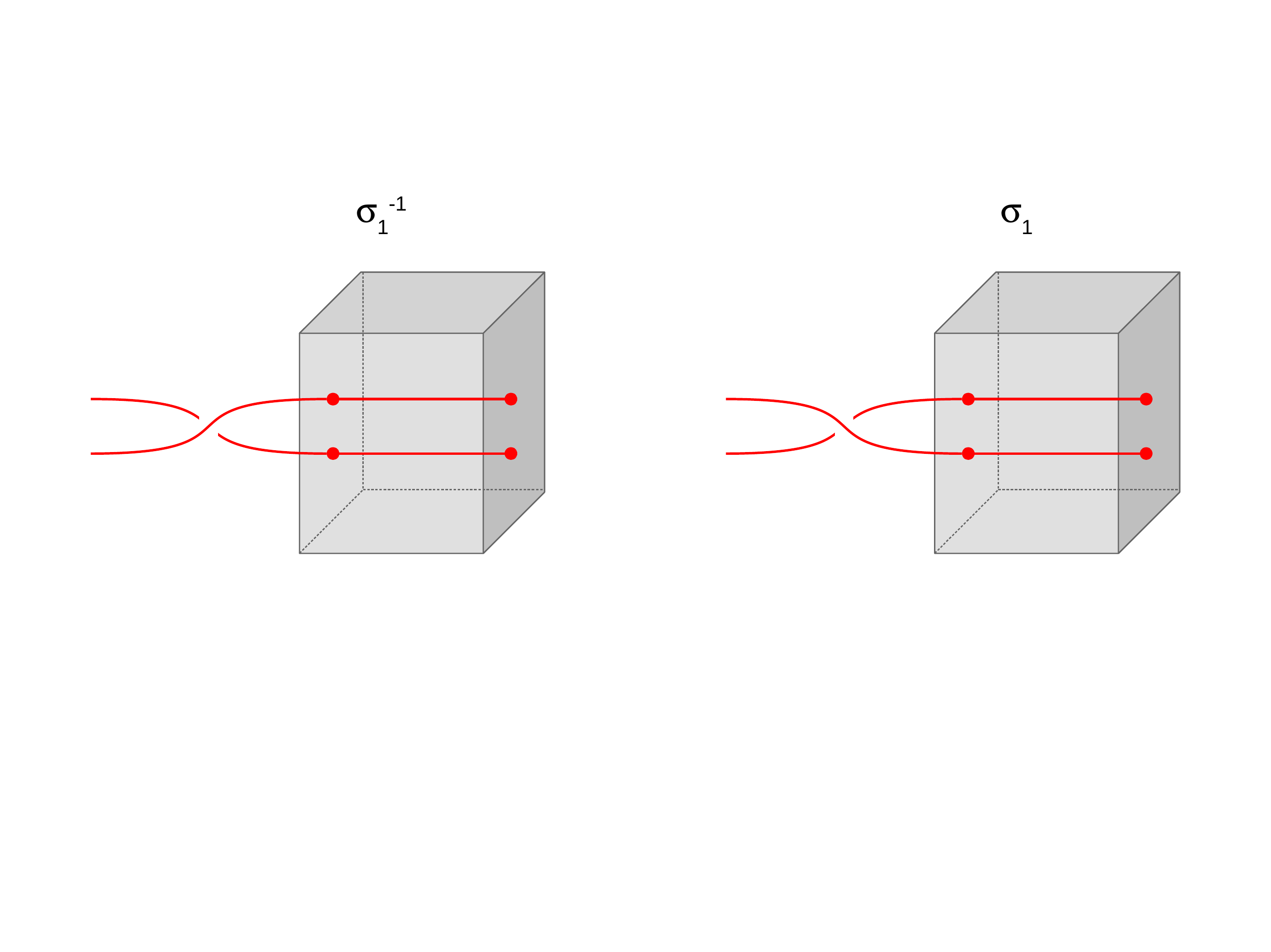}
\vskip-1in
 \caption{The action of the braid group on the Wilson lines. They are distinguised by their over-crossing and 
under-crossing pattern. The first one has a braid group action $\sigma_1^{-1}$, whereas the second one has a braid group action $\sigma_1$.}\label{fig5} \end{figure}

The above discussion raises an interesting question, namely, what is the interpretation of the above story from our M-theory uplift? To answer this,  
recall how we arrived at the Chern-Simons theory \eqref{plaza}. Our starting point was the four-dimensional action \eqref{stotal}, from where 
we derived the Hamiltonian \eqref{hamilbeta}. The electric and the magnetic charges ${\bf Q}_E$ and ${\bf Q}_M$ respectively, when arranged properly by taking care of the subtleties mention in 
section \ref{boundth}, gave rise to the boundary action \eqref{plaza}. There were two crucial ingredients in the discussion: one, the expression \eqref{stwist}, which was 
important in deriving the coupling 
constant $k$ and two, the twisted gauge 
field ${\cal A}_d$ which in turn was composed of the original gauge field ${\cal A}_\mu$ and the twisted scalar field $\phi_\mu$. Looking even further back, both the
ingredients appeared from M-theory: the twisted gauge field from the G-flux ${\cal G}_4$ via \eqref{chorM}; and 
the coupling $k$ (i.e $b_2$ and $c_2$) essentially from the M-theory action via \eqref{mtheorya}\footnote{For the full non-abelian enhancement the readers may refer to section \ref{NAE}.}.  
This means the complexification of the fields that is necessary to analyze \eqref{fireworks} should somehow also appear directly from our M-theory analysis.   

\begin{figure}[t]\centering \includegraphics[width=0.6\textwidth]{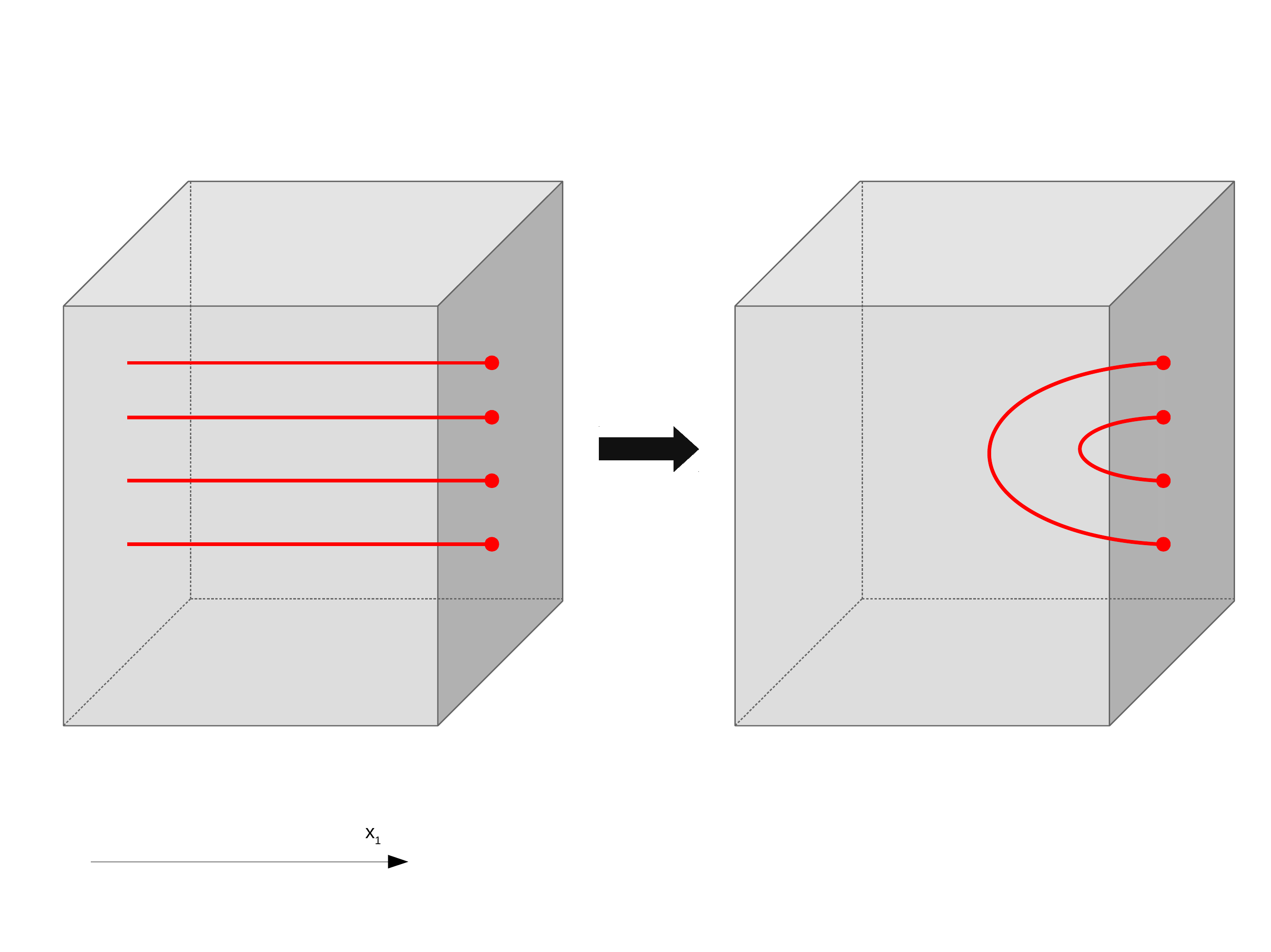} \caption{Four Wilson lines are joined pairwise by identifying the respective monodromies around them.} 
\label{fig6}\end{figure}

The analysis gets harder because in M-theory, or in the eleven-dimensional supergravity, the ingredients enhancing the four-dimensional gauge theory from abelian to non-abelian and creating the 
knots may be the same M2-branes. The distinguishing feature is of course their orientations: the 
non-abelian enhancements appear from M2-branes wrapped on the Taub-NUT two-cycles, whereas the knots appear from 
M2-branes having at most one leg along the Euclideanized boundary ${\bf W}$ (or being a one-dimensional curve in the three-dimensional Minkowskian boundary). 
On the other hand when the knot configurations are dual to the D0-branes, with the worldline of the D0-branes forming a
knot configuration in the three-dimensional boundary ${\bf W}$, the analysis is equally challenging from M-theory. Even at the abelian level, the essential path-integral that we can lay out is
the following:
\bg\label{mesrine1}
Z(a, b) = \int {\cal D}{\cal G}_4 ~{\rm exp}\left[i a \int_{\Sigma_{11}} {\cal G}_4 \wedge \ast{\cal G}_4 + i b\int_{\Sigma_8} {\cal G}_4 \wedge {\cal G}_4 \right] 
{\rm exp}\left(i\oint {\cal C}_3\right). \nd
This is good enough to capture certain aspects of four-dimensional abelian gauge theory as well as the boundary three-dimensional Chern-Simons theory, but definitely not the full story, 
at least not yet in the present incarnation with $a$ providing the gauge coupling and:
\bg\label{mesrine2}
b \equiv {c_2\over \int_{\rm TN} \omega \wedge \omega}, ~~~~~\Sigma_8 = {\bf W} \times {\bf R}_+ \times {\bf TN}, \nd
on the eleven-dimensional manifold $\Sigma_{11} = \Sigma_8 \times {\bf S}^3$, where ${\bf S}^3$ is a three-cycle and $\omega$ is the normalizable harmonic form defined on the warped Taub-NUT space. 
To complete the story, 
we will need a few crucial intermediate steps: one, that converts $b$ in \eqref{mesrine1} to $k$ as in \eqref{lonkeys} via a step similar to \eqref{stwist}; two, that converts 
G-flux ${\cal G}_4$ to three-dimensional twisted gauge field ${\cal A}_d$; and three, that finally converts \eqref{mesrine1} to \eqref{fireworks}. The search then 
is a formalism for a {\it topological M-theory} where calculations of the kind mentioned above may be performed (somewhat along the lines of \cite{topM}). 

\begin{figure}[t]\centering \includegraphics[width=0.8\textwidth]{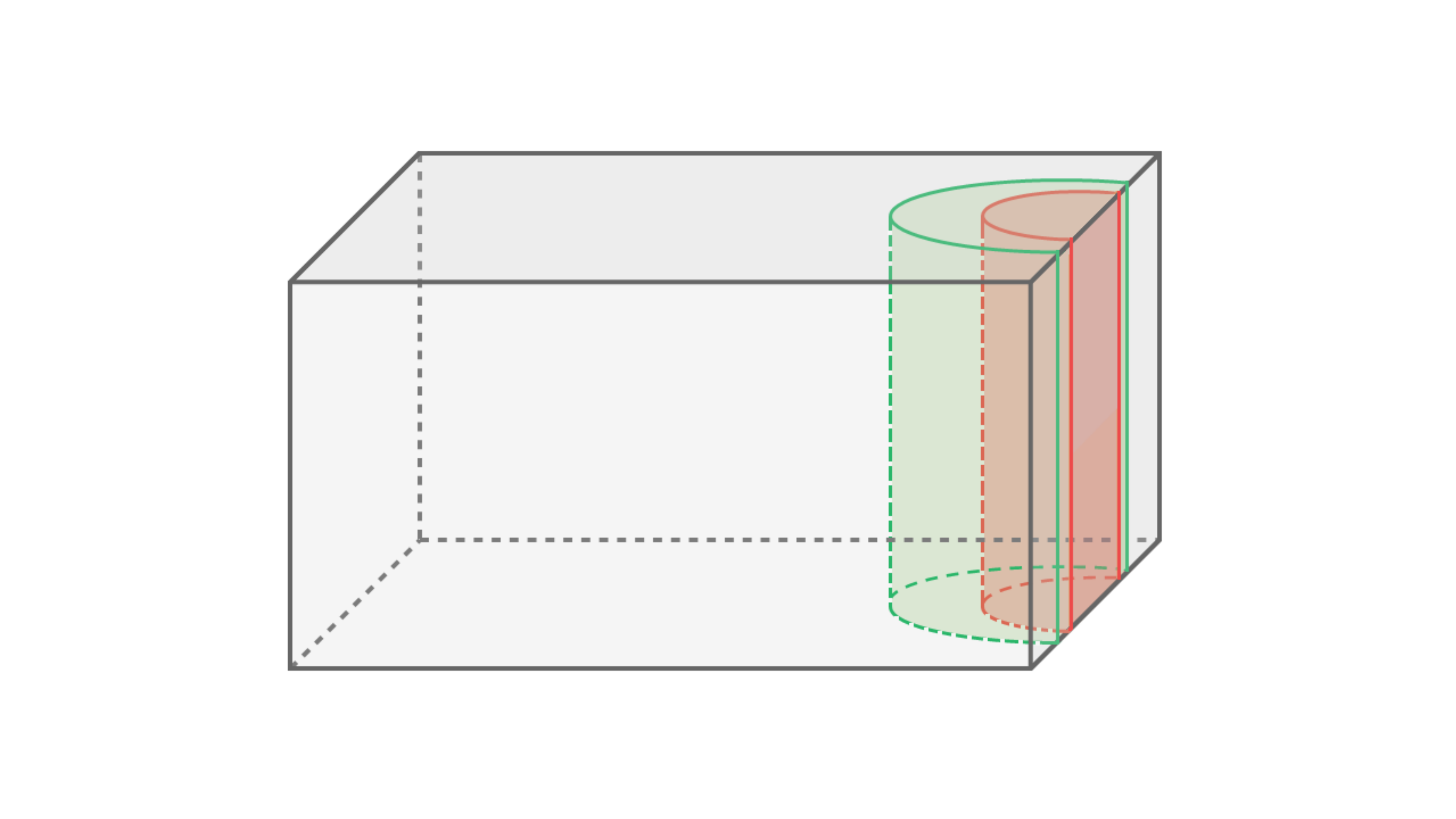}
\caption{Once we identify monodromies of a pair of Wilson lines, the structure of the co-dimension two surface operator in four-dimensional space can be formed out of two-branes. Here two 
such configurations are shown on a Heegaard-split three-manifold base.}
\label{fig7}
\end{figure}

In the absence of such a formalism, simplification occurs when $k$ becomes an integer, so that we can ignore complexification, and when we go to the abelian case, where we can resort to our 
earlier calculations. This then brings us to the following path-integral representation:
\bg\label{ludivin}
Z({\bf K}, k) = \int {\cal D}{\cal A}_d ~{\rm exp}\left({ik\over 4\pi}\int_{\bf W} {\cal A}_d \wedge d{\cal A}_d\right) {\rm exp}\left(iQ_2 \oint_{\bf K} {\cal A}_d\right), \nd
which is simpler than both \eqref{lonkeys} as well as \eqref{fireworks} and where ${k \over 4\pi} = b_2 + c_2$ is now an integer. Additionally, the quadratic form of the Chern-Simons action
implies that \eqref{ludivin} can be simplified further. Defining ${\cal A}_d = \langle {\cal A}_d \rangle + a_d$, where $a_d$ is the fluctuation over the background field 
$\langle {\cal A}_d \rangle$, and using \eqref{darkness} now for the background field strength $\langle {\cal F}_d \rangle$, we can express \eqref{ludivin} equivalently as:
\bg\label{ejong}
Z({\bf K}, k) = Z_0 \int {\cal D}{a}_d ~{\rm exp}\left({ik\over 4\pi}\int_{\bf W} {a}_d \wedge d{a}_d\right), \nd
where $Z_0$ is a number and is given by $Z_0 = {\rm exp}\left({ik\over 4\pi}\int_{\bf W} \langle {\cal A}_d \rangle \wedge d \langle{\cal A}_d\rangle \right)
{\rm exp}\left(\oint_{\bf K}\langle{\cal A}_d\rangle\right)$, 
implying that the quantum computations 
in the presence of a knot may be performed by studying the fluctuations over a classical background as if the knot was absent. This simplification is of course only for the abelian case, as the 
non-abelian case would require more elaboarate computational machinery.  

\begin{figure}[t]\centering \includegraphics[width=0.8\textwidth]{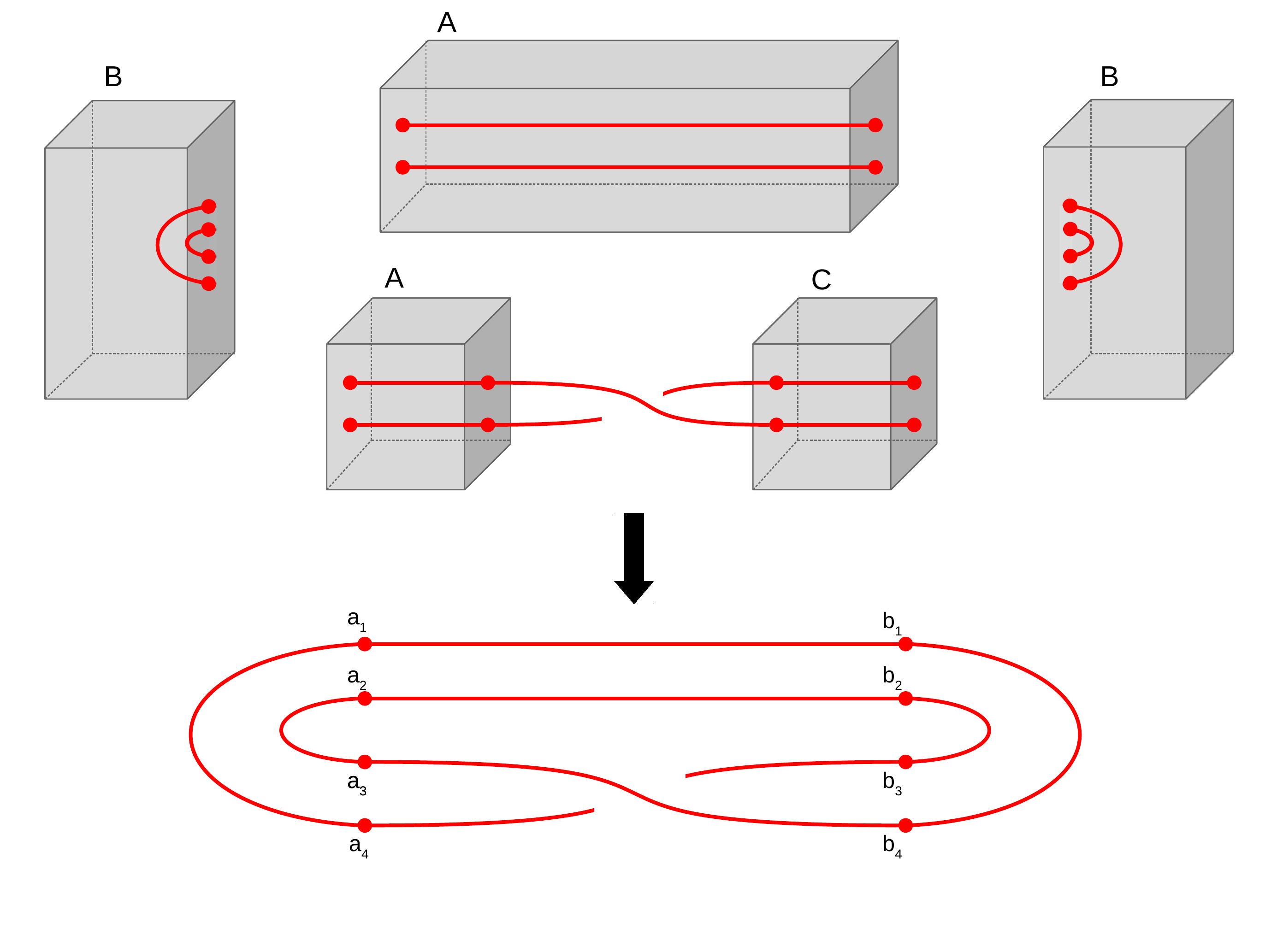}
\caption{Construction of an unknot using all the ingredients that we developed earlier. Boxes ${\bf A}$ represent the Wilson lines parallel to $x_1$ axis, boxes ${\bf B}$ 
denote the curving of the Wilson lines
by identifying pairwise monodromies, and finally box ${\bf C}$ denote the braid group action. Together they form an unknot configuration. The points $a_i$ and $b_i$ are the points where the Wilson lines 
end on the Heegaard-split three manifolds.}
\label{fig8}\end{figure}

There is something puzzling about \eqref{ejong} that we would like to clarify. Rephrasing \eqref{ludivin} to \eqref{ejong} one might worry that all information about the knot ${\bf K}$ is now lost. 
In fact what we have in \eqref{ejong} is the following additional integral:
\bg\label{jojames}
{\rm exp}\left(-iQ_2\int_{\bf W} a_d \wedge \delta_{\bf K}\right) {\rm exp}\left(iQ_2 \oint_{\bf K} a_d \right), \nd
which vanishes classically and so the computations proceeds as though no knot is present in \eqref{ejong}. However \eqref{jojames} imples that the actual quantum mechanical computation should have
{\it another} knot linked to the previous one. In other words there should be a framing anomaly \cite{witten89}. Taking this into account, the information about the knot can thus be recovered in the 
quantum computations. 

Let us elaborate this a bit more. For abelian, Chern-Simons, the cubic interaction term is absent. The expectation value of Wilson loop operator for knot $\mathbf K$ in \eqref{ludivin} can be 
expressed as:
\bg\label{ludivin2}
Z({\bf K}, k) =  \exp\left\{-Q_2^2 \langle \oint_{\mathbf K} dx_{\mu} {\cal A}_d^{\mu}(x) \oint_{\mathbf K} dy_{\nu} {\cal A}_d^{\nu}(y)\rangle\right\}, 
\nd
where $Q_2$, as mentioned earlier, may be computed in M-theory from the dual D0-brane charge \eqref{dragonfly} or from dual M2-brane charge depending on our choice of orientation. 
Using the gauge field two point function:  
\bg\label{4thfloor}
\langle {\cal A}_d^\mu(x) {\cal A}_d^\nu(y)\rangle = \epsilon^{\mu \nu \lambda}~ {(x-y)_{\lambda} \over |x-y|}, \nd  
we see that the above invariant \eqref{ludivin2} will blow up at coincident points 
$x_{\mu} \rightarrow y_{\mu}$. So we will have to regularize the integral. 
This is achieved by choosing a suitable frame with a ${\mathbf K_f}$ knot slightly displaced from the original knot. In other words, we take the coincident points as 
$y_{\mu}= x_{\mu} +\epsilon_{\mu}$, with $\epsilon_\mu$ approaching zero.  
Depending on the choice of frame, we will get the $U(1)$ knot invariant \eqref{ludivin2} to be in terms of a framing number $p$, defined as the  
linking number of knot ${\mathbf K}$ with its frame knot ${\mathbf K_f}$, in the following way:
\bg\label{jullewis}
-{1\over Q_2^2} ~{\rm log}~Z({\bf K}, k) = {}^{\rm lim}_{\epsilon_\mu \to 0}
\langle \oint_{\mathbf K} dx_{\mu} {\cal A}_d^{\mu}(x)  \oint_{\mathbf K_f} dy_{\nu} {\cal A}_d^{\nu}(y)\rangle = - {i \pi p \over k}, \nd
implying that for any knot the result is proportional to $p$. 
However, we can always choose a canonical frame in ${\bf S}^3$ where $p=0$. In other words, this canonical frame does not give any information about knots within 
abelian Chern-Simons theory. This is exactly reflected by perturbing the classical background solution as detailed in \eqref{ejong} and \eqref{jojames}. Thus non-trivial information is 
achieved when we go from one frame to another. For more details see \cite{mmarino}.  

We are now ready to discuss the construction of knots using our surface operators. One of the crucial ingredient is the Heegaard splitting, which states that a three manifold ${\bf W}$ can be 
obtained as a connected sum of three manifolds ${\bf W}_1$ and ${\bf W}_2$ joined along their common boundary ${\bf \Sigma} \equiv \partial {\bf W}$. Thus:
\bg\label{heegaard}
{\bf W} = {\bf W}_1 \cup_{\bf \Sigma} {\bf W}_2. \nd
In the presence of a surface operator, a three manifold can also be split in a similar way as depicted in {\bf Fig \ref{fig4}}, ($A$) and ($B$). Once we extend the Wilson lines along the 
${\bf R}_+$ direction (or alternatively the $\psi$ direction) in {\bf Fig \ref{fig4}}($B$), we can see how the surface operators split. The way we have expressed the surface operators, they 
are parallel to $x_1$ axis as can be seen from the Wilson line representation \eqref{krisisjun25}. This means on the boundary 
${\bf \Sigma}$ of our three manifold ${\bf W}$ the Wilson lines will end on points, and the splitting of the surface operators would imply how the points are distributed on different
boundaries. In a standard quantization of the Chern-Simons theory on ${\bf W}$, where ${\bf W}$ is locally a product of ${\bf \Sigma} \times {\bf R}_1$ with ${\bf R}_1$ representing the
direction $x_1$, the Hilbert space ${\cal H}_{\bf \Sigma}$ associated to the boundary ${\bf \Sigma}$ changes, in the presence of the surface operator, to:
\bg\label{jaazend}  
{\cal H}_{\bf\Sigma} ~ \to ~ {\cal H}_{{\bf \Sigma}; p_i; R_i}, \nd
where $p_i$ are the points on ${\bf \Sigma}$ where the Wilson lines end and $R_i$ are the representations of each points. In the present case the Hilbert space is precisely the gauge theory described
on the D2-brane surface operator that we use here.

\begin{figure}[t]\centering \includegraphics[width=0.8\textwidth]{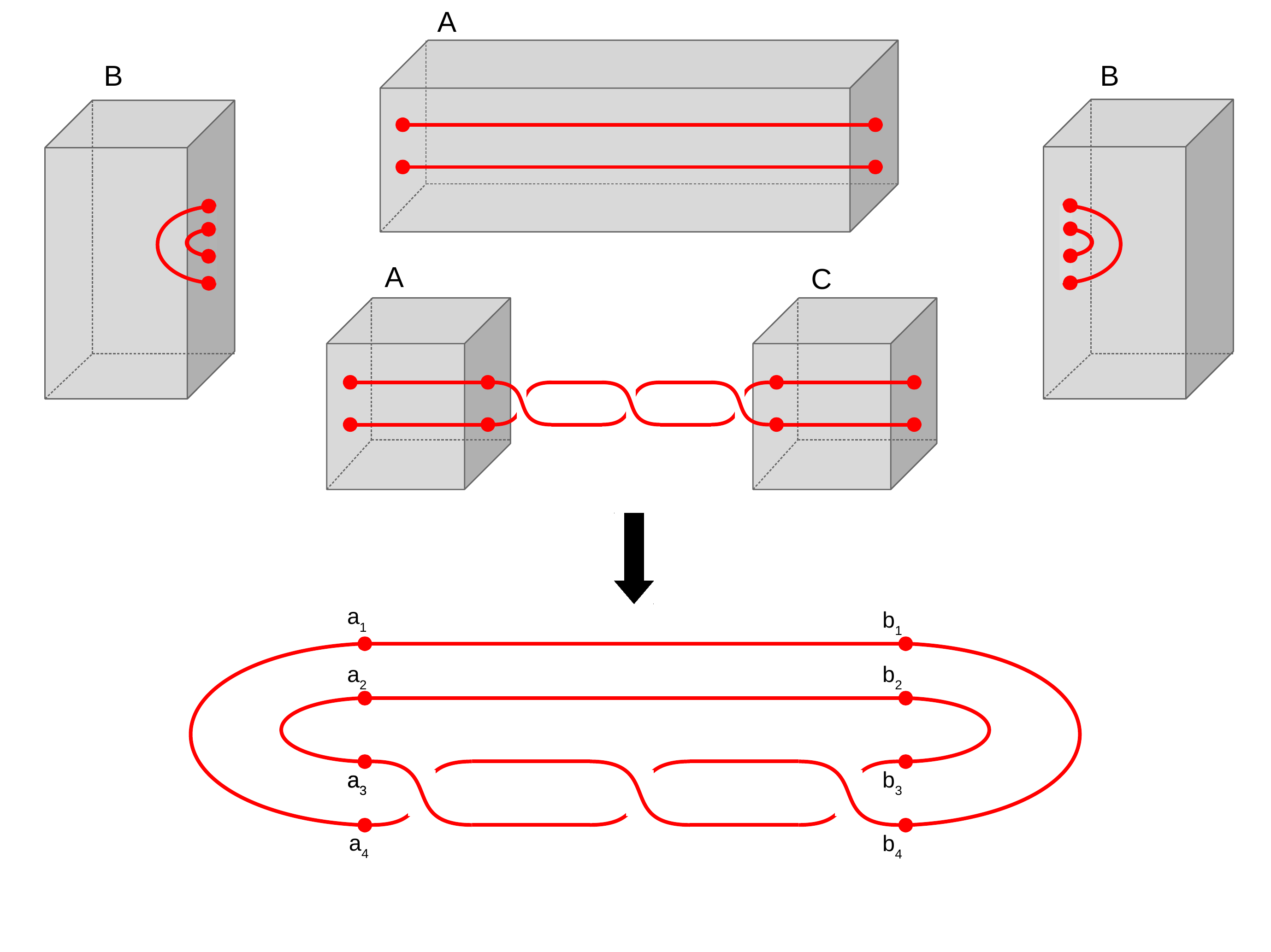}\caption{Construction of a trefoil knot by joining boxes ${\bf A}$, ${\bf B}$ and ${\bf C}$ appropriately. 
The braid group action now acts twice. The points ($a_i, b_i$) still remain the points where the Wilson lines end on the Heegaard-split manifolds.}\label{fig9} \end{figure}

The next ingredient is the monodromy around the surface operator. We already described the case when we have a loop $K$ in the ($x_1, x_2$) plane for the surface operator given in 
{\bf Table \ref{surfaceoper}}. The monodromy therein was given by \eqref{pocket}, which can be re-expressed in the language of ($\alpha, \beta, \gamma$) using the BHN equations \eqref{sandstime},
where ($\alpha, \beta, \gamma$) have in-turn 
been expressed using supergravity variables in \eqref{lateni} and \eqref{f685}. In eq. (6.4) of \cite{wittenknots1}, and also 
in eq. (2.2) and eq. (2.3) of \cite{gukovwitten} with more details, the gauge field ${\cal A}$ and the scalar field $\hat\phi$ have been described using ($\alpha, \beta, \gamma$). Using 
\eqref{lateni} and \eqref{f685}, we now express (${\cal A}, \hat\phi$) using supergravity variables. This is no surprise, of course, as in our earlier sections we used supergravity to write the 
BHN equations for ${\cal F}_{\alpha\beta}$ and  ${\cal F}_{\alpha\psi}$. Thus the monodromy around the $k$-th surface operator \eqref{pocket} can be now written as:
\bg\label{iamlegend}
{\cal M}_k \equiv {\rm exp}\left[-2\pi(\alpha_k - i\gamma_k)\right]. \nd
Since a given surface operator is a solution of the set of equations \eqref{karfish}, \eqref{asnles}, \eqref{rodmoore}, \eqref{mklee} and \eqref{kyalee},  
monodromies around different surface operators depend on their respective choices of the triplets ($\alpha_k, \beta_k, \gamma_k$).  

The gauge field set (${\cal A}, \hat\phi$) that we take 
appears in the boundary Chern-Simons theory as a combined gauge field ${\cal A}_d$ as defined in \eqref{newgf} and in \eqref{copycat}. There are 
three parameters that appear in the definition of ${\cal A}_d$: $b_2$ and $c_2$ from gauge theory coupling constant \eqref{stwist} after twisting, and $d_1$ from \eqref{nest}. It is easy to see that
although $b_2$ and $c_2$ both have to be real, $d_1$ can in principle be {\it complex}\footnote{When $b_2$ and $c_2$ are also complex, we are in the regime where we have to analytically continue 
the Chern-Simons theory. We discussed this briefly earlier and more details are in \cite{jimjam}.}. 
Nothing that we discussed earlier will modify if $d_1$ becomes a complex function. In fact there are two ways
to go about this, with definite advantages in either formalism. Using ${\cal A}_\mu = - iA_\mu$ as in \eqref{moyukh}, we can express ${\cal A}_d$ as:
\bg\label{seccup}
{\cal A}_d ~ = ~ -i\left(A - {id_1\over b_2 + c_2}\hat\phi\right), \nd
which keeps $d_1$ real, but inserts an $i$ in the definition of the gauge field. In this formalism, a boundary {\it flat} connection implies a Hitchin equation of the following form:
\bg\label{6sense}
d{\cal A}_d - i {\cal A}_d \wedge {\cal A}_d ~ = ~ 0 ~ = ~ F + \left({d_1\over b_2 + c_2}\right)^2 \hat\phi \wedge \hat\phi, \nd
where note the relative plus sign\footnote{If we now define $\hat\phi = -i\Phi$, we will get back \eqref{may6baz} as expected. However for the computations at hand, we keep the twisted 
one-form scalar fields unchanged, and only redefine the gauge fields. As noted above, this line of thought has some distinct advantages.}. 
Comparing this with say \eqref{asnles}, which is expressed in variables before twisting, we see that they are similar provided we use ${\cal F}_{12} = -iF_{12}$ as
in \eqref{moyukh}. After twising the coefficients of \eqref{6sense} may be identified with the ones in \eqref{asnles} and this way the value of $d_1$ may be determined.

\begin{figure}[t]\centering \includegraphics[width=0.8\textwidth]{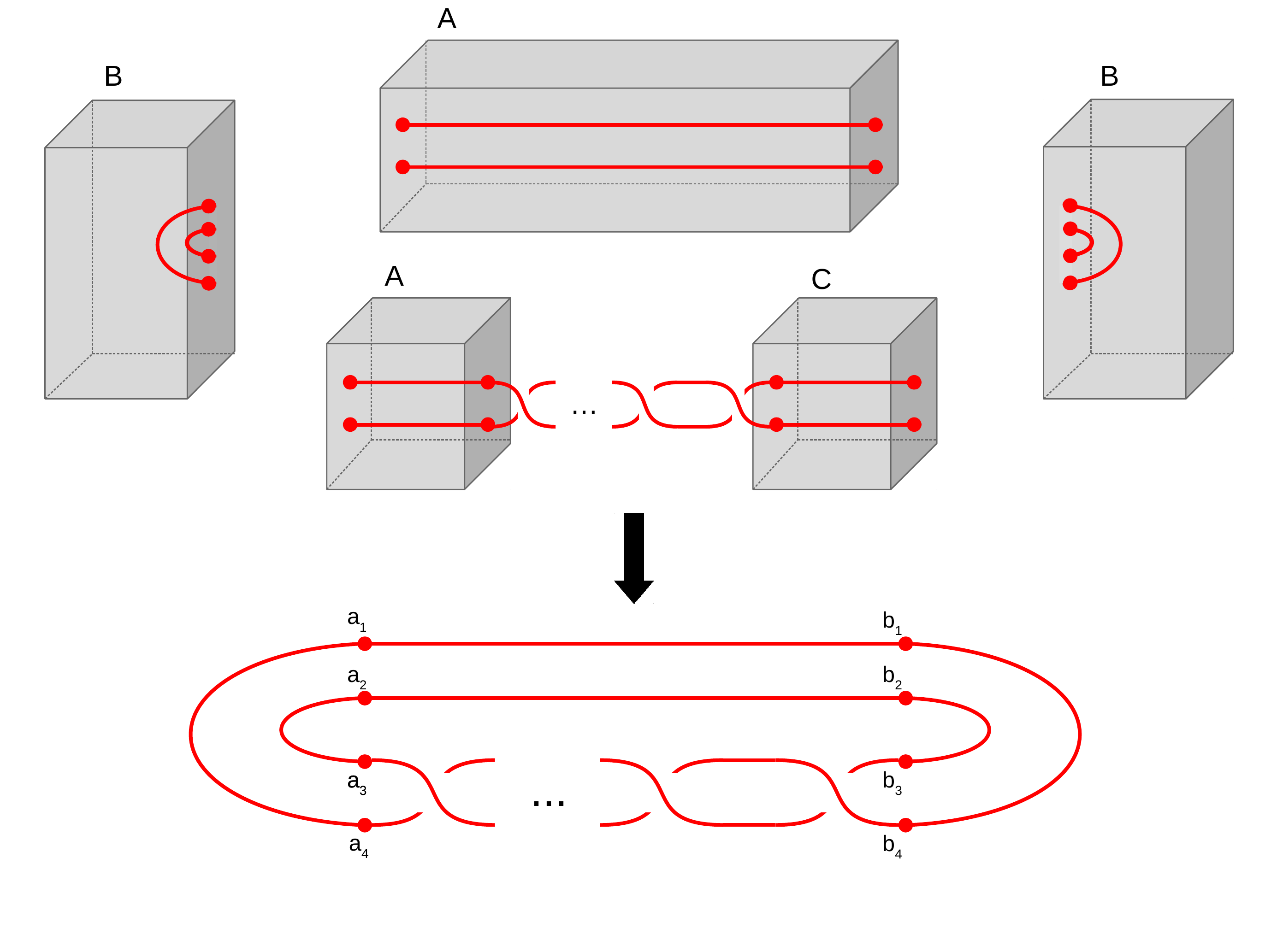}\caption{A specific construction of a ($2, n$) torus knot by joining boxes ${\bf A}$, ${\bf B}$ and ${\bf C}$ appropriately. 
The braid group action now acts $n$ times. The points ($a_i, b_i$) still remain the points where the Wilson lines end on the Heegaard-split manifolds. Once we extend the figure 
along ${\bf R}_+$ (or $\psi$) direction, we will get the configuration of the surface operator.}\label{fig10} \end{figure}

In the second formalism, we keep the gauge field as ${\cal A}_\mu$, but make $d_1$ itself complex.   
If we now map all the 
variables in the action \eqref{stotal} to the ones appearing in say \cite{wittenknots1} using ($\sigma_0, \gamma_0, \kappa_0$) etc in \eqref{bpsi3a} and \eqref{nunchat} respectively, then one can 
show that:
\bg\label{darlent}
d_1 ~\equiv ~ \pm {ic_{11}(b_2 + c_2)\over \sqrt{c_{11}^2 + v_3^2 q^2 ~ \sin^2\theta}}, \nd
where $c_{11}$ is given in \eqref{c11theta}, $v_3$ in \eqref{intig} and $q(\theta)$ in \eqref{chotas} with a NC deformation $\theta$. This definition of $d_1$ doesn't change if we change 
$\phi$, in the absence of a surface operator, to $\hat\phi$, in the presence of one. Additionally it is interesting to note that there are certain values of the NC parameter $\theta$ for which the 
definition of the boundary gauge field ${\cal A}_d$ simplifies to:
\bg\label{nungondho}
{\cal A}_d = {\cal A} \pm i \hat\phi. \nd
The simplest case is of course when $\theta$ vanishes. The other case may arise when $q(\theta)$, as defined in \eqref{qutta}, vanishes for non-zero $\theta$. Clearly for all these cases 
$c_2$ also vanishes, and $t$ becomes $t = \pm i$. However the boundary gauge theory coupling continues to remain non-zero and now takes the value $b_2$ as can be seen from \eqref{stwist}. 

Unfortunately, as it turns out, by doing similar mapping of our variables to the ones in \cite{wittenknots1} as discussed above, $b_2$ becomes infinite when $t$ approaches $\pm i$. In this limit, and
as elaborated in \cite{langland}, $\tau$ defined in \eqref{taude} becomes irrelevant and therfore is not an useful arena to study the boundary theory. Thus 
it seems we should only allow $t \ne \pm i$ cases, 
which then brings us to the question whether the simplification \eqref{nungondho} is any way useful for us. 

A path integral representation sheds some light here. Let us first discuss the non-abelian case in the absence of any knots. The path integral can be written as:
\bg\label{latecasin}
\int_{\cal C} {\cal D}{\cal A}_d~{\rm exp}\Bigg[i(b_2+c_2)\int_{\bf W}{\rm Tr}\left({\cal A}_d \wedge d{\cal A}_d + {2i\over 3} {\cal A}_d \wedge {\cal A}_d \wedge {\cal A}_d \right)\Bigg], \nd 
where ${\cal C}$ is the same integration cycle that we discussed earlier; and 
we see that \eqref{latecasin} only depends on the combination $b_2 + c_2$ but does not depend on the ratio ${d_1\over b_2 + c_2}$, which is another way of saying that ${\cal A}_d$ is a dummy variable 
in the integral \eqref{latecasin}. We can therefore replace ${\cal A}_d$ by {\it any} complex function and the definition \eqref{nungondho} would equally suffice if we view ${\cal A}$ and $\hat\phi$ to be
arbitrary functions appearing in the path integral. All in all, it boils down to the fact that the gauge field appearing in the path integral may be an arbitrary complex one-form, although the 
boundary action is defined with a specific functional form for ${\cal A}_d$. Even in the presence of a knot, for both abelian and non-abelian cases, the arguments presented above go through because 
the Wilson loop is defined with ${\cal A}_d$, and as such could again be replaced by an arbitrary complex one-form in the path integral.   
All these observations resonate well with the ones presented in sec (2.4) of \cite{wittenknots1}.   

\begin{figure}[t]\centering \includegraphics[width=0.8\textwidth]{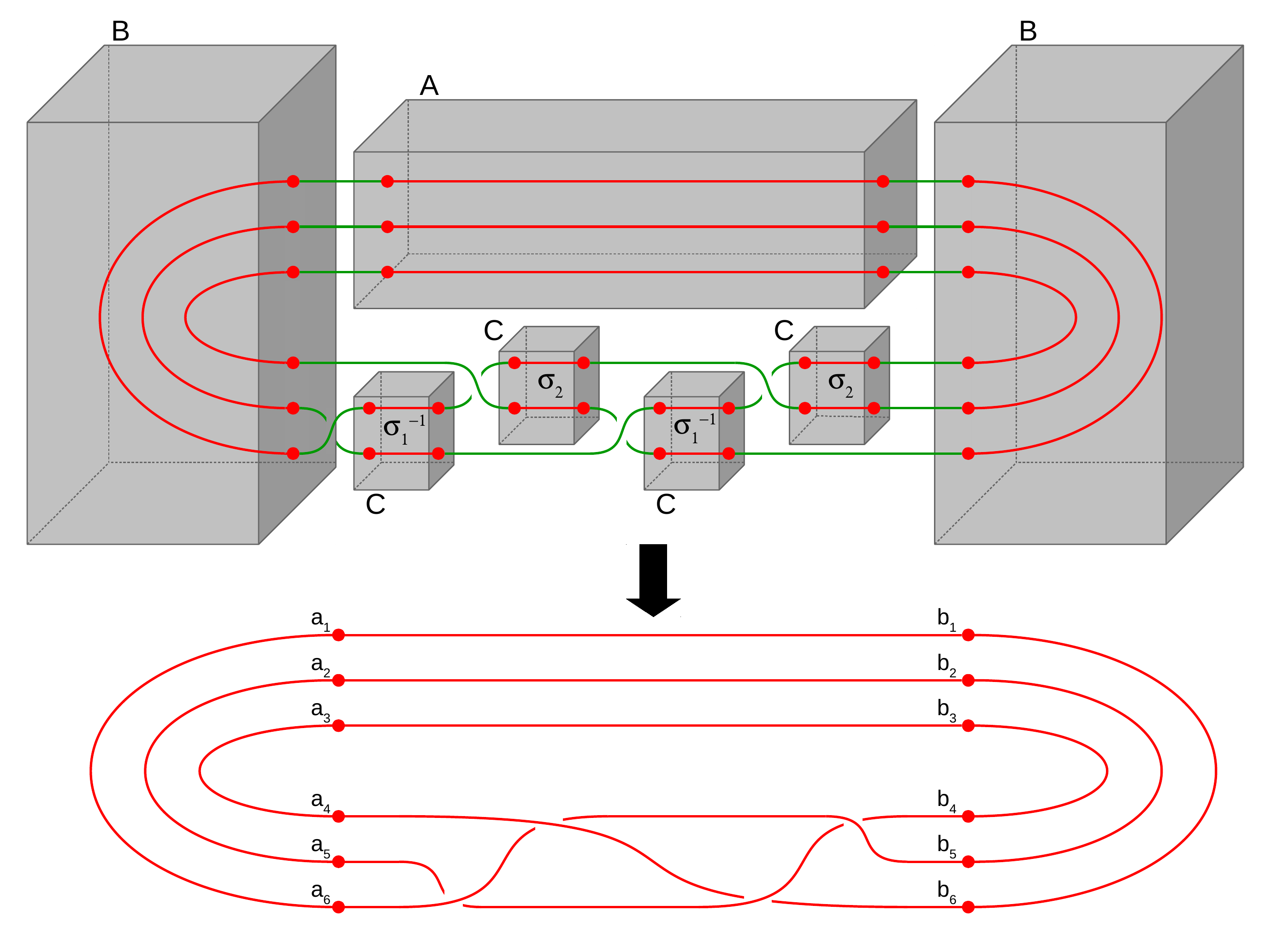}\caption{The construction of a figure 8 knot using ${\bf A}$, ${\bf B}$ and ${\bf C}$ boxes in a 
slightly different way than discussed earlier. The braid group action is now $\sigma_1^{-1}\cdot \sigma_2 \cdot \sigma_1^{-1} \cdot \sigma_2$ acting on the Wilson lines as shown.} \label{fig11} \end{figure}

Further simplification occurs when we look at the BHN equations \eqref{karfish} and \eqref{asnles} on a plane orthogonal to the surface operator. Since $\sigma$, as well as its 
covariant derivatives \eqref{arbitrage} and \eqref{chukmukh}, are localized functions we expect the behavior in a plane away from the center of the surface operator to be:
\bg\label{diamondQ}
{\cal F}_{1\psi} - i \gamma_4 \left[\varphi_1, \varphi_3\right] ~=~  {\cal F}_{2\psi} - i \gamma_4 \left[\varphi_2, \varphi_3\right] ~=~  
{\cal F}_{12} - i \gamma_7 \left[\varphi_1, \varphi_2\right] ~=~ 0, \nd
where $\gamma_7$ is the coefficient of the commutator piece in \eqref{asnles}. Note that we have expressed the BHN equations
without the hats, as the $\sigma$ dependences die off in the orthogonal plane. Converting the gauge fields from ${\cal A}_\mu$ to $-iA_\mu$ using \eqref{moyukh}, the Hitchin equation for ${\cal F}_{12}$ 
in \eqref{diamondQ} match with \eqref{6sense} as noted earlier. Of course the above discussion is good only for the configuration that we study in {\bf Table \ref{surfaceoper}} which is a 
co-dimension {\it one} singularity in the three-dimensional boundary. For a co-dimension {\it two} singularity in the boundary, we will have to study the Hitchin's equations in a plane orthogonal to the 
surface operator. The analysis would be similar to what we did above, although certain specific details might be different now.

The picture that we developed above
leads to the concept of {\it holonomy} of the complex gauge field around a given surface operator. This can typically be represented by $V_k \equiv {\rm Hol}({\cal A}_{d(k)})$ with 
$k$ representing the $k$-th surface operator. For flat connections holonomy and monodromy are related so $V_k$ will be conjugate to the monodromy ${\cal M}_k$ in \eqref{iamlegend}. An interesting 
consequence of having holonomy around a surface operator is the following. Consider four Wilson lines parallel to each other and intersecting at four-points on $\Sigma$ in a Heegaard split 
three-manifold. This is depicted to the left of {\bf Fig \ref{fig6}}, where the Wilson lines are parallel to $x_1$ direction. If we name them as 1, 2, 3 and 4, then by identifying the monodromies:
\bg\label{mormath}
{\cal M}_1 = {\cal M}_4^{-1}, ~~~~~~~~~ {\cal M}_2 = {\cal M}_3^{-1}, \nd
or equivalently the holonomies, we can go to the configuration depicted to the right of {\bf Fig \ref{fig6}}. This operation is useful because it tells us that we can join two Wilson lines 
by identifying monodromies. In terms of surface operators, this procedure will lead to the configuration depicted in {\bf Fig \ref{fig7}}. 
 
In fact we now have two distinct configurations of Wilson lines, or equivalently, surface operators. The first one, we will call it box {\bf A} and is depicted in {\bf Fig \ref{fig4}} (A), is a 
configuration of parallel surface operators. The second one, and we will call it box {\bf B}, is depicted to the right of {\bf Fig \ref{fig6}}: a configuration of curved surface operators. 
Associated to these boxes will be the operators ${\bf A}_k$ and ${\bf B}_k$ where $k$ denote the number of surface operators (or equivalently, Wilson lines).

\begin{figure}[t]\centering \includegraphics[width=0.8\textwidth]{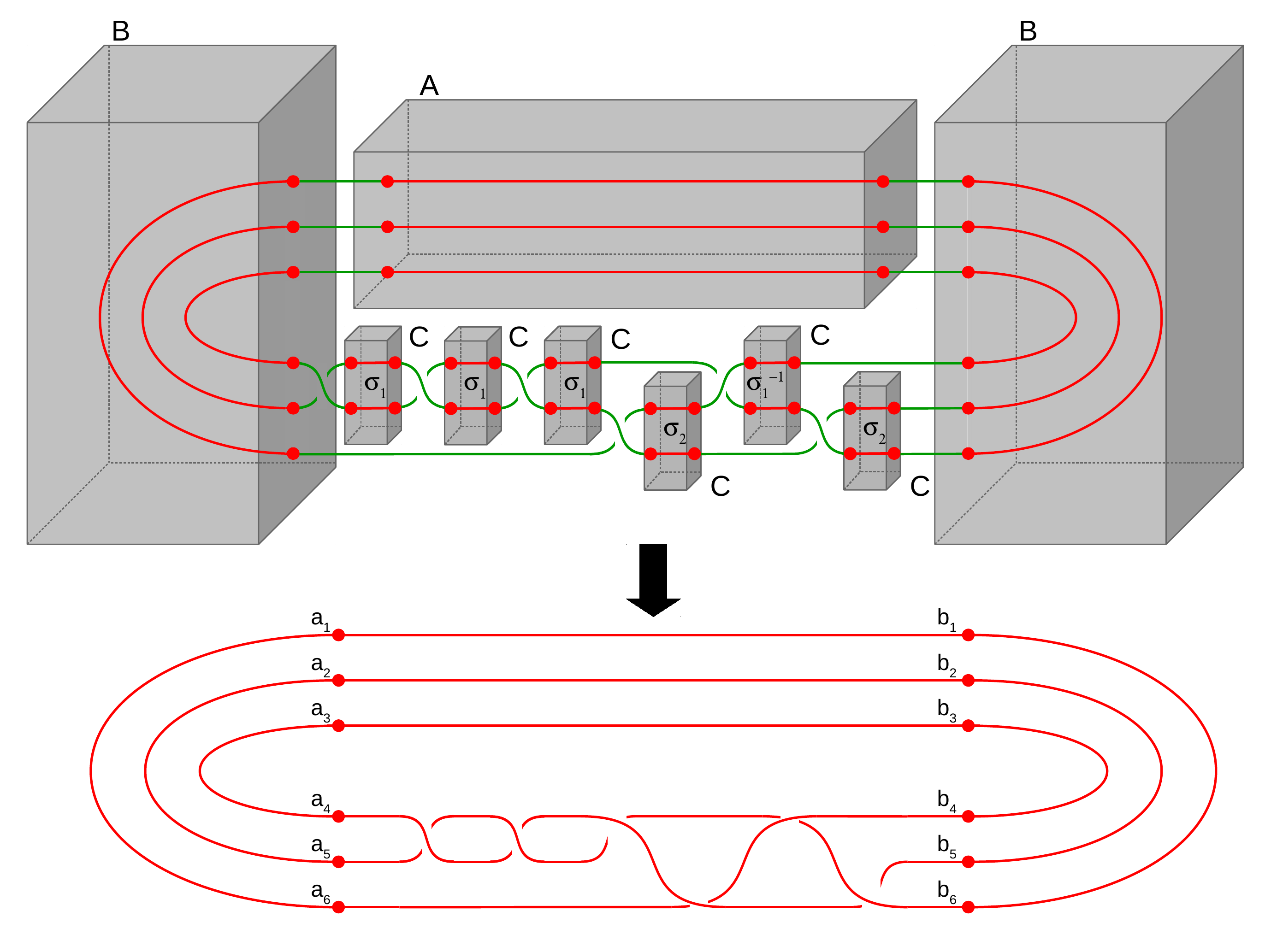}\caption{The construction of $5_2$ knot using the ${\bf A}$, ${\bf B}$ and ${\bf C}$ boxes. 
The braid group action is now $\sigma_1^{3}\cdot \sigma_2 \cdot \sigma_1^{-1} \cdot \sigma_2$ acting on the Wilson lines as shown.} \label{fig12} \end{figure}

There is a third possibility that we can entertain and is depicted in {\bf Fig \ref{fig4}} (C). We will call it box {\bf C}, where the Wilson lines are swapped by a braid group action 
$\sigma_\alpha$. We 
will concentrate on a braid group with two strands, with generators $\sigma_\alpha$ where the subscript $\alpha$ denote which set of two strands, out of a given set of Wilson lines, we choose here. 
The operators associated with the 
braid group action will be ${\bf C}_{(2, \sigma_1)}$ 
and ${\bf C}_{(2, \sigma^{-1}_1)}$ where we take $\alpha = 1$ for illustrative purpose and the two operations are depicted in 
{\bf Fig \ref{fig5}}. We therefore expect:
\bg\label{dechin}
{\bf C}_{(2, \sigma_1)}  {\bf C}_{(2, \sigma^{-1}_1)} ~ = ~ {\bf C}_{(2, 1)}, \nd
where $\sigma_1 = 1$ implies no braid group action. This is therefore topologically equivalent to ${\bf A}^\intercal_2$, with transpose put in to account for the orientations of the Wilson lines. 

We now have more or less all the necessary ingredients to analyze the invariants for various knots. 
Let us start with the simplest case of an unknot as depicted in {\bf Fig \ref{fig8}}. 
Combining the boxes ${\bf A}$, ${\bf B}$ and ${\bf C}$ we can express the invariant (or the linking number) in the following way:
\bg\label{unknot}
Z(q; ~{\bf K}_0)~ = ~ \langle {\rm exp}\oint_{{\bf K}_0}{\cal A}_d \rangle ~ = ~ \sum_{n_2}
\langle n_2 \vert{\bf B}_2^\intercal {\bf C}_{(2, \sigma_1)} {\bf B}_2 {\bf A}_2\vert n_2 \rangle, \nd  
where ${\cal A}_d$, as described above, could be any complex one-form; and ${\bf K}_0$ is the unknot configuration. The action of the operators in the RHS of \eqref{unknot} can be elaborated in the 
following way. Consider box ${\bf A}$ in {\bf Fig \ref{fig8}}
where the Wilson lines intersect the top right two-dimensional surface $\Sigma$ at two points. 
These two points may be considered as a given state $\vert n_2 \rangle$ in the boundary $U(1)$ Chern-Simons
theory. The operator ${\bf A}_2$ evolves the state from right to left (here we take the direction to be parallel to $x_1$, but this is not necessary). The subscript $2$ denotes two strands 
(or the two particle  
state $\vert n_2 \rangle$) in the field theory.
The operator ${\bf B}_2$ then curves the 
Wilson-line states by monodromy identification, much like \eqref{mormath} discussed above. This evolution continues till the braid group operation ${\bf C}_{(2, \sigma_1)}$ acts in the 
way depicted in {\bf Fig \ref{fig8}}. The braided state is then evolved by ${\bf B}_2^\intercal$ where the transpose operation just reverses the orientations of ${\bf B}_2$. Finally we  
sum over all possible two Wilson-line states in the Chern-Simons theory.  

The above, slightly unconventional way, reproduces the invaraint \eqref{ludivin2}
for the unknot case using the operators ${\bf A}_2$, ${\bf B}_2$ and ${\bf C}_{(2, \sigma_1)}$ combined as \eqref{unknot}. All the three operators can be thought of as a $2 \times 2$ matrices whose
components are evolution operators. As such they are expressible in terms of exponentials of generators integrated over the knot configuration, exactly as in \eqref{ludivin2}.  
This can be
normalized to 1, so one might wonder why we went about expressing the unknot in a rather complicated way. 
The answer is that the above way of expressing the unknot using the operators, help us to generalize the picture
to any complicated torus knots. For example, let us consider the trefoil depicted in {\bf Fig \ref{fig9}}, which again uses the three set of operators ${\bf A}_2$, ${\bf B}_2$ 
and ${\bf C}_{(2, \sigma_1)}$. The knot 
invariant associated with the trefoil then is:
\bg\label{trefoil}
Z(q; ~{\bf K}_t)~ = ~ \langle {\rm exp}\oint_{{\bf K}_t}{\cal A}_d \rangle ~ = ~ \sum_{n_2}
\langle n_2 \vert{\bf B}_2^\intercal {\bf C}_{(2, \sigma_1)} {\bf C}_{(2, \sigma_1)} {\bf C}_{(2, \sigma_1)} 
{\bf B}_2 {\bf A}_2\vert n_2 \rangle, \nd   
where ${\bf K}_t$ denotes the trefoil knot. The operators act in the same way as in \eqref{unknot}, except now we have three times the braid group action by the operator ${\bf C}_{(2, \sigma_1)}$. 
This of course distinguishes it from \eqref{unknot}, and
thus the above analysis generalizes easily to the torus knots ($2, n$) as:
\bg\label{torusknot} 
Z(q; ~{\bf K}_\intercal)~ = ~ \langle {\rm exp}\oint_{{\bf K}_\intercal}{\cal A}_d \rangle ~ = ~ \sum_{n_2}
\langle n_2 \vert{\bf B}_2^\intercal {\bf C}^n_{(2, \sigma_1)} {\bf B}_2 {\bf A}_2\vert n_2 \rangle, \nd   
with ${\bf K}_\intercal$ representing the torus ($2, n$) knots. Clearly when $n = 3$ we get our trefoil invariant. 

So far we have been using the operator ${\bf C}_{(2, \sigma_1)}$ 
to represent the braid group action for two Wilson lines. The question is what happens when we have more than two Wilson lines. It turns out we can still use ${\bf C}_{(2, \sigma_1)}$ but represent the
braid group action is a slightly different way. An example of this can be presented for the figure 8 knot, which is the simplest non-torus knot, 
given in {\bf Fig \ref{fig11}}. The knot invariant for this is now:
\bg\label{figure8}
Z(q; ~{\bf K}_8)~ = ~ \langle {\rm exp}\oint_{{\bf K}_8}{\cal A}_d \rangle ~ = ~ \sum_{n_3} \langle n_3 \vert{\bf B}_3^\intercal {\bf C}^{m_{{}_1}}_{(2, \sigma_2)}  
{\bf C}^{m_{{}_2}}_{(2, \sigma^{-1}_1)}   {\bf C}^{m_{{}_3}}_{(2, \sigma_2)}  
{\bf C}^{m_{{}_4}}_{(2, \sigma^{-1}_1)} {\bf B}_3 {\bf A}_3\vert n_3 \rangle, \nonumber\\ \nd   
where ${\bf K}_8$ is the figure 8 knot with $m_1 = m_2 = m_3 = m_4 = 1$; and 
${\bf C}_{(2, \sigma_2)}$ and ${\bf C}_{(2, \sigma^{-1}_1)}$ are the braid group actions ($\sigma_1, \sigma_2$) on two different strands.  The inverse to be understood as the operations depicted in 
{\bf Fig \ref{fig5}}. The rest of the operators act in the way we described earlier.
We can generalize \eqref{figure8} by considering arbitrary values for $n_i$ in \eqref{figure8}. One such 
generalization leads to the $5_2$ knot given in {\bf Fig \ref{fig12}}, whose knot invariant may be written as:  
\bg\label{52}
Z(q; ~{\bf K}_{5_{{}_2}})~ = ~ \langle {\rm exp}\oint_{{\bf K}_{5_{{}_2}}}{\cal A}_d\rangle ~ = ~ \sum_{n_3} \langle n_3 \vert{\bf B}_3^\intercal  
{\bf C}_{(2, \sigma_2)}  {\bf C}_{(2, \sigma^{-1}_1)}  {\bf C}_{(2, \sigma_2)}  {\bf C}^3_{(2, \sigma_1)} {\bf B}_3 {\bf A}_3\vert n_3 \rangle, \nonumber\\ \nd   
where ${\bf K}_{5_{{}_2}}$ is the $5_2$ knot, and we have taken ${\bf C}^3_{(2, \sigma_1)}$ instead of ${\bf C}^3_{(2, \sigma^{-1}_1)}$ action. 
We thus see that the three ingredients, namely (a) the Heegaard splittings, that typically lead to a class of operators 
${\bf A}_k$; (b) Monodromy identifications, that lead to a class of operators ${\bf B}_k$; and (c) braid group actions that lead to a class of operators ${\bf C}_{(2, \sigma_\alpha)}$ and 
${\bf C}_{(2, \sigma^{-1}_\alpha)}$ are sufficient to give us both the surface operator representations as well as the invariants for any given knots. All these are expressible in the 
language of a $U(1)$ Chern-Simons theory with a complex gauge group ${\cal A}_d$ and the invariants that we computed above are proportional to:
\bg\label{dhakka}
\exp \left({i\pi p~ Q_2^2\over k}\right), \nd
which are consistent with the generic argument that we presented for \eqref{ludivin2}. This is not a big surprise, and one might wonder if we can get anything {\it more} out of our elaborate 
constructions beyond the expected result \eqref{dhakka}.
The answer turns out to be affirmative and in fact accommodates 
the polynomial constructions outlined in \cite{surfknot} where the 
monodromies ${\cal M}_k$ in \eqref{iamlegend} are used to construct the variables ($\theta_i$, $x_i$) and the affine cubic $f(x_i, \theta_m) = 0$
 (see for example equations (4.6), (4.7) and (4.9) of \cite{surfknot}). This means the surface operator representations presented for various knot configurations above 
not only give us the knot invariants, 
but also reproduce all the constructions of \cite{surfknot}. Additionally,  
our analysis shows that all the constructions of \cite{surfknot} may be given a supergravity interpretation! 

However once we go to the 
non-abelian extension, we face many issues, and the simple minded analysis that we presented here will have to be modified. This means, for example, a surface operator representations of 
Jones polynomials using the three kinds of operators we used here are not sufficient.  A more detailed framework is then called for, which is unfortunately beyond the scope of the present work.
We will therefore not discuss this further, instead we will elaborate on another set of constructions that generalize easily to the non-abelian case.


\subsubsection{'t Hooft operator, opers and supergravity parameters \label{opar}}

In the previous section we have considered the co-dimension two defect operators in the field theory. The monodromy defect supported on a knot ${\bf K}$ inside the
Chern-Simons boundary was extended in four dimensions to a singularity that the fields had along a two dimensional surface ${\bf K} \times {\bf R}_+$ inside the four dimensional space.

In \cite{wittenknots1} and \cite{wittenknots2}, other defect operators were considered in a four dimensional theory, the co-dimension one Wilson line operators and the co-dimension three 
't Hooft operators. Especially important are the co-dimension three 't Hooft operators which can be related to the Nahm pole solution where the dependence of the co-dimension three object is only on
 $\psi$, the four dimensional
coordinate transversal to the three dimensional boundary. The relevant equations have already appeared in \eqref{bhnwitten}, which are of course the ones of \cite{wittenknots1}.  
Note that, compared to our earlier sections, nothing we say in this subsection will be new. However an attempt will be made to pave a way for possible connections between the results of 
\cite{wittenknots2} and our supergravity analysis.  

Let us first consider $t = 1$ case, where $t$ is given, in our language of supergravity, by \eqref{tkya}.
In this case, a stationary solution (invariant under translations along time direction) with zero $A_\mu$ and $\phi_3$ reduce to Nahm's equations for the components of the field 
$\phi$ tangent to the boundary\footnote{Our analysis here is generic and therefore $\phi_i$ and $\varphi_k$ can be related via any of the two mappings \eqref{4scalars} or \eqref{foursca}. In fact 
our gauge choice could also be generic i.e \eqref{gaugec} or \eqref{gaugec2}. Additionally we will be using the gauge fields $A_\mu$ instead of ${\cal A}_\mu$ so that we can easily compare our results 
to \cite{wittenknots2}.}:
\begin{equation}\label{gemma}
\frac{d\overrightarrow{\phi}}{dy} + \overrightarrow{\phi} \times \overrightarrow{\phi} = 0, 
\end{equation}
where we have identified $y$ as our $\psi$ coordinate. The above equation follows easily from our BHN equation \eqref{prolpart}, and also from \eqref{asnles} which is in the presence of a surface 
operator provided we change $\phi$ to $\hat\phi$. In the language of commutator brackets of \eqref{prolpart} or \eqref{asnles}, it is not too hard to guess the solution of the above equation to be: 
\begin{equation}\label{alicecreed}
\overrightarrow{\phi} = \frac{\overrightarrow{\tau}}{y}, 
\end{equation}
where $\tau_a$ are the three Pauli matrices. The advantage of expressing the equations in terms of three-dimensional vectors, before twisting, allows us to compare with the equations after twisting
when they all become one-forms.  

Once $A_\mu$'s are non-vanishing, the scenario is not so simple as the Nahm equation \eqref{gemma}. From our earlier analysis, we know that we need the full BHN equations. Of course, as expected, the 
solutions to the BHN equation again cannot be as simple as \eqref{alicecreed}. It turns out, there are  
two types of solutions to the BHN equations which may be succintly presented in terms of a complex coordinate $z$ defined as $z = x_1 + i x_2$ (recall that our four-dimensional space is parametrized 
by ($x_0, x_1, x_2, \psi$) where we already identified $y$ with $\psi$). The first type of solutions are independent of $z$ coordinate. Defining:
\bg\label{midnight}
&& D_1 \equiv  \frac{\partial}{\partial x_1} + i \frac{\partial}{\partial x_2} + \left[A_1+iA_2,~.\right] \nonumber\\
&&  D_2 \equiv \frac{\partial}{\partial y} + \left[A_y - i\phi_0,~.\right], ~~ D_3 \equiv  \left[\phi_1 - i \phi_2,~.\right], 
\nd
where $A_y \equiv A_\psi$; and 
as mentioned earlier, depending on the mapping \eqref{4scalars} or \eqref{foursca}, we can identify $\phi_0$ to either ${\cal A}_3$ or $\varphi_3$ respectively. This means, for certain choice of
the gauge (\eqref{gaugec} or \eqref{gaugec2}), $\phi_0$ may vanish and therefore $D_2$ described above may be simplified. However for the present discussion, we will keep things generic.  
The first order differential operators $D_i$ therefore satisfy:
\begin{equation}\label{lajosh}
\left[D_i, D_j\right] = 0,~~i,j = 1, 2, 3;~~~~~~~~~~~\sum_{i=1}^3 \left[D_i, D^{\dagger}_{i}\right] = 0, 
\end{equation}
which are alternative ways to express the BHN equation \eqref{binu} or \eqref{prolpart} once we absorb some factors and signs appropriately. To verify that this is indeed, for example, \eqref{binu} 
we note that the second equation in \eqref{lajosh} is the moment map equation which can be written as:
\begin{equation}
F_{12} - \left[\phi_1, \phi_2\right] - D_{y} \phi_0 = 0. 
\end{equation}
In the gauge $A_1 + i A_2 = 0$, the operator $D_1$ becomes derivative with respect to
$\bar{z}$ and with the gauge choice $A_y = i\phi_0$, the operator $D_2$ becomes derivative with respect to $y$. $D_3$ is proportional to $\phi \equiv \phi_1 - i \phi_2$ as should be 
clear from \eqref{midnight}.
The commutation relation $[D_1, D_3] = 0$  implies that $\phi$ is holomorphic in $z$ and the relation $[D_2, D_3] = 0$ means that $\phi$ is independent of $y$. Near $y=0$, $\phi$ is a constant and a 
complex valued gauge transformation maps it into the Nahm pole solution with $\frac{1}{y}$ dependence.

What about outside the region $y = 0$?  The vanishing of the commutator brackets
$[D_1, D_3]$ and $[D_2, D_3]$ define a Higgs bundle $(E, \phi)$ where $\phi$ is independent of $y$ and holomorphic. The Nahm pole solution (also called the {\it model solution}) 
is trusted around the $y = 0$ boundary but we can extend the model solution as a Higgs bundle $(E, \phi)$ away from $y=0$. In fact, as described in \cite{wittenknots2}, such extension gives a 
Higgs bundle $(E, \phi)$ endowed with a holomorphic line sub-bundle $L$ which is not stabilised by $\phi$. In other words, for any section $s$ of $L$ we expect
$s \wedge \phi s \ne 0$, as described in \cite{wittenknots2}.
 
Let us now consider the second type of solutions that depend on $z$.
The dependence on $z$ is determined by the presence of extra monopoles with extra charges
$k_a$ at points $z = z_a$. Next to $y = 0$, the solution is a simple modification of the Nahm pole solution as
the field $\phi$ has a holomorphic entry with a power of $z$. Away from $y = 0$ the  solution is given again by a triplet $(E, \phi, L)$ of a Higgs bundle with a holomorphic sub-bundle $L$.

How do we now extend this result to the case $t \ne 1$?  A key observation of \cite{wittenknots2} is that the  Higgs bundle (with the key ingredient of a holomorphic scalar field $\phi$) can be obtained 
by starting from a set of Hitchin equations:
\begin{equation}\label{stephk}
F - \phi \wedge \phi = 0;~~~~d *\phi = d \phi = 0, 
\end{equation}
and combining the last two equations to get the holomorphicity condition on $\phi$, namely
$\bar{\partial} \phi = 0$. This is true for $t = 1$. When $t \ne 1$, it is useful to modify the definition of the derivatives with respect to $z, \bar{z}$ by introducing a complex parameter 
$\zeta$ in the following way:
\begin{equation}\label{kege}
D_{z}^{\zeta} = \frac{D}{D z} - \zeta^{-1} \left[\phi,~.\right], ~~~~~D_{\bar{z}}^{\zeta} = \frac{D}{D \bar{z}} + \zeta \left[\bar{\phi},~.\right].
\end{equation}
We have $[D_z^{\zeta}, D_{\bar{z}}^{\zeta}]=0$ which is taken as an equation governing holomorphic data. In fact using 
vector field components $A_{z}^{\zeta} = A_z - \zeta^{-1} \phi$ and $A_{\bar{z}}^{\zeta} = A_{\bar{z}} + \zeta \bar{\phi}$ makes \eqref{kege} 
holomorphic in these variables. Additionally, the holomorphicity condition on $\phi$ is mapped into a holomorphicity condition on $A_z^\zeta$
and the Higgs bundle condition is now replaced by a complex flat connection.  The Nahm pole solution around $y = 0$ now describes a singularity in $A_{z}^{\zeta}$ and $A_y$. 
Away from $y \rightarrow 0$
region, the solution is a complex flat bundle $E$ with a holomorphic bundle $L$ defined such that its holomorphic sections are not annihilated by $D_{z}$. 
Such a pair $(E, L)$ is called an {\it oper} \cite{wittenknots2}.

In the Appendix A of  \cite{wittenknots2}, the reduction of a  four dimensional stationary solution to a topological
theory in three dimensions was a function of a rotational angle $\theta$ where the parameter
$t$ was set to $\tan\left({3 \theta\over 2} + {\pi\over 4}\right)$ and $\zeta$ to $\tan~\theta$. This relation between
$t$ and $\zeta$  should also appear from our M-theory reduction. As $t$ is related to the supergravity parameters via \eqref{tkya}, we expect
$\zeta$ to also be represented by our supergravity parameter. From here we conclude  
that the oper solution is automatically fixed once we have determined the supergravity parameters. This is somewhat along the lines of the discussion 
in the previous subsections where we saw that many of the results discussed in \cite{wittenknots1} automatically appear from our supergravity analysis. 
More details on this will be presented in the sequel to this paper.















\section{Model B: The type IIB dual description and non-K$\ddot{a}$hler resolved cone  \label{nonkah}}

In section \ref{nonkah2} most of our analysis revolved around the uplift of the brane configuration given in {\bf Table \ref{wittenbranes}} to M-theory, and the subsequent physics associated to the 
presence of a knot in $2+1$ dimensional boundary ${\bf W}$. The existence of a Coulomb branch, as well as dipole (or RR) deformation, helped us to study the knots and their localization 
to the boundary ${\bf W}$. Many of the details, that were studied exclusively from the boundary point of view in \cite{wittenknots1}, appeared very naturally in our set-up from the bulk 
dynamics in M-theory. The starting point 
of all our discussion was the Hamiltonian \eqref{hamilbeta} from where, and in the presence of surface operators, we were led to the detailed study of knots and knots invariants.

At this stage it is interesting to ask if we can repeat the success using the second brane configuration given in {\bf Table \ref{modbranes}}. One immediate difference from the 
earlier brane configuration in {\bf Table \ref{wittenbranes}} (or its T-dual type IIA version) is the {\it absence} of the Coulomb branch. Recall that the existence of the Coulomb branch earlier was
responsible in constructing the twisted gauge field ${\cal A}_d$ in \eqref{newgf} which eventually led us to the boundary Chern-Simons theory \eqref{plaza}. Once we lose the Coulomb branch, 
restricting the knot to the three-dimensional boundary ${\bf W}$ is more subtle. In fact the whole boundary picture developed from four-dimensional space 
${\bf V} = {\bf W} \times {\bf R}^+$ {\it a la} \cite{wittenknots1} will need to be re-interpreted differently now. Problems lie in restricting the knots to three-dimensions, constructing the 
twisted gauge field and resolving the conundrum addressed earlier in section \ref{TRC}.     

We will start by discussing, in series of steps, a 
way out of the conundrum for {\bf Model B} by analyzing the picture from M-theory in a slightly different way from what is discussed in section 5 of \cite{wittenknots1}. 
In the process we will get some understanding how to address the other two issues namely,
restricting knots to 3d and topological twisting of the scalar fields. But we make only the barest beginnings in this direction, and leave most of the details for the sequel. 

\subsection{Second look at the gravity and the topological gauge theory \label{second}} 

We saw, from our earlier discussion in section \ref{TRC}, that an appropriate 
duality to the brane configuration of {\bf Model B}
leads to a type IIB picture with wrapped D5-branes on the two-cycle of a resolved conifold. According to \cite{DEM} the metric on the
resolved conifold should be non-K\"ahler. Ignoring the dipole deformation for the time being (we will insert this soon), the supergravity background for the configuration is 
given by \eqref{sugra1} as before
with $\phi$ being the dilaton and the Hodge star and the fundamental form $J$ are wrt to the dilaton deformed metric $e^{2\phi} ds_2^6$. The metric $ds_6^2$ is now different from 
\eqref{chote} as its a non-K\"ahler
resolved conifold metric written as:
\bg\label{ds6}
ds^2_6 = F_1~ dr^2 + F_2 (d\psi + {\rm cos}~\theta_1 d\phi_1 + {\rm cos}~\theta_2 d\phi_2)^2  + \sum_{i = 1}^2 F_{2+i}
(d\theta_i^2 + {\rm sin}^2\theta_i d\phi_i^2), \nonumber\\
\nd
where $F_i(r)$ are warp factors that are functions of the radial coordinate $r$ only\footnote{One may generalize this to make the warp factors $F_i$ functions of 
($r, \theta_1, \theta_2$) but we will not do so here.}. The above background \eqref{ds6} can be easily converted to a background with both ${\cal H}_3$ and ${\cal F}_3$ fluxes by
a series of duality specified in \cite{MM, DEM}. The duality converts \eqref{ds6} to:
\bg\label{iibform}
&&ds^2 = {1\over e^{2\phi/3} \sqrt{e^{2\phi/3} + \Delta}} ~ds^2_{0123} + e^{2\phi/3} \sqrt{e^{2\phi/3} + \Delta} ~ds^2_6 \\
&& {\cal F}_3 = - e^{2\phi} {\rm cosh}~\beta \sqrt{F_2\over F_1}\left(g_1 ~e_\psi \wedge e_{\theta_1} \wedge e_{\phi_1}  +
g_2~e_\psi \wedge e_{\theta_2} \wedge e_{\phi_2}\right)\nonumber\\ 
&& {\widetilde{\cal F}}_5 = -{\rm sinh}~\beta~{\rm cosh}~\beta\left(1 + \ast_{10}\right) {\cal C}_5(r)~d\psi \wedge \prod_{i=1}^2~\sin~\theta_i~ d\theta_i \wedge d\phi_i\nonumber\\
&& {\cal H}_3 = {\rm sinh}~\beta \Big[\left(\sqrt{F_1 F_2} - F_{3r}\right) e_r \wedge e_{\theta_1} \wedge e_{\phi_1} 
+ \left(\sqrt{F_1 F_2} - F_{4r}\right) e_r \wedge e_{\theta_2} \wedge e_{\phi_2}\Big] \nonumber \nd
with a dilaton $e^{\phi_B} = e^{-\phi}$ and a $\Delta$ defined as:
\bg\label{linfri}
\Delta = {\rm sinh}^2\beta \left(e^{2\phi/3} - e^{-4\phi/3}\right)\nd
and $\beta$ is a parameter related to certain boost that is explained in \cite{DEM} while
the others, namely ($g_1, g_2, {\cal C}_5$)  are given by:
\bg\label{jmar}
&&g_1(r) = F_3\left(\sqrt{F_1 F_2} - F_{4r} \over F_4\right), ~~~g_2(r) = F_4\left(\sqrt{F_1 F_2} - F_{3r} \over F_3\right)\\ 
&&{\cal C}_5(r) = \int^r~{e^{2\phi} F_3 F_4 \sqrt{F_1 F_2}\over F_1}\left[\left({\sqrt{F_1F_2} - F_{3r}\over F_3}\right)^2 
+ \left({\sqrt{F_1F_2} - F_{4r}\over F_4}\right)^2\right] dr .\nonumber\nd

\subsubsection{Revisiting the topologically twisted theory \label{toptwist}}

Before moving further, let us ask how does finding the type IIB background \eqref{ds6} and \eqref{iibform}
helps us in understanding the topologically twisted theory. Recall what we did in section \ref{nonkah2}. We mapped the
type IIB brane configuration of {\bf Table \ref{wittenbranes}} to a configuration of wrapped D5-${\overline{\rm D5}}$ branes on two-cycle of a warped Taub-NUT space. An M-theory uplift then 
gave us the required action \eqref{stotal} and the Hamiltonian \eqref{hamilbeta} from where we extracted our boundary three-dimensional Chern-Simons action \eqref{plaza}.  

The situation now is a bit different as has been hinted above. The Ooguri-Vafa model \cite{OV} has two different realizations that are connected via large $N$ dualities.
On one hand the $SU(N)$  
Chern-Simons theory is defined on ${\bf S}_{(2)}^3$, the subscript 2 is for later convenience,  
with the {\it dual} closed topological string theory of $A$-type defined on the ${\bf S}^2$ blown-up of a conifold geometry (i.e on a 
resolved conifold). On the other hand, we have $N$ D6-branes wrapped on the ${\bf S}_{(2)}^3$ 
of a deformed conifold giving us ${\cal N} = 1$ SYM theory in four spacetime dimensions that is {\it dual} 
to closed type IIA string theory on a resolved conifold with fluxes and no branes.

\begin{figure}[t]\centering \includegraphics[width=0.8\textwidth]{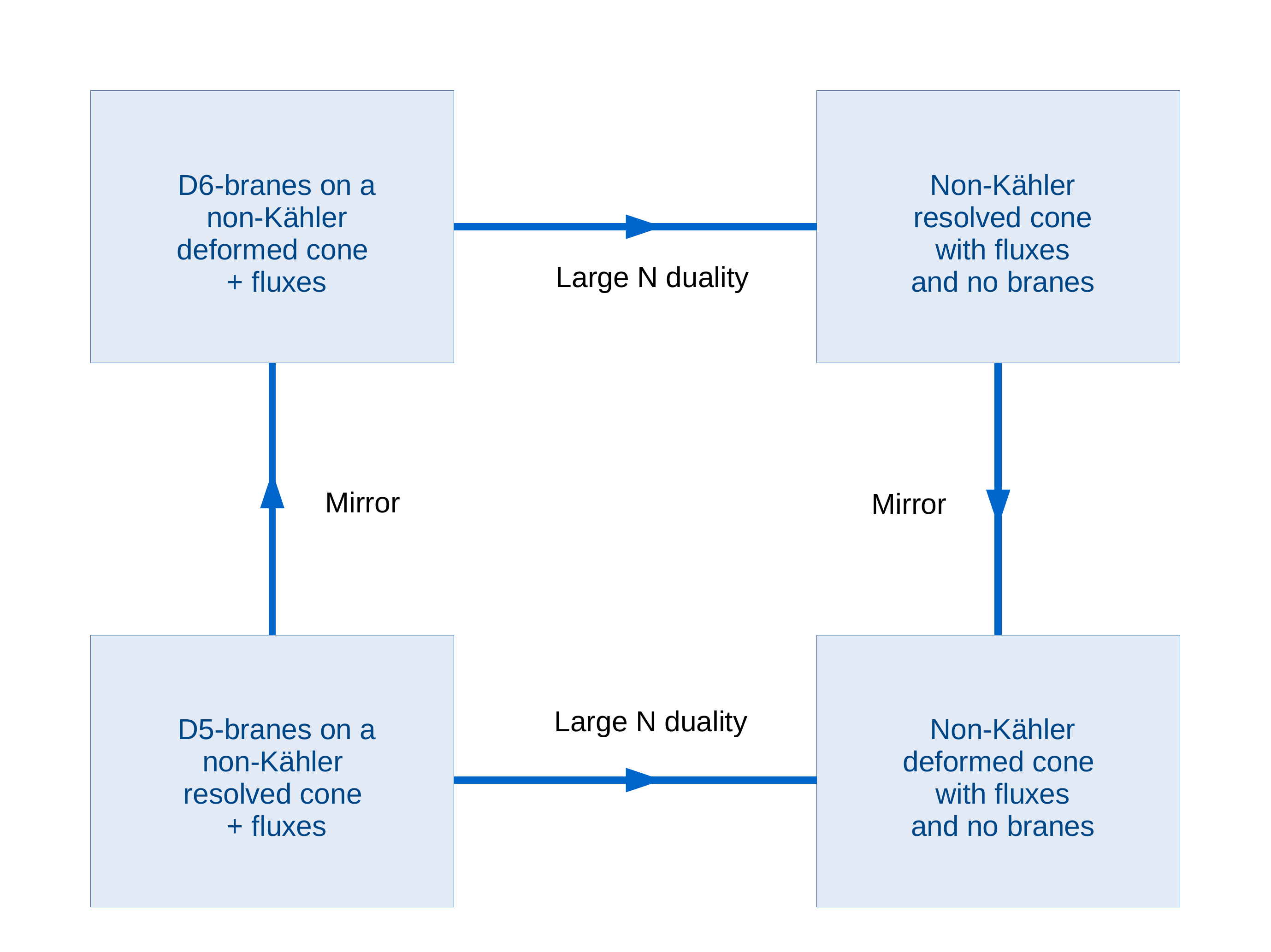}\caption{The web of dualities that connect various configurations in type IIB and type IIA theories. Here we will concentrate 
mostly on the lower left hand box that captures the physics of D5-branes wrapped on the two-cycle of a non-K\"ahler resolved conifold.}
\label{fig13} \end{figure}

There appears to be some mismatch between the locations of four-dimensional gauge theory and the three-dimensional Chern-Simons theory. The four-dimensional ${\cal N} = 1$ gauge theory is 
defined along the space-time directions ($x_0, x_1, x_2, x_3$). Although this is slightly different from our earlier case, where the four-dimensional gauge theory was located along 
($x_0, x_1, x_2, \psi$) directions, it is nevertheless consistent with both the brane configurations in {\bf Table \ref{modbranes}} as well as the configuration after a duality to  
a non-K\"ahler resolved conifold with wrapped D5-branes. However what is {\it different} now is the location of the Chern-Simons theory. Previously the Chern-Simons theory was localized to the 
boundary ${\bf W}$ of the four-dimensional space. For the present case the Chern-Simons theory is most succinctly described on the three-cycle ${\bf S}_{(2)}^3$ of a deformed conifold got by taking the 
mirror of the resolved conifold picture with wrapped D5-branes.   

This apparent mismatch of the location of the Chern-Simons theory is not just a relocalization of the topological theory, but lies at the heart of the problem.
To see this, 
first note that the partition function of the 
Chern-Simons theory on ${\bf S}_{(2)}^3$ in the large $N$ limit, takes the following form \cite{GoV}:
\bg\label{kothabaaz}
Z\left[{\bf S}_{(2)}^3\right] = {\rm exp}\left[-\sum_{g = 0}^\infty \lambda^{2g-2} F_g(t)\right], \nd
where $\lambda$ is the string coupling and $t = i\lambda N$ is the K\"ahler modulus of the blown-up ${\bf S}^2$ of a resolved conifold. This resolved conifold is not the same one studied in
\eqref{ds6} above. Rather it is the one that appears to the top right of {\bf Fig \ref{fig13}}. The factor $g$ in \eqref{kothabaaz} is the genus $g$ of Riemann surfaces that parametrize the moduli 
space ${\cal M}_g$ with Euler characteristics $\chi_g$. Together they can be used to define $F_g(t)$, for $g \ge 2$, appearing in \eqref{kothabaaz} as (see \cite{rahul}, and \cite{OV} for details):
\bg\label{peterpa}
F_g(t) ~ \equiv~ \int_{{\cal M}_g} c_{g-1}^3 - {\chi_g\over (2g-3)!} \sum_{n = 1}^\infty n^{2g-3} e^{-nt}, \nd
where the first term denotes the Chern class of the Hodge bundle over the moduli space ${\cal M}_g$, derived in \cite{rahul}. As noted in \cite{GoV}, \eqref{peterpa} is very suggestive of a $g$-loop
topological string amplitude. 
 
Secondly, there are two different ways we can study knots here as mentioned above. 
The first is with intersecting 
D4-branes where a set of $N$ D4-branes wrap ${\bf S}_{(2)}^3 \times {\bf R}^2$ and another set of $M$ D4-branes intersect the first set on the knot ${\bf K}$ and are stretched along 
the remaining directions ${\bf R}^2 \times {\bf D}^2$, where ${\bf D}^2$ is a two-dimensional subspace in $T^\ast {\bf S}_{(2)}^3$. The second is with $N$ D6-branes 
wrapping ${\bf R}^4 \times {\bf S}_{(2)}^3$. 
 Once we go to Euclidean space, the knots appearing on ${\bf S}_{(2)}^3$ may be constructed using D2- or D4-branes intersecting the D6-branes on ${\bf K}$.
Clearly it is the second case that is more relevant to us because the brane configuration given in {\bf Table \ref{modbranes}} take us directly to this set-up via a series of T and 
SYZ \cite{SYZ, syz2} dualities as shown in {\bf fig \ref{fig13}}, at least in the absence of knots. Knots can then be inserted in the type IIA picture by surface operators\footnote{In section 5 of 
\cite{wittenknots1} the Ooguri-Vafa \cite{OV} model with intersecting D4-branes is derived using a different route. 
The D4-branes are oriented in a way that the four-dimensional gauge theory and the three-dimensional Chern-Simons theory have similar representations as before. 
We thank Johannes Walcher for explaining the construction to us \cite{walcher}.}. 

It turns out, for the case that we are most interested in, the topological string amplitude $F_{g, h}$ with $g = 0$ and $h = 1$ computes the {\it superpotential} terms for 
the ${\cal N} = 1$ theories in four-dimensions. The superpotential terms are in general harder to compute in type IIA language, but become easier in the mirror type IIB language. The 
mirror is of course our configuration of D5-branes wrapped on the two-cycle of a non-K\"ahler resolved conifold, bringing us back to the analysis performed in section \ref{second}.  

The above discussion should hopefully suggest the usefulness of the type IIB analysis. However we haven't yet reconconciled with all the steps of our earlier analysis performed in section 
\ref{nonkah2}. For example, if we want to localize the knots to the three-cycle ${\bf S}_{(2)}^3$ of the deformed conifold, what is the usefulness of the boundary ${\bf W}$ used earlier?   

The answer can be given in a few steps. First, 
let us go back to the type IIB D5-$\overline{\rm D5}$ branes wrapped on the two-cycle of our Taub-NUT space discussed in section \ref{flook1}. We can move the 
$\overline{\rm D5}$-branes away on the Coulomb branch so that we are left with only D5-branes wrapped on the two-cycle of the Taub-NUT space. The geometry is discussed in \eqref{chote} before. 
To go from this geometry to the one studied above in \eqref{ds6}, we will assume that our Taub-NUT space is fibered over a ${\bf P}^1$, in other words, a resolved conifold geometry may be 
viewed as a Taub-NUT space fibered over a ${\bf P}^1$.  
The precise relationship between the two geometries is 
studied in section 3.1 of \cite{lapan} (see equations (3.10) to (3.13) in \cite{lapan}). The only difference\footnote{The discussion in \cite{lapan} is for a resolved conifold with a Calabi-Yau metric on 
it. It can be easily generalized for a resolved conifold with a non-K\"ahler metric on it.} 
here is now that the two-cycle, on which we have our wrapped D5-branes, should be along 
($\theta_1, \phi_1$). This is of course just a renaming of coordinates from section \ref{nonkah2}. The fibration breaks the four-dimensional 
supersymmetry down to ${\cal N} =1$, but for the time being we will 
not be too concerned with the supersymmetry.  
The above manipulation tells us how we can channel our earlier calculations for the new set-up. Locally, at every point on the base ${\bf P}^1$, parametrized by ($\theta_2, \phi_2$), we have 
D-branes wrapped on the two-cycle of a Taub-NUT space. 

Secondly, we go to Euclidean space and assume that the spacetime directions with Minkowskian coordinates ($x_0, x_1, x_2$) are now on an Euclidean ${\bf S}_{(1)}^3$. Thus the four-dimensional space 
${\bf V} = {\bf W} \times {\bf R}^+$ previously, now becomes ${\bf S}_{(1)}^3$ that represents ${\bf W}$ and the half coordinate $x_3$ that parametrizes ${\bf R}^+$. Further,  
the four-dimensional theory that we have on ${\bf S}_{(1)}^3\times {\bf R}^+$ can also be got from the mirror construction of D6-branes wrapped on three-cycle of a non-K\"ahler deformed conifold. 
Since they are connected by SYZ transformations \cite{SYZ, syz2}, the theories on ${\bf V}$, and therefore also on ${\bf W}$, are identical.   

The above discussions suggest that we can perform similar computations in type IIB theory as in section \ref{nonkah2}, 
but now appropriately modified to incorporate D5-branes wrapped on two-cycle of a non-K\"ahler resolved conifold.
This is easier than the mirror computations with D6-branes, and one may now insert the knots using surface operators on ${\bf S}^3$. Since the mirror picture is identical, we can view the 
theory on ${\bf S}_{(1)}^3$, got from our IIB computations, to be exactly the same in the type IIA side. 

In the type IIA side, as shown on {\bf fig \ref{fig13}}, the D6-branes are wrapped on the three-cycle ${\bf S}_{(2)}^3$ of a non-K\"ahler deformed conifold. In fact the world-volume of the D6-branes
is oriented along ${\cal M}_7$ where:
\bg\label{whistle}
{\cal M}_7 ~ \equiv ~ {\bf S}_{(1)}^3 ~ \times ~ {\bf S}_{(2)}^3 ~ \times ~ {\bf R}^+, \nd
and the physics on the first three-cycle ${\bf S}_{(1)}^3$ is directly imported from our type IIB analysis. Since the deformed conifold is non-compact, Gauss' law is not violated and the wrapped 
D6-branes continue to be a valid supergravity solution there. We can now perform the following {\it flop} operation:
\bg\label{sebeast}
{\bf S}_{(1)}^3   ~ \leftrightarrow ~ {\bf S}_{(2)}^3, \nd
transferring all the physics on ${\bf S}_{(1)}^3$ to the three-cycle of the non-K\"ahler deformed conifold\footnote{One may also look up section 5.5 of \cite{lapan} where somewhat similar kind of 
flop operation is discussed. Note that D6-branes continue to remain D6-branes under the flop operation \eqref{sebeast}
because the flop is performed inside the manifold ${\cal M}_7$ given in \eqref{whistle}.}. 
This is exactly the D6-brane realization of the Ooguri-Vafa \cite{OV} model! Our construction 
differs from the intersecting D4-branes' realization of the Ooguri-Vafa model in \cite{wittenknots1, walcher}. 

The above discussions suggest the power of the IIB analysis: we can continue working on the type IIB side, albeit with a different background, and perform similar manipulations as in 
section \ref{nonkah2}. Of course subtleties appear because of the underlying supersymmetry, twisting etc, but presumably none too unsurmountable. Remarkably,
once we have the full IIB analysis at hand, we can {\it transfer} the physics to the 
type IIA side by a mirror transformation followed by a flop operation \eqref{sebeast} giving us the full realization of the Ooguri-Vafa \cite{OV} model. Therefore
in the following we will elaborate on the type IIB side, by analyzing the background with and 
without dipole deformation and then discuss how to extract the four-dimensional physics similar to what we did in section \ref{nonkah2}. Most of the other details regarding the subtleties 
coming from reduced supersymmetry, twisting and the exact boundary theory on ${\bf S}_{(1)}^3$; including the type IIA mirror and the flop operation \eqref{sebeast} will only be briefly touched upon here, 
and detailed elaborations will be relegated to the sequel.

\subsection{Five branes on a resolved conifold: Exact results \label{5brane}}

Let us now consider specific choices of the warp factors $F_i(r)$ that would not only solve the EOMs but also preserve supersymmetry. One solution that was briefly mentioned 
in \cite{DEM} and studied in some details in \cite{lapan} is:
\bg\label{suramon} 
F_1 = {e^{-\phi}\over 2F}, ~~~~~ F_2 = {r^2 e^{-\phi} F\over 2}, ~~~~~F_3 = {r^2 e^{-\phi}\over 4} + a^2(r), ~~~~~ 
F_4 = {r^2 e^{-\phi}\over 4}, \nd
where $a^2 \equiv a_0^2 + a_1(r)$ and $a_0^2$ 
is the resolution parameter, $F(r)$ is some function of $r$ whose value will determined soon and $\phi$, as usual, is related to the type IIB dilaton. 
The function $F(r)$ has to be related to the dilaton $\phi(r)$ because any arbitrary 
choice of $F$ and $\phi$ will break supersymmetry. We will determine the equation relating $F$ and $\phi$ using torsion classes \cite{chiossi, luest, louis, gauntlett}. 
In the process we will also argue for supersymmetry. 

\subsubsection{Analysis of the background fluxes \label{5brane1}}

Before we go about determing the functional form for $r$, let us work out the three-form fluxes from \eqref{iibform}. Plugging \eqref{suramon} into \eqref{iibform}, they are 
given by:
\bg\label{tota}
&&{\cal H}_3 = {1\over 4} {\rm sinh}~\beta~e^{-\phi}~r^2 \left[\left(\phi_r - {8a\over r^2} e^{\phi} a_r\right) e_r \wedge e_{\theta_1} \wedge e_{\phi_1} + 
\phi_r~e_r \wedge e_{\theta_2} \wedge e_{\phi_2}\right]\\
&& {\cal F}_3 = -{1\over 4} {\rm cosh}~\beta~e^{\phi}~r^3 F(r)\left[\left(1 + {4a^2\over r^2} e^{\phi}\right) \phi_r~e_\psi \wedge e_{\theta_1} \wedge e_{\phi_1} + 
\left({r^2\phi_r - 8 a a_r e^{\phi}\over r^2 + 4a^2 e^{\phi}}\right)e_\psi \wedge e_{\theta_2} \wedge e_{\phi_2}\right]. \nonumber \nd
Looking carefully at the three-forms we see that ${\cal H}_3$ is closed but ${\cal F}_3$ is not. This is good because non-closure of ${\cal F}_3$ is related to the wrapped 
five-brane sources. Recall that the five-branes are wrapped on the two-cycle ($\theta_1, \phi_1$) and stretched along the space-time directions $x_{0, 1, 2, 3}$, which will be later 
converted to Euclidean ${\bf S}^3_{(1)} \times {\bf R}^+$. This means the
source equation should have delta function like singularity along the orthogonal directions of the brane, namely the ($\theta_2, \phi_2, \psi$) and the radial direction $r$. In
the limit when both $a^2$ as well as $a_r$ are smaller than some chosen scale in the theory, ${\cal F}_3$ can be expressed in the following suggestive way:
\bg\label{calf3}
{\cal F}_3 & = & -{1\over 4}~{\rm cosh}~\beta ~e^\phi r^3 F ~\phi_r ~e_\psi \wedge \left(e_{\theta_1} \wedge e_{\phi_1}  - e_{\theta_2} \wedge e_{\phi_2}\right) \\
&& - {\rm cosh}~\beta ~e^{2\phi} r F ~e_\psi \wedge \left[{\hat a}^2~e_{\theta_1} \wedge e_{\phi_1}  - {\hat a}^2~e_{\theta_2} \wedge e_{\phi_2} - 
\left(2 a a_r - {1\over 2} e^{-\phi} r^2 \phi_r\right) e_{\theta_2} \wedge e_{\phi_2}\right]\nonumber \nd
where the implications of the relative sign between the vielbein products will become clear soon. We have also defined:
\bg\label{hatp}
{\hat a}^2 = a^2 \phi_r. \nd
As mentioned earlier, ${\cal F}_3$ is not closed, and therefore $d{\cal F}_3$ should be related to localized or delocalized sources along the ($\theta_2, \phi_2$)
and ($r, \psi$) 
directions. Using the fact that the three-form:
\bg\label{3form}
\eta_3 \equiv e_\psi \wedge \left(e_{\theta_1} \wedge e_{\phi_1}  - e_{\theta_2} \wedge e_{\phi_2}\right) \nd
is closed we can find some relations between the three unknown functions $F(r), \phi(r)$ and $a(r)$ that appear in \eqref{calf3}. All we need is to express the dilaton $\phi(r)$ and
the resolution parameter $a(r)$ in terms of the function $F(r)$ that appears in our ansatze \eqref{suramon}. One simple relation between the three variables is given by:
\bg\label{onesol}
{de^\phi\over dr}\left({1\over 4} + {e^\phi ~a^2\over r^2}\right) = {c_0\over r^3 F}, \nd
where $c_0$ is a  constant whose value could be determined from the boundary condition. Note that this is an additional constraint compared to \cite{lapan}.
Pluging in \eqref{onesol} in \eqref{calf3}, we get:
\bg\label{calf3now}
{{\cal F}_3\over {\rm cosh}~\beta} = - c_0~\eta_3 + \left(2 a a_r - {1\over 2} e^{-\phi} r^2 \phi_r\right) e^{2\phi} r F~e_\psi \wedge e_{\theta_2} \wedge e_{\phi_2}. \nd
The source equation is now easy to determine from \eqref{calf3now}. It is clear that the first term does not contribute, and the contribution therefore solely comes from the 
second term of \eqref{calf3now}:
\bg\label{sour}
d{\cal F}_3 = G_r(r) ~e_r \wedge e_\psi \wedge e_{\theta_2} \wedge e_{\phi_2} - G(r)~ e_{\theta_1} \wedge e_{\phi_1} \wedge e_{\theta_2} \wedge e_{\phi_2} \nd 
with $G(r)$ defined as:
\bg\label{grdef}
G(r) =   \left(2 a a_r - {1\over 2} e^{-\phi} r^2 \phi_r\right) e^{2\phi} r F~{\rm cosh}~\beta. \nd
Looking at \eqref{sour} we see that we have two terms. The first term of \eqref{sour} captures the Gauss' charge along the orthogonal directions of the wrapped D5-branes i.e
the ($r, \psi, \theta_2, \phi_2$) directions. The second term, that is proportional to the volume of the four-cycle, captures the Gauss' charge 
along the ($\theta_2, \phi_2$) directions. In fact this term tells us that even if $G(r)$ is a constant, the D5-branes' charge would be calculable.

We see that there are two constraint equations, \eqref{onesol} and \eqref{sour}, for three functions $F(r), e^\phi$ and $a(r)$. The third equation will be determined soon when we 
will demand supersymmety in the system. We could also go for more generic solution to the system. Constraint on D5-brane charges impose the following relation between the 
four warp factors $F_i(r)$ and the dilaton $e^{\phi}$:
\bg\label{cons1}
{dF_4\over dr} = \sqrt{F_1 F_2}\left(1- {e^{-2\phi}F_4\over F_2 F_3}\right). \nd
One may compare this with the recently found constraint relations in \cite{DEM}. Since we are not imposing integrable complex structures, we don't have additional constraint
equations as in \cite{DEM}. Note also that an equation like \eqref{cons1} is not required in the heterotic theory as the anomalous Bianchi identity is enough \cite{lapan}. 
Thus plugging in \eqref{cons1} in \eqref{iibform} we get:
\bg\label{fata3}
{{\cal F}_3\over {\rm cosh}~\beta} = -\eta_3 - \left[1 + e^{2\phi} {F_4\over F_3}\sqrt{F_2\over F_1}\left(\sqrt{F_1F_2} - F_{3r}\right)\right] ~e_\psi \wedge e_{\theta_2} \wedge 
e_{\phi_2}. \nd
The second constraint would come from \eqref{fata3} if we demand charge quantization. Of course if the D5-brane charges are {\it delocalized} there is no strong constraint 
being imposed by \eqref{fata3}. However demanding supersymmetry does introduce new constraint on the warp factors. 
In the following section we will use the 
powerful machinery of the torsion classes ${\cal W}_i$ \cite{chiossi} to analyze this. 

\subsubsection{Finding the warp factors using torsion classes \label{5brane2}}

To study the constraint on the warp factors one may use the technique of the torsion classes \cite{chiossi}.
For us the relevant torsion classes are the ${\cal W}_4$ and ${\cal W}_5$ classes, defined as:  
\bg\label{w4w5}
&&\mathcal{W}_4  =  {F_{3r} - \sqrt{F_1 F_2}\over 4 F_3}  + {F_{4r} -  \sqrt{F_1 F_2}\over 4 F_4} + \phi_r , \nonumber\\
&&{\rm Re}~\mathcal{W}_5  =  {F_{3r} \over 12 F_3} + {F_{4r} \over 12 F_4} + {F_{2r} - 2\sqrt{F_1F_2} \over 12 F_2} + 
{\phi_r\over 2}, \nd  
where one may look at the detailed derivations from \cite{lapan, DEM} or some of the earlier papers in the series namely \cite{fangpaul, anke} etc. Plugging in the warp factor choice
\eqref{suramon}, it is easy to see that:
\bg\label{tota}
&& {\cal W}_4 = {\phi_r\over 2} + {\cal O}(a^2) \nonumber\\
&& {\cal W}_5 \equiv {\rm Re}~{\cal W}_5 = {1\over 12} \left({6\over r} + 3\phi_r + {F_r\over F} - {2\over r F}\right). \nd
Depending on how to define our dilaton, 
\bg\label{dil}
{\rm Re}~{\cal W}_5 = \pm \phi_r + {\cal O}(a^2), \nd
such that the supersymmetry condition will take the following well-known form in terms of the torsion classes \cite{luest, louis, gauntlett}:
\bg\label{susycond}
2{\cal W}_4 \pm {\rm Re}~{\cal W}_5 = 0. \nd
For us we will choose the minus sign in \eqref{dil} such that \eqref{susycond} will appear with a relative plus sign\footnote{The overall behavior of fluxes etc do not change if
we go from one convention to another as shown in \cite{lapan}.}. This gives the following equation for the variables $F(r)$ and $e^\phi$ upto ${\cal O}(a^2)$:
\bg\label{orda2}
r{d\phi\over dr} + {r\over 15 F} {dF\over dr} - {2\over 15 F} + {2\over 5} + {\cal O}(a^2) = 0. \nd
The above is the simplified version where the dependence of the resolution parameter is not shown. If we insert $a^2$, the EOM becomes more involved and takes the following form:
\bg\label{refeq}
\left(15 + {88 a^2 e^\phi \over r^2}\right) {d\phi\over dr} + {56 e^\phi a \over r^2} {da\over dr} + \left({4\over r} + {1\over F} {dF\over dr} - {2\over rF}\right)
\left(1 + {4a^2 e^\phi\over r^2}\right) + {2\over r} = 0, \nonumber\\  \nd
and reduces to \eqref{orda2} in the limit where $a^2$ as well as $da/dr$ are small. In this limit we can combine \eqref{onesol} and \eqref{orda2} to eliminate $F(r)$ and express 
everything in terms of the following dilaton equation:
\bg\label{dilaeqn}
r{d^2 Z\over dr^2} - 3{dZ\over dr} + r\left({r^2\over 2 c_0} - {15\over Z}\right) \left({dZ\over dr}\right)^2 = 0 \nd  
where $Z = e^\phi$ and $c_0$ is a constant appearing in \eqref{onesol}. To solve the above equation let us take the following ansatze for $Z$:
\bg\label{Zdef}
Z(r) ~ = ~ {\alpha(r)\over r^2}, \nd
with $\alpha(r)$ a positive definite function for all $r$. Plugging \eqref{Zdef} in \eqref{dilaeqn}, we see that $\alpha(r)$ satisfies the following second-order differential 
equation:
\bg\label{secord}
c_0 {d^2\alpha \over dr^2} + \left({53c_0 - 2\alpha \over r}\right) {d\alpha \over dr} 
+   \left({1\over 2} - {15 c_0 \over \alpha}\right) \left({d\alpha\over dr}\right)^2 
+ {2\alpha(\alpha - 24 c_0) \over r^2} = 0. \nd
One simple solution for the system is given by a constant $\alpha$, i.e:
\bg\label{simsol}
\alpha ~ = ~ 24 c_0. \nd
Other solutions to \eqref{secord} could be entertained but we will not do so here. Plugging \eqref{simsol} in \eqref{Zdef} and \eqref{onesol}, and using the definition of $Z$, we
find that:
\bg\label{dilbac}
e^\phi = {24 c_0\over r^2}, ~~~~~~~~ F = -{1\over 12}. \nd
The careful reader will be alarmed by seeing the negative value for $F$ because $F$ goes into the definition for the warp-factors in \eqref{suramon}. However if we look at 
\eqref{suramon} carefully, we see that $F$ appears in the definitions of $F_1$ and $F_2$ but not in the definitions of $F_3$ and $F_4$. This is good because ($F_1, F_2$) appear
in the three-form fluxes ${\cal H}_3$ and ${\cal F}_3$ {\it only} in the combinations $F_1F_2$ and $F_2/F_1$. Thus we can change the sign of ($F_1, F_2$) simultaneously without
changing the fluxes or the constraint equation \eqref{onesol}! The consequence of this invariance is simply the following changes to the definition of the warp factors:
\bg\label{redeff}
F_1 ~ \to ~ \vert F_1 \vert, ~~~~~~~ F_2 ~ \to ~ \vert F_2 \vert , \nd
without changing $F_3$ and $F_4$. This means, after the dust settles, the internal six-dimensional manifold in type IIB theory will be given by the following metric:
\bg\label{intmet}
ds_6^2 & = &{r^2\over 4 c_0}\Big[dr^2 + {r^2\over 144}\left(d\psi + \cos~\theta_1~d\phi_1 + \cos~\theta_2~d\phi_2\right)^2 \nonumber\\
&& + \left({r^2\over 24} + {\cal O}(a^2)\right)(d\theta_1^2 + \sin^2\theta_1 ~d\phi_1^2) + {r^2\over 24} (d\theta_2^2 + \sin^2\theta_2 ~d\phi_2^2)\Big]. \nd
The above metric is a non-K\"ahler metric on the resolved conifold, and can be compared to the recently studied examples in \cite{DEM}. If we change our initial ansatze \eqref{suramon},
we can allow for a different non-K\"ahler metric on the resolved conifold. There is of course an infinite class of possible non-K\"ahler metric that we can allow for a given
complex structure and satisfying the constraint equation \eqref{cons1} and the supersymmetry condition \eqref{susycond} with a relative plus sign between the ${\cal W}_4$  and 
${\cal W}_5$ torsion classes. The generic solution for the metric and the three-form fluxes with these constraints will then be \eqref{iibform}. 
For the specific choice \eqref{intmet} of the internal metric, the three-form fluxes are given by: 
\bg\label{inth3f3}
&&{\cal F}_3 = + c_0~{\rm cosh}~\beta~e_\psi \wedge \left(e_{\theta_1} \wedge e_{\phi_1} + e_{\theta_2} \wedge e_{\phi_2}\right) \nonumber\\
&&{\cal H}_3 = - {r^3\over 48 c_0} ~{\rm sinh}~\beta~e_r \wedge \left(e_{\theta_1} \wedge e_{\phi_1} + e_{\theta_2} \wedge e_{\phi_2}\right), \nd 
with the five-form flux derivable from \eqref{inth3f3} and \eqref{iibform}. The IIB dilaton, on the other hand, is $e^{\phi_B} = e^{-\phi}$ and so for 
\bg\label{rval}
r ~ \ge  ~ \sqrt{24 c_0}, \nd
classical supergravity solution will not capture the full dynamics and one has to go to it S-dual, or weakly coupled version of the theory. Combining the two patches, one should be
able to study the sugra limit of the theory. 

On the other hand if dilaton is slowly varying from its weak coupling value then one may express \eqref{refeq} as:
\bg\label{a2eq}
{da^2\over dr} + {1\over 28}\left(4r + {r^2\over F}{dF\over dr} - {2r\over F}\right)\left(e^{-\phi} + {4a^2 \over r^2}\right) + {r e^{-\phi}\over 14}  = 0.\nd
To solve \eqref{a2eq}, let us assume that the dilaton is given by the following expression in terms of a slowly varying function $f(r)$:
\bg\label{dilan}
e^\phi ~= ~ e^{\phi_0} + f(r), \nd
where the constant factor is the weak coupling limit. To proceed, let us define two functions $H(x)$ and $G(x)$ using the function $F(x)$ appearing in \eqref{a2eq}, in the following way:
\bg\label{bkcase}
G(x) = {1\over 7}\left({4\over x} - {2\over x F(x)} + {F'(x)\over F(x)}\right), ~~~~~ H(x) = x e^{-\phi_0} \left({x G(x)  \over 4} + {1 \over 14}\right), \nd
where the prime is defined as the derivative of $x$. Using \eqref{dilan} and \eqref{bkcase}, we can solve for the resolution parameter $a^2$ in terms of the functions $G(x)$ and $H(x)$ as:
\bg\label{a2sol}
a^2(r) = -\int_0^r dy~H(y) ~{\rm exp}\left(\int_r^y dx~G(x)\right) + {\cal O}(f), \nd
where the overall negative sign shouldn't be a concern because the functional form for $F(x)$ will be chosen so that $a^2$ remains positive definite.

\subsection{A four-fold from the $G_2$ structure  manifold in M-theory \label{g2man}}

In the previous section we discussed possible ways to construct the metric of D5-branes wrapped on two-cycle of a non-K\"ahler resolved conifold. We discussed a class of these 
solutions satisfying the charge constraint \eqref{cons1} and the supersymmetry constraint \eqref{susycond}. The M-theory uplift of these solutions can be done by first 
T-dualizing along $\psi$ direction to allow for D6-branes in type IIA theory oriented along ($\theta_1, \phi_1, \psi$) and spanning the space-time directions $x_{0, 1, 2, 3}$. 
We can then lift this configuration to M-theory on a $G_2$ structure manifold. The way we constructed our scenario, T-duality of the IIB configuration will lead to D6-branes and 
not D4-branes as in \cite{DM1, DM2}. At low energies, and as discussed around \eqref{c7}, we do get the D4-branes configuration (see also \cite{DEM}). Furthermore, we will start
by studying a single D6-brane and insert the dipole deformation of the T-dual wrapped D5-brane. Later on we will generalize this to multiple D6-branes. 

\subsubsection{First look at the $G_2$ structure manifold \label{g21}}

The D6-brane configuration, without dipole deformation of the T-dual wrapped D5-brane on non-K\"ahler resolved conifold, is given by the following metric structure on an internal
six-dimensional space:
\bg\label{d6brane2}
ds^2 &= &{1\over \sqrt{h}}~ds^2_{012} + {1\over \sqrt{h}}\left(dx_3^2 + {1\over F_2}~d\psi^2\right) \nonumber\\
& + & \sqrt{h}\left[F_1 dr^2 + F_3(d\theta_1^2 + \sin^2\theta_1~d\phi_1^2) + F_4(d\theta_2^2 + \sin^2\theta_1~d\phi_2^2)\right], \nd 
where we have separated the compact directions ($x_3, \psi$) in anticipation of the dipole deformations along those directions in the type IIB side. The type IIA dilaton 
$e^{\phi_A}$ and
the warp factor $h$ are defined in the following way:
\bg\label{dilhdef} 
e^{\phi_A} \equiv e^{-\phi} F_2^{-1/2} h^{-1/4}, ~~~~
h \equiv  e^{2\phi}~{\rm cosh}^2\beta - {\rm sinh}^2\beta, \nd
such that when $\beta = 0$ we get back the standard picture. Combining the IIA metric \eqref{d6brane2} with the dilaton \eqref{dilhdef} we can easily get the M-theory 
manifold as:
\bg\label{mmetric}
ds^2_{11} = {e^{2\phi/3} F_2^{1/3}\over h^{1/3}}~ds^2_{012} + ds^2_8 \nd
where $ds^2_8$ is a eight-dimensional manifold that, in the absence of the dipole deformation in the type IIB side, is simply a $S^1$ fibration over a $G_2$ structure 
seven-manifold expressed as:
\bg\label{8fold}   
ds^2_8 &=& {e^{2\phi/3} F_2^{1/3}\over h^{1/3}}\left(dx^2_3 + {1\over F_2} d\psi^2\right) + {1\over e^{4\phi/3}F_2^{2/3}h^{1/3}} (dx_{11} + {\bf A}_{1\mu} dx^\mu)^2 \\
&+& e^{2\phi/3}F_2^{1/3} h^{2/3}\left[F_1 dr^2 + F_3(d\theta_1^2 + \sin^2\theta_1~d\phi_1^2) + F_4(d\theta_2^2 + \sin^2\theta_2~d\phi_2^2)\right]. \nonumber \nd
The ${\bf A}_1$ appearing above is the type IIA gauge field whose value will be determined soon.  
As discussed in details in \cite{DEM}, the $G_2$ structure seven-manifold in-turn is a four-dimensional 
warped Taub-NUT manifold $ds^2_{\rm TN}$ fibered over a three-dimensional base $ds_3^2$ parametrized by 
($\theta_1, \phi_1, \psi$): 
\bg\label{g2}
ds^2_7 &= & ds^2_3 + ds^2_{\rm TN} \nonumber\\ 
&=& G_2\left(d\theta_1^2 + {\rm sin}^2\theta_1~d\phi_1^2 + {G_1\over G_2 F_2}d\psi^2 \right)\nonumber\\
&+& G_3 dr^2 + G_4 \left(d\theta_2^2 + {G_5\over G_4} d\phi_2^2\right) + G_6\left(dx_{11} + {\bf A}_{1\mu} dx^\mu\right)^2 , \nd
where $G_i$ are the warp factors that can be read up from \eqref{8fold} or from \cite{DEM} and the third line of \eqref{g2} is the metric of the warped Taub-NUT space.  

\subsubsection{Dipole deformation and the M-theory uplift \label{g22}}

It is now time to see what effect would the type IIB dipole deformation have on our M-theory manifold. Dipole deformation of four-dimensional Yang-Mills theory was
first introduced from gauge theory side in \cite{GB} and from type IIB gravity dual in \cite{bergman1, bergman2}. Elaborate study was performed in \cite{jabbari, shmakov}.
Essentially the simplest dipole deformation amounts to switching on a NS B-field with one component along the brane and the other component orthogonal to the brane. Generalization
of this picture exists, but we will not discuss this here. The B-field for our case will have component $B_{3\psi}$ as we mentioned before, which of course
has the required property in the presence of a D5-brane along ($x_{0, 1, 2, 3}, \theta_1, \phi_1$). 
However as before this B-field cannot be a constant otherwise it will be gauged away. Thus again
we expect a field strength of the form $dB$, which in turn will then back-react on our original type IIB background \eqref{iibform} and change the metric to the following:
\bg\label{iibform1}
ds^2 & = & {1\over \sqrt{h}}\left(-dt^2 + dx_1^2 + dx_2^2 + {dx_3^2\over \cos^2~\theta + F_2~\sin^2~\theta}\right)\\
&+& \sqrt{h}\left[F_1 dr^2 + {F_2({d\psi \over \cos~\theta} + \cos~\theta_1~d\phi_1 + \cos~\theta_2~d\phi_2)^2\over 1 + F_2~{\rm tan}^2~\theta} + 
\sum_{i=1}^2 F_{2+i}\left(d\theta_i^2 + \sin^2~\theta_i~d\phi_i^2\right)\right], \nonumber  \nd
where $\theta$ is the dipole deformation parameter. The three-form fluxes also change from their values in \eqref{iibform} to the following:
\bg\label{iibform2} 
{\cal F}_3 &= & - e^{2\phi}~{\rm cosh}~\beta \sqrt{F_2\over F_1}\left(g_1~{\widetilde e_\psi} \wedge e_{\theta_1} \wedge e_{\phi_1} + 
g_2~{\widetilde e_\psi} \wedge e_{\theta_2} \wedge e_{\phi_2}\right) \nonumber\\
{\cal H}_3 &=& {\rm sinh}~\beta \left[\left(\sqrt{F_1F_2} - F_{3r}\right) e_r \wedge e_{\theta_1} \wedge e_{\phi_1} + 
\left(\sqrt{F_1F_2} - F_{4r}\right) e_r \wedge e_{\theta_2} \wedge e_{\phi_2}\right] \nonumber\\ 
&& + {F_{2r}~{\rm sin}~2\theta\over 2\left(\cos^2~\theta + F_2~\sin^2~\theta\right)^2}~e_r \wedge 
{\widetilde e_\psi} \wedge e_3 + {F_2~\sin~\theta\over \cos^2~\theta + F_2~\sin^2~\theta} \sum_{i=1}^2 e_{\phi_i} \wedge e_{\theta_i}\wedge e_3
\nonumber\\ \nd 
where as before we note that the dipole deformation has appeared as an additional term in the definition of the three-form flux ${\cal H}_3$, and helped to break the Lorentz invariance 
between $x_{0, 1, 2}$ and $x_3$ directions.   
The type IIB dilaton $e^{\phi_B}$ and 
${\widetilde e_\psi}$ are defined in the following way:
\bg\label{dilpsi}
e^{\phi_B} = {e^{-\phi}\over \sqrt{\cos^2~\theta + F_2 ~\sin^2~\theta}}, ~~~~ {\widetilde e_\psi} = d\psi + \cos~\theta_1~\cos~\theta~d\phi_1 
+ \cos~\theta_2~\cos~\theta~d\phi_2. \nonumber\\ \nd
The M-theory uplift of the dipole-deformed type IIB set-up is now easy to perform once we get the type IIA configuration. The type IIA dilaton does not change from its value 
\eqref{dilhdef}, and the only change in the metric \eqref{d6brane2} is:
\bg\label{d6now}
{1\over \sqrt{h}}\left(dx_3^2 + {1\over F_2}~d\psi^2\right) ~~ \to ~~ {1\over \sqrt{h}}\left[{dx_3^2\over \cos^2~\theta} + 2{\rm tan}~\theta~dx_3 d\psi +
\left(\sin^2~\theta + {\cos^2~\theta\over F_2}\right)d\psi^2\right], \nonumber\\ \nd
which means the M-theory metric retain its form \eqref{mmetric} except the metric of the eight manifold changes slightly from \eqref{8fold} to the following metric:
\bg\label{8foldnow}
ds^2_8 &=& {e^{2\phi/3} F_2^{1/3}\over h^{1/3}\cos^2~\theta} ~{\bf \vert} dx_3 + \tau_1~d\psi {\bf \vert}^2
+ {1\over e^{4\phi/3}F_2^{2/3}h^{1/3}} (dx_{11} + {\bf A}_{1\mu} dx^\mu)^2 \\
&+& e^{2\phi/3}F_2^{1/3} h^{2/3}\left[F_1 dr^2 + F_3(d\theta_1^2 + \sin^2\theta_1~d\phi_1^2) + F_4(d\theta_2^2 + \sin^2\theta_2~d\phi_2^2)\right], \nonumber \nd
where the complex structure $\tau_1$ of the ($x_3, \psi$) torus is given by:
\bg\label{cmplx}
\tau_1 ~= ~ \sin~\theta~\cos~\theta + {i\cos^2~\theta\over \sqrt{F_2}}.\nd
Note that the warped Taub-NUT space doesn't change from what we had earlier in \eqref{8fold} without dipole deformation. The gauge field ${\bf A}_1$ in the Taub-NUT fibration
structure also doesn't change, and is given by the following field-strength:
\bg\label{f2form}
{{\cal F}_2 \over {\rm cosh}~\beta} &= & - e_{\theta_1} \wedge e_{\phi_1} 
- e^{2\phi}\sqrt{F_2\over F_1}\cdot {F_4\over F_3}\left(\sqrt{F_1F_2} - F_{3r}\right)~e_{\theta_2} \wedge e_{\phi_2}
\nonumber\\
& = & {d{\bf A}_1\over {\rm cosh}~\beta} 
+ \left[1 - e^{2\phi}\sqrt{F_2\over F_1}\cdot {F_4\over F_3}\left(\sqrt{F_1F_2} - F_{3r}\right)\right]~e_{\theta_2} \wedge 
e_{\phi_2}, \nd
using the constraint \eqref{cons1} and defining the gauge field ${\bf A}_1$ in the following way:
\bg\label{gaugo}
{\bf A}_1 = {\rm cosh}~\beta \left(\cos~\theta_1 ~d\phi_1 + \cos~\theta_2 ~d\phi_2\right), \nd
which would appear in the fibration \eqref{8foldnow}. However expressing the gauge field as \eqref{gaugo} does not introduce any additional constraint on the warp-factors 
in the metric (see discussion in \cite{DEM}). The ${\cal G}_4$ flux in M-theory can now be expressed as:
\bg\label{gfla}
{{\cal G}_4\over {\rm sinh}~\beta} &=& \left(\sqrt{F_1 F_2} - F_{3r}\right) e_r \wedge e_{\theta_1} \wedge e_{\phi_1} \wedge e_{11} 
+ {\rm cosech}~\beta~d\psi \wedge e_{\theta_1} \wedge e_{\phi_1}\wedge \widetilde{e}_{11} \nonumber\\
&+& \left(\sqrt{F_1 F_2} - F_{4r}\right) e_r \wedge e_{\theta_2} \wedge e_{\phi_2} \wedge e_{11} 
+ {\rm cosech}~\beta~d\psi \wedge e_{\theta_2} \wedge e_{\phi_2}\wedge \widetilde{e}_{11}, \nonumber\\ \nd
where we see that the dipole deformation appears in an appropriate way in the ${\cal G}_4$ flux. In the absence of the type IIB dipole deformation the form of \eqref{gfla} is almost 
similar to what we had in \cite{DEM} except the vielbeins 
$e_{11}$ and $\widetilde{e}_{11}$ are defined in a slightly different way as:
\bg\label{e11def}
&&e_{11} = dx_{11} + \cos~\theta~{\rm cosh}~\beta(\cos~\theta_1~d\phi_1 + g_o~\cos~\theta_2~d\phi_2) \nonumber\\
&& \widetilde{e}_{11} = dx_{11} + \cos~\theta~{\rm cosh}~\beta(g_o~\cos~\theta_1~d\phi_1 + \cos~\theta_2~d\phi_2), \nd
using the following functional form for $g_o(r)$:
\bg\label{g1now}
g_o(r) = e^{2\phi}\sqrt{F_2\over F_1}\cdot {F_4\over F_3}\left(\sqrt{F_1F_2} - F_{3r}\right).\nd

\subsubsection{Revisiting gauge theory from M-theory \label{g23}}

We have by now developed all the machinery needed for determining the gauge field on the wrapped D5-branes from M-theory. If we take a single wrapped D5-brane on the 
non-K\"ahler resolved conifold, the M-theory manifold \eqref{8foldnow} will be a warped single-centered Taub-NUT space fibered over a four-dimensional base parametrized by
($x_3, \psi, \theta_1, \phi_1$) coordinates. The gauge-field in the type IIB side will appear as localized G-flux in M-theory, similar to what we had earlier in section \ref{jahangir} (see also
the discussion 
in \cite{DEM}). For the single centered Taub-NUT case in \eqref{8foldnow}, at any given point on four-dimensional base, the localized G-flux can be expressed as:
\bg\label{locg}
{\cal G}_4^{\rm loc} = {\cal F} \wedge \omega, \nd
where ${\cal F}$ is the world-volume gauge field that, in the language of the wrapped D6-brane, will be along four-dimensional spacetime parametrized by $x_{0, 1, 2, 3}$ coordinates. This of course
parallels the story we discussed in great details in section \ref{fstep}. There is also an option to define the gauge theory 
along the compact ($\psi, \theta_1, \phi_1$) directions, or even along all compact and non-compact directions. Each of these possibilities will lead to interesting interpretations
for the knot invariants once we extend this to the non-abelian case. We will however only concentrate on the gauge theory along the spacetime directions so that comparison with 
earlier sections like \ref{fstep}, \ref{inclu} and \ref{NAE} as well as with \cite{wittenknots1} may be made easily. In fact we will follow similar logic as in sections \ref{fstep} and \ref{inclu}, namely, 
study the {\it abelian} theory and then proceed to discuss the non-abelian case (which is the large $N$ limit here). 

The abelian case is succinctly represented by $\omega$
in \eqref{locg}, which is a normalizable harmonic two-form, expressed as $\omega = d \zeta$. The procedure is similar to 
what we had in \eqref{chokranto}, \eqref{gsaisfy} and \eqref{gchakka}, so we will avoid the details. Once the dust settles,  $\zeta$ is given by the 
following expression\footnote{Note that at any given point on the four-dimensional 
base, $\phi_1$ is a constant and therefore the eleven-dimensional fibration structure is the correct form for a warped Taub-NUT space.}:
\bg\label{zeta}
\zeta(r, \theta_2) &=& g_0~{\rm exp}\left[-\int^r dr ~{e^{-\phi}\over F_4}\sqrt{F_1\over h F_2}\right]
 \left(d\Psi + \cos~\theta_2~d\phi_2\right)\\
& = & g_0~{\rm exp}\left[-\int_0^r {48~dx\over x\sqrt{576 c_0^2~{\rm cosh}^2~\beta - x^4~{\rm sinh}^2~\beta}}\right]\left(d\Psi + \cos~\theta_2~d\phi_2\right), \nonumber \nd
where $d\Psi = dx_{11}/{\rm cosh}~\beta$ and the second line is from using the background \eqref{dilbac} and \eqref{intmet}. Note that the harmonic form tells us that 
for:
\bg\label{rvalnow}
r ~> ~ \sqrt{24 c_0~{\rm coth}~\beta}, \nd
new description has to be devised as the harmonic form will become oscillatory. This bound should be compared to \eqref{rval} where strong coupling sets in for the radius 
equals $\sqrt{24 c_0}$.   

The non-abelian enhancement now follows similar procedure as outlined in section \ref{NAE}. The M2-brane states wrap around the Taub-NUT singularities to enhance the gauge symmetry to 
$SU(N)$. This way we will have ${\cal N} = 1$ supersymmetric $SU(N)$ Yang-Mills theory in four spacetime dimensions appearing from $N$ D5-branes wrapped on the two-cycle of a non-K\"ahler 
resolved conifold.

\subsection{Comparing knots from branes and from gravity duals \label{coknots}}

In the previous sections we have developed most of the machinery needed to study the abelian and the non-abelian theories on the wrapped D5-branes on a 
resolved conifold from M-theory point of view. Our aim is to concentrate on
the non-abelian case with two goals in mind: the first is to study the connection between the model of Witten \cite{wittenknots1} 
using five-branes 
and the model of Ooguri-Vafa \cite{OV} using geometric transition picture to study knots invariants and Khovanov homology. The second goal is to use our M-theory picture to 
actually compute some of these invariants and develop the picture in more generic direction. A discussion of the first goal, namely connecting the two models: \cite{wittenknots1} and 
\cite{OV}, is presented in section \ref{toptwist} and in the following we will elaborate the story a bit more. 

Our starting point, which is the configuration of $N$
D5-branes wrapped on a two-cycle of a non-K\"ahler resolved conifold, may look a bit different from the configuration that we used before in section \ref{nonkah2}, namely, a finite 
number of D5-branes wrapped on the two-cycles of a 
warped Taub-NUT space. Additionally, the supersymmetry is now no longer ${\cal N} = 4$, but is the minimal ${\cal N} =1$. The latter tells us that we have no Coulomb branch, implying that the vector 
multiplet is devoid of any scalar fields. Thus the {\it twisting} that we performed in section \ref{boundth} to determine the boundary theory cannot be done in a similar way now. Additionally, we see
that there are apparently {\it two} realizations of the Ooguri-Vafa model in M-theory from the type IIB configuration.

\vskip.1in

\noindent {\bf Using one T-duality}: This will lead to the D6-branes that we studied above. Subsequent lift to M-theory results in 
the localized G-flux that has two legs along the spacetime  $x_{0, 1, 2, 3}$ directions and two legs along the Taub-NUT directions leading to gauge fields in the spacetime directions. 
The other components of the
gauge fields in the internal directions will appear as non-abelian scalars in the non-compact three-dimensions. Together they will generate the ${\cal N} = 1$ non-abelian vector multiplet
with scalar fields forming the chiral multiplets. 

\vskip.1in

\noindent {\bf Using three T-dualities}: Instead of making one T-duality to go to the D6-brane picture, we can make three T-dualities to go to the mirror 
picture\footnote{One encounters various subtleties in the duality procedure, which have been explained in details in \cite{syz2}.} \cite{SYZ, syz2}.  
Here we will again get D6-branes but 
wrapped on the three-cycle of a non-K\"ahler {\it deformed} conifold. Lifting this to M-theory this will lead to another $G_2$ structure manifold which is yet again a warped Taub-NUT space fibered over a 
three-dimensional base \cite{lapan}. The localized G-flux can now be used to compute the four-dimensional theory as before. 

\vskip.1in

\noindent As explained in section \ref{toptwist}, despite appearance, the physics in four spacetime dimensions for both cases are identical. This is not a surprise 
because T-dualities generally do not change 
the four-dimensional physics. Thus either of the two configurations $-$ D5-branes wrapped on two-cycle of a resolved conifold or 
D6-branes wrapped on three-cycle of a deformed conifold $-$ may be used to study the Ooguri-Vafa \cite{OV} model. However since the latter is technically harder, we have used the type IIB model to 
study the four-dimensional physics above. Additionally since a non-K\"ahler resolved cone may be expressed as a warped Taub-NUT fibered over a ${\bf P}^1$ base \cite{lapan}, locally at a 
given point on ${\bf P}^1$, the D5-branes can be thought of as wrapping the two-cycle of the Taub-NUT space. We now see some resemblance with \cite{wittenknots1} locally, although the global picture 
is different. Unfortunately we cannot extend the similarity too far because, in the Ooguri-Vafa case, the absence of the Coulomb branch will not allow us to make similar manipulations as we did in 
section \ref{boundth}.   

Despite this, the gauge theory derivation from M-theory in the previous section helps us to at least get the topological piece in a way similar to what we had in \eqref{mtheorya} before. Let us
concentrate on the second piece in \eqref{mtheorya}, namely the topological term. For the present case, it is more instructive to Euclideanize everthing, as we hinted in section \ref{toptwist}. 
Assuming this, we get: 
\bg\label{mtheoryb}
\int_{\Sigma_{11}} {\cal C}_3 \wedge {\cal G}_4 \wedge {\cal G}_4 = {\widetilde c}_2 \int_{\Sigma_4} {\cal F} \wedge {\cal F}, \nd
where both $\Sigma_{11}$ and $\Sigma_4$ are eleven and four-dimensional Euclidean spaces respectively, and the coupling constant ${\widetilde c}_2$ is defined as:
\bg\label{vietduck}
{\widetilde c}_2 \equiv \int_{\Sigma_7} \langle {\cal C}_3 \rangle \wedge \omega \wedge \omega, \nd
with $\omega = d\zeta$ as described in \eqref{zeta} above, $\Sigma_7$ is the $G_2$ structure manifold in M-theory and $\langle {\cal C}_3 \rangle$ is the expectation value of the three-form potential
$\left({\cal C}_3\right)_{r\psi\phi_1}$ which may be extracted from the four-form ${\cal G}_4$ in \eqref{gfla} using the vielbeins \eqref{e11def}. 

One of the key difference between ${\widetilde c}_2$ in \eqref{vietduck} and $c_2$ in \eqref{c1c2c1c2} is the orientations of $\langle {\cal C}_3 \rangle$ appearing in both. Previously 
we needed three-form potential of the form $\left({\cal C}_3\right)_{3 r\phi_1}$  \eqref{chotas} to determine $c_2$ in \eqref{c1c2c1c2}. 
Such a component was generated from the subtle flux arrangement on the two-cycle of the warped 
Taub-NUT space to stabilize the D5-${\overline{\rm D5}}$ pairs against tachyonic instabilities. Now we don't have such instabilities, and the three-form potential does appear more naturally 
from \eqref{gfla}. 

Once we allow for the non-abelian extension, the coefficient of the topological term ${\widetilde c}_2$ will remain the same as \eqref{vietduck} with a $SU(N)$ trace inserted in the 
action \eqref{vietduck}, 
similar to what we had in section \ref{NAE}. 
The 
boundary theory may now be derived in a much simpler way that what we had in section \ref{boundth}. To proceed, we will first assume that the Euclidean space $\Sigma_4$ may be 
written as $\Sigma_4 = {\bf S}^3_{(1)} \times {\bf R}^+$ where ${\bf R}^+$ is parametrized by $x_3$ in either the M-theory or the type IIB metrics. Taking $x_3$ or ${\bf R}^+$ to be the 
half-line, we can easily infer the boundary theory to be:
\bg\label{ajapple}
S_{ov} = \left({\widetilde b}_2 + {\widetilde c}_2\right) \int_{{\bf S}^3_{(1)}} {\rm Tr}\left({\cal A} \wedge d{\cal A} + {2i\over 3} {\cal A} \wedge {\cal A} \wedge {\cal A}\right), \nd
where the trace is in the adjoint representation of $SU(N)$ and ${\cal A}$ is the non-abelian gauge field derived from ${\cal F}$ once we allow for the full non-abelian extension in M-theory (this is
similar to what we had in section \ref{NAE}). The coefficient ${\widetilde c}_2$ is of course the one in \eqref{vietduck}, however ${\widetilde b}_2$ is new. We expect ${\widetilde b}_2$ to appear 
in somewhat similar way as $b_2$ appearing in \eqref{stwist} earlier. In other words, in the presence of a boundary, the kinetic term is not completely ${\bf Q}$ invariant, and a piece proportional to 
\eqref{mtheoryb} should appear as described in \eqref{stwist}. Considering this, reproduces \eqref{ajapple}.

\begin{figure}[t]\centering \includegraphics[width=0.8\textwidth]{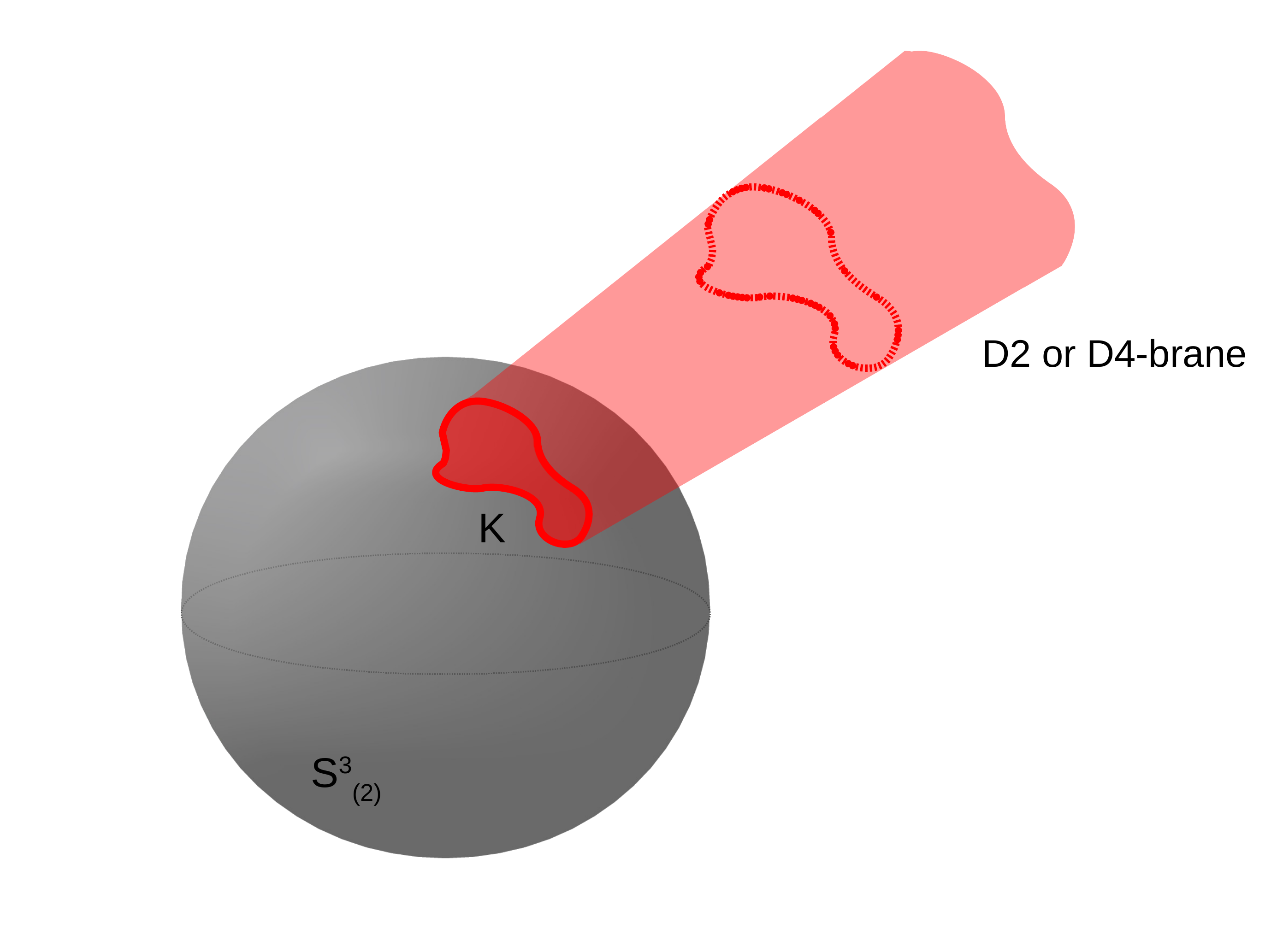}\caption{Knot ${\bf K}$ on D6-branes wrapped on ${\bf S}^3_{(2)}$ of a deformed conifold is represented by a
D2-brane (or D4-brane) surface operator that intersects the D6-branes on ${\bf K}$. This picture is before geometric transition. After geometric transition, the D6-branes disappear and are replaced by 
fluxes on a non-K\"ahler resolved conifold, but the D2-brane (or D4-brane) state survives on the dual side retaining all information of the knot ${\bf K}$.}
\label{fig14}\end{figure}

The attentive reader must have noticed the key difference between \eqref{plaza} and \eqref{ajapple}. The former is constructed from a modified gauge field ${\cal A}_d$ by combining the original gauge 
field ${\cal A}$ and the Coulomb branch scalars $\phi$ as in \eqref{newgf}. For the present case, the vector multiplet has no scalars, and assuming we keep vanishing expectation values of the 
scalars in the chiral multiplets, the boundary theory will be constructed solely using the non-abelian gauge field ${\cal A}$, leading to \eqref{ajapple}. Quantum mechanically however the 
difference is only in the choices of the coupling constants for the boundary theories \eqref{plaza} and \eqref{ajapple}. This is because of the following path integral equivalence in the 
Euclidean formalism:
\bg\label{effsit}
\int_{\cal C} {\cal D}{\cal A}_d ~{\rm exp}\left[-S_{bnd}({\cal A}_d)\right] ~F({\cal A}_d)  = 
\int_{\cal C} {\cal D}{\cal A} ~{\rm exp}\left[-S_{bnd}({\cal A})\right] ~F({\cal A}), \nd
where $F({\cal A})$ is any observable in the theory and ${\cal C}$ is the integration cycle. Therefore in the path integral ${\cal A}_d$ is just a dummy variable and can be replaced by the 
gauge field ${\cal A}$. Although our discussion above is a bit sloppy as we are ignoring many subtle points, the essential physics is captured in \eqref{effsit}. 
For more details on the equivalence of two path integrals for both real and complex 
gauge fields, one may refer to section (2.4) of \cite{wittenknots1}. 

The three-dimensional boundary theory \eqref{ajapple}, defined on ${\bf S}^3_{(1)}$, remains the same when we go to the {\it mirror} type IIA side. Since the SYZ transformations \cite{SYZ, syz2} 
do not change the spacetime metric, the three-cycle ${\bf S}^3_{(1)}$ on the type IIB side goes unchanged to the type IIA side. However the D5-branes wrapped on the two-cycle of the non-K\"ahler resolved 
confold become D6-branes wrapped on the three-cycle ${\bf S}^3_{(2)}$
of the non-K\"ahler deformed conifold. The world-volme of the D6-branes is now \eqref{whistle}, and therefore a flop operation \eqref{sebeast} 
will transfer the boundary theory \eqref{ajapple} defined on the three-cycle ${\bf S}^3_{(1)}$ to the three-cycle ${\bf S}^3_{(2)}$ of the deformed conifold, giving us:
\bg\label{pasoli}
S_{ov} = \left({\widetilde b}_2 + {\widetilde c}_2\right) \int_{{\bf S}^3_{(2)}} {\rm Tr}\left({\cal A} \wedge d{\cal A} + {2i\over 3} {\cal A} \wedge {\cal A} \wedge {\cal A}\right), \nd 
where, although we use the same notation of \eqref{ajapple}, ${\cal A}$ should be thought of as the gauge field defined on ${\bf S}^3_{(2)}$.
Knots may now be inserted on ${\bf S}^3_{(2)}$ using D2-brane (or D4-brane) surface operators as shown in {\bf fig \ref{fig14}}. 
The construction parallels the discussion in section \ref{bathle} in spirit only as specific details differ. The 
difference of course stems from the construction of the Ooguri-Vafa model \cite{OV} starting with {\bf Table \ref{modbranes}} compared to the construction in section \ref{nonkah2} starting with
{\bf Table \ref{wittenbranes}}. The flop operation \eqref{sebeast} with the added complication of geometric transition, as well as the absence of the Coulomb branch scalars, 
in fact makes it harder to implement similar procedure as in section \ref{bathle}. We will therefore not analyze the story further and only make few observations
keeping most of the details for the sequel. 

The first observation is the M-theory lift of the knot configurations on ${\bf S}^3_{(2)}$. The uplift leads to M2-brane states\footnote{We can also entertain M5-brane states related to 
D4-branes in type IIA. This is allowed because we only require co-dimension two singularities in ${\bf S}^3_{(2)} \times {\bf R}^+$ space, and as such can come from both D2 and D4-branes. This 
is depicted in {\bf fig \ref{fig14}}.} 
in the $G_2$ structure manifold of the second kind associated with 
{\it three} T-dualities (see discussion above). These M2-brane states do not wrap the eleven-dimensional circle, so are distinct from the ones leading to non-abelian enhancement discussed for the $G_2$ 
structure manifold of the first kind associated with {\it one} T-duality. This would then be the uplift of the surface operators in M-theory. 

The second observation is that 
the knots appearing from the surface operators do not follow similar pathway that we developed earlier in section \ref{bathle} and \ref{knotty} for Witten's model \cite{wittenknots1}. 
This is because we cannot study the abelian version now as the model is only defined for large $N$, implying that 
our earlier analysis of the knots using operators ${\bf A}_k$, ${\bf B}_k$  and ${\bf C}_{(2, \sigma_j)}$ in section \ref{knotty} may not be possible now. 
Secondly, similar manipulations to the BHN equations that we did in section \ref{bathle} now cannot be performed. 

What {\it can} be done here? There is one well known procedure that we can follow. We can use
the canonical quantization approach by slicing the three-cycle  ${\bf S}^3_{(2)}$  containing the knot ${\bf K}$  into many pieces so that  each piece
appears locally as  ${\bf S}^2_{(2)} \times {\bf R}$ where ${\bf S}^2_{(2)}$ is a two-dimensional sphere with punctures $p_i$'s.  On every piece, the action \eqref{pasoli} in gauge ${\cal A}_0=0$
gives classical solution ${\cal F}_{ij}=0$. One may compare this to the classical solution ${\cal F}_{12} = 0$ that we get from \eqref{ajapple} $-$ which in turn may be assumed to be the special 
case of \eqref{f12bhn} with the scalar fields switched off. 
The constraint implies that the physical space $\{{\cal A}\}$ to be  moduli space of flat connections on the punctured sphere 
${\bf S}^2_{(2)}$ (modulo gauge transformation) which has a finite volume.  After imposing the constraint and then quantizing gives a finite 
dimensional Hilbert space $\mathcal{H}_{({\bf S}^2_{(2)}, p_i)}$, with $i = 1, 2, ...r$,  
whose states are related to the $r$-point correlation functions of the Wess-Zumino-Novikov-Witten conformal field theory (WZNW model) in the
two dimensional sphere ${\bf S}^2_{(2)}$ \cite{WZNW}. The WZNW model possesses  level $k$  current  algebra symmetry $G_k$ besides the conformal symmetry, where  
the Chern-Simons coupling $k \equiv 2\pi \left({\widetilde b}_2 + {\widetilde c}_2\right)$ is identified with the level $k$ of WZNW models.

\begin{figure}[t]\centering \includegraphics[width=0.6\textwidth]{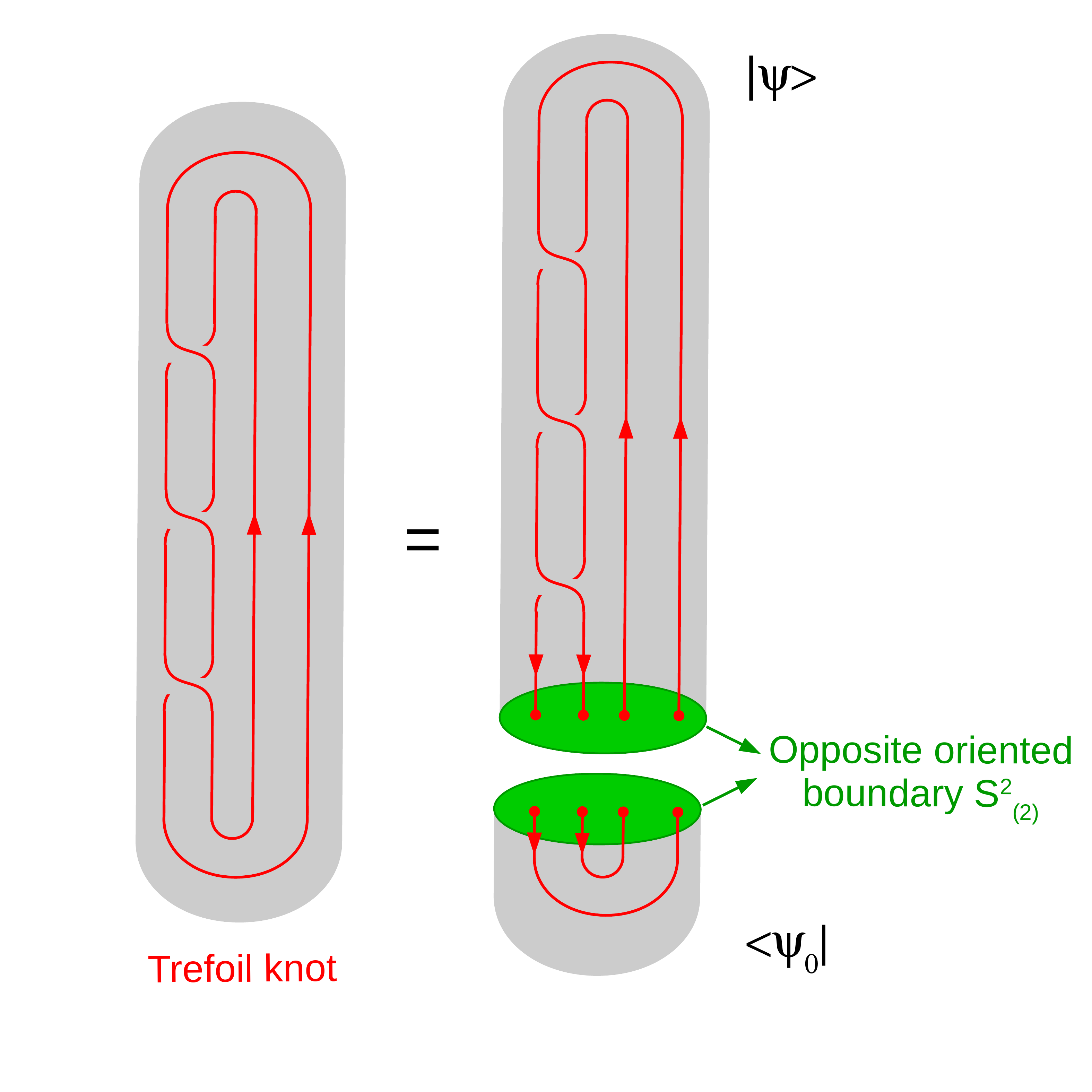}\caption{An example of a trefoil knot computation in the Ooguri-Vafa model. The knot invariant is now proportional to 
$\langle \Psi_0 \vert \Psi\rangle$, which is somewhat similar in spirit with the knot invariants computed earlier. The details however differ.}
\label{fig15} \end{figure}

This connection between Chern-Simons theory \eqref{pasoli} and WZNW model \cite{witten89, kaul} brings us to the familiar playground where a path integral of the form \eqref{effsit}, now defined with 
\eqref{pasoli}, may be identified with a quantum state in the Hilbert space of WZNW model with $r$ punctures. The story can be elaborated by working out the link invariants, one example is 
shown in {\bf fig \ref{fig15}}, but we will 
not do so here. Our aim is to find a supergravity link to this construction, and we leave this for the sequel. 

The third observation is related to geometric transition in the wrapped D6-branes' picture. Under geometric transition, the D6-branes wrapped on the three-cycle ${\bf S}^3_{(2)}$ of a non-K\"ahler deformed
conifold disappear and are
replaced by a non-K\"ahler resolved conifold with fluxes and no branes. What happens to the knot configurations on ${\bf S}^3_{(2)}$? This was the conundrum that we started off with in 
section \ref{TRC}. Introducing the D2-brane surface operators (or equivalently D4-brane surface operators)
in the wrapped D6-branes' picture now resolves the conundrum. After geometric transition, even though the D6-branes disappear, the D2-brane (or D4-brane)
configurations that are responsible for the knots, as shown in {\bf fig \ref{fig14}}, continue to survive on the resolved conifold side. Thus the {\it gravity dual}, which is our 
non-K\"ahler resolved conifold with fluxes, now equipped with the 
D2-brane (or D4-brane) states, continues to retain all the informations of knots and knot invariants and may be extracted with high fidelity.

\section{Discussions and conclusions \label{disco}}

In recent times we have understood that knot invariants like  Jones polynomial in three-dimensional space ${\bf W}$ 
can be computed by understanding the solutions of certain elliptic partial differential equations in four-dimensional space ${\bf V}$, where ${\bf W}$ is the boundary of ${\bf V}$.  
These equations were originally derived in a topologically modified ${\cal N}=4$ Super Yang-Mills by imposing a localization condition into the 
Chern-Simons theory in the three-dimensional boundary ${\bf W}$ \cite{wittenknots1}. The restriction to the three-dimensional boundary was realized by switching on an axionic field in the 
four-dimensional gauge theory defined on ${\bf V} = {\bf W} \times {\bf R}^+$. This way various details about knot configurations may be addressed directly using the dynamics of four-dimensional 
gauge theory. 

In a parallel development, Ooguri-Vafa \cite{OV} studied $SU(N)$ knot invariants using a topological theory generated by wrapping D6-branes on three-cycle of a deformed conifold. Here the 
knot invariants may be associated to counting certain BPS configurations that have origins in the gravity {\it dual} of the wrapped D6-branes' configuration. The gravity dual is 
given by resolved conifold with topological fluxes.  

In the first part of 
our work we present an alternative derivation of the results of \cite{wittenknots1}. We show that the physics studied in both ${\bf W}$ and ${\bf V}$ can be derived from a configuration in 
M-theory on a certain seven-dimensional manifold with fluxes and no branes other than the M2-branes. These M2-branes serve dual purpose: one set of configurations lead to non-abelian gauge theory
in ${\bf V}$; and another set of configurations lead to surface operators in ${\bf V}$ that are responsible for knots in ${\bf W}$. 

Restricting the knots to the boundary ${\bf W}$ is achieved by switching on a dipole or a RR deformation in ${\bf V}$ that can be parametrized from supergravity. The M-theory uplifts leads to 
a seven-dimensional manifold, as mentioned above, of the form of a warped Taub-NUT space fibered over a three-dimensional base. Supergravity analysis leads to a four-dimensional 
Hamiltonian \eqref{hamilbeta}, from where a series of BPS equations are derived. A sets of these BPS equations are exactly the localization equations of \cite{wittenknots1} and \cite{wittenknots2}, and 
we call them the BHN equations (the acronym stands for Bogomolnyi, Hitchin and Nahm). The remaining sets of the BPS equations are shown to be solved exactly using supergravity variables. 
Therefore one of our results was to show that such equations emerge from M-theory compactifications and their coefficients are succinctly interpreted in terms of supergravity parameters. 
 
We also considered various types of solutions of such BHN equations alongwith
their deformations. One possibility is to have codimension three solutions denoted by 't Hooft operators.  These solutions 
appear as opers, and we discuss them briefly here attempting a supergravity interpretation. Another possibility is to have codimension two solutions denoted by surface operators. We make a detailed study 
of this in our work and show how the surface operators, which we interpret as certain configuration of M2-branes, modify the BHN equations. These modifications are given by introducing delta function 
sources whose coefficients can be traced to the supergravity parameters in our model. Additionally we argue how the M2-brane surface operators help us to study the link invariants for various 
knot configurations in the abelian case. 

In the second part of the paper we argue how the Ooguri-Vafa model may also be derived from a configuration in M-theory defined on a {\it different} seven-dimensional manifold that is given 
by another warped Taub-NUT fibered over a three-dimensional base. The warping and fluxes now are such that the supersymmetry is reduced to ${\cal N} = 1$, and the seven-dimensional manifold has a 
$G_2$ structure. Nevertheless, many of the physics discussed in the first part of the paper follow a similar route in the second part too. There are crucial differences of course, which we 
point out in our paper. For example the topological theory is simpler now, but the analysis of knots using surface operators are harder because there is no abelian simplification that can be
performed now. There is also a {\it relocation} of the knots on the three-cycle of the deformed conifold instead on the spacetime boundary ${\bf W}$ earlier. This relocation is associated to a 
flop transition that can be performed on the mirror type IIA side. In our opinion these are all new results. 

There are a number of future directions. For example, 
in the first part we only studied the link invariants for the abelian case, so a natural question would to investigate the non-abelian scenario. This is harder
because, as we discussed in the text, the effect of the non-abelian configuration of the surface operators on the BHN equations are difficult to handle. Thus solving the BHN equations 
and interpreting the knots in terms of solutions of the BHN equations in the 
non-abelian case will be more challenging. 

For the second part we only make the barest beginnings in this direction, and leave most of the details for the sequel. For example the configuration of the surface operators in terms of M2 or 
M5-branes, details about the flop transition and the subsequent analysis of knot invariants still remain to be elaborated. Other connection to A-polynomial of \cite{AV}, 
Khovanov homology \cite{kovan0} etc have 
not been touched here at all, and we expect to study them in the sequel.  
Thus we see that the two connections to M-theory seven-manifolds explored in this paper lead to a rich spectrum of ideas that can allow us to have a fruitful dialogue 
between M-theory supergravity on one hand and topological field theory and mathematics on the other.
 
\vskip.2in

\centerline{\bf Acknowledgements}

\vskip.1in

\noindent We would like to specifically thank Edward Witten for helpful correspondences clarifying many subtleties of \cite{wittenknots1}; to Johannes Walcher for initial 
collaboration and for \cite{walcher}; to Ori Ganor for explaining the cubic term in \eqref{toptheory} and for other helpful comments via \cite{gan0}; 
and to Sergei Gukov for discussions and for a helpful reference. 
V. E. D. is grateful to Maxim Emelin, Avinash Raju, Sam Selmani and Bala Subramanian for illuminating discussions, to Chethan Krishnan for pointing out useful references and to Jatin Panwar for invaluable help with the figures. We would also like to thank Siddharth Dwivedi, Sam Selmani, Jihye Seo and Zodinmawia for initial collaborations. 
The work of K. D and V. E. D is supported in part by National Science and Engineering Research Council of Canada.


{}
 \end{document}


\newpage

\begin{thebibliography}{}


\bibitem{skein}
 J.~W. ~Alexander, {``Topological invariants of knots and links''}, Trans. Am. Math. Soc. {\bf 30}, 275 (1928); 
V.~F.~R. ~Jones, {``A polynomial invariant for knots via von Neumann algebras''}, Bull. AMS, 103 (1985); 
P.~ Freyd, D. ~Yetter, J. ~Hoste, W.~B.~R. ~Lickorish, K.~ Millet and A.~ Ocneanu, {``A new polynomial invariant of knots and links''}, Bull.
AMS.12(1985) 239; J.~H.~ Przytycki and K.~P. ~Traczyk, {``Invariants of links of Conway Type''}, Kobe J. Math. {\bf 4}, 115 (1987);
L.~H. ~Kauffman, {``An invariant of regular isotopy''}, Trans. Amer. Math. Soc. {\bf 318}, 417 (1990).


\bibitem{witten89}  E.~Witten, ``Quantum Field Theory and the Jones Polynomial,'' Commun.\ Math.\ Phys.\  {\bf 121}, 351 (1989). 

\bibitem{kaul} P. ~Ramadevi, T.~R. ~Govindarajan  and R.~K. ~Kaul, 
   {``Three-dimensional Chern-Simons theory as a theory of
                   knots and links. 3. Compact semisimple group''}, Nucl. Phys. B {\bf 402}, 548 (1993) [hep-th/9212110]      

\bibitem{kovan0} M.~Khovanov, {``A categorification of the Jones polynomial''},
    Duke Math. J. {\bf 3}, 359 (2000).

\bibitem{khovan1}  M.~ Khovanov, {``$sl(3)$ link homology''},
Algebr. Geom. Topol. {\bf 4}, 1045 (2004).

\bibitem{kovrosan} M.~ Khovanov and L.~ Rozansky, 
  {``Matrix factorizations and link homology''}, Fund. Math. 
     {\bf 199}, 1 (2008).


\bibitem{GoV} 
  R.~Gopakumar and C.~Vafa,
  ``On the gauge theory / geometry correspondence,''
  Adv.\ Theor.\ Math.\ Phys.\  {\bf 3}, 1415 (1999)
  [hep-th/9811131].


\bibitem{OV}  H.~Ooguri and C.~Vafa, ``Knot invariants and topological strings,'' Nucl.\ Phys.\ B {\bf 577}, 419 (2000) [hep-th/9912123].


\bibitem{integ} 
  J.~M.~F.~Labastida and M.~Marino,
  ``Polynomial invariants for torus knots and topological strings,''
  Commun.\ Math.\ Phys.\  {\bf 217}, 423 (2001)
    [hep-th/0004196];
  P.~Ramadevi and T.~Sarkar,
  ``On link invariants and topological string amplitudes,''
  Nucl.\ Phys.\ B {\bf 600}, 487 (2001)
    [hep-th/0009188].


\bibitem{sgv} 
S.~Gukov, A.~S.~Schwarz and C.~Vafa,
  ``Khovanov-Rozansky homology and topological strings,''
  Lett.\ Math.\ Phys.\  {\bf 74}, 53 (2005)
    [hep-th/0412243].


\bibitem{wittenknots1}  E.~Witten, ``Fivebranes and Knots,'' arXiv:1101.3216 [hep-th].


\bibitem{langland} 
  A.~Kapustin and E.~Witten,
  ``Electric-Magnetic Duality And The Geometric Langlands Program,''
  Commun.\ Num.\ Theor.\ Phys.\  {\bf 1}, 1 (2007)
  [hep-th/0604151].




\bibitem{wittenknots2} 
  D.~Gaiotto and E.~Witten,
  ``Knot Invariants from Four-Dimensional Gauge Theory,''
  Adv.\ Theor.\ Math.\ Phys.\  {\bf 16}, no. 3, 935 (2012)
  [arXiv:1106.4789 [hep-th]].



\bibitem{AV} 
  M.~Aganagic and C.~Vafa,
  ``Large N Duality, Mirror Symmetry, and a Q-deformed A-polynomial for Knots,''
  arXiv:1204.4709 [hep-th].

\bibitem{veronica}
V.~Errasti D\'{i}ez,
 ``A companion to Knot Invariants and M-Theory I: Proofs and Derivations,"
 To appear.


\bibitem{DOT1} 
  K.~Dasgupta, K.~Oh and R.~Tatar,
  ``Geometric transition, large N dualities and MQCD dynamics,''
  Nucl.\ Phys.\ B {\bf 610}, 331 (2001)
  [hep-th/0105066].

\bibitem{DOT2} 
  K.~Dasgupta, K.~Oh and R.~Tatar,
  ``Open / closed string dualities and Seiberg duality from geometric transitions in M theory,''
  JHEP {\bf 0208}, 026 (2002)
  [hep-th/0106040].


\bibitem{bvs} 
  M.~Bershadsky, C.~Vafa and V.~Sadov,
  ``D strings on D manifolds,''
  Nucl.\ Phys.\ B {\bf 463}, 398 (1996)
  [hep-th/9510225].

\bibitem{DM2} 
  K.~Dasgupta and S.~Mukhi,
  ``Brane constructions, fractional branes and Anti-de Sitter domain walls,''
  JHEP {\bf 9907}, 008 (1999)
  [hep-th/9904131].

\bibitem{uranga} 
  A.~M.~Uranga,
  ``Brane configurations for branes at conifolds,''
  JHEP {\bf 9901}, 022 (1999)
  [hep-th/9811004].


\bibitem{DM1} 
  K.~Dasgupta and S.~Mukhi,
  ``Brane constructions, conifolds and M theory,''
  Nucl.\ Phys.\ B {\bf 551}, 204 (1999)
  [hep-th/9811139].


\bibitem{leeyee} 
  K.~M.~Lee and H.~U.~Yee,
  ``BPS String Webs in the 6-dim (2,0) Theories,''
  JHEP {\bf 0703}, 057 (2007)
  [hep-th/0606150].


\bibitem{gan0}
O.~Ganor,
  ``A derivation of the $\phi^3$ term in eqn. \eqref{toptheory} and in \cite{wittenknots1},'' Nov 2015.


\bibitem{GB} 
  A.~Bergman and O.~J.~Ganor,
  ``Dipoles, twists and noncommutative gauge theory,''
  JHEP {\bf 0010}, 018 (2000)
  [hep-th/0008030].

\bibitem{bergman1} 
  K.~Dasgupta, O.~J.~Ganor and G.~Rajesh,
  ``Vector deformations of N=4 superYang-Mills theory, pinned branes, and arched strings,''
  JHEP {\bf 0104}, 034 (2001)
  [hep-th/0010072].

\bibitem{bergman2} 
  A.~Bergman, K.~Dasgupta, O.~J.~Ganor, J.~L.~Karczmarek and G.~Rajesh,
  ``Nonlocal field theories and their gravity duals,''
  Phys.\ Rev.\ D {\bf 65}, 066005 (2002)
  [hep-th/0103090].

\bibitem{imamura} 
  Y.~Imamura,
  ``Born-Infeld action and Chern-Simons term from Kaluza-Klein monopole in M theory,''
  Phys.\ Lett.\ B {\bf 414}, 242 (1997)
  [hep-th/9706144]; A.~Sen,
  ``Dynamics of multiple Kaluza-Klein monopoles in M and string theory,''
  Adv.\ Theor.\ Math.\ Phys.\  {\bf 1}, 115 (1998)
  [hep-th/9707042]; ``A Note on enhanced gauge symmetries in M and string theory,''
  JHEP {\bf 9709}, 001 (1997)
  [hep-th/9707123]; K.~Becker, M.~Becker, K.~Dasgupta and P.~S.~Green,
  ``Compactifications of heterotic theory on nonKahler complex manifolds. 1.,''
  JHEP {\bf 0304}, 007 (2003)
  [hep-th/0301161].

\bibitem{rajaraman} 
  W.~Fischler and A.~Rajaraman,
  ``M(atrix) string theory on K3,''
  Phys.\ Lett.\ B {\bf 411}, 53 (1997)
  [hep-th/9704123].

\bibitem{hanany} 
  A.~Hanany and G.~Lifschytz,
  ``M(atrix) theory on T**6 and a m(atrix) theory description of K K monopoles,''
  Nucl.\ Phys.\ B {\bf 519}, 195 (1998)
  [hep-th/9708037].

\bibitem{myers}
R.~C.~Myers,
  ``Dielectric branes,''
  JHEP {\bf 9912}, 022 (1999)
  [hep-th/9910053].

\bibitem{bagger} 
  J.~Bagger and N.~Lambert,
  ``Modeling Multiple M2's,''
  Phys.\ Rev.\ D {\bf 75}, 045020 (2007)
  [hep-th/0611108]; 
``Gauge symmetry and supersymmetry of multiple M2-branes,''
  Phys.\ Rev.\ D {\bf 77}, 065008 (2008)
  [arXiv:0711.0955 [hep-th]];
J.~Bagger, N.~Lambert, S.~Mukhi and C.~Papageorgakis,
  ``Multiple Membranes in M-theory,''
  Phys.\ Rept.\  {\bf 527}, 1 (2013)
  [arXiv:1203.3546 [hep-th]].

\bibitem{gukovwitten2} 
  S.~Gukov and E.~Witten,
  ``Gauge Theory, Ramification, And The Geometric Langlands Program,''
  hep-th/0612073.

\bibitem{gukovwitten}
S.~Gukov and E.~Witten,
  ``Rigid Surface Operators,''
  Adv.\ Theor.\ Math.\ Phys.\  {\bf 14}, no. 1, 87 (2010)
  [arXiv:0804.1561 [hep-th]].

\bibitem{gukovsurf} 
  S.~Gukov,
  ``Surface Operators,''
  arXiv:1412.7127 [hep-th].

\bibitem{smallins} 
  E.~Witten,
  ``Small instantons in string theory,''
  Nucl.\ Phys.\ B {\bf 460}, 541 (1996)
    [hep-th/9511030].
  

\bibitem{surfknot} 
  S.~Gukov,
  ``Gauge theory and knot homologies,''
  Fortsch.\ Phys.\  {\bf 55}, 473 (2007)
    [arXiv:0706.2369 [hep-th]].

\bibitem{jimjam} 
  E.~Witten,
  ``Analytic Continuation Of Chern-Simons Theory,''
  AMS/IP Stud.\ Adv.\ Math.\  {\bf 50}, 347 (2011)
  [arXiv:1001.2933 [hep-th]].

\bibitem{topM} 
  R.~Dijkgraaf, S.~Gukov, A.~Neitzke and C.~Vafa,
  ``Topological M-theory as unification of form theories of gravity,''
  Adv.\ Theor.\ Math.\ Phys.\  {\bf 9}, no. 4, 603 (2005)
    [hep-th/0411073].





\bibitem{mmarino} 
  M.~Marino,
  ``Chern-Simons theory, Matrix Models, and Topological Strings,''
  Int.\ Ser.\ Monogr.\ Phys.\  {\bf 131}, 1 (2005).



\bibitem{DEM} 
  K.~Dasgupta, M.~Emelin and E.~McDonough,
  ``Non-Kahler Resolved Conifold, Localized Fluxes in M-Theory and Supersymmetry,''
JHEP {\bf 1502}, 179 (2015),   
[arXiv:1412.3123 [hep-th]].

\bibitem{MM} 
  J.~Maldacena and D.~Martelli,
  ``The Unwarped, resolved, deformed conifold: Fivebranes and the baryonic branch of the Klebanov-Strassler theory,''
  JHEP {\bf 1001}, 104 (2010)
  [arXiv:0906.0591 [hep-th]].


\bibitem{rahul}
C.~Faber and R.~Pandharipande,
``Hodge Integrals and Gromov-Witten Theory'', math.AG/9810173.


\bibitem{walcher}
J.~Walcher, ``Goods to those who wait'', private email correspondence, April 2013.


\bibitem{SYZ} 
  A.~Strominger, S.~T.~Yau and E.~Zaslow,
  ``Mirror symmetry is T duality,''
  Nucl.\ Phys.\ B {\bf 479}, 243 (1996)
  [hep-th/9606040].

\bibitem{syz2} 
  M.~Becker, K.~Dasgupta, A.~Knauf and R.~Tatar,
  ``Geometric transitions, flops and nonKahler manifolds. I.,''
  Nucl.\ Phys.\ B {\bf 702}, 207 (2004)
    [hep-th/0403288];  S.~Alexander, K.~Becker, M.~Becker, K.~Dasgupta, A.~Knauf and R.~Tatar,
  ``In the realm of the geometric transitions,''
  Nucl.\ Phys.\ B {\bf 704}, 231 (2005)
    [hep-th/0408192]; M.~Becker, K.~Dasgupta, S.~H.~Katz, A.~Knauf and R.~Tatar,
  ``Geometric transitions, flops and non-Kahler manifolds. II.,''
  Nucl.\ Phys.\ B {\bf 738}, 124 (2006)
    [hep-th/0511099].


\bibitem{lapan} 
  F.~Chen, K.~Dasgupta, J.~M.~Lapan, J.~Seo and R.~Tatar,
  ``Gauge/Gravity Duality in Heterotic String Theory,''
  Phys.\ Rev.\ D {\bf 88}, 066003 (2013)
  [arXiv:1303.4750 [hep-th]].


\bibitem{fangpaul} 
  F.~Chen, K.~Dasgupta, P.~Franche, S.~Katz and R.~Tatar,
  ``Supersymmetric Configurations, Geometric Transitions and New Non-Kahler Manifolds,''
  Nucl.\ Phys.\ B {\bf 852}, 553 (2011)
  [arXiv:1007.5316 [hep-th]].

\bibitem{anke} 
  K.~Dasgupta, M.~Grisaru, R.~Gwyn, S.~H.~Katz, A.~Knauf and R.~Tatar,
  ``Gauge-Gravity Dualities, Dipoles and New Non-Kahler Manifolds,''
  Nucl.\ Phys.\ B {\bf 755}, 21 (2006)
  [hep-th/0605201].

\bibitem{chiossi} 
  S.~Chiossi and S.~Salamon,
  ``The Intrinsic torsion of SU(3) and G(2) structures,''
  [math/0202282 [math-dg]].


\bibitem{luest} 
  G.~Lopes Cardoso, G.~Curio, G.~Dall'Agata, D.~Lust, P.~Manousselis and G.~Zoupanos,
  ``Non-K\"ahler string backgrounds and their five torsion classes,''
  Nucl.\ Phys.\ B {\bf 652}, 5 (2003)
  [hep-th/0211118].

\bibitem{louis} 
  S.~Gurrieri, J.~Louis, A.~Micu and D.~Waldram,
  ``Mirror symmetry in generalized Calabi-Yau compactifications,''
  Nucl.\ Phys.\ B {\bf 654}, 61 (2003)
  [hep-th/0211102].

\bibitem{gauntlett} 
  J.~P.~Gauntlett, D.~Martelli and D.~Waldram,
  ``Superstrings with intrinsic torsion,''
  Phys.\ Rev.\ D {\bf 69}, 086002 (2004)
  [hep-th/0302158].

\bibitem{jabbari} 
  K.~Dasgupta and M.~M.~Sheikh-Jabbari,
  ``Noncommutative dipole field theories,''
  JHEP {\bf 0202}, 002 (2002)
  [hep-th/0112064].

\bibitem{shmakov} 
  K.~Dasgupta and M.~Shmakova,
  ``On branes and oriented B fields,''
  Nucl.\ Phys.\ B {\bf 675}, 205 (2003)
  [hep-th/0306030].

\bibitem{WZNW}
J.~Wess and B.~Zumino,
  ``Consequences of anomalous Ward identities,''
  Phys.\ Lett.\ B {\bf 37}, 95 (1971);  E.~Witten,
  ``Global Aspects of Current Algebra,''
  Nucl.\ Phys.\ B {\bf 223}, 422 (1983); S.~P.~Novikov,
  ``The Hamiltonian formalism and a many valued analog of Morse theory,''
  Usp.\ Mat.\ Nauk {\bf 37N5}, no. 5, 3 (1982)
  [Russ.\ Math.\ Surveys {\bf 37}, no. 5, 1 (1982)].
  
\end{thebibliography}
\end{document}